\documentclass[10pt,reprint,aip,amsmath,amssymb]{revtex4-1}
\usepackage{graphicx}
\input epsf

\newcommand\Alfven{Alfv\'en }
\newcommand\Alfvenic{Alfv\'enic }
\newcommand{\V}[1]{\mathbf{#1}}

\newcommand{\secref}[1]{\S\ref{#1}}

\begin{document}

\title{Laboratory Space Physics: Investigating the Physics of Space
  Plasmas in the Laboratory}

 \author{Gregory~G. Howes}
\email[]{gregory-howes@uiowa.edu}
\affiliation{Department of Physics and Astronomy, University of Iowa, Iowa City, 
Iowa 52242, USA.}

\date{\today}

\begin{abstract}
Laboratory experiments provide a valuable complement to explore the
fundamental physics of space plasmas without the limitations inherent
to spacecraft measurements. Specifically, experiments overcome the
restriction that spacecraft measurements are made at only one (or a
few) points in space, enable greater control of the plasma conditions
and applied perturbations, can be reproducible, and are orders of
magnitude less expensive than launching spacecraft. Here I highlight
key open questions about the physics of space plasmas and identify the
aspects of these problems that can potentially be tackled in
laboratory experiments. Several past successes in laboratory space
physics provide concrete examples of how complementary experiments can
contribute to our understanding of physical processes at play in the
solar corona, solar wind, planetary magnetospheres, and outer boundary
of the heliosphere. I present developments on the horizon of
laboratory space physics, identifying velocity space as a key new
frontier, highlighting new and enhanced experimental facilities, and
showcasing anticipated developments to produce improved diagnostics
and innovative analysis methods.  A strategy for future laboratory
space physics investigations will be outlined, with explicit
connections to specific fundamental plasma phenomena of interest.
\end{abstract}

\pacs{}

\maketitle 

\section{Introduction}

On January 31, 1958, the United States entered the space age with the
launch of \emph{Explorer I}, the first U.S. satellite to orbit the
Earth. This historic event also marked the birth of an entirely new
field of science: experimental space physics. \emph{In situ}
measurements of high-energy particles by James Van Allen's Geiger
counter instrument on \emph{Explorer I} led to the first major
scientific discovery of the space age---that the Earth's magnetic field
traps high-energy particles in regions encircling the Earth,
\citep{VanAllen:1959} now known as the Van Allen Radiation Belts.

The ability to launch spacecraft to make direct measurements of the
plasma and electromagnetic fields beyond the Earth's atmosphere has
stimulated tremendous progress in our understanding of the
heliosphere, our home in the universe. The heliosphere is the realm of
influence of the Sun, within which the planets of our solar system
orbit.  Heliophysics, more commonly known as space physics, is
dedicated to the study of the physical mechanisms that govern the
dynamics and evolution of the space environment from the Sun to outer
boundary of the heliosphere, beyond which the field of astrophysics
takes over.

The international space physics community tackles major open questions
about the flow of matter and energy from the Sun, to the Earth and
other planets, and out to the edge of the heliosphere, where the
heliosphere interacts with the surrounding interstellar medium. At the
heart of heliophysics are fundamental questions about how the solar
magnetic dynamo draws on the energy released by fusion reactions in
the Sun's core to generate strong magnetic fields. As these magnetic
fields intensify, they become buoyant, rising through the solar
interior and emerging through the photosphere into the atmosphere of
the sun, driving the frenzied and occasionally explosive dynamics that
occur in active regions on the Sun. How magnetic reconnection governs
the eruptions of coronal mass ejections into space, and how plasma
physics processes determine whether such explosive activity is
accompanied by a solar flare or by the acceleration of solar energetic
particles, remain poorly understood at present. Another long-standing
open question in heliophysics is what role plasma turbulence and
magnetic reconnection play in the heating of the solar corona to more
than a million Kelvin---nearly a thousand times hotter than the
approximately six thousand Kelvin temperature of the solar photosphere
below.

Extreme space weather is caused by these violent events on the
Sun---solar flares, coronal mass ejections, and the acceleration of
solar energetic particles---and can impact the technological
infrastructure upon which our society depends daily, such as satellite
communication and navigation, radar and radio communication, air
travel, and even the electrical power grid. Predicting how coronal
mass ejections and solar energetic particles propagate through the
turbulent interplanetary medium is critical to our ability to predict
their detrimental effect at Earth, and to unravel how the Earth's
magnetosphere responds to forcing by the variable solar wind and by
these sporadic outbursts from the Sun.  The conditions in the
near-Earth space environment, including the Van Allen Radiation Belts,
react to this external forcing in complicated ways that are not well
understood. Challenges at the forefront of space physics include
discovering the processes responsible for the acceleration of
high-energy particles trapped in the radiation belts and for the
eventual loss of those confined particles.  Furthermore, unraveling
the complicated interactions and feedback mechanisms that couple the
ionosphere and thermosphere to the magnetosphere will require an
improved understanding of ion-neutral coupling and of the physics of
multi-ion plasmas.

The solar wind carries the influence of the Sun beyond the orbit of
the planets in our solar system, where its supersonic and
super-Alfv\'enic flow is abruptly slowed at the collisionless
termination shock.  The shocked heliosheath plasma beyond this point
slows further until it finally halts at the heliopause, the final
boundary separating the heliosphere from the surrounding interstellar
medium.  Numerous new questions have arisen as we obtain new
information about the interaction of the heliosphere with the
interstellar medium through remote sampling by energetic neutral atoms
\citep{McComas:2009a,McComas:2009b} and direct measurements by
\emph{Voyager I}. \citep{Gurnett:2013} Although distant stellar
systems and galaxies almost certainly lie beyond the reach of \emph{in
  situ} measurement in our lifetimes, the lessons learned about the
fundamental physical mechanisms that govern the evolution of the
plasmas that fill the heliosphere can be applied to better interpret
remote measurements of these distant and fascinating astrophysical
systems.

Being able to understand, and ultimately to predict, the evolution of
the heliosphere requires not only learning how different regions of
the system---such as the sun, the solar wind, the planetary
magnetospheres, and the interface with the surrounding interstellar
medium---interact and feedback upon each other, but also discovering
how the fundamental dynamics of the heliospheric plasma mediates those
interactions.  In other words, it requires a ``systems-science'' study
of the heliosphere as a coupled system of different interacting
elements complemented by detailed investigations of the underlying
plasma physics that governs those interactions. \citep{NRCspace:2013}
The Heliophysics System Observatory (HSO) comprises the current fleet
of thirty-three spacecraft \footnote{As of summer 2017, the NASA
  missions comprising the Heliophysics System Observatory (HSO)
  include the single-spacecraft missions ACE, AIM, Geotail, GOLD, Hinode,
  IBEX, IRIS, RHESSI, SDO, SOHO, TIMED, and Wind, as well as the
  multiple-spacecraft missions ARTEMIS (2), Cluster (4), MMS (4)
  STEREO (2), THEMIS(3), TWINS (2), Van Allen Probes (2), and Voyager
  (2).}  that simultaneously sample many different regions of the
heliosphere, enabling the complex interplay of the different elements
to be observed.  Individual missions return, from each spacecraft's
location, direct measurements of the plasma and electromagnetic field
fluctuations that are critical to unravel the fundamental plasma
mechanisms at play in the heliosphere. In many cases, the macroscopic
evolution of the heliosphere depends upon the physical processes at
play in the plasma on microscopic scales. For example, the heating of
the solar corona is widely believed to depend on the dissipation of plasma
turbulence at small scales.

The ability to make direct measurements of the plasma dynamics using
the spacecraft of the HSO, however, does not guarantee that we will
understand the physical processes governing what we observe.  In
particular, the HSO spacecraft missions make measurements of the
plasma at a single point---or, for multi-spacecraft missions, a few
points---in space.  But to unravel the complicated evolution of
fundamental plasma mechanisms, such as magnetic reconnection, the
restriction to single-point or few-point measurements significantly
limits our ability to discern the underlying plasma physics.  It is
here that laboratory experiments can provide an invaluable complement
in the effort to understand the physics of space plasmas, motivating
an approach denoted here as \emph{laboratory space physics}.

In particular, laboratory experiments enable detailed investigations
of the fundamental physical mechanisms that govern the transport of
mass and energy within heliospheric environments of interest. Plasma
turbulence, magnetic reconnection, particle acceleration, and kinetic
instabilities are four physical processes that play key roles in the
evolution of the plasmas that fill the heliosphere, constituting four
grand challenge problems at the frontier of heliophysics. Terrestrial
experiments enable us to measure in the laboratory what is often
difficult or impossible to measure in space, facilitating our study of
these mechanisms to gain much deeper insight into the fundamental
plasma physics that governs the dynamics of the heliosphere.

The natural synergy between spacecraft observations and laboratory
experiments has only become stronger as improved capabilities in both
arenas are leading to a convergence of their regimes of applicability.
In particular, the cadence of measurements by modern spacecraft
instruments has finally decreased to a point that the kinetic length
and time scales of the plasma can be directly probed.  For example,
the \emph{Magnetospheric Multiscale} (\emph{MMS})
mission\citep{Burch:2016a,Pollock:2016} measures the three-dimensional proton
distribution function at a sampling rate of 150~ms, shorter than the
typical timescales of 1~s associated with ion cyclotron motion and
with the Doppler shift of fluctuations on the kinetic scale of the ion
Larmor radius. The three-dimensional electron distribution function is
sampled by \emph{MMS} at 30~ms, nearing the typical timescales of
25~ms for the Doppler-shifted electron Larmor radius.

In contrast, laboratory experiments have generally suffered from the
inability to model the large scales (relative to kinetic length
scales) characteristic of many physical processes at play in space
plasmas.  But intermediate-scale experiments now, often operated as
national user facilities, can generate sufficiently large plasmas that
there exists a substantial dynamic range above the typical kinetic
length scales.  For example, the \emph{Large Plasma Device}
(\emph{LAPD}) \citep{Gekelman:1991} at UCLA produces a 17~m long,
60~cm diameter cylindrical plasma confined radially by a strong axial
magnetic field around 1~kG; with an ion Larmor radius around 0.2~cm
and a long-enough parallel length to contain MHD waves with
frequencies below the ion cyclotron frequency, the \emph{LAPD} enables
exploration of the large-scale, MHD dynamics of magnetospheric and
heliospheric plasmas.  Another example is newly constructed
\emph{Facility for Laboratory Reconnection Experiments}
(\emph{FLARE}), which can produce plasmas with a Lundquist number
above $10^5$ and a current sheet length relative to the ion inertial
length or ion Larmor radius of around $10^3$, enabling studies of the
plasmoid instability in the both the collisional and collisionless
regimes of magnetic reconnection.\citep{Ji:2011}

This convergence of the regimes accessible to spacecraft measurements
and laboratory experiments is a key reason why the study of space
physics in the laboratory is now particularly timely, with the
potential to make a transformative contribution to the study of
fundamental plasma mechanisms in the heliosphere.

Previous works by F{\"a}lthammar \citep{Falthammar:1974} in 1974 and
Koepke\citep{Koepke:2008} in 2008 have reviewed the use of laboratory
experiments to investigate space plasma physics.  In this review, I
discuss the application of appropriately scaled laboratory experiments
to study space and astrophysical phenomena and emphasize the
advantages of laboratory experiments in \secref{sec:how}. In
\secref{sec:rev}, I identify key questions in space physics and
astrophysics and discuss aspects of these questions that can be
tackled in the laboratory, briefly reviewing previous successful
laboratory investigations.  Next, I outline in \secref{sec:horizon}
developments on the horizon of laboratory investigations of space
physics, identifying velocity space as a key new frontier,
highlighting new and enhanced experimental facilities, and showcasing
anticipated developments to produce improved diagnostics and
innovative analysis methods.  Finally, in \secref{sec:future}, I wrap
up the discussion with a perspective on why laboratory experiments are
likely to play an increasingly important role in the study of space
physics, with a call for the strategic alignment of experimental
efforts with upcoming spacecraft missions.

\section{How Do You Study the Physics of Space Plasmas in the Laboratory?}
\label{sec:how}

Laboratory experiments provide a valuable complement to the
observational, theoretical, and numerical study of the physics of space
plasmas because they make possible the study of fundamental plasma
physics processes in a controlled and well-diagnosed environment. Here
I discuss how to design experiments relevant to space or astrophysical
environments of interest.  I also highlight the key advantages that
laboratory experiments have over measurements by spacecraft, arguing
that they provide a unique complement to observational studies of
space plasmas.

\subsection{Scaling}
In the effort to explore the physics of space and astrophysical
plasmas in the laboratory, an obvious question arises: how can one
reconcile the enormous discrepancy in the length and time scales of
astrophysical phenomena with what can be achieved in a terrestrial
laboratory?  The answer is that the physical equations governing the
evolution of a system of interest remain invariant through a careful
scaling of the length scales, time scales, and other dimensional
parameters of the system. Through dimensional analysis, the physical
behavior of a given system can be found to depend on a minimal number
of dimensionless parameters, as shown by Buckingham in 1914, commonly
known as the Buckingham ``Pi Theorem.''  \citep{Buckingham:1914} This
type of similarity analysis \citep{Barenblatt:1996} represents a
powerful approach to understand the fundamental physical behavior of a
system which remains invariant upon appropriate scaling of dimensional
quantities.

One of the most famous applications of similarity analysis occurred in
1950 when G.~I.~Taylor used declassified photos of the first atomic
explosion in New Mexico to estimate accurately the yield of that
atomic weapon test. \citep{Taylor:1950a,Taylor:1950b} Similarity
analysis has been applied successfully to a wide variety of physical
systems, from turbulent fluids \citep{Frisch:1995} to avalanching
systems that exhibit self-organized criticality. \citep{Chapman:2009}

In the effort to employ laboratory experiments to explore the physics
of space and astrophysical plasmas, the high-energy density physics
community has taken full advantage of similarity analysis to design
experiments using laser-generated plasmas to explore the physics of
supernova explosions and the nonlinear evolution of their
remnants.\citep{Ryutov:1999,Drake:2000,Ryutov:2000,Ryutov:2001,Remington:2006}
Specifically, a careful analysis has been carried out to identify the
``Euler similarity'' for compressible hydrodynamic fluids
\citep{Ryutov:1999} and ideal compressible magnetohydrodynamic fluids,
\citep{Ryutov:2000} with significant attention to the limits of
validity of such a mathematical description
\citep{Ryutov:1999,Ryutov:2000,Ryutov:2001} and to the constraints
inherent to the laboratory environment. \citep{Drake:2000} A thorough
review of the application of similarity analysis to produce
appropriately scaled high-energy density physics laboratory
experiments can be found in Remington, Drake, and
Ryutov. \citep{Remington:2006}

In this review, I focus not on the high-energy density environments
relevant to supernova explosions that can be explored using
facility-class laser plasmas, typically denoted using the term
\emph{laboratory astrophysics}, but rather on scaled laboratory
experiments at the lower energy densities that are typical of many
heliospheric and less extreme astrophysical environments. Here I
choose to refer to the experimental study of those moderate-energy
density plasma environments as \emph{laboratory space physics},
although many such studies indeed may be applicable to processes
occuring beyond the limits of the heliosphere. A wide range of
cutting-edge experiments in laboratory space physics can be
successfully conducted at universities on small-scale to
moderate-scale laboratory devices.

\subsection{Advantages of Laboratory Experiments}
Laboratory experiments enjoy a number of significant advantages over
the spacecraft exploration of the physics of space plasmas: (i)
many-point measurements, (ii) reproducibility, (iii) controlled
conditions, and (iv) much lower cost.  On the other hand, spacecraft
missions can probe the plasma dynamics without perturbing the plasma
substantially, three-dimensional velocity distributions are much
easier to measure in space than in the laboratory, and it is often not
possible, even with appropriately scaled experiments, to reproduce the
large dynamic range of time and length scales found in space.

One of the most restrictive limitations of spacecraft measurements is
that they are limited to a single point in space for the majority of
missions which consist of only one spacecraft, or a few points in
space for multi-spacecraft missions.  Since  heliospheric plasmas are
often moving at a large velocity relative to the sampling spacecraft,
the spacecraft obtains a time series of measurements along a
trajectory slicing through the plasma being studied.  For single-point
measurements made in the spacecraft frame, in relative motion with
respect to the plasma rest frame of reference, it is not possible to
definitively separate temporal variations in the plasma frame from
spatial variations that are being advected past the spacecraft at the
plasma flow velocity. If the plasma flow is much faster than typical
wave velocities in the plasma, as is generally the case in the solar
wind, observers often adopt the Taylor hypothesis, \citep{Taylor:1938}
assuming that the temporal fluctuations in the spacecraft frame are
solely due to the spatial fluctuations in the plasma frame sweeping
over the spacecraft with the plasma
flow. \citep{Matthaeus:1982b,Perri:2010a} But if the condition of fast
plasma flow is not satisfied, such as in the solar corona or planetary
magnetospheres, detailed calculations must be used to estimate when
the Taylor hypothesis is violated.
\citep{Howes:2014a,Klein:2014b,Perri:2017}

In contrast to the case with spacecraft, an array of diagnostic probes
can relatively easily inserted into a laboratory experiment to sample
the plasma at many points simultaneously, for example in the
\emph{Magnetic Reconnection Experiment} (\emph{MRX})
\citep{Yamada:1997b} for studies of magnetic reconnection or in the
\emph{Swarthmore Spheromak Experiment} (\emph{SSX}) MHD wind tunnel
for studies of plasma turbulence. \citep{Brown:2014,Brown:2015} Such
diagnostic access makes possible the sampling across a region of
interest in the plasma, enabling a more complete characterization of
the dynamics that is extremely valuable for the illumination of the
underlying plasma physical processes.

A second key advantage of many laboratory investigations over
spacecraft measurements is the ability to design experiments that are
reproducible. At the Basic Plasma Science Facility (BAPSF) at UCLA,
the \emph{Large Plasma Device} (\emph{LAPD}) \citep{Gekelman:1991} takes
reproducibility to new limits, with experimental shots in the machine
fired at a cadence of 1~Hz, meaning 86,400 separate experiments can be
performed in a single day.  In typical \emph{LAPD} experiments, this enables
a single probe (rather than a probe array, which may interfere with
the plasma dynamics under investigation) to be moved throughout the
plasma, making possible the measurement of the evolution of the plasma
throughout the plasma volume. Furthermore, for such reproducible
experiments, measurements may be made at the same location for many
independent shots, allowing the signal-to-noise ratio to be improved
through averaging.

Another major advantage of laboratory experiments over spacecraft
measurements is the ability to exert some measure of control over the
plasma conditions and dynamics.  For example, in the \emph{Space Physics
Simulation Chamber} (\emph{SPSC}) at that Naval Research Laboratory and in the
\emph{Auburn Linear Experiment for Instability Studies} (\emph{ALEXIS}) at Auburn
University, \citep{Wallace:2004,Eadon:2011,Dubois:2013a} studies of the
relaxation of stressed plasma boundary layers measured unstable
fluctuations with a frequency that varied over five orders of
magnitude as the ratio of the boundary layer width to the ion
gyroradius was varied over nearly an order of magnitude.
\citep{Dubois:2013b,Dubois:2014}


A final major advantage is that laboratory experiments can be orders
of magnitude less costly than spacecraft missions. For example,
expenses for five years of operation at the Basic Plasma Science
Facility (BAPSF) at UCLA, a national user facility supported jointly
by the National Science Foundation and the Department of Energy, total
approximately \$14M.  For comparison, the life-cycle cost of the twin
\emph{Van Allen Probes} spacecraft mission totals
\$686M.\citep{GAO:2013} The cost for the launch vehicle for the
\emph{Van Allen Probes} was approximately \$135M, so just getting the
instruments into space represents a major investment of research
funding. The life-cycle cost of the upcoming \emph{Parker Solar Probe}
mission, the first spacecraft to visit the Sun, is \$1,553M,
\citep{GAO:2017} more than 100 times the operating expenses of the
BAPSF.  Naturally, if we want to understand in detail the physics of
space plasmas, then the sampling of those plasmas directly with
spacecraft is an essential endeavor. But to develop a complete
understanding of the plasma physics mechanisms at play in space
plasmas, laboratory experiments represent a cost-effective means of
complementing spacecraft missions.

It is worthwhile mentioning, however, that spacecraft observations do
enjoy some specific advantages over laboratory measurements, besides
the obvious fact that spacecraft are directly sampling the space
plasma of interest.  First, the size of the spacecraft, typically on
the order of meters, is typically much smaller than the characteristic
length scales in a plasma: in the near-Earth solar wind, the proton
gyroradius is around 100~km, the electron gyroradius is around 1~km,
and the Debye length is around 10~m.  Therefore, the plasma dynamics
can be sampled without perturbing the plasma substantially, although
near-spacecraft effects generally need to be taken into account to
calibrate measurements.  Second, although the measurement of particle
velocity distribution functions---a key measurement to probe the
kinetic plasma physics---is rather difficult to achieve in the
laboratory, spacecraft instruments have been capable of measuring
three-dimensional velocity distributions for decades, although the
cadence of these measurements has only recently improved to the point
that kinetic timescales are accessible. Finally, even with
appropriately scaled laboratory experiments, it is not always possible
to reproduce the dynamic range of time and length scales occurring in
the space environment.  Notwithstanding these limitations in the
laboratory, terrestrial experiments are likely to play an increasingly
important role in illuminating the physics of space and astrophysical
plasmas.

\section{Key Questions in Space Physics and  Past Successes in Laboratory Space Physics}
\label{sec:rev}

Plasma physics governs the fundamental interactions that influence the
evolution of plasmas throughout the heliosphere, from the depths of
the solar interior to the far reaches of the heliosphere, where the
solar wind slows and ultimately comes to a halt at the heliopause,
marking the boundary between the heliosphere and the surrounding
galactic interstellar medium. Appropriately scaled laboratory
experiments provide a uniquely accessible platform to explore in
detail the same physical processes that are believed to control the
dynamics of space plasmas. Here I review these fundamental mechanisms
of plasma physics and identify the space environments in which they
arise, briefly highlighting existing experimental facilities where
these problems can be studied and reviewing previous suceessful
studies of the physics of space plasmas in the laboratory.

\subsection{Plasma Turbulence}
Turbulence arises in nearly every plasma environment throughout the
heliosphere: the solar convection zone, chromosphere, transition zone,
and corona; the solar wind; the planetary magnetospheres and their
interaction with their underlying thermospheres and ionospheres; and
the heliospheric termination shock and heliosheath. Fundamentally,
turbulence influences the transport of plasma particles, momentum, and
energy. Specifically, turbulence governs the interpenetration of
distinct plasmas, affects the propagation of energetic particles, and
mediates the conversion of the energy of large-scale plasma flows and
electromagnetic fields into plasma heat or other non-thermal forms of
particle energization. In many space and astrophysical environments,
the microphysics of turbulence and its dissipation at small scales can
strongly influence the macroscopic evolution of the system.

In the solar convection zone, turbulent magnetoconvection is widely
believed to play a key role in the magnetic dynamo that drives the
solar cycle. \citep{Parker:1955,Brun:2004} As dynamo activity enhances
the magnetic field strength, strong magnetic flux tubes buoyantly rise
through the solar convection zone and emerge through the solar
photosphere into the sun's atmosphere, generating sunspots that often
gather together in areas of intense magnetic activity called Active
Regions. The granular and supergranular networks of the convecting
photosphere buffet the solar magnetic fields that pass through the
surface, leading to a persistently turbulent state of the solar
chromosphere; high resolution imaging by \emph{Hinode} satellite has
shown that even the quiescent regions of the solar atmosphere abound
with turbulent motion. \citep{Berger:2008,Berger:2010}

As you move up through the sun's atmosphere from the chromosphere, the
narrow transition region marks a drastic increase in temperature and
decrease in density from the cool, dense, partially ionized, and
collisional conditions of the chromosphere to the hot, diffuse, fully
ionized, and collisionless conditions of the lower solar corona. Ever
since coronal emission lines were used to demonstrate in the 1940s
that the temperature of the coronal plasma is more than a million
Kelvin, \citep{Edlen:1943,Edlen:1945} the question of how the solar
corona is heated to a temperature nearly three orders of magnitude
higher than the photosphere below has persisted as a major unanswered
question in heliophysics.  Although the detailed mechanism by which
the coronal plasma is heated remains poorly understood, plasma
turbulence plays a crucial role in many of the proposed mechanisms.
\citep{Withbroe:1977,Heyvaerts:1983,Parker:1988,Klimchuk:2006,Cranmer:2009b,Chandran:2010b}

A major problem at the forefront of heliophysics research is to
determine the detailed plasma physics governing the removal of energy
from turbulent fluctuations and the consequent heating of the plasma
particles.  Under the weakly collisional conditions relevant to most
heliospheric environments, such as the solar corona, solar wind, and
planetary magnetospheres, the proposed collisionless energy transfer
mechanisms fall into three broad categories: (i) resonant
wave-particle interactions, such as Landau damping, transit-time
damping, or cyclotron damping;
\cite{Landau:1946,Barnes:1966,Coleman:1968,Denskat:1983,Isenberg:1983,Goldstein:1994,Leamon:1998a,Leamon:1998b,Quataert:1998,Gary:1999a,Leamon:1999,Quataert:1999,Leamon:2000,Isenberg:2001,Hollweg:2002,Howes:2008b,Schekochihin:2009,TenBarge:2013a,Howes:2015b,TCLi:2016,Howes:2018a}
(ii) nonresonant wave-particle interactions, primarily leading to
stochastic ion heating;
\cite{McChesney:1991,Johnson:2001,Chen:2001,White:2002,Voitenko:2004,Bourouaine:2008,Chandran:2010a,Chandran:2010b,Chandran:2011,Bourouaine:2013}
and (iii) dissipation in coherent structures, such as collisionless
magnetic reconnection occurring in small-scale current sheets.
\cite{Ambrosiano:1988,Dmitruk:2004,Markovskii:2011,Matthaeus:2011,Osman:2011,Servidio:2011a,Osman:2012a,Osman:2012b,Wan:2012,Karimabadi:2013,Zhdankin:2013,Dalena:2014,Osman:2014a,Osman:2014b,Zhdankin:2015a,Zhdankin:2015b}

Which of these mechanisms dominates in a given turbulent plasma likely
depends on the character of the turbulence and on the plasma
parameters. For example, the strength of resonant wave-particle
interactions depends on where the phase velocity of a given wave falls
within the particle velocity distribution, and the location of this
resonant velocity for both ions and electrons depends on the ion
plasma beta, $\beta_i = 8 \pi n_i T_i/B^2$, and the ion-to-electron
temperature ratio, $T_i/T_e$.  Thus, to understand fully the
mechanisms that govern how turbulent energy is converted into energy
of the plasma particles, it is important to study turbulence under a
wide range of turbulent and plasma conditions: in the solar
chromosphere, the plasma is partially ionized and strongly
collisional, so the impact of ion-neutral collisions will have a
significant impact on how the turbulent energy is dissipated; in the
solar corona, the turbulent dynamics are weakly collisional and the
plasma beta is low, $\beta_i \ll 1$; in the solar wind near the Earth,
the plasma beta increases to unity, $\beta_i \sim 1$; much further out in the
heliosphere, the magnetic field magnitude decreases
with the radially expanding solar wind, but \emph{in situ} heating of
the plasma leads to a slower radial decline in plasma temperature,
leading to high beta conditions, $\beta_i \gg 1$.
\citep{Cranmer:2009a,Howes:2011c} A key motivation for the upcoming
\emph{Parker Solar Probe} mission \citep{Fox:2016} is to explore the
energization of the plasma in the unexplored inner heliosphere within
the orbit of Mercury, sampling the turbulent conditions down to low
beta values, $\beta_i \ll 1$, so that we can discover how the solar
corona is heated and how the solar wind is accelerated.

The interplanetary medium, which is dominated by the radially outward
flow of the supersonic and super-\Alfvenic solar wind, is always
observed to be in a turbulent state. This turbulence naturally leads
to a very tangled interplanetary magnetic field,
\citep{Similon:1989,Maron:2001,Howes:2017b,Bourouaine:2017} and this
tangled magnetic field has important implications for the transport of
energetic particles through the heliosphere to the
Earth,\citep{Parker:1965,Jokipii:1966} including solar energetic
particles generated by violent activity at the Sun, \cite{Reames:2013}
anomalous cosmic rays generated through poorly understood mechanisms
in the outer reaches of our heliosphere,
\citep{Cummings:1996,Cummings:2007} and galactic cosmic
rays.\citep{Jokipii:1977,Kota:1983} For particularly extreme space
weather events, the copious solar energetic particles that are often
generated represent a significant hazard for robotic and human assets
in space, so accurate prediction of their propagation through the
turbulent solar wind toward the Earth is critical to prevent damage to
spaceborne technology and harm to astronauts.

Turbulence plays a role in a number of other important physical
processes that occur in various regions of the heliosphere. Turbulence
arises as a response of the Earth's coupled
magnetosphere-ionosphere-thermosphere (MIT) system to forcing by the
variable solar wind and to major impulses during extreme space weather
events. For example, ionospheric turbulence driven by the MIT system
response to variations in the solar wind can drive density
irregularities that scatter radio waves, causing amplitude and phase
scintillations in radio signals and thereby affecting satellite
communication and GPS navigation systems. \cite{Basu:2002} In
addition, many proposed mechanisms for particle acceleration at
collisionless shocks, such as diffusive shock acceleration, require
turbulent fluctuations upstream of the shock to scatter particles back
toward the shock. \citep{Bell:1978,Malkov:2001}

Laboratory experiments open up valuable new avenues to explore the
dynamics of plasma turbulence over a wide range of turbulence and
plasma parameters.  In the \emph{Large Plasma Device} (\emph{LAPD}) at UCLA,
\citep{Gekelman:1991} experiments have been conducted to understand
the nonlinear evolution of \Alfven wave collisions
\citep{Howes:2013a}---the nonlinear interactions among
counterpropagating \Alfven waves, proposed as the fundamental
mechanism mediating turbulent energy transfer to small scales in early
studies of MHD turbulence. \cite{Iroshnikov:1963,Kraichnan:1965}
Asymptotic analytical solutions for the evolution of \Alfven wave
collisions in the weakly nonlinear limit \cite{Howes:2013a} have been
confirmed numerically with gyrokinetic numerical simulations in the
MHD regime \cite{Nielson:2013a} and verified experimentally in the
laboratory,
\cite{Howes:2012b,Howes:2013b,Drake:2013,Drake:2014,Drake:2016}
establishing \Alfven wave collisions as the fundamental building block
of astrophysical plasma turbulence. The success of this experimental
investigation of \Alfven wave collisions has laid the foundation for
subsequent advances in our theoretical understanding of how current
sheets arise self-consistently in plasma
turbulence,\citep{Howes:2015b,Howes:2016b} the role played by resonant
wave-particle interactions in the dissipation of these current sheets,
\citep{Howes:2018a} and how collisions between localized \Alfven
wavepackets in the strongly nonlinear limit mediate the turbulent
cascade of energy to small
scales.\citep{Verniero:2018a,Verniero:2018b}

Observations of enhanced perpendicular ion temperatures in the solar
corona \citep{Kohl:1998} have lead to the important question of how
ions are energized perpendicular to the magnetic field under low
plasma beta conditions. Under similar low plasma beta and weakly
collisional conditions, a broadband spectrum of anisotropic magnetic
turbulence has been observed to arise during magnetic relaxation
events in the \emph{Madison Symmetric Torus} (\emph{MST}) experiment;
\citep{Ren:2011} application of a Rutherford scattering diagnostic
\citep{Reardon:2001} has shown there to be anomalous ion heating
coincident with this turbulence. \citep{Fiksel:2009} Such laboratory
experiments, using devices that have primarily been used to explore
magnetic confinement fusion, provide an alternative path to understand
the physics underlying ion heating mechanisms in the low beta,
collisionless plasma turbulence relevant to the solar corona.

In the near-Earth solar wind, in the outer heliosphere, and in many
astrophysical systems such as accretion disks around black holes,
turbulence occurs in plasmas with plasma beta of unity or higher.
Experimental facilities capable of generating plasma turbulence with
beta of order unity or higher enable uniquely detailed investigations.
The \emph{Swarthmore Spheromak Experiment} (\emph{SSX}) MHD plasma
wind tunnnel \citep{Brown:2014,Brown:2015} is used to launch a
spheromak which immediately relaxes its magnetic configuration,
generating broadband plasma turbulence with $0.1 \lesssim \beta_i
\lesssim 1$, enabling studies of intermittency \citep{Schaffner:2014}
and statistical complexity \citep{Weck:2015} in the resulting
turbulence. In the \emph{Big Red Plasma Ball}\citep{Cooper:2014} at
the Wisconsin Plasma Astrophysics Laboratory
(WiPAL),\citep{Forest:2015} a new national user facility for frontier
plasma science, a multi-cusp magnetic field configuration using a
spherical array of permanent magnets can be used to confine a
turbulent plasma at high plasma beta, $\beta_i \gg 1$. The
\emph{Plasma Liner Experiment}\citep{Hsu:2015} at Los Alamos National
Laboratory uses coaxial plasma guns mounted to a spherical chamber to
launch colliding plasma jets that can generate flow-dominated
turbulence with target plasma beta conditions over the range $0.01
\lesssim \beta_i \lesssim 10$.

Together, these unique facilities make possible laboratory experiments
that can overcome the limitations of spacecraft measurements to
understand in more detail how plasma turbulence influences the
evolution of the space environment.

\subsection{Magnetic Reconnection}
Another fundamental phenomenon of plasma physics that influences the
evolution of many diverse environments in space and astrophysical
plasmas is magnetic reconnection, mediating the rapid release of
magnetic energy through the topological rearrangement of magnetic field
lines. Although the field lines reconnect in spatially localized
places, the process often causes fundamental changes in macroscopic
configurations, such as in the evolution of solar flares or in the
dynamics of the Earth's magnetosphere during geomagnetic storms and
substorms. Over the years, numerous reviews have thoroughly explored
the implications of magnetic reconnection for heliospheric,
astrophysical, and laboratory
plasmas.\citep{Vasyliunas:1975,Biskamp:2000,Priest:2000,Zweibel:2009,Yamada:2010,Zweibel:2016}
Here I highlight the different space environments in which magnetic
reconnection plays a key role, with an emphasis on experimental
efforts in the laboratory that are helping to illuminate the
properties of this fundamental physical mechanism under various
conditions.

Magnetic reconnection under the collisional plasma conditions within
the solar convection zone plays an essential role in the incompletely
understood dynamo mechanism \citep{Parker:1955,Brun:2004} that
generates the Sun's magnetic field and dictates the 22-year magnetic
solar cycle.  The magnetic flux generated by the solar dynamo emerges
through the photosphere into the solar atmosphere, generating sunspots
and driving intense activity that can lead to stress in the
configuration of the magnetic field that pervades the solar
chromosphere and solar corona, ultimately building up energy in the
magnetic field. In this stressed magnetic field topology, the sudden
onset of magnetic reconnection can trigger explosive activity by
tapping the magnetic energy to power intense solar
flares\citep{Masuda:1994} or cause the eruption of a coronal mass
ejection. \citep{Antiochos:1999,Forbes:2006} Unraveling the complex
evolution of these poorly understood events that drive extreme space
weather is a major goal of the heliophysics community. Beyond the
lower solar atmosphere, the plasma conditions become collisionless,
where kinetic effects rather than resistivity lead to the breaking of
the magnetic field lines. A key frontier in heliophysics and
astrophysics is to understand the details of collisionless magnetic
reconnection. \citep{Birn:2001,Shay:2001,Ricci:2004,Drake:2008,Ji:2011}
In the solar corona, for example, the collective effect of many
magnetic small magnetic reconnection events, known as nanoflares, has
been suggested to be a candidate process for the heating of the
coronal plasma to more than a million Kelvin.  \citep{Parker:1988}

Near the Earth, if the interplanetary magnetic field carried by the
solar wind or by a coronal mass ejection has a significant southward
component, asymmetric magnetic reconnection can occur at the dayside
magnetopause,\citep{Pritchett:2008} leading to a sweeping of the reconnected
magnetic field toward the tail of the Earth's magnetosphere, driving
magnetic substorms or geomagnetic storms, as first proposed by Dungey
in the 1960s.  \citep{Dungey:1961} This dayside magnetic reconnection
drives a two-cell convective pattern at the polar caps
\citep{Crooker:1979} and sweeps magnetic flux into the
magnetotail. Symmetric magnetic reconnection in the magnetotail leads
to a shift in the topology of the magnetospheric field which is
transmitted to Earth via \Alfvenic fluctuations that propagate along
the magnetic field line to the polar ionosphere, triggering a magnetic
substorm \citep{Angelopoulos:2008} and leading to the glowing of
discrete auroral arcs. \citep{Stasiewicz:2000} Furthermore, the
ionosphere and thermosphere respond to shifts in the Earth's
magnetosphere caused by magnetic reconnection, and this coupling of the
magnetosphere-ionosphere-thermosphere (MIT) system represents an
important aspect of how the Earth reacts to forcing by the variable
solar wind and extreme space weather events.

Of course, these same collisional and collisionless magnetic
reconnection processes play an important role in a wide range of
plasma systems throughout the Universe.\citep{Zweibel:2009} In
addition, magnetic reconnection also occurs in partially ionized
plasmas, where it impacts the evolution of a number of important
astrophysical systems, such as in the interstellar medium,
protostellar and protoplanetary disks, and the outer envelopes of cool
stars.  \citep{Zweibel:2011}

The inherent geometry of the magnetic reconnection process---with
spatially separated inflow, x-point, and outflow regions---presents
severe a challenge to develop a full understanding of the mechanism
using observations in space plasmas due to the limitation to
single-point or few-point measurements that are possible with
spacecraft missions.  The laboratory provides an ideal complement to
observational studies, enabling multi-point measurements along with
some measure of control over the plasma conditions and the driving of
the reconnection process.  A number of experimental facilities have
pursued studies of magnetic reconnection, including the \emph{Large
  Plasma Device} (\emph{LAPD}) at UCLA, \citep{Gekelman:1991} the
\emph{Magnetic Reconnection Experiment} (\emph{MRX}) at Princeton
University, \citep{Yamada:1997b} the \emph{Swarthmore Spheromak
  Experiment} (\emph{SSX}), \citep{Brown:1999} the \emph{Versatile
  Toroidal Facility}\citep{Egedal:2000} at the Massachusetts Institute
of Technology, and the recently completed \emph{Facility for
  Laboratory Reconnection Experiments} (\emph{FLARE})\cite{Ji:2011} at
Princeton University.  There have also been studies of magnetic
reconnection in plasmas generated by facility-class
lasers,\citep{Nilson:2006,Fiksel:2014} but the broad range of
laboratory astrophysical studies of high-energy density plasmas using
these facility-class lasers---such as the \emph{National Ignition
  Facility} at Lawrence Livermore National Laboratory, the
\emph{Omega} laser at the University of Rochester, and the
\emph{Vulcan} laser Rutherford Laboratory in the UK---is beyond the
scope of this review.

By enabling some measure of control over plasma conditions, laboratory
experiments make possible studies of how magnetic reconnection
influences the plasma evolution across the broad range of space and
astrophysical environments discussed above.  Although recent reviews
\citep{Zweibel:2009,Yamada:2010,Zweibel:2016} provide a more thorough
discussion of past successes in the laboratory, I showcase here the
flexibility of laboratory experiments with the range of reconnection
problems examined on \emph{MRX}, from
collisional\citep{Ji:1998,Kuritsyn:2007,Jara-Almonte:2016} to
collisionless reconnection,\citep{Ren:2005,Yamada:2006} from
electron-scale \citep{Ji:2008,Ren:2008} to ion-scale
physics,\citep{Yoo:2013} from periodic to line-tied boundary
conditions, \citep{Oz:2011,Myers:2015} from zero-guide-field to
finite-guide-field reconnection,\citep{Tharp:2012,Fox:2017} from
two-dimensional to three-dimensional physics,
\citep{Carter:2002a,Ji:2004,Dorfman:2013a} from fully ionized to
partially ionized plasmas, \citep{Lawrence:2013} from symmetric
\citep{Yamada:2014,Yamada:2015} to asymmetric
reconnection,\citep{Yoo:2014,Yoo:2017} and from single X-line to
multiple X-line geometries.\citep{Dorfman:2013a,Jara-Almonte:2016}

A current frontier in the study of magnetic reconnection is to
understand how the plasmoid instability arises under high Lundquist
number conditions, leading to breakup of a thin current sheet with a
single x-point into multiple
x-points.\citep{Shibata:2001,Loureiro:2007,Bhattacharjee:2009,Samtaney:2009,Uzdensky:2010,Loureiro:2012}
To access in the laboratory the parameter regime in which the plasmoid
instability arises, \citep{Daughton:2009b,Ji:2011} new experimental
facilities must developed that can achieve a larger dynamic range of
plasma size relative to current sheet thickness.  The recently
completed \emph{Facility for Laboratory Reconnection Experiments}
(\emph{FLARE})\cite{Ji:2011} at Princeton has been constructed specifically
to provide an experimental platform for understanding in detail how
the plasmoid instability influences the microscopic dynamics and
macroscopic structure of magnetic reconnection that governs the
evolution of the space environments enumerated above. Another new
experimental facility for magnetic reconnection operates at the
Wisconsin Plasma Astrophysics Laboratory (WiPAL), where the
\emph{Terrestrial Reconnection Experiment} (\emph{TREX}) is a set of coils that can
be inserted into the \emph{Big Red Ball} to drive
reconnection,\citep{Forest:2015} with sufficient flexibility to access
numerous key regimes, including the collisionless plasmoid
instability,\citep{Olson:2016} anti-parallel reconnection, strong
guide-field reconnection, and 3D reconnection.

\subsection{Particle Acceleration}
Another poorly understood fundamental process in plasmas is how a
small fraction of particles in weakly collisional plasmas can be
accelerated to very high energies, seemingly defying the laws of
thermodynamics.  The acceleration of particles arises in many
different space and astrophysical plasma environments, creating
populations of energetic particles that impact regions far beyond the
region of acceleration. Unknown particle acceleration mechanisms lead
to a spectrum of galactic cosmic rays up to more than 10$^{20}$~eV per
particle, and anomalous cosmic rays up to about 100 MeV/nucleon are
believed to be accelerated in the outer regions of our
heliosphere.\citep{Cummings:1996,Cummings:2007} Near the sun, both
impulsive and gradual solar energetic particle events
\citep{Reames:2013} shower near-Earth space with high-energy particles
that pose a threat to manned space exploration and spaceborne assets
for communication and navigation by satellite.  The upcoming
\emph{Interstellar Mapping and Acceleration Probe} (\emph{IMAP}) mission,
recommended by the National Research Council's (NRC) 2013 Decadal
Strategy for Solar and Space Physics,\cite{NRCspace:2013} is intended
to make observations needed to tackle the problem of particle
acceleration throughout the heliosphere.

Within the Earth's magnetosphere, protons and electrons trapped in
Earth's dipolar magnetic field can be accelerated to high energies,
populating the Earth's Van Allen radiation belts. Particle
acceleration also plays a part in the coupling of the
magnetosphere-ionosphere (MI) system,\cite{Keiling:2009}
where shifts in the magnetic field in the distant magnetotail---due to
magnetic reconnection driven by substorms or geomagnetic storms---are
transmitted down towards the Earth by propagating \Alfven waves. Under
the very low plasma $\beta$ conditions of the polar magnetosphere,
these downward traveling \Alfven waves are believed to accelerate
electrons which precipitate into the ionosphere and cause the glowing
of discrete auroral arcs.
\cite{Hasegawa:1976c,Goertz:1979,Kletzing:1994,Stasiewicz:2000,Kletzing:2001}
Finally, turbulent electromagnetic fluctuations driven in the inner
magnetosphere by geomagnetic storms have been proposed to accelerate
ionospheric ions and drive ion outflows.
\cite{Chaston:2014,Chaston:2015,Chaston:2016}

In these various space environments, acceleration is believed to be
caused either by the interaction of particles with collisionless
shocks, as a consequence of collisionless magnetic reconnection, or
via resonant wave-particle interactions.  Common shock acceleration
mechanisms include shock surfing acceleration (SSA),
\cite{Sagdeev:1966,Lever:2001,Shapiro:2003a} shock drift acceleration
(SDA), \citep{Anagnostopoulos:1998a, Park:2013a} diffusive shock
acceleration (DSA), \citep{Bell:1978,Malkov:2001} and the
single-bounce ``fast Fermi'' mechanism.  \citep{Wu:1984b,Leroy:1984a}
Magnetic reconnection has also been proposed as an effective
acceleration mechanism for ions\citep{Drake:2009c} and electrons
\citep{Hoshino:2001,Drake:2006a,Dahlin:2014,Dahlin:2015} and for the
generation of anomalous cosmic rays. \citep{Drake:2010} Resonant
wave-particle interactions can also lead to the acceleration of a
small number of particles if the resonant velocity occurs in the tail
of the velocity distribution, thus enabling a small fraction of the
total distribution of particles to gain significant energy, as is
believed to occur in the case of auroral electron acceleration.
\citep{Kletzing:1994,Stasiewicz:2000}

The observational study of particle acceleration in space is hampered
by the frequent case that we measure the accelerated particles away
from the region of acceleration, such as the cases of impulsive solar
energetic particles associated with solar flares \citep{Masuda:1994,Aschwanden:2002,Lin:2003,Reames:2013}
and anomalous cosmic rays accelerated in the outer
heliosphere.\citep{Cummings:1996,Cummings:2007} Although reproducing
in the laboratory the highly energetic conditions of acceleration
regions in space is a major challenge, the ability to control the
location and properties of shock interactions and magnetic
reconnection in experiments provides exciting new possibilities for
illuminating the fundamental plasma physics of particle acceleration.

The capability to collide multiple plasma jets to generate shocks in
the \emph{Plasma Liner Experiment}\citep{Hsu:2012,Hsu:2015} at Los
Alamos National Laboratory makes possible a new path to study the
fundamental particle energization and acceleration processes that
occur in space plasmas, with ongoing work pushing from the collisional
\citep{Merritt:2013,Merritt:2014} to the collisionless
regime. \citep{Moser:2015} Laboratory experiments may also provide a
new avenue for exploring the acceleration of particles due to magnetic
reconnection, for example recent measurements of the generation of an
anisotropic non-thermal tail in the electron velocity distribution
during magnetic reconnection events occurring in the \emph{Madison
  Symmetric Torus} (\emph{MST}) experiment at the University of
Wisconsin.  \citep{DuBois:2017} Furthermore, expanding plasmas
generated by moderate-power lasers in a pre-existing plasma
environment make possible the investigation of magnetized shocks, and
recent work in the \emph{Large Plasma Device} (\emph{LAPD})
\citep{Gekelman:1991} at UCLA has made possible the first experimental
creation in the laboratory of sub-critical \citep{Niemann:2014} and
super-critical \citep{Schaeffer:2017} perpendicular shocks, as well as
the first observation of collisionless Larmor coupling.
\citep{Bondarenko:2016} Future experiments will be able to explore
also the formation of quasi-parallel shocks, \citep{Weidl:2016} and
possibly the physics of diffusive particle acceleration and injection.
\citep{Revile:2013}

The \emph{LAPD} has also been used to explore two proposed mechanisms for the
downward acceleration of the magnetospheric electrons that precipitate
into the ionosphere and lead to the glowing of discrete auroral
arcs.\citep{Sandahl:2008} Long-lived discrete auroral arcs have been
proposed to be caused by stationary inertial \Alfven waves on the
Earth's magnetic field, \citep{Knudsen:1996,Finnegan:2008} and recent
\emph{LAPD} experiments have successfully generated a stationary inertial
\Alfven wave in the laboratory frame.\citep{Koepke:2016} An
alternative idea is that propagating inertial \Alfven waves, generated
along the plasma sheet boundary layer by magnetotail reconnection,
propagate downward along the Earth's magnetic field toward to the
high-latitude ionosphere and accelerate auroral electrons along their
way.\cite{Hasegawa:1976c,Goertz:1979,Kletzing:1994,Stasiewicz:2000,Kletzing:2001}
To test this hypothesis, recent experiments on the \emph{LAPD} have launched
inertial \Alfven waves down the 16~m cylindrical plasma column and
measured the resulting perturbations of the parallel electron velocity
distribution function using a novel whistler wave absorption
diagnostic.\citep{Thuecks:2012} These auroral electron acceleration
experiments have found good agreement with the predicted linear
perturbation of the parallel electron distribution function,
\citep{Schroeder:2015,Schroeder:2016,Schroeder:2017} establishing the
critical foundation for future experiments to directly measure the
acceleration of the electrons by inertial \Alfven waves.  With
enhanced capabilities, laboratory experiments are likely to contribute
increasingly to our understanding of the mechanisms for particle
acceleration in space and astrophysical plasmas.

\subsection{Collisional and Collisionless Shocks}

In addition to their role in particle acceleration, collisionless
shocks in space plasmas establish the macroscopic boundaries
separating distinct regions of the heliosphere, decelerating
super-magnetosonic flows while heating and compressing the downstream
plasma. \citep{Treumann:2009} Such important heliospheric boundaries
include planetary bowshocks, \citep{Russell:1979} shocks associated
with solar flares and coronal mass ejections, \citep{Burlaga:2001} and
the heliospheric termination shock that abruptly slows the outward
flow of the solar wind. \citep{Richardson:2008}  In these rarefied
environments, the Coulomb collisional mean free paths exceed the
observed interaction length scales by many orders of magnitude,
signifying that the shocked plasma exchanges momentum and energy with
the ambient plasma via collisionless, collective, electromagnetic
effects.

Achieving astrophysically relevant collisionless shocks in the
laboratory is extremely challenging, but it can be achieved using
scaled experiments, as shown in a detailed analysis by
Drake. \cite{Drake:2000} Unlike in space shocks, important
dimensionless parameters in laboratory experiments are of order unity
and only marginally satisfy the shock formation criteria.
\citep{Schaeffer:2016} But laboratory experiments enable a detailed
study of the poorly understood physics of the collisionless coupling
between the shocked and ambient plasma that mediates the acceleration
and heating of the upstream plasma, including micro-instabilities and
ion reflection.

In addition to the studies of shocks in high-energy density plasmas
relevant to extreme astrophysical environments that are generated by
facility-class lasers or imploding theta-pinch or Z-pinch experiments,
a number of facilities are tackling shock conditions relevant to the
lower-energy-density environment of the heliosphere.  The \emph{Plasma
  Liner Experiment}\citep{Hsu:2012,Hsu:2015} at Los Alamos National
Laboratory uses colliding supersonic plasma jets, providing a unique
experimental platform to explore key issues in space plasma shock
physics, including the identification of the two-scale structure and
ambipolar electric fields predicted by two-fluid plasma theory
\citep{Jaffrin:1964} as well as the subtle effects of interspecies ion
separation arising in plasmas consisting of multiple ion species.
\citep{Kagan:2012,Hsu:2016} In the strongly magnetized plasma of
the \emph{LAPD}, \citep{Gekelman:1991} magnetized
plasma shocks relevant to the heliosphere can be studied by generating
expanding laser-produced plasmas in the pre-existing plasma
environment. Successful experiments have studied perpendicular shocks,
where the magnetic field is perpendicular to the shock normal,
observing both sub-critical \citep{Niemann:2014} and super-critical
cases,\citep{Schaeffer:2017} and future experiments will tackle 
fundamental questions about the formation of quasi-parallel shocks.
\citep{Weidl:2016}

\subsection{Kinetic and Fluid Instabilities}
\label{sec:instab}
Kinetic and fluid instabilities arise in a wide variety of space and
astrophysical plasma environments, regulating the thermodynamic state
of the plasma by limiting temperature anisotropies, differential drift
among species, and heat flux; governing eruptive behavior driven by
magnetic buoyancy that drives the solar dynamo and triggers extreme
space weather events; impacting the relaxation of stressed boundary
layers; and controlling the linear and nonlinear response of space
plasmas to applied perturbations.

Temperature anisotropy instabilities are a critical class of kinetic
instabilities that both regulate the thermodynamic state of space
plasmas and lead to the generation of unstable fluctuations that can
significantly impact the evolution of different space environments.
For a plasma with a bi-Maxwellian proton distribution, there exist
four potential proton temperature anisotropy instabilities: the
parallel (or whistler) firehose
instability,\citep{Kennel:1966,Gary:1976} the \Alfven (or oblique)
firehose instability,\citep{Hellinger:2000} the mirror
instability,\citep{Chandrasekhar:1958,Vedenov:1958,Barnes:1966,Tajiri:1967,Southwood:1993}
and the proton cyclotron
instability.\citep{Kennel:1966,Davidson:1975,Smith:1984,Gary:1976}
Spacecraft measurements in the near-Earth solar wind demonstrate that
the observed proton temperature anisotropy $T_{\perp p}/T_{\parallel
  p}$ is constrained, as a function of the parallel proton beta
$\beta_{\parallel p}$, by the marginal stability boundaries of kinetic
proton temperature anisotropy
instabilities,\citep{Kasper:2002,Hellinger:2006,Bale:2009,Maruca:2011,Chen:2016}
and these instabilities are likely to affect the dynamics and
dissipation of turbulent fluctuations in space
plasmas. \citep{Klein:2015,Kunz:2015,Kunz:2018} For example, for
sufficiently large plasma $\beta$, the anisotropic velocity
distributions self-consistently generated by \Alfven waves of
sufficiently large amplitude can trigger the parallel firehose
instability, effectively eliminating the magnetic tension that serves
as the restoring force for the wave, thereby interrupting the \Alfven
wave dynamics.\citep{Squire:2016,Squire:2017a,Squire:2017b}

Experimental validation of the instability predictions for $\beta_{\|
  p} > 1$ plasmas is an area of active research. At West Virginia
University, laboratory studies of instability growth and the limits of
ion temperature anisotropy in $\beta_i \sim 1$, space-relevant plasmas
were conducted in the \emph{Large Experiment on Instabilities and
  Anisotropies} (\emph{LEIA}) facility. Those experiments demonstrated
an upper bound on the ion temperature anisotropy $T_{\perp
  i}/T_{\parallel i}$ that scales inversely with the parallel ion beta in low
collisionality plasmas. \citep{Keiter:2000, Scime:2000} Additional
measurements provided direct evidence of enhanced electromagnetic,
ion-cyclotron-like, fluctuations for the same plasma conditions.
\citep{Keiter:2000,Scime:2000} Current experiments in the \emph{Large
  Plasma Device} at UCLA aim to measure the parallel firehose
instability by generating a plasma with $T_{\perp i}/T_{\parallel
  i}<1$ and $\beta_{\parallel i} \sim 1$.

Of course, more general non-Maxwellian velocity distributions, such as
loss-cone or ring distributions, can also lead to instabilities that
generate electromagnetic fluctuations.  In the Earth's magnetosphere
and Van Allen radiation belts, the acceleration and loss of energetic
particles trapped in the Earth's dipolar magnetic field
\citep{Summers:2007a,Summers:2007b} is affected by different
instability-driven waves: whistler-mode
chorus,\citep{Summers:1998,Horne:2005,Omura:2008} electromagnetic ion
cyclotron (EMIC)
waves,\citep{Cornwall:1965,Horne:1994,Anderson:1996,Fraser:2001,Min:2012}
extremely low frequency (ELF) magnetosonic equatorial noise,
\citep{Russell:1970,Gurnett:1976,Olsen:1987,Liu:2011} and
plasmaspheric hiss.
\citep{Thorne:1973,Church:1983,Sonwalkar:1989,Hayakawa:1992,Abel:1998,Green:2005,Meredith:2006,Bortnik:2008}
Laboratory investigations are a unique tool that can be used to
improve our understanding of the instabilities that drive these
different waves and their effect on the particles trapped in the
magnetosphere.

At altitudes below 2,000~km in the auroral ionosphere, laboratory
experiments have played a critical role in understanding a
velocity-shear driven instability,\citep{Kintner:1992} the
Inhomogeneous Energy Density Driven Instability
(IEDDI),\citep{Ganguli:1985a,Ganguli:1988a,Penano:1999} that leads to
intense ion heating that drives the observed outflow of heavy oxygen
ions to higher altitude.\citep{Pollock:1990} Experiments in the
\emph{Q machine} at West Virginia University and in the \emph{Space
  Plasma Simulation Chamber} at the Naval Research Laboratory (NRL)
showed the generation of unstable modes,
\citep{Koepke:1994,Amatucci:1994} and later experiments confirmed the
broadband electrostatic emissions \citep{Amatucci:1996} and ion
heating \citep{Walker:1997} as well as electromagnetic
emission\citep{Tejero:2011} of the IEDDI. This experimental program
verified key aspects of a comprehensive ionospheric heating model,
\citep{Ganguli:1994a} inspiring numerous sounding rocket missions to
look for corroborating signatures in the
ionosphere.\citep{Earle:1989,Bonnell:1996,Bonnell:1997,Lundberg:2012}

Laboratory experiments have also made valuable contributions to our
understanding of the Electron-Ion Hybrid Instability
\citep{Romero:1992,Romero:1993,Romero:1994} that arises in the
relaxation of stressed boundary layers in the Earth's magnetotail
between the high-pressure plasma sheet and low-pressure
lobe. \citep{Forbes:1981,Parks:1984} A clever means was devised to
establish a scaled laboratory experiment of the plasma sheet-lobe
boundary\citep{Amatucci:2003} in the \emph{Space Physics Simulation
  Chamber} at the NRL, demonstrating the unstable generation of waves
with similar properties to those observed in space.  Subsequent
experiments using the \emph{Auburn Linear Experiment for Instability
  Studies} (\emph{ALEXIS})\citep{Dubois:2013a} at Auburn University
characterized these broadband electrostatic emissions over five orders
of magnitude in frequency, \citep{Dubois:2013b,Dubois:2014} providing
an experimental confirmation of theoretical
predictions.\citep{Ganguli:1994a} Subsequent experiments at NRL have
also measured electromagnetic emissions from a similar stressed
boundary layer experiment, \citep{Tejero:2011,Enloe:2017} but at a
much smaller energy density than the electrostatic emission.

Parametric instabilities have long been suggested as a potential
nonlinear mechanism underlying the turbulent transfer of energy from
large-scale, large-amplitude unidirectional \Alfven waves into a
turbulent cascade.\citep{Schmidt:1995,DelZanna:2001} Recent
experimental work in the \emph{Large Plasma Device} at UCLA has
focused on exploring the physics of parametric instabilities. Initial
experiments demonstrated the resonant excitation of acoustic modes by
large-amplitude \Alfven waves,\citep{Dorfman:2013b,Dorfman:2015}
followed by the successful measurement of an Alfv{\'e}n wave
parametric instability in the laboratory.\citep{Dorfman:2016}

Finally, a number of other important instabilities in space and
astrophysical plasmas are susceptible to investigation in the
laboratory, including the magnetic buoyancy
instabilities\citep{Parker:1966} that play a key role in the solar
dynamo and drive space weather eruptions, the gradient drift coupling
(GDC) instability that can potentially enhance magnetic reconnection
in astrophysical plasmas, \citep{Pueschel:2015,Pueschel:2017} and
current-driven instabilities of coronal arches. \citep{Moser:2012} In
summary, a host of kinetic and fluid instabilities are believed to
significantly influence the evolution of space and astrophysical
plasmas, and carefully devised laboratory experiments can be used to
explore the physics of these instabilities in great detail and confirm
their effect on the evolution of these plasmas.

\subsection{Self-Organization}
An important fundamental process that occurs in heliospheric plasmas
is the self-organization of turbulent motions to generate ordered
magnetic fields through a dynamo mechanism, the details of which
remain incompletely understood. Such turbulent magnetic dynamos
dominate the evolution of the heliosphere, generating the strong solar
magnetic fields \citep{Parker:1955,Brun:2004} that drive explosive
dynamics on the Sun's surface and are eventually swept out with the
supersonic solar wind to form the heliosphere.  Furthermore, turbulent
dynamos also operate in the liquid metal cores of the Earth
\cite{Bullard:1954,Parker:1955,Kuang:1997} and other planets that
generate their own protective magnetospheres.

To understand better the nature of magnetic dynamos, many laboratories
around the world have performed magnetic dynamo experiments using
liquid sodium, including constrained flows
\citep{Gailitis:2001,Stieglitz:2001} and unconstrained turbulent
flows.
\citep{Reighard:2001,Noguchi:2002,Nornberg:2006,Monchaux:2007,Zimmerman:2014}
More recently, a plasma dynamo experiment \cite{Cooper:2014} has been
designed in the \emph{Big Red Ball} at the University of Wisconsin, employing
a cusp magnetic field confinement scheme to make possible the
investigation of the dynamo mechanism under high-$\beta$ (\emph{i.e.}, weak
magnetic field) conditions within the interior of the
plasma. \citep{Weisberg:2017} Magnetic field generation by the Weibel
instability \citep{Weibel:1959} driven by interpenetrating plasma flows
at collisionless shocks has also been studied successfully in facility-class laser
plasmas.\citep{Huntington:2015}

Another example of self-organization in space plasmas is the
relaxation of a magnetized, low-$\beta$ plasma toward a minimum-energy
state under the constraint of constant magnetic helicity, occurring
due to the faster decay of magnetic energy compared to the decay of
magnetic helicity in low-$\beta$, MHD-unstable
plasmas.\citep{Bellan:2000} In the \emph{Caltech Spheromak
  Experiment}, experiments using a simple, planar magnetized coaxial
plasma gun \citep{Hsu:2003, Hsu:2005} have revealed vivid details of
the 3D, dynamical processes involved in magnetic relaxation, including
the formation and collimation of a plasma jet, kink instability of the
jet, and subsequent poloidal flux amplification associated with
spheromak formation and evolution toward a minimum energy
state.\citep{Stenson:2012,Moser:2012} The relationship of these
experiments to jet formation and morphology in astrophysical
environments has also been described in detail. \citep{Hsu:2002} Such
magnetic relaxation of a spheromak plasma injected into a larger
plasma volume also generates vigorous plasma turbulence, an approach
exploited in the \emph{Swarthmore Spheromak Experiment} (\emph{SSX})
wind tunnel to study MHD
turbulence. \citep{Brown:2014,Schaffner:2014,Brown:2015,Weck:2015}

Magnetic self-organization also underlies the development of
collimated astrophysical jets, where laboratory experiments have been
used to generate and explore the evolution of magnetized plasma jets
both at Caltech \citep{Stenson:2012,Moser:2012} and at the University
of Washington.\citep{You:2018} Laser plasmas at the \emph{ELFIE} laser
facility at Ecole Polytechnique in France have also been used to
produce scaled experiments of a collimated plasma outflow relevant to
the physics of young stellar
objects. \citep{Albertazzi:2013,Ciardi:2013,Albertazzi:2014}

\subsection{Physics of Multi-Ion and Dusty Plasmas}
Due to the inherent complexity of plasma physics, many studies of
space and astrophysical plasmas treat the idealized case of a fully
ionized, proton-electron plasma, ignoring the presence of neutrals,
additional ion species, or charged microparticles (dust).  But the
effects of these other species cannot always be treated as a higher
order correction, and for some space and astrophysical environments
the effects of neutrals, heavier ions, or dust fundamentally alters
the macroscopic evolution.  Early studies recognized the potentially
important role played by charged dust in star formation,
\citep{Mestel:1956} the formation of planetary rings,
\citep{Goertz:1983,Gurnett:1983} and other processes throughout the
solar system.\citep{Goertz:1989} In the ionosphere, ion-neutral
interactions and the presence of a significant density of minority
ions, such as oxygen, influence how the ionosphere responds to various
impulses from the
magnetosphere.\citep{Ahn:1983,Luhmann:1983,Lysak:1983}

The study of strongly coupled, dusty (complex) plasmas has emerged as
a unique discipline within the larger community of the basic plasma
physics.  Beyond the fundamental investigation of strongly coupled
plasma physics, applications to the dynamics of the solar
system\citep{Goertz:1989,Horanyi:1996} and more distant astrophysical
environments, \citep{Mendis:1994} for example the study of waves and
instabilities in dusty plasmas,\citep{Verheest:1996} have defined a
new frontier of space plasma physics. In addition to a number of
small-scale laboratory experiments for the study of dusty plasmas at
different universities, experimental facilities included experiments
onboard the Mir space station \citep{Fortov:1998} and the
international space station
\citep{Nefedov:2003,Thomas:2008,Fortov:2005} to explore the dynamics
under microgravity conditions.  Currently, the \emph{Magnetized Dusty
  Plasma Experiment} (\emph{MPDX}) \citep{Thomas:2015a} is a user
facility at Auburn University devised to explore how the presence of a
magnetic field alters the physics of a dusty plasma, applicable to
many space and astrophysical environments.  The Caltech
\emph{Water-Ice Dusty Plasma Experiment} \citep{Bellan:2015} is
dedicated to understanding the accretion and grain growth of ice
crystals in a cold, magnetized dusty plasma environment relevant to
protoplanetary disks and molecular clouds.\citep{Marshall:2017}

In contrast to the wide range of laboratory experiments dedicated to
the study of the physics of dusty plasmas, there have yet been
relatively few experimental studies of the physics of multi-ion
plasmas. But the partially ionized, multi-ion nature of plasmas in the
Earth's ionosphere and magnetosphere significantly influences the
topside current instabilities \citep{Kindel:1971} that play a role in
the dynamics of the magnetosphere-ionosphere coupling, the resonant
acceleration and scattering of particles in the radiation belts,
\citep{Papadopoulos:1980,Summers:1998,Hu:2010,Saikin:2015} the
stability of ion-ring distributions in multi-ion plasmas,
\citep{Mithaiwala:2010} the acceleration and heating of particles in
the solar wind\citep{Hu:1999,Perrone:2013} including when ions are
differentially streaming, \citep{Isenberg:1982,McKenzie:1994} the
physics of \Alfven waves in the multi-ion plasma of the chromosphere
and lower solar corona,\citep{Ofman:2005} and the impulsive
acceleration of different ions in solar flares.\citep{Steinacker:1997}

Improving our understanding of the physics of multi-ion plasmas and
ion-neutral coupling will take on an enhanced strategic importance
with plans for NASA to launch a number of missions in the near future
that will focus on probing magnetosphere-ionosphere-thermosphere (MIT)
coupling. The \emph{Global-Scale Observations of the Limb and Disk}
(\emph{GOLD}) mission\citep{Eastes:2017} was launched in January 2018
and the \emph{Ionospheric Connection Explorer} (\emph{ICON})
mission\citep{Immel:2018} is due for launch later in 2018. In
addition, the National Research Council's (NRC) 2013 Decadal Strategy
for Solar and Space Physics\cite{NRCspace:2013} proposes three
upcoming missions: the \emph{Magnetosphere Energetics, Dynamics, and
  Ionospheric Coupling Investigation} (\emph{MEDICI}) mission, the
\emph{Dynamical Neutral Atmosphere-Ionosphere Coupling}
(\emph{DYNAMIC}) mission, and the \emph{Geospace Dynamics Coupling}
(\emph{GDC}) mission.  Targeted laboratory experiments to elucidate
the physics of multi-ion plasmas and ion-neutral coupling have the
potential to complement these scientific missions, making a valuable
contribution to our knowledge of MIT coupling.




\subsection{Astrophysical Connections}
The physical mechanisms governing the evolution of the heliosphere are
broadly expected to apply also to more distant astrophysical
environments that are, for the foreseeable future, beyond the reach of
measurement by \emph{in situ} instrumentation.  But the universe also
harbors more extreme astrophysical environments without analogue in
the heliosphere, with many open scientific questions about the
physical processes governing the evolution of those
systems. Appropriately scaled laboratory experiments can provide
unique opportunities for  detailed examination of the physics of
these astrophysical plasmas.

A key question in supernova physics is the strong-shock-driven
instabilities that generate turbulence and the impact of those effects
on the supernova blast wave and the resulting supernova
remnant.\citep{Arnett:1996} Appropriately scaled laboratory
experiments\citep{Ryutov:1999,Drake:2000,Ryutov:2000,Ryutov:2001,Remington:2006}
using intense facility-class lasers provide unique opportunities to
study the nonlinear evolution of instabilities, such as the Weibel
instability,\citep{Weibel:1959} that arise in these extreme
environments.\citep{Huntington:2015} Laser plasmas can also be used to
create the hot, dense conditions of stellar interiors to enable
improved measurements of iron opacity\citep{Bailey:2015} that are
needed to refine models of stellar structure.\citep{Cox:1970}

Exploring the interaction between plasma flows and magnetic fields
remains a fundamental plasma physics problem necessary to understand
the formation and stability of astrophysical
jets.\citep{Shibata:1985,Shibata:1986,Blandford:1982,Begelman:1984,McKinney:2006}
Laser plasma experiments have been used to investigate the physics of
radiative jets \citep{Farley:1999} and protostellar
jets.\citep{Albertazzi:2014} Some aspects of MHD models of the
formation and collimation of astrophysical
jets\citep{Shibata:1985,Shibata:1986,Blandford:1982,Begelman:1984,McKinney:2006}
can be tested in laboratory experiments, \citep{Hsu:2002} where
experiments at Caltech have generated a collimated plasma jet and
observed its unstable evolution due to the kink
instability.\citep{Hsu:2003,Hsu:2005} At the University of Washington,
the \emph{MoCHI} device has been recently constructed with the aim to
produce long, collimated, stable, magnetized plasma jets by mimicking
an accretion disk threaded by a poloidal magnetic field with
concentric planar electrodes in front of a solenoidal
coil. \citep{You:2018}

The physics of astrophysical accretion
disks\cite{Pringle:1981,Narayan:1994,Blandford:1999,Frank:2002}
represents a major topic at the frontier of astrophysics, particularly
the attempt to understand the generation of turbulence and its impact
on the rate of accretion under the weakly collisional conditions
typical of accretion disks around compact objects, such black holes
and neutron stars. The magnetorotational instability
\citep{Balbus:1991,Hawley:1991,Balbus:1998} explains how MHD
turbulence arises in a weakly magnetized disk with a Keplerian
rotation profile, and numerous efforts are underway to understand the
physics of the magnetorotational instability under the weakly
collisional conditions of hot accretion
disks\citep{Quataert:1999,Quataert:2002,Sharma:2006,Sharma:2007,Riquelme:2012,Hoshino:2013,Heinemann:2014,Hoshino:2015,Sironi:2015a,Sironi:2015b,Kunz:2016}
as well as under the poorly ionized conditions relevant to
protoplanetary disks.\citep{Kunz:2013} In particular, the shearing due
to the differential rotation in the accretion disk can drive firehose
and mirror instabilities in the collisionless plasma, significantly
impacting the turbulence and thermodynamic evolution of the
disk. \citep{Kunz:2014,Quataert:2015,Riquelme:2015,Melville:2016} The
ability to study the plasma turbulence under high plasma $\beta_i$
conditions experimentally, possible in the \emph{Big Red
  Ball}\citep{Forest:2015} at the University of Wisconsin, may provide
new opportunities to illuminate the physics of turbulent astrophysical
plasmas.

Furthermore, exploring the physics of kinetic temperature anisotropy
instabilities, such as the firehose and mirror instabilities (see
\secref{sec:instab}), in the laboratory may contribute to a better
understanding of how these instabilities affect the turbulent dynamics
in astrophysical systems.  For example, carefully devised laboratory
experiments may be able to probe how \Alfven waves can be interrupted
by these instabilities under sufficient high $\beta_i$ plasma
conditions.\citep{Squire:2016,Squire:2017a,Squire:2017b}

Another major topic at the frontier of astrophysical research is the
plasma physics of the intracluster medium in galaxy clusters,
including the generation of magnetic fields by a turbulent dynamo in
the weakly collisional intracluster plasma
\citep{Colgate:2001,Schekochihin:2006,Parrish:2008,McCourt:2011,Santos-Lima:2014,Rincon:2016}
and the thermodynamics of cooling flows in galaxy
clusters. \citep{Fabian:1984,Fabian:1994,Carilli:2002} In particular,
the dynamics and instabilities associated with thermal conduction in
the turbulent intracluster
medium\citep{Chandran:1998,Narayan:2001,Parrish:2009,Komarov:2016}
remains an active area of research, with a focus on recently
discovered instabilities in the collisionless intracluster plasma,
such as the magnetothermal
instability\citep{Balbus:2000,Parrish:2005,Parrish:2007} and the heat
flux buoyancy instability.\citep{Quataert:2008,Parrish:2008}
Laboratory experiments in the \emph{Large Plasma Device} are being
designed to explore electron heat conduction through a series of
magnetic mirrors, connecting to the frontier issue of thermal
conduction in a mirror-unstable plasma. \citep{Komarov:2016}

\section{What's on the Horizon for Laboratory Space Physics?}
\label{sec:horizon}
The value of complementary efforts between plasma physics experiments
in the laboratory and satellite measurements in space is dramatically
increasing as the accessible length and time scales of spacecraft
observations and laboratory measurements are converging. In the past,
limitations on the size of the vacuum chambers and the magnitude of
magnetic fields and other applied perturbations limited the
experimental investigation of space-relevant plasma phenomena to
length scales at or below the typical ion kinetic length
scales. Similarly, the cadence of plasma and field measurements by
older spacecraft missions limited the accessible length scales---when
using Taylor hypothesis to use the plasma flow relative to the
spacecraft to convert from the temporal sampling cadence to a spatial
resolution\citep{Taylor:1938}---to scales typically in the MHD regime,
much larger than the ion kinetic length scales.  Improved experimental
facilities and modern spacecraft instrumentation have closed this gap,
enabling the two complementary approaches to tackle the same plasma
phenomena in an overlapping regime of applicability, exponentiating
the impact of the synergy between laboratory investigations and
spacecraft missions on our understanding of the physics of space and
astrophysical plasmas.

In this section, I stress specific ways in which laboratory
experiments can make unique contributions to potentially
transformative progress in our knowledge of space plasmas. A
valuable new frontier for developing an improved understanding of
space plasmas using laboratory experiments is to exploit fully the
vast store of information about the plasma dynamics stored within the
fluctuations of the particle velocity distributions.  New experimental
facilities, improved diagnostic capabilities, and novel analysis
methods will open new avenues of investigation and revitalize
established approaches, with the promise to revolutionize our
understanding of the physics of space plasmas, which can be sampled
directly by spacecraft missions, and of astrophysical plasmas, which
are presently beyond the reach of \emph{in situ} probes.

\subsection{Velocity Space: A New Frontier }
Although spacecraft missions have been able to measure
three-dimensional particle velocity distributions since the 1970s
(\emph{e.g.}, the \emph{Helios} missions provided unique measurements
of 3D proton velocity distributions at 90~s
cadence\citep{Marsch:1982,Marsch:1991,Marsch:2004,Marsch:2006}), only
with modern instrumentation are spacecraft now able to sample the
fluctuations of the plasma particle velocity distributions at the
kinetic time and length scales of the plasma dynamics.  For example,
the \emph{Magnetospheric MultiScale} mission (\emph{MMS}) provides 3D
velocity distributions for protons at a cadence of 150~ms and for
electrons at a cadence of 30~ms. \citep{Pollock:2016} By comparison,
in the solar wind, the period (Doppler-shifted by the solar wind flow
relative to the spacecraft \citep{Taylor:1938}) of fluctuations with
typical ion kinetic length scales is approximately 0.4~s and with
typical electron kinetic length scales is approximately 50~ms. Since
the collisionless mechanisms of energy transfer between the
electromagnetic fields and the particles dominantly operate on the
corresponding particle kinetic length scales, only now with such recent
improvements in instrumental capabilities is it possible to
explore properly these important dynamical processes in space plasmas.

Fundamentally, collisionless interactions among the electromagnetic
fields and the individual plasma particles generate characteristic
signatures in the particle velocity distributions, \citep{Howes:2017c}
although novel means of interpreting these velocity-space signatures
are in their relative infancy. Nonetheless, the fluctuations of the
particle velocity distributions provide a largely untapped potential
source for discovery science in space physics. In the effort to
diagnose and interpret the velocity-space dynamics of space plasmas,
laboratory experiments provide an invaluable complement to spacecraft
missions.  Although three-dimensional velocity distributions are
extremely challenging to measure in the laboratory (see
\secref{sec:diag} for more detail), the combination of multi-point
diagnostic access and reproducibility in experiments makes the
conjunction of laboratory and spacecraft investigations a powerful
tool for illuminating the fundamental physics governing space plasmas,
in particular the kinetic plasma physics of turbulence, magnetic
reconnection, particle acceleration, and kinetic instabilities.
Improvements in experimental diagnostics (which, in some cases, may
contribute to refinements of spacecraft instrumentation) and analysis
methods will enable heliophysicists and astrophysicists to exploit
fully the velocity-space dynamics to understand the plasma physics
that governs the evolution of space and astrophysical environments.

\subsection{New and Enhanced Experimental Capabilities}
The potential for discovery science using laboratory experiments is
directly related to the accessibility of ground-breaking experimental
facilities. The immediate future for space physics in the laboratory
looks especially promising at present with number of new and enhanced
experimental facilities that are available across the plasma physics
community.

National user facilities for laboratory investigations provide a
frontline for the collaborative investigation of fundamental space
physics phenomena.  The \emph{Basic Plasma Science Facility}
(\emph{BAPSF}) at UCLA is an established national user facility for
the experimental investigation of space plasma physics and fundamental
plasma physics.  Funded over its lifetime by multiple federal
agencies, including the National Science Foundation, Department of
Energy, and Office of Naval Research, this facility boasts as its
primary experimental platform the unique \emph{Large Plasma Device}
(\emph{LAPD}), \citep{Gekelman:1991} a long cylindrical plasma chamber
capable of generating a plasma of length 17~m and diameter 60~cm, with
sufficient flexibility in axial magnetic field, plasma temperature,
and density to access plasma betas over the range $10^{-5} \lesssim
\beta \lesssim 0.1$. The shot repetition rate is 1~Hz, enabling up to
$86,400$ shots per day. The flexibility of the \emph{LAPD} plasma has
enabled a broad range of experiments, from \Alfvenic space plasma
turbulence\citep{} to magnetized collisionless shocks\citep{} to the
interaction of energetic particles with plasma waves\citep{} to
magnetic reconnection between flux ropes.\citep{}

In 2017, the \emph{Wisconsin Plasma Astrophysics Laboratory}
(\emph{WiPAL})\citep{Forest:2015} was named a new national user
facility, boasting the combined experimental capabilities of the
\emph{Madison Symmetric Torus} (\emph{MST}), the \emph{Big Red Ball}
(\emph{BRB}), \citep{Cooper:2014} and the \emph{Terrestrial Magnetic
  Reconnection Experiment} (\emph{TREX}).  The \emph{MST} device is
reverse field pinch that can confine plasmas with $0.03 \lesssim \beta
\lesssim 0.15$ and, in addition to the numerous magnetic confinement
fusion studies it has hosted, has also provided a unique platform for
the investigation of anomalous ion\citep{Ren:2011} and
electron\citep{DuBois:2017} heating in broadband plasma
turbulence. The \emph{BRB} is a spherical, multi-cusp magnetic field
confinement device that can be used to confine a turbulent plasma at
high plasma beta, $\beta \gg 1$.  By adjusting external magnetic field
coils, it can access plasma beta over the range $10^{-3}\lesssim \beta
\lesssim \infty$.  \emph{TREX} is a set of magnetic field control
coils that can be inserted into the \emph{BRB} to enable studies of
magnetic reconnection under a wide variety of background plasma
conditions, with an effective plasma beta range of $10^{-4}\lesssim
\beta \lesssim 4$. Together, \emph{MST}, \emph{BRB}, and \emph{TREX}
provide a wide range of experimental platforms for the exploration of
space and astrophysical plasma phenomena available to the national
research community.

In 2014, the \emph{Magnetized Dusty Plasma Experiment} (\emph{MPDX})
\citep{Thomas:2015a} at Auburn University began operating as a user
facility for the study of the physics of dusty, or complex, plasmas in
the magnetized environment that is applicable to many space plasmas.
Completed experiments have explored the physics particle
motion\citep{Thomas:2015b} and ordering in different
dimensionality\citep{Thomas:2016} as well as the effect of the
magnetic field on phase transitions in a dusty
plasma.\citep{Jaiswal:2017}

The recently constructed \emph{Facility for Laboratory Reconnection
  Experiments} (\emph{FLARE}) at Princeton University
is intended to be a user facility open to worldwide users from
multiple communities, with a focus on exploring new reconnection
phases modified by the plasmoid
instability.\citep{Shibata:2001,Loureiro:2007,Bhattacharjee:2009}
Experimental studies of the plasmoid instability necessarily require a
larger dynamic range of accessible Lundquist number and normalized
plasma size.\citep{Ji:2011} The physics of the plasmoid instability
may provide the much needed multi-scale solution to couple global MHD
scales to local kinetic scales.\citep{Daughton:2009b,Ji:2011}

In addition to these user-class experimental facilities, the
collective capabilities of numerous moderate-scale experimental
platforms at national laboratories and universities around the country
provide a powerful complement to spacecraft missions to discover the
fundamental plasma physics mechanisms that govern the evolution of
space and astrophysical plasmas. The \emph{Space Physics Simulation
  Chamber} at the Naval Research Laboratory is a laboratory device
dedicated to the investigation of near-Earth space plasma phenomena,
under scaled ionospheric and magnetospheric conditions. The large
plasma size of approximately 150 ion gyroradii across the plasma
column enables a range of experimental studies, from exploring the
plasma response to strongly sheared flow
\citep{Amatucci:1996,Amatucci:1998,Amatucci:1999} to understanding the
triggering of the Electron-Ion Hybrid Instability
\citep{Romero:1992,Romero:1993,Romero:1994} in stressed boundary
layers, \citep{Amatucci:2003} such as that found in the plasma sheet
boundary layer in the Earth's magnetotail. The physics of the
relaxation of stressed boundary layers was further explored using the
\emph{Auburn Linear Experiment for Instability Studies}
(\emph{ALEXIS})\citep{Dubois:2013a} at Auburn University, where
varying the width of the boundary layer to the ion gyroradius lead to
the generation of unstable electrostatic fluctuations which varied
over five orders of magnitude in
frequency. \citep{Dubois:2013b,Dubois:2014}

At Los Alamos National Laboratory, collisional plasma shocks are
generated in the \emph{Plasma Liner Experiment} using colliding plasma
jets driven by pulsed-power-driven plasma guns,\citep{Hsu:2012,
  Hsu:2015} with future experiments aiming to form collisionless
shocks. At Swarthmore College, the \emph{Swarthmore Spheromak
  Experiment} (\emph{SSX}) MHD plasma wind tunnel
\citep{Brown:2014,Brown:2015} launches a spheromak which immediately
relaxes to a lower energy magnetic configuration, enabling studies of
magnetic reconnection \citep{Brown:1999} and plasma turbulence in the
ion plasma beta parameter range $0.1 \lesssim \beta_i \lesssim
1$.\citep{Schaffner:2014,Weck:2015} At Bryn Mawr College, the
\emph{Bryn Mawr Magnetohydrodynamic Experiment} (\emph{BMX}) is a
larger scale version of the \emph{SSX} wind tunnel that is currently
under construction, with the aim to produce a continuous injection of
plasma and to allow extensive diagnostic access to the turbulent
flowing plasma.  In the \emph{Caltech Spheromak Experiment},
experimental campaigns are conducted to explore the physics of
magnetic relaxation \citep{Bellan:2000,Hsu:2003, Hsu:2005} and
magnetic self-organization related to solar coronal arches and
astrophysical jets.\citep{Bellan:2005,Moser:2012,Stenson:2012} At West
Virginia University, an experimental platform combining the \emph{Hot
  Helicon Experiment} and the \emph{Large Experiment on Instabilities
  and Anisotropies} (\emph{HELIX-LEIA}) together  enables studies of
ion temperature anisotropy instabilities in space
plasmas,\citep{Klein:2015} in particular the mirror instability
\citep{Chandrasekhar:1958,Barnes:1966} and the ion cyclotron
instability \citep{Kennel:1966,Davidson:1975,Smith:1984} that both
occur for ion temperature anisotropies $T_{\perp i}/ T_{\parallel i} >
1$. For $\beta_i \sim 1$ conditions relevant to the near-Earth solar
wind, \emph{HELIX-LEIA} experiments have demonstrated the
theoretically predicted\citep{Gary:1994} inverse scaling of the ion
temperature anisotropy with the parallel ion beta in low
collisionality plasmas.  \citep{Keiter:2000,Scime:2000} Finally, the
\emph{Colorado Solar Wind Experiment}\citep{Ulibarri:2017} at the
University of Colorado, Boulder explores the interaction of supersonic
and super-Alfv\'enic plasma flow with small-scale magnetic anomalies at
unmagnetized bodies, such as the Moon.

Even facilities that have been constructed for fusion energy research
are beginning to make available a small fraction of runtime for
science at the frontier of plasma physics, including the physics of
space and astrophysical plasmas.  The Department of Energy supports the
Frontier Science Campaign on the \emph{DIII-D} tokamak at General
Atomics in San Diego, dedicating one week of runtime annually to
non-fusion plasma physics experiments.  During the initial year of the
campaign in 2017, experiments observed whistler waves driven by
runaway electrons, relevant to the physics of the Van Allen radiation
belts.\citep{Spong:2018}

Laboratory experiments also contribute to the investigation of other
aspects of space and planetary environments beyond the physics of
plasmas, including the study of the physics of ice under the cryogenic
vacuum conditions of space, \citep{Berisford:2016,Berisford:2017}
investigation the formation of ice in dusty plasma environments,
\citep{Bellan:2015,Marshall:2017} the ablation of micrometeoroids in
Earth's atmosphere, \citep{Thomas:2017} the effect of micrometeoroid
impact on icy surfaces, \citep{Nelson:2016} and the determination of
the surface chemistry of planets and moons, such as
Titan.\citep{Malaska:2017,Vu:2017,Mahjoub:2017}

Viewed collectively as a national resource for the laboratory
investigation of the physics of space plasmas, akin to the collection
of spacecraft missions that comprise the Heliophysics System
Observatory, these facilities have the capability the contribute
uniquely to progress in our understanding of space physics. Increased
coordination and collaboration among these distinct facilities,
through an emerging organized \emph{laboratory space physics}
community, will enable us to maximize the scientific return from our
investments in spacecraft missions and in these facilities and
contribute to transformative progress in our knowledge of the
physics of the heliosphere, as well as of more remote astrophysical
systems.


\subsection{Improved Diagnostic Capabilities}
\label{sec:diag}
One of the primary advantages of laboratory experiments is the ability
to overcome the practical limitation that spacecraft missions can
measure the plasma at only a single point (or a few points) in space.
On the other hand, the measurement of three-dimensional particle
velocity distributions is routinely performed by spacecraft but
remains extremely difficult to accomplish in the laboratory.  Thus,
the development of enhanced diagnostic capabilities for laboratory
experiments will increase their impact on our understanding of the
physics of space and astrophysical plasmas.

One of the key challenges in the laboratory is to sample the plasma
behavior without substantially perturbing its dynamics. In particular,
obtaining three-dimensional particle velocity measurements in
laboratory plasmas is challenging primarily due to the size of
diagnostic instrumentation relative to the plasma kinetic length
scales. In the solar wind, for example, the ion gyroradius is around
100~km, whereas the size of the spacecraft is on the order of
magnitude of meters.  In a laboratory plasma, such as the \emph{Large
  Plasma Device} (\emph{LAPD}), the ion gyroradius is of order one
centimeter, approximately the same size as physical probes inserted
into the plasma. One approach used to overcome this problem is to
exploit the reproducibility of experimental shots in the \emph{LAPD}
plasma, enabling a single small probe to sample the plasma at a
different location during each repeated shot. But a more widely
applicable approach is the develop miniaturized diagnostics for use in
the laboratory.

Retarding potential analyzers have long been used in the laboratory to
sample the velocity distribution of electrons
\citep{Henderson:1939,Young:1959} and ions,
\citep{Conway:1998,Rudakov:1999,Charles:1992,Charles:1993,Charles:2000}
and technological developments are continually improving the
performance and robustness of such diagnostics. \citep{Enloe:2015} At
West Virginia University, researchers are working on the development
of an ultra-compact plasma spectrometer (UCPS)\citep{Scime:2016} for
the measurement of particle velocity distributions in both laboratory
and space plasmas.  Ongoing innovation will improve the ability to
measure the particle velocity distributions, making possible the
illumination of the weakly collisional dynamics and energy transfer
mechanisms that govern the evolution of many space plasma
environments.

Besides the miniaturization of diagnostics, another valuable approach
is the application of advanced non-perturbative techniques,
particularly for the measurement of particle velocity distributions.
For example, laser induced fluorescence (LIF) is a sophisticated means
of making spatially resolved measurements of ion velocity
distributions in the
laboratory.\citep{Skiff:1987,Sarfaty:1996,Scime:1998,Boivin:2003,Keesee:2004,Biloiu:2005,Scime:2005,Biloiu:2006,UzunKaymak:2006,Mattingly:2013,Thompson:2017}
Several studies have compared the performance of retarding field
energy analyzers to that of LIF systems in measuring ion velocity
distributions.\citep{Harvey:2008,Gulbrandsen:2015} Note, however, that
because LIF depends on the existence of suitable atomic emission
lines, it cannot be used to probe the velocities of protons (in
ionized hydrogen plasmas) or electrons.  To circumvent this
limitation, the wave absorption technique is an innovative approach
that has been developed and refined in the laboratory relatively
recently.  \citep{Skiff:1993,Skiff:2006,Thuecks:2012} Whistler wave
absorption, for example, has been used in \emph{LAPD} experiments to
study the physics of electron acceleration by \Alfven waves in the
auroral regions. \citep{Schroeder:2015,Schroeder:2016,Schroeder:2017}
A number of other non-perturbative approaches have also been applied
in the laboratory to determine various aspects of the velocity space
dynamics of plasmas, including Rutherford scattering diagnostics
applied in the \emph{Madison Symmetric Torus}, \cite{Reardon:2001}
charge-exchange recombination spectroscopy,
\citep{Fonck:1984,DenHartog:2006,Podesta:2008,Heidbrink:2008,Heidbrink:2010}
continuous wave cavity ring-down spectroscopy,
\citep{ChakrabortyThakur:2012,McCarren:2015} and nonlinear optical
tagging.\citep{Skiff:1995,Claire:2001}

Improvements in conventional diagnostic techniques and the development
and refinement of innovative ideas for the measurement of laboratory
plasmas will no doubt boost the impact of laboratory experiments on
studies at the frontier of heliophysics and astrophysics.

\subsection{Novel Analysis Methods}
To make the most of the improved measurements of laboratory plasmas by
innovative diagnostics, in particular detailed measurements of ion and
electron velocity distributions, the development and refinement of
novel analysis methods is essential. Most valuable are methods that
can be applied to the analysis of both laboratory experiments and
spacecraft observations, enabling the controlled nature of the
laboratory to be exploited to interpret the dynamics of the
uncontrolled space environment.

Nonlinear kinetic theory dictates that the collisionless interactions
between the electromagnetic fields and charged particles in weakly
collisional heliospheric plasmas necessarily lead to correlations
between the fields and fluctuations in the particle velocity
distributions.  Based on this fundamental insight, a novel
field-particle correlation technique
\citep{Klein:2016a,Howes:2017a,Klein:2017b} has been developed that
employs single-point measurements of the electromagnetic fields and
particle velocity distributions to determine the net energy transfer
between the fields and particles. Furthermore, the technique can be
used to identify which particles in velocity space take part in this
energy transfer, enabling different kinetic physical mechanisms for
particle energization to be distinguished.  Previous applications of
this technique have successfully explored particle energization by
resonant collisionless wave-particle interactions,
\citep{Klein:2016a,Howes:2017a,Howes:2017c} by kinetic instabilities,
\citep{Klein:2017a} by the damping of strong plasma turbulence,
\citep{Klein:2017b} and in current sheets generated self-consistently in
strong \Alfven wave collisions.\citep{Howes:2018}

To understand how energy is cascaded by turbulence through the
six-dimensional phase-space of weakly collisional heliospheric
plasmas, recent studies have employed a Hermite spectral
representation of the structures in velocity
space. \citep{Watanabe:2004,Zocco:2011,Hatch:2013,Loureiro:2013,Hatch:2014,Kanekar:2015b,Parker:2015,Pezzi:2016,Schekochihin:2016,Parker:2016,Groselj:2017,White:2017,Servidio:2017}
Such an optimal spectral representation of the deviations from
equilibrium in the particle velocity distribution functions has lead
to the discovery of an unanticipated process, called
\emph{anti-phase-mixing}, that may inhibit collisionless damping in a
turbulent environment. \cite{Schekochihin:2016,Parker:2016} Such
elegant spectral methods maximize the scientific return from the
detailed measurements of fluctuations in velocity-space that can be
made both in the laboratory and by modern spacecraft instrumentation.

Other innovative methods have been primarily developed for application
to the study of the kinetic plasma physics of the heliosphere, but may
also find use in the analysis of laboratory experiments.  One such
techniques is the determination of the wave vector $\V{k}$, assuming a
single dominant plane-wave mode, using single-point measurements of
the magnetic field and current density fluctuations,
\citep{Bellan:2012,Bellan:2016} a technique recently used to analyze
\emph{MMS} measurements to show the collisionless transfer of energy
between the electric field and plasma particles.\citep{Gershman:2017}
Another novel technique is the application of the Nyquist stability
analysis to solar wind plasma measurements to make possible the
determination of whether kinetic instabilities can tap free energy in
observed non-thermal particle velocity distributions to drive
turbulent fluctuations. \cite{Klein:2017c}

\section{The Future of Laboratory Space Physics}
\label{sec:future}
The convergence and overlap of the plasma length and time scales that
can be accessed by spacecraft missions and by terrestrial laboratory
experiments presents a new, unique opportunity for synergy in the
study of the physics of space and astrophysical plasmas.  By
overcoming the limitations of single-point or few-point measurements
and affording reproducibility and control of conditions, laboratory
experiments represent a powerful complementary approach to explore
space plasma physics, here denoted by the term \emph{laboratory space
  physics}.  One particularly promising application is to determine
the kinetic mechanisms responsible for particle energization and
plasma heating in space plasmas.  Being able to sample the
fluctuations in velocity distributions at cadences associated with the
time scales of kinetic dynamics and energy transfer processes in space
plasmas provides a new opportunity to answer key questions
definitively.  Laboratory platforms enable a sufficient spatial
sampling of the plasma, over a controllable range of parameters, that
will help to facilitate the development of a predictive capability of
the heliospheric plasma evolution, the ultimate goal of heliophysics.
With space measurements alone, due to their inherently uncontrolled
conditions, it would be very difficult to develop predictive theories
and test them thoroughly over a range of parameters.

In the effort to employ laboratory studies to examine the mechanisms
at play in the heliosphere, it is critical to initiate active
collaborations between laboratory plasma physicists and space
physicists. The effort to make direct contact between laboratory
measurements and spacecraft observations, even when an appropriately
scaled experiment has been devised, is decidedly non-trivial.  Often,
the work needed to connect laboratory measurements to spacecraft
observations reveals unforeseen issues, the resolution of which
ultimately leads to a more complete understanding of the underlying
fundamental plasma physics, enriching both the observational and
experimental perspectives on the problem at hand.  Numerical
simulations, which have intentionally not been reviewed here, can
provide a critical bridge between spacecraft observations and
experimental measurements and can establish a direct connection to
idealized theoretical models.
 
How can laboratory space physics experiments make the biggest possible
impact on the frontier of space physics and astrophysics? The
laboratory space physics community can maximize their scientific
contribution by strategically aligning their efforts to current and
upcoming spacecraft missions.  The current \emph{Magnetospheric
  Multiscale} (\emph{MMS}) mission\citep{Burch:2016a} focuses
primarily on exploring the physics of magnetic reconnection and aligns
well with the recently completed \emph{Facility for Laboratory
  Reconnection Experiments} (\emph{FLARE}) at Princeton
University.\citep{Ji:2011} Understanding turbulent heating and
particle acceleration are key science targets of the upcoming
\emph{Parker Solar Probe}\citep{Fox:2016,Bale:2016,Kasper:2016} and
\emph{Solar Orbiter} \citep{Muller:2013} missions as well as the
proposed \emph{Turbulent Heating ObserveR} (\emph{THOR})
\citep{Vaivads:2016} and \emph{Interstellar Mapping and Acceleration
  Probe} (\emph{IMAP})\citep{NRCspace:2013} missions. Designing
well-diagnosed laboratory experiments to explore plasma turbulence and
particle acceleration will complement the single-point measurements
returned by each of these four spacecraft missions. In addition,  the
development and refinement of novel analysis methods that make full use of
particle velocity distribution data, a key goal advocated by this
review, will enable us to maximize the scientific return from the
costly investment in these strategic missions.

The physics of multi-ion plasmas and ion-neutral coupling, critical to
improving our knowledge of magnetosphere-ionosphere-thermosphere
(MIT) coupling, is a topic that has scarcely been investigated using
laboratory experiments in the past.  Presently, there is a significant
need for synergistic laboratory experiments that can complement a host
of upcoming missions aimed at developing a better understanding of the
dynamics and feedback of the MIT system.  The poorly understood
processes controlling how the Earth responds to variable forcing from
the Sun will be addressed by several new spacecraft missions, including
the recently launched \emph{Global-Scale Observations of the Limb and
  Disk} (\emph{GOLD}) mission\citep{Eastes:2017} and upcoming
\emph{Ionospheric Connection Explorer} (\emph{ICON})
mission,\citep{Immel:2018} as well as the proposed \emph{Magnetosphere
  Energetics, Dynamics, and Ionospheric Coupling Investigation}
(\emph{MEDICI}), \emph{Dynamical Neutral Atmosphere-Ionosphere
  Coupling} (\emph{DYNAMIC}), and \emph{Geospace Dynamics Coupling}
(\emph{GDC}) missions. \cite{NRCspace:2013} Timely experiments
exploring multi-ion effects in plasmas and the impacts of ion-neutral
coupling can make valuable contributions to our understanding of the
MIT system.

Ultimately, laboratory space physics experiments present a cost
effective means to complement spacecraft observations in our quest to
better understand the physics of space and astrophysical
plasmas. Collectively, the user-class experimental facilities and
moderate-scale experimental platforms at national laboratories and
universities around the country provide a powerful complement to
spacecraft missions to discover the fundamental plasma physics
mechanisms that govern the evolution of the heliosphere.  Improved
diagnostics and innovative analysis techniques, in particular those
that make full use of the measured dynamics of the particle velocity
distributions, are likely to spur a significant leap in our study of
the physics of space plasmas. Ultimately, the convergence of the
regimes accessible to spacecraft measurements and laboratory
experiments makes the pursuit of laboratory space physics particularly
timely, with the potential to make a transformative contribution to
the study of fundamental plasma mechanisms in the heliosphere.  All
signs indicate a bright future for laboratory space physics.

\begin{acknowledgments}
  I thank all of the participants of the Bringing Space Down to Earth
  Workshop, held at UCLA in April 2017, for their contributions on the
  topic of laboratory space physics. The workshop was supported by the
  Basic Plasma Science Facility at UCLA through DOE award
  DE‐FC02‐07ER54918 and NSF award 1561912. This work was supported by
  DOE grant DE-SC0014599.
\end{acknowledgments}

%


\begin{thebibliography}{489}%
\makeatletter
\providecommand \@ifxundefined [1]{%
 \@ifx{#1\undefined}
}%
\providecommand \@ifnum [1]{%
 \ifnum #1\expandafter \@firstoftwo
 \else \expandafter \@secondoftwo
 \fi
}%
\providecommand \@ifx [1]{%
 \ifx #1\expandafter \@firstoftwo
 \else \expandafter \@secondoftwo
 \fi
}%
\providecommand \natexlab [1]{#1}%
\providecommand \enquote  [1]{``#1''}%
\providecommand \bibnamefont  [1]{#1}%
\providecommand \bibfnamefont [1]{#1}%
\providecommand \citenamefont [1]{#1}%
\providecommand \href@noop [0]{\@secondoftwo}%
\providecommand \href [0]{\begingroup \@sanitize@url \@href}%
\providecommand \@href[1]{\@@startlink{#1}\@@href}%
\providecommand \@@href[1]{\endgroup#1\@@endlink}%
\providecommand \@sanitize@url [0]{\catcode `\\12\catcode `\$12\catcode
  `\&12\catcode `\#12\catcode `\^12\catcode `\_12\catcode `\%12\relax}%
\providecommand \@@startlink[1]{}%
\providecommand \@@endlink[0]{}%
\providecommand \url  [0]{\begingroup\@sanitize@url \@url }%
\providecommand \@url [1]{\endgroup\@href {#1}{\urlprefix }}%
\providecommand \urlprefix  [0]{URL }%
\providecommand \Eprint [0]{\href }%
\providecommand \doibase [0]{http://dx.doi.org/}%
\providecommand \selectlanguage [0]{\@gobble}%
\providecommand \bibinfo  [0]{\@secondoftwo}%
\providecommand \bibfield  [0]{\@secondoftwo}%
\providecommand \translation [1]{[#1]}%
\providecommand \BibitemOpen [0]{}%
\providecommand \bibitemStop [0]{}%
\providecommand \bibitemNoStop [0]{.\EOS\space}%
\providecommand \EOS [0]{\spacefactor3000\relax}%
\providecommand \BibitemShut  [1]{\csname bibitem#1\endcsname}%
\let\auto@bib@innerbib\@empty
\bibitem [{\citenamefont {{van Allen}}\ and\ \citenamefont
  {{Frank}}(1959)}]{VanAllen:1959}%
  \BibitemOpen
  \bibfield  {author} {\bibinfo {author} {\bibfnamefont {J.~A.}\ \bibnamefont
  {{van Allen}}}\ and\ \bibinfo {author} {\bibfnamefont {L.~A.}\ \bibnamefont
  {{Frank}}},\ }\bibfield  {title} {\enquote {\bibinfo {title} {{Radiation
  Around the Earth to a Radial Distance of 107,400 km.}}}\ }\href {\doibase
  10.1038/183430a0} {\bibfield  {journal} {\bibinfo  {journal} {Nature}\
  }\textbf {\bibinfo {volume} {183}},\ \bibinfo {pages} {430--434} (\bibinfo
  {year} {1959})}\BibitemShut {NoStop}%
\bibitem [{\citenamefont {{McComas}}\ \emph
  {et~al.}(2009{\natexlab{a}})\citenamefont {{McComas}}, \citenamefont
  {{Allegrini}}, \citenamefont {{Bochsler}}, \citenamefont {{Bzowski}},
  \citenamefont {{Collier}}, \citenamefont {{Fahr}}, \citenamefont
  {{Fichtner}}, \citenamefont {{Frisch}}, \citenamefont {{Funsten}},
  \citenamefont {{Fuselier}}, \citenamefont {{Gloeckler}}, \citenamefont
  {{Gruntman}}, \citenamefont {{Izmodenov}}, \citenamefont {{Knappenberger}},
  \citenamefont {{Lee}}, \citenamefont {{Livi}}, \citenamefont {{Mitchell}},
  \citenamefont {{M{\"o}bius}}, \citenamefont {{Moore}}, \citenamefont
  {{Pope}}, \citenamefont {{Reisenfeld}}, \citenamefont {{Roelof}},
  \citenamefont {{Scherrer}}, \citenamefont {{Schwadron}}, \citenamefont
  {{Tyler}}, \citenamefont {{Wieser}}, \citenamefont {{Witte}}, \citenamefont
  {{Wurz}},\ and\ \citenamefont {{Zank}}}]{McComas:2009a}%
  \BibitemOpen
  \bibfield  {author} {\bibinfo {author} {\bibfnamefont {D.~J.}\ \bibnamefont
  {{McComas}}}, \bibinfo {author} {\bibfnamefont {F.}~\bibnamefont
  {{Allegrini}}}, \bibinfo {author} {\bibfnamefont {P.}~\bibnamefont
  {{Bochsler}}}, \bibinfo {author} {\bibfnamefont {M.}~\bibnamefont
  {{Bzowski}}}, \bibinfo {author} {\bibfnamefont {M.}~\bibnamefont
  {{Collier}}}, \bibinfo {author} {\bibfnamefont {H.}~\bibnamefont {{Fahr}}},
  \bibinfo {author} {\bibfnamefont {H.}~\bibnamefont {{Fichtner}}}, \bibinfo
  {author} {\bibfnamefont {P.}~\bibnamefont {{Frisch}}}, \bibinfo {author}
  {\bibfnamefont {H.~O.}\ \bibnamefont {{Funsten}}}, \bibinfo {author}
  {\bibfnamefont {S.~A.}\ \bibnamefont {{Fuselier}}}, \bibinfo {author}
  {\bibfnamefont {G.}~\bibnamefont {{Gloeckler}}}, \bibinfo {author}
  {\bibfnamefont {M.}~\bibnamefont {{Gruntman}}}, \bibinfo {author}
  {\bibfnamefont {V.}~\bibnamefont {{Izmodenov}}}, \bibinfo {author}
  {\bibfnamefont {P.}~\bibnamefont {{Knappenberger}}}, \bibinfo {author}
  {\bibfnamefont {M.}~\bibnamefont {{Lee}}}, \bibinfo {author} {\bibfnamefont
  {S.}~\bibnamefont {{Livi}}}, \bibinfo {author} {\bibfnamefont
  {D.}~\bibnamefont {{Mitchell}}}, \bibinfo {author} {\bibfnamefont
  {E.}~\bibnamefont {{M{\"o}bius}}}, \bibinfo {author} {\bibfnamefont
  {T.}~\bibnamefont {{Moore}}}, \bibinfo {author} {\bibfnamefont
  {S.}~\bibnamefont {{Pope}}}, \bibinfo {author} {\bibfnamefont
  {D.}~\bibnamefont {{Reisenfeld}}}, \bibinfo {author} {\bibfnamefont
  {E.}~\bibnamefont {{Roelof}}}, \bibinfo {author} {\bibfnamefont
  {J.}~\bibnamefont {{Scherrer}}}, \bibinfo {author} {\bibfnamefont
  {N.}~\bibnamefont {{Schwadron}}}, \bibinfo {author} {\bibfnamefont
  {R.}~\bibnamefont {{Tyler}}}, \bibinfo {author} {\bibfnamefont
  {M.}~\bibnamefont {{Wieser}}}, \bibinfo {author} {\bibfnamefont
  {M.}~\bibnamefont {{Witte}}}, \bibinfo {author} {\bibfnamefont
  {P.}~\bibnamefont {{Wurz}}}, \ and\ \bibinfo {author} {\bibfnamefont
  {G.}~\bibnamefont {{Zank}}},\ }\bibfield  {title} {\enquote {\bibinfo {title}
  {{IBEX---Interstellar Boundary Explorer}},}\ }\href {\doibase
  10.1007/s11214-009-9499-4} {\bibfield  {journal} {\bibinfo  {journal} {Space
  Sci.~Rev.}\ }\textbf {\bibinfo {volume} {146}},\ \bibinfo {pages} {11--33}
  (\bibinfo {year} {2009}{\natexlab{a}})}\BibitemShut {NoStop}%
\bibitem [{\citenamefont {{McComas}}\ \emph
  {et~al.}(2009{\natexlab{b}})\citenamefont {{McComas}}, \citenamefont
  {{Allegrini}}, \citenamefont {{Bochsler}}, \citenamefont {{Bzowski}},
  \citenamefont {{Christian}}, \citenamefont {{Crew}}, \citenamefont
  {{DeMajistre}}, \citenamefont {{Fahr}}, \citenamefont {{Fichtner}},
  \citenamefont {{Frisch}}, \citenamefont {{Funsten}}, \citenamefont
  {{Fuselier}}, \citenamefont {{Gloeckler}}, \citenamefont {{Gruntman}},
  \citenamefont {{Heerikhuisen}}, \citenamefont {{Izmodenov}}, \citenamefont
  {{Janzen}}, \citenamefont {{Knappenberger}}, \citenamefont {{Krimigis}},
  \citenamefont {{Kucharek}}, \citenamefont {{Lee}}, \citenamefont
  {{Livadiotis}}, \citenamefont {{Livi}}, \citenamefont {{MacDowall}},
  \citenamefont {{Mitchell}}, \citenamefont {{M{\"o}bius}}, \citenamefont
  {{Moore}}, \citenamefont {{Pogorelov}}, \citenamefont {{Reisenfeld}},
  \citenamefont {{Roelof}}, \citenamefont {{Saul}}, \citenamefont
  {{Schwadron}}, \citenamefont {{Valek}}, \citenamefont {{Vanderspek}},
  \citenamefont {{Wurz}},\ and\ \citenamefont {{Zank}}}]{McComas:2009b}%
  \BibitemOpen
  \bibfield  {author} {\bibinfo {author} {\bibfnamefont {D.~J.}\ \bibnamefont
  {{McComas}}}, \bibinfo {author} {\bibfnamefont {F.}~\bibnamefont
  {{Allegrini}}}, \bibinfo {author} {\bibfnamefont {P.}~\bibnamefont
  {{Bochsler}}}, \bibinfo {author} {\bibfnamefont {M.}~\bibnamefont
  {{Bzowski}}}, \bibinfo {author} {\bibfnamefont {E.~R.}\ \bibnamefont
  {{Christian}}}, \bibinfo {author} {\bibfnamefont {G.~B.}\ \bibnamefont
  {{Crew}}}, \bibinfo {author} {\bibfnamefont {R.}~\bibnamefont
  {{DeMajistre}}}, \bibinfo {author} {\bibfnamefont {H.}~\bibnamefont
  {{Fahr}}}, \bibinfo {author} {\bibfnamefont {H.}~\bibnamefont {{Fichtner}}},
  \bibinfo {author} {\bibfnamefont {P.~C.}\ \bibnamefont {{Frisch}}}, \bibinfo
  {author} {\bibfnamefont {H.~O.}\ \bibnamefont {{Funsten}}}, \bibinfo {author}
  {\bibfnamefont {S.~A.}\ \bibnamefont {{Fuselier}}}, \bibinfo {author}
  {\bibfnamefont {G.}~\bibnamefont {{Gloeckler}}}, \bibinfo {author}
  {\bibfnamefont {M.}~\bibnamefont {{Gruntman}}}, \bibinfo {author}
  {\bibfnamefont {J.}~\bibnamefont {{Heerikhuisen}}}, \bibinfo {author}
  {\bibfnamefont {V.}~\bibnamefont {{Izmodenov}}}, \bibinfo {author}
  {\bibfnamefont {P.}~\bibnamefont {{Janzen}}}, \bibinfo {author}
  {\bibfnamefont {P.}~\bibnamefont {{Knappenberger}}}, \bibinfo {author}
  {\bibfnamefont {S.}~\bibnamefont {{Krimigis}}}, \bibinfo {author}
  {\bibfnamefont {H.}~\bibnamefont {{Kucharek}}}, \bibinfo {author}
  {\bibfnamefont {M.}~\bibnamefont {{Lee}}}, \bibinfo {author} {\bibfnamefont
  {G.}~\bibnamefont {{Livadiotis}}}, \bibinfo {author} {\bibfnamefont
  {S.}~\bibnamefont {{Livi}}}, \bibinfo {author} {\bibfnamefont {R.~J.}\
  \bibnamefont {{MacDowall}}}, \bibinfo {author} {\bibfnamefont
  {D.}~\bibnamefont {{Mitchell}}}, \bibinfo {author} {\bibfnamefont
  {E.}~\bibnamefont {{M{\"o}bius}}}, \bibinfo {author} {\bibfnamefont
  {T.}~\bibnamefont {{Moore}}}, \bibinfo {author} {\bibfnamefont {N.~V.}\
  \bibnamefont {{Pogorelov}}}, \bibinfo {author} {\bibfnamefont
  {D.}~\bibnamefont {{Reisenfeld}}}, \bibinfo {author} {\bibfnamefont
  {E.}~\bibnamefont {{Roelof}}}, \bibinfo {author} {\bibfnamefont
  {L.}~\bibnamefont {{Saul}}}, \bibinfo {author} {\bibfnamefont {N.~A.}\
  \bibnamefont {{Schwadron}}}, \bibinfo {author} {\bibfnamefont {P.~W.}\
  \bibnamefont {{Valek}}}, \bibinfo {author} {\bibfnamefont {R.}~\bibnamefont
  {{Vanderspek}}}, \bibinfo {author} {\bibfnamefont {P.}~\bibnamefont
  {{Wurz}}}, \ and\ \bibinfo {author} {\bibfnamefont {G.~P.}\ \bibnamefont
  {{Zank}}},\ }\bibfield  {title} {\enquote {\bibinfo {title} {{Global
  Observations of the Interstellar Interaction from the Interstellar Boundary
  Explorer (IBEX)}},}\ }\href {\doibase 10.1126/science.1180906} {\bibfield
  {journal} {\bibinfo  {journal} {Science}\ }\textbf {\bibinfo {volume}
  {326}},\ \bibinfo {pages} {959} (\bibinfo {year}
  {2009}{\natexlab{b}})}\BibitemShut {NoStop}%
\bibitem [{\citenamefont {Gurnett}\ \emph {et~al.}(2013)\citenamefont
  {Gurnett}, \citenamefont {Kurth}, \citenamefont {Burlaga},\ and\
  \citenamefont {Ness}}]{Gurnett:2013}%
  \BibitemOpen
  \bibfield  {author} {\bibinfo {author} {\bibfnamefont {D.~A.}\ \bibnamefont
  {Gurnett}}, \bibinfo {author} {\bibfnamefont {W.~S.}\ \bibnamefont {Kurth}},
  \bibinfo {author} {\bibfnamefont {L.~F.}\ \bibnamefont {Burlaga}}, \ and\
  \bibinfo {author} {\bibfnamefont {N.~F.}\ \bibnamefont {Ness}},\ }\bibfield
  {title} {\enquote {\bibinfo {title} {In situ observations of interstellar
  plasma with voyager 1},}\ }\href {\doibase 10.1126/science.1241681}
  {\bibfield  {journal} {\bibinfo  {journal} {Science}\ }\textbf {\bibinfo
  {volume} {341}},\ \bibinfo {pages} {1489} (\bibinfo {year}
  {2013})}\BibitemShut {NoStop}%
\bibitem [{\citenamefont {on~a Decadal Strategy~for Solar}\ and\ \citenamefont
  {(Heliophysics)}(2013)}]{NRCspace:2013}%
  \BibitemOpen
  \bibfield  {author} {\bibinfo {author} {\bibfnamefont {C.}~\bibnamefont {on~a
  Decadal Strategy~for Solar}}\ and\ \bibinfo {author} {\bibfnamefont {S.~P.}\
  \bibnamefont {(Heliophysics)}},\ }\href
  {http://www.nap.edu/openbook.php?record_id=13060} {\emph {\bibinfo {title}
  {Solar and Space Physics: A Science for a Technological Society}}}\ (\bibinfo
   {publisher} {The National Academies Press},\ \bibinfo {year}
  {2013})\BibitemShut {NoStop}%
\bibitem [{Note1()}]{Note1}%
  \BibitemOpen
  \bibinfo {note} {As of summer 2017, the NASA missions comprising the
  Heliophysics System Observatory (HSO) include the single-spacecraft missions
  ACE, AIM, Geotail, GOLD, Hinode, IBEX, IRIS, RHESSI, SDO, SOHO, TIMED, and
  Wind, as well as the multiple-spacecraft missions ARTEMIS (2), Cluster (4),
  MMS (4) STEREO (2), THEMIS(3), TWINS (2), Van Allen Probes (2), and Voyager
  (2).}\BibitemShut {Stop}%
\bibitem [{\citenamefont {{Burch}}\ \emph {et~al.}(2016)\citenamefont
  {{Burch}}, \citenamefont {{Moore}}, \citenamefont {{Torbert}},\ and\
  \citenamefont {{Giles}}}]{Burch:2016a}%
  \BibitemOpen
  \bibfield  {author} {\bibinfo {author} {\bibfnamefont {J.~L.}\ \bibnamefont
  {{Burch}}}, \bibinfo {author} {\bibfnamefont {T.~E.}\ \bibnamefont
  {{Moore}}}, \bibinfo {author} {\bibfnamefont {R.~B.}\ \bibnamefont
  {{Torbert}}}, \ and\ \bibinfo {author} {\bibfnamefont {B.~L.}\ \bibnamefont
  {{Giles}}},\ }\bibfield  {title} {\enquote {\bibinfo {title} {{Magnetospheric
  Multiscale Overview and Science Objectives}},}\ }\href {\doibase
  10.1007/s11214-015-0164-9} {\bibfield  {journal} {\bibinfo  {journal} {Space
  Sci.~Rev.}\ }\textbf {\bibinfo {volume} {199}},\ \bibinfo {pages} {5--21}
  (\bibinfo {year} {2016})}\BibitemShut {NoStop}%
\bibitem [{\citenamefont {{Pollock}}\ \emph {et~al.}(2016)\citenamefont
  {{Pollock}}, \citenamefont {{Moore}}, \citenamefont {{Jacques}},
  \citenamefont {{Burch}}, \citenamefont {{Gliese}}, \citenamefont {{Saito}},
  \citenamefont {{Omoto}}, \citenamefont {{Avanov}}, \citenamefont {{Barrie}},
  \citenamefont {{Coffey}}, \citenamefont {{Dorelli}}, \citenamefont
  {{Gershman}}, \citenamefont {{Giles}}, \citenamefont {{Rosnack}},
  \citenamefont {{Salo}}, \citenamefont {{Yokota}}, \citenamefont {{Adrian}},
  \citenamefont {{Aoustin}}, \citenamefont {{Auletti}}, \citenamefont {{Aung}},
  \citenamefont {{Bigio}}, \citenamefont {{Cao}}, \citenamefont {{Chandler}},
  \citenamefont {{Chornay}}, \citenamefont {{Christian}}, \citenamefont
  {{Clark}}, \citenamefont {{Collinson}}, \citenamefont {{Corris}},
  \citenamefont {{De Los Santos}}, \citenamefont {{Devlin}}, \citenamefont
  {{Diaz}}, \citenamefont {{Dickerson}}, \citenamefont {{Dickson}},
  \citenamefont {{Diekmann}}, \citenamefont {{Diggs}}, \citenamefont
  {{Duncan}}, \citenamefont {{Figueroa-Vinas}}, \citenamefont {{Firman}},
  \citenamefont {{Freeman}}, \citenamefont {{Galassi}}, \citenamefont
  {{Garcia}}, \citenamefont {{Goodhart}}, \citenamefont {{Guererro}},
  \citenamefont {{Hageman}}, \citenamefont {{Hanley}}, \citenamefont
  {{Hemminger}}, \citenamefont {{Holland}}, \citenamefont {{Hutchins}},
  \citenamefont {{James}}, \citenamefont {{Jones}}, \citenamefont {{Kreisler}},
  \citenamefont {{Kujawski}}, \citenamefont {{Lavu}}, \citenamefont {{Lobell}},
  \citenamefont {{LeCompte}}, \citenamefont {{Lukemire}}, \citenamefont
  {{MacDonald}}, \citenamefont {{Mariano}}, \citenamefont {{Mukai}},
  \citenamefont {{Narayanan}}, \citenamefont {{Nguyan}}, \citenamefont
  {{Onizuka}}, \citenamefont {{Paterson}}, \citenamefont {{Persyn}},
  \citenamefont {{Piepgrass}}, \citenamefont {{Cheney}}, \citenamefont
  {{Rager}}, \citenamefont {{Raghuram}}, \citenamefont {{Ramil}}, \citenamefont
  {{Reichenthal}}, \citenamefont {{Rodriguez}}, \citenamefont {{Rouzaud}},
  \citenamefont {{Rucker}}, \citenamefont {{Saito}}, \citenamefont {{Samara}},
  \citenamefont {{Sauvaud}}, \citenamefont {{Schuster}}, \citenamefont
  {{Shappirio}}, \citenamefont {{Shelton}}, \citenamefont {{Sher}},
  \citenamefont {{Smith}}, \citenamefont {{Smith}}, \citenamefont {{Smith}},
  \citenamefont {{Steinfeld}}, \citenamefont {{Szymkiewicz}}, \citenamefont
  {{Tanimoto}}, \citenamefont {{Taylor}}, \citenamefont {{Tucker}},
  \citenamefont {{Tull}}, \citenamefont {{Uhl}}, \citenamefont {{Vloet}},
  \citenamefont {{Walpole}}, \citenamefont {{Weidner}}, \citenamefont
  {{White}}, \citenamefont {{Winkert}}, \citenamefont {{Yeh}},\ and\
  \citenamefont {{Zeuch}}}]{Pollock:2016}%
  \BibitemOpen
  \bibfield  {author} {\bibinfo {author} {\bibfnamefont {C.}~\bibnamefont
  {{Pollock}}}, \bibinfo {author} {\bibfnamefont {T.}~\bibnamefont {{Moore}}},
  \bibinfo {author} {\bibfnamefont {A.}~\bibnamefont {{Jacques}}}, \bibinfo
  {author} {\bibfnamefont {J.}~\bibnamefont {{Burch}}}, \bibinfo {author}
  {\bibfnamefont {U.}~\bibnamefont {{Gliese}}}, \bibinfo {author}
  {\bibfnamefont {Y.}~\bibnamefont {{Saito}}}, \bibinfo {author} {\bibfnamefont
  {T.}~\bibnamefont {{Omoto}}}, \bibinfo {author} {\bibfnamefont
  {L.}~\bibnamefont {{Avanov}}}, \bibinfo {author} {\bibfnamefont
  {A.}~\bibnamefont {{Barrie}}}, \bibinfo {author} {\bibfnamefont
  {V.}~\bibnamefont {{Coffey}}}, \bibinfo {author} {\bibfnamefont
  {J.}~\bibnamefont {{Dorelli}}}, \bibinfo {author} {\bibfnamefont
  {D.}~\bibnamefont {{Gershman}}}, \bibinfo {author} {\bibfnamefont
  {B.}~\bibnamefont {{Giles}}}, \bibinfo {author} {\bibfnamefont
  {T.}~\bibnamefont {{Rosnack}}}, \bibinfo {author} {\bibfnamefont
  {C.}~\bibnamefont {{Salo}}}, \bibinfo {author} {\bibfnamefont
  {S.}~\bibnamefont {{Yokota}}}, \bibinfo {author} {\bibfnamefont
  {M.}~\bibnamefont {{Adrian}}}, \bibinfo {author} {\bibfnamefont
  {C.}~\bibnamefont {{Aoustin}}}, \bibinfo {author} {\bibfnamefont
  {C.}~\bibnamefont {{Auletti}}}, \bibinfo {author} {\bibfnamefont
  {S.}~\bibnamefont {{Aung}}}, \bibinfo {author} {\bibfnamefont
  {V.}~\bibnamefont {{Bigio}}}, \bibinfo {author} {\bibfnamefont
  {N.}~\bibnamefont {{Cao}}}, \bibinfo {author} {\bibfnamefont
  {M.}~\bibnamefont {{Chandler}}}, \bibinfo {author} {\bibfnamefont
  {D.}~\bibnamefont {{Chornay}}}, \bibinfo {author} {\bibfnamefont
  {K.}~\bibnamefont {{Christian}}}, \bibinfo {author} {\bibfnamefont
  {G.}~\bibnamefont {{Clark}}}, \bibinfo {author} {\bibfnamefont
  {G.}~\bibnamefont {{Collinson}}}, \bibinfo {author} {\bibfnamefont
  {T.}~\bibnamefont {{Corris}}}, \bibinfo {author} {\bibfnamefont
  {A.}~\bibnamefont {{De Los Santos}}}, \bibinfo {author} {\bibfnamefont
  {R.}~\bibnamefont {{Devlin}}}, \bibinfo {author} {\bibfnamefont
  {T.}~\bibnamefont {{Diaz}}}, \bibinfo {author} {\bibfnamefont
  {T.}~\bibnamefont {{Dickerson}}}, \bibinfo {author} {\bibfnamefont
  {C.}~\bibnamefont {{Dickson}}}, \bibinfo {author} {\bibfnamefont
  {A.}~\bibnamefont {{Diekmann}}}, \bibinfo {author} {\bibfnamefont
  {F.}~\bibnamefont {{Diggs}}}, \bibinfo {author} {\bibfnamefont
  {C.}~\bibnamefont {{Duncan}}}, \bibinfo {author} {\bibfnamefont
  {A.}~\bibnamefont {{Figueroa-Vinas}}}, \bibinfo {author} {\bibfnamefont
  {C.}~\bibnamefont {{Firman}}}, \bibinfo {author} {\bibfnamefont
  {M.}~\bibnamefont {{Freeman}}}, \bibinfo {author} {\bibfnamefont
  {N.}~\bibnamefont {{Galassi}}}, \bibinfo {author} {\bibfnamefont
  {K.}~\bibnamefont {{Garcia}}}, \bibinfo {author} {\bibfnamefont
  {G.}~\bibnamefont {{Goodhart}}}, \bibinfo {author} {\bibfnamefont
  {D.}~\bibnamefont {{Guererro}}}, \bibinfo {author} {\bibfnamefont
  {J.}~\bibnamefont {{Hageman}}}, \bibinfo {author} {\bibfnamefont
  {J.}~\bibnamefont {{Hanley}}}, \bibinfo {author} {\bibfnamefont
  {E.}~\bibnamefont {{Hemminger}}}, \bibinfo {author} {\bibfnamefont
  {M.}~\bibnamefont {{Holland}}}, \bibinfo {author} {\bibfnamefont
  {M.}~\bibnamefont {{Hutchins}}}, \bibinfo {author} {\bibfnamefont
  {T.}~\bibnamefont {{James}}}, \bibinfo {author} {\bibfnamefont
  {W.}~\bibnamefont {{Jones}}}, \bibinfo {author} {\bibfnamefont
  {S.}~\bibnamefont {{Kreisler}}}, \bibinfo {author} {\bibfnamefont
  {J.}~\bibnamefont {{Kujawski}}}, \bibinfo {author} {\bibfnamefont
  {V.}~\bibnamefont {{Lavu}}}, \bibinfo {author} {\bibfnamefont
  {J.}~\bibnamefont {{Lobell}}}, \bibinfo {author} {\bibfnamefont
  {E.}~\bibnamefont {{LeCompte}}}, \bibinfo {author} {\bibfnamefont
  {A.}~\bibnamefont {{Lukemire}}}, \bibinfo {author} {\bibfnamefont
  {E.}~\bibnamefont {{MacDonald}}}, \bibinfo {author} {\bibfnamefont
  {A.}~\bibnamefont {{Mariano}}}, \bibinfo {author} {\bibfnamefont
  {T.}~\bibnamefont {{Mukai}}}, \bibinfo {author} {\bibfnamefont
  {K.}~\bibnamefont {{Narayanan}}}, \bibinfo {author} {\bibfnamefont
  {Q.}~\bibnamefont {{Nguyan}}}, \bibinfo {author} {\bibfnamefont
  {M.}~\bibnamefont {{Onizuka}}}, \bibinfo {author} {\bibfnamefont
  {W.}~\bibnamefont {{Paterson}}}, \bibinfo {author} {\bibfnamefont
  {S.}~\bibnamefont {{Persyn}}}, \bibinfo {author} {\bibfnamefont
  {B.}~\bibnamefont {{Piepgrass}}}, \bibinfo {author} {\bibfnamefont
  {F.}~\bibnamefont {{Cheney}}}, \bibinfo {author} {\bibfnamefont
  {A.}~\bibnamefont {{Rager}}}, \bibinfo {author} {\bibfnamefont
  {T.}~\bibnamefont {{Raghuram}}}, \bibinfo {author} {\bibfnamefont
  {A.}~\bibnamefont {{Ramil}}}, \bibinfo {author} {\bibfnamefont
  {L.}~\bibnamefont {{Reichenthal}}}, \bibinfo {author} {\bibfnamefont
  {H.}~\bibnamefont {{Rodriguez}}}, \bibinfo {author} {\bibfnamefont
  {J.}~\bibnamefont {{Rouzaud}}}, \bibinfo {author} {\bibfnamefont
  {A.}~\bibnamefont {{Rucker}}}, \bibinfo {author} {\bibfnamefont
  {Y.}~\bibnamefont {{Saito}}}, \bibinfo {author} {\bibfnamefont
  {M.}~\bibnamefont {{Samara}}}, \bibinfo {author} {\bibfnamefont {J.-A.}\
  \bibnamefont {{Sauvaud}}}, \bibinfo {author} {\bibfnamefont {D.}~\bibnamefont
  {{Schuster}}}, \bibinfo {author} {\bibfnamefont {M.}~\bibnamefont
  {{Shappirio}}}, \bibinfo {author} {\bibfnamefont {K.}~\bibnamefont
  {{Shelton}}}, \bibinfo {author} {\bibfnamefont {D.}~\bibnamefont {{Sher}}},
  \bibinfo {author} {\bibfnamefont {D.}~\bibnamefont {{Smith}}}, \bibinfo
  {author} {\bibfnamefont {K.}~\bibnamefont {{Smith}}}, \bibinfo {author}
  {\bibfnamefont {S.}~\bibnamefont {{Smith}}}, \bibinfo {author} {\bibfnamefont
  {D.}~\bibnamefont {{Steinfeld}}}, \bibinfo {author} {\bibfnamefont
  {R.}~\bibnamefont {{Szymkiewicz}}}, \bibinfo {author} {\bibfnamefont
  {K.}~\bibnamefont {{Tanimoto}}}, \bibinfo {author} {\bibfnamefont
  {J.}~\bibnamefont {{Taylor}}}, \bibinfo {author} {\bibfnamefont
  {C.}~\bibnamefont {{Tucker}}}, \bibinfo {author} {\bibfnamefont
  {K.}~\bibnamefont {{Tull}}}, \bibinfo {author} {\bibfnamefont
  {A.}~\bibnamefont {{Uhl}}}, \bibinfo {author} {\bibfnamefont
  {J.}~\bibnamefont {{Vloet}}}, \bibinfo {author} {\bibfnamefont
  {P.}~\bibnamefont {{Walpole}}}, \bibinfo {author} {\bibfnamefont
  {S.}~\bibnamefont {{Weidner}}}, \bibinfo {author} {\bibfnamefont
  {D.}~\bibnamefont {{White}}}, \bibinfo {author} {\bibfnamefont
  {G.}~\bibnamefont {{Winkert}}}, \bibinfo {author} {\bibfnamefont {P.-S.}\
  \bibnamefont {{Yeh}}}, \ and\ \bibinfo {author} {\bibfnamefont
  {M.}~\bibnamefont {{Zeuch}}},\ }\bibfield  {title} {\enquote {\bibinfo
  {title} {{Fast Plasma Investigation for Magnetospheric Multiscale}},}\ }\href
  {\doibase 10.1007/s11214-016-0245-4} {\bibfield  {journal} {\bibinfo
  {journal} {Space Sci.~Rev.}\ }\textbf {\bibinfo {volume} {199}},\ \bibinfo
  {pages} {331--406} (\bibinfo {year} {2016})}\BibitemShut {NoStop}%
\bibitem [{\citenamefont {{Gekelman}}\ \emph {et~al.}(1991)\citenamefont
  {{Gekelman}}, \citenamefont {{Pfister}}, \citenamefont {{Lucky}},
  \citenamefont {{Bamber}}, \citenamefont {{Leneman}},\ and\ \citenamefont
  {{Maggs}}}]{Gekelman:1991}%
  \BibitemOpen
  \bibfield  {author} {\bibinfo {author} {\bibfnamefont {W.}~\bibnamefont
  {{Gekelman}}}, \bibinfo {author} {\bibfnamefont {H.}~\bibnamefont
  {{Pfister}}}, \bibinfo {author} {\bibfnamefont {Z.}~\bibnamefont {{Lucky}}},
  \bibinfo {author} {\bibfnamefont {J.}~\bibnamefont {{Bamber}}}, \bibinfo
  {author} {\bibfnamefont {D.}~\bibnamefont {{Leneman}}}, \ and\ \bibinfo
  {author} {\bibfnamefont {J.}~\bibnamefont {{Maggs}}},\ }\bibfield  {title}
  {\enquote {\bibinfo {title} {{Design, construction, and properties of the
  large plasma research device - The LAPD at UCLA}},}\ }\href {\doibase
  10.1063/1.1142175} {\bibfield  {journal} {\bibinfo  {journal} {Rev. Sci.
  Instrum.}\ }\textbf {\bibinfo {volume} {62}},\ \bibinfo {pages} {2875--2883}
  (\bibinfo {year} {1991})}\BibitemShut {NoStop}%
\bibitem [{\citenamefont {{Ji}}\ and\ \citenamefont
  {{Daughton}}(2011)}]{Ji:2011}%
  \BibitemOpen
  \bibfield  {author} {\bibinfo {author} {\bibfnamefont {H.}~\bibnamefont
  {{Ji}}}\ and\ \bibinfo {author} {\bibfnamefont {W.}~\bibnamefont
  {{Daughton}}},\ }\bibfield  {title} {\enquote {\bibinfo {title} {{Phase
  diagram for magnetic reconnection in heliophysical, astrophysical, and
  laboratory plasmas}},}\ }\href {\doibase 10.1063/1.3647505} {\bibfield
  {journal} {\bibinfo  {journal} {Phys.~Plasmas}\ }\textbf {\bibinfo {volume}
  {18}},\ \bibinfo {pages} {111207--111207} (\bibinfo {year} {2011})},\ \Eprint
  {http://arxiv.org/abs/1109.0756} {arXiv:1109.0756 [astro-ph.IM]} \BibitemShut
  {NoStop}%
\bibitem [{\citenamefont {{F{\"a}lthammar}}(1974)}]{Falthammar:1974}%
  \BibitemOpen
  \bibfield  {author} {\bibinfo {author} {\bibfnamefont {C.-G.}\ \bibnamefont
  {{F{\"a}lthammar}}},\ }\bibfield  {title} {\enquote {\bibinfo {title}
  {{Laboratory Experiments of Magnetospheric Interest}},}\ }\href {\doibase
  10.1007/BF00241062} {\bibfield  {journal} {\bibinfo  {journal} {Space
  Sci.~Rev.}\ }\textbf {\bibinfo {volume} {15}},\ \bibinfo {pages} {803--825}
  (\bibinfo {year} {1974})}\BibitemShut {NoStop}%
\bibitem [{\citenamefont {{Koepke}}(2008)}]{Koepke:2008}%
  \BibitemOpen
  \bibfield  {author} {\bibinfo {author} {\bibfnamefont {M.~E.}\ \bibnamefont
  {{Koepke}}},\ }\bibfield  {title} {\enquote {\bibinfo {title} {{Interrelated
  laboratory and space plasma experiments}},}\ }\href {\doibase
  10.1029/2005RG000168} {\bibfield  {journal} {\bibinfo  {journal} {Rev.
  Geophys.}\ }\textbf {\bibinfo {volume} {46}},\ \bibinfo {eid} {RG3001}
  (\bibinfo {year} {2008})}\BibitemShut {NoStop}%
\bibitem [{\citenamefont {{Buckingham}}(1914)}]{Buckingham:1914}%
  \BibitemOpen
  \bibfield  {author} {\bibinfo {author} {\bibfnamefont {E.}~\bibnamefont
  {{Buckingham}}},\ }\bibfield  {title} {\enquote {\bibinfo {title} {{On
  Physically Similar Systems; Illustrations of the Use of Dimensional
  Equations}},}\ }\href {\doibase 10.1103/PhysRev.4.345} {\bibfield  {journal}
  {\bibinfo  {journal} {Phys.~Rev.}\ }\textbf {\bibinfo {volume} {4}},\
  \bibinfo {pages} {345--376} (\bibinfo {year} {1914})}\BibitemShut {NoStop}%
\bibitem [{\citenamefont {{Barenblatt}}(1996)}]{Barenblatt:1996}%
  \BibitemOpen
  \bibfield  {author} {\bibinfo {author} {\bibfnamefont {G.~I.}\ \bibnamefont
  {{Barenblatt}}},\ }\href@noop {} {\emph {\bibinfo {title} {Scaling,
  Self-similarity, and Intermediate Asymptotics, by Grigory Isaakovich
  Barenblatt, pp.~408.~ISBN 0521435226.~Cambridge, UK: Cambridge University
  Press, December 1996.}}}\ (\bibinfo  {publisher} {Cambridge University
  Press},\ \bibinfo {year} {1996})\ p.\ \bibinfo {pages} {408}\BibitemShut
  {NoStop}%
\bibitem [{\citenamefont {{Taylor}}(1950{\natexlab{a}})}]{Taylor:1950a}%
  \BibitemOpen
  \bibfield  {author} {\bibinfo {author} {\bibfnamefont {G.}~\bibnamefont
  {{Taylor}}},\ }\bibfield  {title} {\enquote {\bibinfo {title} {{The Formation
  of a Blast Wave by a Very Intense Explosion. I. Theoretical Discussion}},}\
  }\href {\doibase 10.1098/rspa.1950.0049} {\bibfield  {journal} {\bibinfo
  {journal} {Proc.~Roy.~Soc.}\ }\textbf {\bibinfo {volume} {201}},\ \bibinfo
  {pages} {159--174} (\bibinfo {year} {1950}{\natexlab{a}})}\BibitemShut
  {NoStop}%
\bibitem [{\citenamefont {{Taylor}}(1950{\natexlab{b}})}]{Taylor:1950b}%
  \BibitemOpen
  \bibfield  {author} {\bibinfo {author} {\bibfnamefont {G.}~\bibnamefont
  {{Taylor}}},\ }\bibfield  {title} {\enquote {\bibinfo {title} {{The Formation
  of a Blast Wave by a Very Intense Explosion. II. The Atomic Explosion of
  1945}},}\ }\href {\doibase 10.1098/rspa.1950.0050} {\bibfield  {journal}
  {\bibinfo  {journal} {Proc.~Roy.~Soc.}\ }\textbf {\bibinfo {volume} {201}},\
  \bibinfo {pages} {175--186} (\bibinfo {year}
  {1950}{\natexlab{b}})}\BibitemShut {NoStop}%
\bibitem [{\citenamefont {{Frisch}}(1995)}]{Frisch:1995}%
  \BibitemOpen
  \bibfield  {author} {\bibinfo {author} {\bibfnamefont {U.}~\bibnamefont
  {{Frisch}}},\ }\href@noop {} {\emph {\bibinfo {title} {Turbulence, by Uriel
  Frisch, Cambridge, UK: Cambridge University Press, 1995}}}\ (\bibinfo
  {publisher} {Cambridge University Press},\ \bibinfo {year}
  {1995})\BibitemShut {NoStop}%
\bibitem [{\citenamefont {{Chapman}}, \citenamefont {{Rowlands}},\ and\
  \citenamefont {{Watkins}}(2009)}]{Chapman:2009}%
  \BibitemOpen
  \bibfield  {author} {\bibinfo {author} {\bibfnamefont {S.~C.}\ \bibnamefont
  {{Chapman}}}, \bibinfo {author} {\bibfnamefont {G.}~\bibnamefont
  {{Rowlands}}}, \ and\ \bibinfo {author} {\bibfnamefont {N.~W.}\ \bibnamefont
  {{Watkins}}},\ }\bibfield  {title} {\enquote {\bibinfo {title} {{Macroscopic
  control parameter for avalanche models for bursty transport}},}\ }\href
  {\doibase 10.1063/1.3057392} {\bibfield  {journal} {\bibinfo  {journal}
  {Phys.~Plasmas}\ }\textbf {\bibinfo {volume} {16}},\ \bibinfo {eid} {012303}
  (\bibinfo {year} {2009})},\ \Eprint {http://arxiv.org/abs/0806.1133}
  {arXiv:0806.1133 [math-ph]} \BibitemShut {NoStop}%
\bibitem [{\citenamefont {{Ryutov}}\ \emph {et~al.}(1999)\citenamefont
  {{Ryutov}}, \citenamefont {{Drake}}, \citenamefont {{Kane}}, \citenamefont
  {{Liang}}, \citenamefont {{Remington}},\ and\ \citenamefont
  {{Wood-Vasey}}}]{Ryutov:1999}%
  \BibitemOpen
  \bibfield  {author} {\bibinfo {author} {\bibfnamefont {D.}~\bibnamefont
  {{Ryutov}}}, \bibinfo {author} {\bibfnamefont {R.~P.}\ \bibnamefont
  {{Drake}}}, \bibinfo {author} {\bibfnamefont {J.}~\bibnamefont {{Kane}}},
  \bibinfo {author} {\bibfnamefont {E.}~\bibnamefont {{Liang}}}, \bibinfo
  {author} {\bibfnamefont {B.~A.}\ \bibnamefont {{Remington}}}, \ and\ \bibinfo
  {author} {\bibfnamefont {W.~M.}\ \bibnamefont {{Wood-Vasey}}},\ }\bibfield
  {title} {\enquote {\bibinfo {title} {{Similarity Criteria for the Laboratory
  Simulation of Supernova Hydrodynamics}},}\ }\href {\doibase 10.1086/307293}
  {\bibfield  {journal} {\bibinfo  {journal} {Astrophys.~J.}\ }\textbf
  {\bibinfo {volume} {518}},\ \bibinfo {pages} {821--832} (\bibinfo {year}
  {1999})}\BibitemShut {NoStop}%
\bibitem [{\citenamefont {Drake}(2000)}]{Drake:2000}%
  \BibitemOpen
  \bibfield  {author} {\bibinfo {author} {\bibfnamefont {R.~P.}\ \bibnamefont
  {Drake}},\ }\bibfield  {title} {\enquote {\bibinfo {title} {The design of
  laboratory experiments to produce collisionless shocks of cosmic
  relevance},}\ }\href {\doibase http://dx.doi.org/10.1063/1.1314625}
  {\bibfield  {journal} {\bibinfo  {journal} {Phys.~Plasmas}\ }\textbf
  {\bibinfo {volume} {7}},\ \bibinfo {pages} {4690--4698} (\bibinfo {year}
  {2000})}\BibitemShut {NoStop}%
\bibitem [{\citenamefont {{Ryutov}}, \citenamefont {{Drake}},\ and\
  \citenamefont {{Remington}}(2000)}]{Ryutov:2000}%
  \BibitemOpen
  \bibfield  {author} {\bibinfo {author} {\bibfnamefont {D.~D.}\ \bibnamefont
  {{Ryutov}}}, \bibinfo {author} {\bibfnamefont {R.~P.}\ \bibnamefont
  {{Drake}}}, \ and\ \bibinfo {author} {\bibfnamefont {B.~A.}\ \bibnamefont
  {{Remington}}},\ }\bibfield  {title} {\enquote {\bibinfo {title} {{Criteria
  for Scaled Laboratory Simulations of Astrophysical MHD Phenomena}},}\ }\href
  {\doibase 10.1086/313320} {\bibfield  {journal} {\bibinfo  {journal}
  {Astrophys.~J.~Supp.}\ }\textbf {\bibinfo {volume} {127}},\ \bibinfo {pages}
  {465--468} (\bibinfo {year} {2000})}\BibitemShut {NoStop}%
\bibitem [{\citenamefont {{Ryutov}}\ \emph {et~al.}(2001)\citenamefont
  {{Ryutov}}, \citenamefont {{Remington}}, \citenamefont {{Robey}},\ and\
  \citenamefont {{Drake}}}]{Ryutov:2001}%
  \BibitemOpen
  \bibfield  {author} {\bibinfo {author} {\bibfnamefont {D.~D.}\ \bibnamefont
  {{Ryutov}}}, \bibinfo {author} {\bibfnamefont {B.~A.}\ \bibnamefont
  {{Remington}}}, \bibinfo {author} {\bibfnamefont {H.~F.}\ \bibnamefont
  {{Robey}}}, \ and\ \bibinfo {author} {\bibfnamefont {R.~P.}\ \bibnamefont
  {{Drake}}},\ }\bibfield  {title} {\enquote {\bibinfo {title}
  {{Magnetohydrodynamic scaling: From astrophysics to the laboratory}},}\
  }\href {\doibase 10.1063/1.1344562} {\bibfield  {journal} {\bibinfo
  {journal} {Phys.~Plasmas}\ }\textbf {\bibinfo {volume} {8}},\ \bibinfo
  {pages} {1804--1816} (\bibinfo {year} {2001})}\BibitemShut {NoStop}%
\bibitem [{\citenamefont {{Remington}}, \citenamefont {{Drake}},\ and\
  \citenamefont {{Ryutov}}(2006)}]{Remington:2006}%
  \BibitemOpen
  \bibfield  {author} {\bibinfo {author} {\bibfnamefont {B.~A.}\ \bibnamefont
  {{Remington}}}, \bibinfo {author} {\bibfnamefont {R.~P.}\ \bibnamefont
  {{Drake}}}, \ and\ \bibinfo {author} {\bibfnamefont {D.~D.}\ \bibnamefont
  {{Ryutov}}},\ }\bibfield  {title} {\enquote {\bibinfo {title} {{Experimental
  astrophysics with high power lasers and Z pinches}},}\ }\href {\doibase
  10.1103/RevModPhys.78.755} {\bibfield  {journal} {\bibinfo  {journal} {Rev.
  Mod. Phys.}\ }\textbf {\bibinfo {volume} {78}},\ \bibinfo {pages} {755--807}
  (\bibinfo {year} {2006})}\BibitemShut {NoStop}%
\bibitem [{\citenamefont {{Taylor}}(1938)}]{Taylor:1938}%
  \BibitemOpen
  \bibfield  {author} {\bibinfo {author} {\bibfnamefont {G.~I.}\ \bibnamefont
  {{Taylor}}},\ }\bibfield  {title} {\enquote {\bibinfo {title} {{The Spectrum
  of Turbulence}},}\ }\href@noop {} {\bibfield  {journal} {\bibinfo  {journal}
  {{Proc. Roy. Soc. A}}\ }\textbf {\bibinfo {volume} {164}},\ \bibinfo {pages}
  {476--490} (\bibinfo {year} {1938})}\BibitemShut {NoStop}%
\bibitem [{\citenamefont {{Matthaeus}}\ and\ \citenamefont
  {{Goldstein}}(1982)}]{Matthaeus:1982b}%
  \BibitemOpen
  \bibfield  {author} {\bibinfo {author} {\bibfnamefont {W.~H.}\ \bibnamefont
  {{Matthaeus}}}\ and\ \bibinfo {author} {\bibfnamefont {M.~L.}\ \bibnamefont
  {{Goldstein}}},\ }\bibfield  {title} {\enquote {\bibinfo {title}
  {{Measurement of the rugged invariants of magnetohydrodynamic turbulence in
  the solar wind}},}\ }\href@noop {} {\bibfield  {journal} {\bibinfo  {journal}
  {J.~Geophys.~Res.}\ }\textbf {\bibinfo {volume} {87}},\ \bibinfo {pages}
  {6011--6028} (\bibinfo {year} {1982})}\BibitemShut {NoStop}%
\bibitem [{\citenamefont {{Perri}}\ and\ \citenamefont
  {{Balogh}}(2010)}]{Perri:2010a}%
  \BibitemOpen
  \bibfield  {author} {\bibinfo {author} {\bibfnamefont {S.}~\bibnamefont
  {{Perri}}}\ and\ \bibinfo {author} {\bibfnamefont {A.}~\bibnamefont
  {{Balogh}}},\ }\bibfield  {title} {\enquote {\bibinfo {title} {{Stationarity
  in Solar Wind Flows}},}\ }\href {\doibase 10.1088/0004-637X/714/1/937}
  {\bibfield  {journal} {\bibinfo  {journal} {Astrophys.~J.}\ }\textbf
  {\bibinfo {volume} {714}},\ \bibinfo {pages} {937--943} (\bibinfo {year}
  {2010})}\BibitemShut {NoStop}%
\bibitem [{\citenamefont {{Howes}}, \citenamefont {{Klein}},\ and\
  \citenamefont {{TenBarge}}(2014)}]{Howes:2014a}%
  \BibitemOpen
  \bibfield  {author} {\bibinfo {author} {\bibfnamefont {G.~G.}\ \bibnamefont
  {{Howes}}}, \bibinfo {author} {\bibfnamefont {K.~G.}\ \bibnamefont
  {{Klein}}}, \ and\ \bibinfo {author} {\bibfnamefont {J.~M.}\ \bibnamefont
  {{TenBarge}}},\ }\bibfield  {title} {\enquote {\bibinfo {title} {{Validity of
  the Taylor Hypothesis for Linear Kinetic Waves in the Weakly Collisional
  Solar Wind}},}\ }\href {\doibase 10.1088/0004-637X/789/2/106} {\bibfield
  {journal} {\bibinfo  {journal} {Astrophys.~J.}\ }\textbf {\bibinfo {volume}
  {789}},\ \bibinfo {eid} {106} (\bibinfo {year} {2014})},\ \Eprint
  {http://arxiv.org/abs/1405.5460} {arXiv:1405.5460 [astro-ph.SR]} \BibitemShut
  {NoStop}%
\bibitem [{\citenamefont {{Klein}}, \citenamefont {{Howes}},\ and\
  \citenamefont {{TenBarge}}(2014)}]{Klein:2014b}%
  \BibitemOpen
  \bibfield  {author} {\bibinfo {author} {\bibfnamefont {K.~G.}\ \bibnamefont
  {{Klein}}}, \bibinfo {author} {\bibfnamefont {G.~G.}\ \bibnamefont
  {{Howes}}}, \ and\ \bibinfo {author} {\bibfnamefont {J.~M.}\ \bibnamefont
  {{TenBarge}}},\ }\bibfield  {title} {\enquote {\bibinfo {title} {{The
  Violation of the Taylor Hypothesis in Measurements of Solar Wind
  Turbulence}},}\ }\href {\doibase 10.1088/2041-8205/790/2/L20} {\bibfield
  {journal} {\bibinfo  {journal} {Astrophys.~J.~Lett.}\ }\textbf {\bibinfo
  {volume} {790}},\ \bibinfo {eid} {L20} (\bibinfo {year} {2014})},\ \Eprint
  {http://arxiv.org/abs/1406.5470} {arXiv:1406.5470 [physics.space-ph]}
  \BibitemShut {NoStop}%
\bibitem [{\citenamefont {{Perri}}\ \emph {et~al.}(2017)\citenamefont
  {{Perri}}, \citenamefont {{Servidio}}, \citenamefont {{Vaivads}},\ and\
  \citenamefont {{Valentini}}}]{Perri:2017}%
  \BibitemOpen
  \bibfield  {author} {\bibinfo {author} {\bibfnamefont {S.}~\bibnamefont
  {{Perri}}}, \bibinfo {author} {\bibfnamefont {S.}~\bibnamefont {{Servidio}}},
  \bibinfo {author} {\bibfnamefont {A.}~\bibnamefont {{Vaivads}}}, \ and\
  \bibinfo {author} {\bibfnamefont {F.}~\bibnamefont {{Valentini}}},\
  }\bibfield  {title} {\enquote {\bibinfo {title} {{Numerical Study on the
  Validity of the Taylor Hypothesis in Space Plasmas}},}\ }\href {\doibase
  10.3847/1538-4365/aa755a} {\bibfield  {journal} {\bibinfo  {journal}
  {Astrophys.~J.~Supp.}\ }\textbf {\bibinfo {volume} {231}},\ \bibinfo {eid}
  {4} (\bibinfo {year} {2017})}\BibitemShut {NoStop}%
\bibitem [{\citenamefont {Yamada}\ \emph {et~al.}(1997)\citenamefont {Yamada},
  \citenamefont {Ji}, \citenamefont {Hsu}, \citenamefont {Carter},
  \citenamefont {Kulsrud}, \citenamefont {Bretz}, \citenamefont {Jobes},
  \citenamefont {Ono},\ and\ \citenamefont {Perkins}}]{Yamada:1997b}%
  \BibitemOpen
  \bibfield  {author} {\bibinfo {author} {\bibfnamefont {M.}~\bibnamefont
  {Yamada}}, \bibinfo {author} {\bibfnamefont {H.}~\bibnamefont {Ji}}, \bibinfo
  {author} {\bibfnamefont {S.}~\bibnamefont {Hsu}}, \bibinfo {author}
  {\bibfnamefont {T.}~\bibnamefont {Carter}}, \bibinfo {author} {\bibfnamefont
  {R.}~\bibnamefont {Kulsrud}}, \bibinfo {author} {\bibfnamefont
  {N.}~\bibnamefont {Bretz}}, \bibinfo {author} {\bibfnamefont
  {F.}~\bibnamefont {Jobes}}, \bibinfo {author} {\bibfnamefont
  {Y.}~\bibnamefont {Ono}}, \ and\ \bibinfo {author} {\bibfnamefont
  {F.}~\bibnamefont {Perkins}},\ }\bibfield  {title} {\enquote {\bibinfo
  {title} {Study of driven magnetic reconnection in a laboratory plasma},}\
  }\href@noop {} {\bibfield  {journal} {\bibinfo  {journal} {Phys.~Plasmas}\
  }\textbf {\bibinfo {volume} {4}},\ \bibinfo {pages} {1936} (\bibinfo {year}
  {1997})}\BibitemShut {NoStop}%
\bibitem [{\citenamefont {{Brown}}\ and\ \citenamefont
  {{Schaffner}}(2014)}]{Brown:2014}%
  \BibitemOpen
  \bibfield  {author} {\bibinfo {author} {\bibfnamefont {M.~R.}\ \bibnamefont
  {{Brown}}}\ and\ \bibinfo {author} {\bibfnamefont {D.~A.}\ \bibnamefont
  {{Schaffner}}},\ }\bibfield  {title} {\enquote {\bibinfo {title} {{Laboratory
  sources of turbulent plasma: a unique MHD plasma wind tunnel}},}\ }\href
  {\doibase 10.1088/0963-0252/23/6/063001} {\bibfield  {journal} {\bibinfo
  {journal} {Plasma Sources Sci. Tech.}\ }\textbf {\bibinfo {volume} {23}},\
  \bibinfo {eid} {063001} (\bibinfo {year} {2014})}\BibitemShut {NoStop}%
\bibitem [{\citenamefont {{Brown}}\ and\ \citenamefont
  {{Schaffner}}(2015)}]{Brown:2015}%
  \BibitemOpen
  \bibfield  {author} {\bibinfo {author} {\bibfnamefont {M.~R.}\ \bibnamefont
  {{Brown}}}\ and\ \bibinfo {author} {\bibfnamefont {D.~A.}\ \bibnamefont
  {{Schaffner}}},\ }\bibfield  {title} {\enquote {\bibinfo {title} {{SSX MHD
  plasma wind tunnel}},}\ }\href {\doibase 10.1017/S0022377815000227}
  {\bibfield  {journal} {\bibinfo  {journal} {J.~Plasma Phys.}\ }\textbf
  {\bibinfo {volume} {81}},\ \bibinfo {eid} {345810302} (\bibinfo {year}
  {2015})}\BibitemShut {NoStop}%
\bibitem [{\citenamefont {{Wallace}}\ \emph {et~al.}(2004)\citenamefont
  {{Wallace}}, \citenamefont {{Thomas}}, \citenamefont {{Eadon}},\ and\
  \citenamefont {{Jackson}}}]{Wallace:2004}%
  \BibitemOpen
  \bibfield  {author} {\bibinfo {author} {\bibfnamefont {E.}~\bibnamefont
  {{Wallace}}}, \bibinfo {author} {\bibfnamefont {E.}~\bibnamefont {{Thomas}}},
  \bibinfo {author} {\bibfnamefont {A.}~\bibnamefont {{Eadon}}}, \ and\
  \bibinfo {author} {\bibfnamefont {J.~D.}\ \bibnamefont {{Jackson}}},\
  }\bibfield  {title} {\enquote {\bibinfo {title} {{Design and initial
  operation of the Auburn Linear Experiment for Instability Studies: A new
  plasma experiment for studying shear driven flows}},}\ }\href {\doibase
  10.1063/1.1818491} {\bibfield  {journal} {\bibinfo  {journal} {Rev. Sci.
  Instrum.}\ }\textbf {\bibinfo {volume} {75}},\ \bibinfo {pages} {5160--5165}
  (\bibinfo {year} {2004})}\BibitemShut {NoStop}%
\bibitem [{\citenamefont {{Eadon}}\ \emph {et~al.}(2011)\citenamefont
  {{Eadon}}, \citenamefont {{Tejero}}, \citenamefont {{DuBois}},\ and\
  \citenamefont {{Thomas}}}]{Eadon:2011}%
  \BibitemOpen
  \bibfield  {author} {\bibinfo {author} {\bibfnamefont {A.~C.}\ \bibnamefont
  {{Eadon}}}, \bibinfo {author} {\bibfnamefont {E.}~\bibnamefont {{Tejero}}},
  \bibinfo {author} {\bibfnamefont {A.}~\bibnamefont {{DuBois}}}, \ and\
  \bibinfo {author} {\bibfnamefont {E.}~\bibnamefont {{Thomas}}},\ }\bibfield
  {title} {\enquote {\bibinfo {title} {{Upgrades to the Auburn linear
  experiment for instability studies}},}\ }\href {\doibase 10.1063/1.3594102}
  {\bibfield  {journal} {\bibinfo  {journal} {Rev. Sci. Instrum.}\ }\textbf
  {\bibinfo {volume} {82}},\ \bibinfo {eid} {063511-063511-9} (\bibinfo {year}
  {2011})}\BibitemShut {NoStop}%
\bibitem [{\citenamefont {{DuBois}}\ \emph {et~al.}(2013)\citenamefont
  {{DuBois}}, \citenamefont {{Arnold}}, \citenamefont {{Thomas}}, \citenamefont
  {{Tejero}},\ and\ \citenamefont {{Amatucci}}}]{Dubois:2013a}%
  \BibitemOpen
  \bibfield  {author} {\bibinfo {author} {\bibfnamefont {A.~M.}\ \bibnamefont
  {{DuBois}}}, \bibinfo {author} {\bibfnamefont {I.}~\bibnamefont {{Arnold}}},
  \bibinfo {author} {\bibfnamefont {E.}~\bibnamefont {{Thomas}}}, \bibinfo
  {author} {\bibfnamefont {E.}~\bibnamefont {{Tejero}}}, \ and\ \bibinfo
  {author} {\bibfnamefont {W.~E.}\ \bibnamefont {{Amatucci}}},\ }\bibfield
  {title} {\enquote {\bibinfo {title} {{Electron-ion hybrid instability
  experiment upgrades to the Auburn Linear Experiment for Instability
  Studies}},}\ }\href {\doibase 10.1063/1.4799288} {\bibfield  {journal}
  {\bibinfo  {journal} {Rev. Sci. Instrum.}\ }\textbf {\bibinfo {volume}
  {84}},\ \bibinfo {eid} {043503-043503-6} (\bibinfo {year}
  {2013})}\BibitemShut {NoStop}%
\bibitem [{\citenamefont {DuBois}\ \emph {et~al.}(2013)\citenamefont {DuBois},
  \citenamefont {Thomas}, \citenamefont {Amatucci},\ and\ \citenamefont
  {Ganguli}}]{Dubois:2013b}%
  \BibitemOpen
  \bibfield  {author} {\bibinfo {author} {\bibfnamefont {A.~M.}\ \bibnamefont
  {DuBois}}, \bibinfo {author} {\bibfnamefont {E.}~\bibnamefont {Thomas}},
  \bibinfo {author} {\bibfnamefont {W.~E.}\ \bibnamefont {Amatucci}}, \ and\
  \bibinfo {author} {\bibfnamefont {G.}~\bibnamefont {Ganguli}},\ }\bibfield
  {title} {\enquote {\bibinfo {title} {Plasma response to a varying degree of
  stress},}\ }\href {\doibase 10.1103/PhysRevLett.111.145002} {\bibfield
  {journal} {\bibinfo  {journal} {Phys.~Rev.~Lett.}\ }\textbf {\bibinfo
  {volume} {111}},\ \bibinfo {pages} {145002} (\bibinfo {year}
  {2013})}\BibitemShut {NoStop}%
\bibitem [{\citenamefont {DuBois}\ \emph {et~al.}(2014)\citenamefont {DuBois},
  \citenamefont {Thomas}, \citenamefont {Amatucci},\ and\ \citenamefont
  {Ganguli}}]{Dubois:2014}%
  \BibitemOpen
  \bibfield  {author} {\bibinfo {author} {\bibfnamefont {A.~M.}\ \bibnamefont
  {DuBois}}, \bibinfo {author} {\bibfnamefont {E.}~\bibnamefont {Thomas}},
  \bibinfo {author} {\bibfnamefont {W.~E.}\ \bibnamefont {Amatucci}}, \ and\
  \bibinfo {author} {\bibfnamefont {G.}~\bibnamefont {Ganguli}},\ }\bibfield
  {title} {\enquote {\bibinfo {title} {Experimental characterization of
  broadband electrostatic noise due to plasma compression},}\ }\href {\doibase
  10.1002/2014JA020198} {\bibfield  {journal} {\bibinfo  {journal}
  {J.~Geophys.~Res.}\ }\textbf {\bibinfo {volume} {119}},\ \bibinfo {pages}
  {5624--5637} (\bibinfo {year} {2014})},\ \bibinfo {note}
  {2014JA020198}\BibitemShut {NoStop}%
\bibitem [{\citenamefont {Office}(2013)}]{GAO:2013}%
  \BibitemOpen
  \bibfield  {author} {\bibinfo {author} {\bibfnamefont {U.~S. G.~A.}\
  \bibnamefont {Office}},\ }\href@noop {} {\enquote {\bibinfo {title} {Nasa:
  Assessments of selected large-scale projects},}\ }\bibinfo {type} {Tech.
  Rep.}\ (\bibinfo  {institution} {United States Government Accountability
  Office},\ \bibinfo {address} {Washington, D.C.},\ \bibinfo {year}
  {2013})\BibitemShut {NoStop}%
\bibitem [{\citenamefont {Office}(2017)}]{GAO:2017}%
  \BibitemOpen
  \bibfield  {author} {\bibinfo {author} {\bibfnamefont {U.~S. G.~A.}\
  \bibnamefont {Office}},\ }\href@noop {} {\enquote {\bibinfo {title} {Nasa:
  Assessments of major projects},}\ }\bibinfo {type} {Tech. Rep.}\ (\bibinfo
  {institution} {United States Government Accountability Office},\ \bibinfo
  {address} {Washington, D.C.},\ \bibinfo {year} {2017})\BibitemShut {NoStop}%
\bibitem [{\citenamefont {{Parker}}(1955)}]{Parker:1955}%
  \BibitemOpen
  \bibfield  {author} {\bibinfo {author} {\bibfnamefont {E.~N.}\ \bibnamefont
  {{Parker}}},\ }\bibfield  {title} {\enquote {\bibinfo {title} {{Hydromagnetic
  Dynamo Models.}}}\ }\href {\doibase 10.1086/146087} {\bibfield  {journal}
  {\bibinfo  {journal} {Astrophys.~J.}\ }\textbf {\bibinfo {volume} {122}},\
  \bibinfo {pages} {293} (\bibinfo {year} {1955})}\BibitemShut {NoStop}%
\bibitem [{\citenamefont {{Brun}}, \citenamefont {{Miesch}},\ and\
  \citenamefont {{Toomre}}(2004)}]{Brun:2004}%
  \BibitemOpen
  \bibfield  {author} {\bibinfo {author} {\bibfnamefont {A.~S.}\ \bibnamefont
  {{Brun}}}, \bibinfo {author} {\bibfnamefont {M.~S.}\ \bibnamefont
  {{Miesch}}}, \ and\ \bibinfo {author} {\bibfnamefont {J.}~\bibnamefont
  {{Toomre}}},\ }\bibfield  {title} {\enquote {\bibinfo {title} {{Global-Scale
  Turbulent Convection and Magnetic Dynamo Action in the Solar Envelope}},}\
  }\href {\doibase 10.1086/423835} {\bibfield  {journal} {\bibinfo  {journal}
  {Astrophys.~J.}\ }\textbf {\bibinfo {volume} {614}},\ \bibinfo {pages}
  {1073--1098} (\bibinfo {year} {2004})},\ \Eprint
  {http://arxiv.org/abs/astro-ph/0610073} {astro-ph/0610073} \BibitemShut
  {NoStop}%
\bibitem [{\citenamefont {{Berger}}\ \emph {et~al.}(2008)\citenamefont
  {{Berger}}, \citenamefont {{Shine}}, \citenamefont {{Slater}}, \citenamefont
  {{Tarbell}}, \citenamefont {{Title}}, \citenamefont {{Okamoto}},
  \citenamefont {{Ichimoto}}, \citenamefont {{Katsukawa}}, \citenamefont
  {{Suematsu}}, \citenamefont {{Tsuneta}}, \citenamefont {{Lites}},\ and\
  \citenamefont {{Shimizu}}}]{Berger:2008}%
  \BibitemOpen
  \bibfield  {author} {\bibinfo {author} {\bibfnamefont {T.~E.}\ \bibnamefont
  {{Berger}}}, \bibinfo {author} {\bibfnamefont {R.~A.}\ \bibnamefont
  {{Shine}}}, \bibinfo {author} {\bibfnamefont {G.~L.}\ \bibnamefont
  {{Slater}}}, \bibinfo {author} {\bibfnamefont {T.~D.}\ \bibnamefont
  {{Tarbell}}}, \bibinfo {author} {\bibfnamefont {A.~M.}\ \bibnamefont
  {{Title}}}, \bibinfo {author} {\bibfnamefont {T.~J.}\ \bibnamefont
  {{Okamoto}}}, \bibinfo {author} {\bibfnamefont {K.}~\bibnamefont
  {{Ichimoto}}}, \bibinfo {author} {\bibfnamefont {Y.}~\bibnamefont
  {{Katsukawa}}}, \bibinfo {author} {\bibfnamefont {Y.}~\bibnamefont
  {{Suematsu}}}, \bibinfo {author} {\bibfnamefont {S.}~\bibnamefont
  {{Tsuneta}}}, \bibinfo {author} {\bibfnamefont {B.~W.}\ \bibnamefont
  {{Lites}}}, \ and\ \bibinfo {author} {\bibfnamefont {T.}~\bibnamefont
  {{Shimizu}}},\ }\bibfield  {title} {\enquote {\bibinfo {title} {{Hinode SOT
  Observations of Solar Quiescent Prominence Dynamics}},}\ }\href {\doibase
  10.1086/587171} {\bibfield  {journal} {\bibinfo  {journal}
  {Astrophys.~J.~Lett.}\ }\textbf {\bibinfo {volume} {676}},\ \bibinfo {eid}
  {L89} (\bibinfo {year} {2008})}\BibitemShut {NoStop}%
\bibitem [{\citenamefont {{Berger}}\ \emph {et~al.}(2010)\citenamefont
  {{Berger}}, \citenamefont {{Slater}}, \citenamefont {{Hurlburt}},
  \citenamefont {{Shine}}, \citenamefont {{Tarbell}}, \citenamefont {{Title}},
  \citenamefont {{Lites}}, \citenamefont {{Okamoto}}, \citenamefont
  {{Ichimoto}}, \citenamefont {{Katsukawa}}, \citenamefont {{Magara}},
  \citenamefont {{Suematsu}},\ and\ \citenamefont {{Shimizu}}}]{Berger:2010}%
  \BibitemOpen
  \bibfield  {author} {\bibinfo {author} {\bibfnamefont {T.~E.}\ \bibnamefont
  {{Berger}}}, \bibinfo {author} {\bibfnamefont {G.}~\bibnamefont {{Slater}}},
  \bibinfo {author} {\bibfnamefont {N.}~\bibnamefont {{Hurlburt}}}, \bibinfo
  {author} {\bibfnamefont {R.}~\bibnamefont {{Shine}}}, \bibinfo {author}
  {\bibfnamefont {T.}~\bibnamefont {{Tarbell}}}, \bibinfo {author}
  {\bibfnamefont {A.}~\bibnamefont {{Title}}}, \bibinfo {author} {\bibfnamefont
  {B.~W.}\ \bibnamefont {{Lites}}}, \bibinfo {author} {\bibfnamefont {T.~J.}\
  \bibnamefont {{Okamoto}}}, \bibinfo {author} {\bibfnamefont {K.}~\bibnamefont
  {{Ichimoto}}}, \bibinfo {author} {\bibfnamefont {Y.}~\bibnamefont
  {{Katsukawa}}}, \bibinfo {author} {\bibfnamefont {T.}~\bibnamefont
  {{Magara}}}, \bibinfo {author} {\bibfnamefont {Y.}~\bibnamefont
  {{Suematsu}}}, \ and\ \bibinfo {author} {\bibfnamefont {T.}~\bibnamefont
  {{Shimizu}}},\ }\bibfield  {title} {\enquote {\bibinfo {title} {{Quiescent
  Prominence Dynamics Observed with the Hinode Solar Optical Telescope. I.
  Turbulent Upflow Plumes}},}\ }\href {\doibase 10.1088/0004-637X/716/2/1288}
  {\bibfield  {journal} {\bibinfo  {journal} {Astrophys.~J.}\ }\textbf
  {\bibinfo {volume} {716}},\ \bibinfo {pages} {1288--1307} (\bibinfo {year}
  {2010})}\BibitemShut {NoStop}%
\bibitem [{\citenamefont {{Edl{\'e}n}}(1943)}]{Edlen:1943}%
  \BibitemOpen
  \bibfield  {author} {\bibinfo {author} {\bibfnamefont {B.}~\bibnamefont
  {{Edl{\'e}n}}},\ }\bibfield  {title} {\enquote {\bibinfo {title} {{Die
  Deutung der Emissionslinien im Spektrum der Sonnenkorona. Mit 6
  Abbildungen.}}}\ }\href@noop {} {\bibfield  {journal} {\bibinfo  {journal}
  {Zeitschrift für Astrophysik}\ }\textbf {\bibinfo {volume} {22}},\ \bibinfo
  {pages} {30} (\bibinfo {year} {1943})}\BibitemShut {NoStop}%
\bibitem [{\citenamefont {{Edl{\'e}n}}(1945)}]{Edlen:1945}%
  \BibitemOpen
  \bibfield  {author} {\bibinfo {author} {\bibfnamefont {B.}~\bibnamefont
  {{Edl{\'e}n}}},\ }\bibfield  {title} {\enquote {\bibinfo {title} {{The
  identification of the coronal lines (George Darwin Lecture)}},}\ }\href
  {\doibase 10.1093/mnras/105.6.323} {\bibfield  {journal} {\bibinfo  {journal}
  {Mon.~Not.~Roy.~Astron.~Soc.}\ }\textbf {\bibinfo {volume} {105}},\ \bibinfo
  {pages} {323} (\bibinfo {year} {1945})}\BibitemShut {NoStop}%
\bibitem [{\citenamefont {{Withbroe}}\ and\ \citenamefont
  {{Noyes}}(1977)}]{Withbroe:1977}%
  \BibitemOpen
  \bibfield  {author} {\bibinfo {author} {\bibfnamefont {G.~L.}\ \bibnamefont
  {{Withbroe}}}\ and\ \bibinfo {author} {\bibfnamefont {R.~W.}\ \bibnamefont
  {{Noyes}}},\ }\bibfield  {title} {\enquote {\bibinfo {title} {{Mass and
  energy flow in the solar chromosphere and corona}},}\ }\href {\doibase
  10.1146/annurev.aa.15.090177.002051} {\bibfield  {journal} {\bibinfo
  {journal} {Ann.~Rev.~Astron.~Astrophys.}\ }\textbf {\bibinfo {volume} {15}},\
  \bibinfo {pages} {363--387} (\bibinfo {year} {1977})}\BibitemShut {NoStop}%
\bibitem [{\citenamefont {{Heyvaerts}}\ and\ \citenamefont
  {{Priest}}(1983)}]{Heyvaerts:1983}%
  \BibitemOpen
  \bibfield  {author} {\bibinfo {author} {\bibfnamefont {J.}~\bibnamefont
  {{Heyvaerts}}}\ and\ \bibinfo {author} {\bibfnamefont {E.~R.}\ \bibnamefont
  {{Priest}}},\ }\bibfield  {title} {\enquote {\bibinfo {title} {{Coronal
  heating by phase-mixed shear Alfven waves}},}\ }\href@noop {} {\bibfield
  {journal} {\bibinfo  {journal} {Astron.~Astrophys.}\ }\textbf {\bibinfo
  {volume} {117}},\ \bibinfo {pages} {220--234} (\bibinfo {year}
  {1983})}\BibitemShut {NoStop}%
\bibitem [{\citenamefont {{Parker}}(1988)}]{Parker:1988}%
  \BibitemOpen
  \bibfield  {author} {\bibinfo {author} {\bibfnamefont {E.~N.}\ \bibnamefont
  {{Parker}}},\ }\bibfield  {title} {\enquote {\bibinfo {title} {{Nanoflares
  and the solar X-ray corona}},}\ }\href {\doibase 10.1086/166485} {\bibfield
  {journal} {\bibinfo  {journal} {Astrophys.~J.}\ }\textbf {\bibinfo {volume}
  {330}},\ \bibinfo {pages} {474--479} (\bibinfo {year} {1988})}\BibitemShut
  {NoStop}%
\bibitem [{\citenamefont {{Klimchuk}}(2006)}]{Klimchuk:2006}%
  \BibitemOpen
  \bibfield  {author} {\bibinfo {author} {\bibfnamefont {J.~A.}\ \bibnamefont
  {{Klimchuk}}},\ }\bibfield  {title} {\enquote {\bibinfo {title} {{On Solving
  the Coronal Heating Problem}},}\ }\href {\doibase 10.1007/s11207-006-0055-z}
  {\bibfield  {journal} {\bibinfo  {journal} {Sol.~Phys.}\ }\textbf {\bibinfo
  {volume} {234}},\ \bibinfo {pages} {41--77} (\bibinfo {year} {2006})},\
  \Eprint {http://arxiv.org/abs/astro-ph/0511841} {astro-ph/0511841}
  \BibitemShut {NoStop}%
\bibitem [{\citenamefont {{Cranmer}}(2009)}]{Cranmer:2009b}%
  \BibitemOpen
  \bibfield  {author} {\bibinfo {author} {\bibfnamefont {S.~R.}\ \bibnamefont
  {{Cranmer}}},\ }\bibfield  {title} {\enquote {\bibinfo {title} {{Coronal
  Holes}},}\ }\href {\doibase 10.12942/lrsp-2009-3} {\bibfield  {journal}
  {\bibinfo  {journal} {Living Reviews in Solar Physics}\ }\textbf {\bibinfo
  {volume} {6}},\ \bibinfo {eid} {3} (\bibinfo {year} {2009})},\ \Eprint
  {http://arxiv.org/abs/0909.2847} {arXiv:0909.2847 [astro-ph.SR]} \BibitemShut
  {NoStop}%
\bibitem [{\citenamefont {{Chandran}}(2010)}]{Chandran:2010b}%
  \BibitemOpen
  \bibfield  {author} {\bibinfo {author} {\bibfnamefont {B.~D.~G.}\
  \bibnamefont {{Chandran}}},\ }\bibfield  {title} {\enquote {\bibinfo {title}
  {{Alfv{\'e}n-wave Turbulence and Perpendicular Ion Temperatures in Coronal
  Holes}},}\ }\href {\doibase 10.1088/0004-637X/720/1/548} {\bibfield
  {journal} {\bibinfo  {journal} {Astrophys.~J.}\ }\textbf {\bibinfo {volume}
  {720}},\ \bibinfo {pages} {548--554} (\bibinfo {year} {2010})}\BibitemShut
  {NoStop}%
\bibitem [{\citenamefont {{Landau}}(1946)}]{Landau:1946}%
  \BibitemOpen
  \bibfield  {author} {\bibinfo {author} {\bibfnamefont {L.~D.}\ \bibnamefont
  {{Landau}}},\ }\bibfield  {title} {\enquote {\bibinfo {title} {{On the
  Vibrations of the Electronic Plasma}},}\ }\href@noop {} {\bibfield  {journal}
  {\bibinfo  {journal} {Journal of Physics}\ }\textbf {\bibinfo {volume}
  {10}},\ \bibinfo {pages} {25} (\bibinfo {year} {1946})}\BibitemShut {NoStop}%
\bibitem [{\citenamefont {{Barnes}}(1966)}]{Barnes:1966}%
  \BibitemOpen
  \bibfield  {author} {\bibinfo {author} {\bibfnamefont {A.}~\bibnamefont
  {{Barnes}}},\ }\bibfield  {title} {\enquote {\bibinfo {title} {{Collisionless
  Damping of Hydromagnetic Waves}},}\ }\href@noop {} {\bibfield  {journal}
  {\bibinfo  {journal} {Phys.~Fluids}\ }\textbf {\bibinfo {volume} {9}},\
  \bibinfo {pages} {1483--1495} (\bibinfo {year} {1966})}\BibitemShut {NoStop}%
\bibitem [{\citenamefont {{Coleman}}(1968)}]{Coleman:1968}%
  \BibitemOpen
  \bibfield  {author} {\bibinfo {author} {\bibfnamefont {P.~J.}\ \bibnamefont
  {{Coleman}}, \bibfnamefont {Jr.}},\ }\bibfield  {title} {\enquote {\bibinfo
  {title} {{Turbulence, Viscosity, and Dissipation in the Solar-Wind
  Plasma}},}\ }\href@noop {} {\bibfield  {journal} {\bibinfo  {journal}
  {Astrophys.~J.}\ }\textbf {\bibinfo {volume} {153}},\ \bibinfo {pages}
  {371--388} (\bibinfo {year} {1968})}\BibitemShut {NoStop}%
\bibitem [{\citenamefont {{Denskat}}, \citenamefont {{Beinroth}},\ and\
  \citenamefont {{Neubauer}}(1983)}]{Denskat:1983}%
  \BibitemOpen
  \bibfield  {author} {\bibinfo {author} {\bibfnamefont {K.~U.}\ \bibnamefont
  {{Denskat}}}, \bibinfo {author} {\bibfnamefont {H.~J.}\ \bibnamefont
  {{Beinroth}}}, \ and\ \bibinfo {author} {\bibfnamefont {F.~M.}\ \bibnamefont
  {{Neubauer}}},\ }\bibfield  {title} {\enquote {\bibinfo {title}
  {{Interplanetary magnetic field power spectra with frequencies from 2.4 X 10
  to the -5th HZ to 470 HZ from HELIOS-observations during solar minimum
  conditions}},}\ }\href@noop {} {\bibfield  {journal} {\bibinfo  {journal}
  {J.~Geophys.~Zeit.~Geophys.}\ }\textbf {\bibinfo {volume} {54}},\ \bibinfo
  {pages} {60--67} (\bibinfo {year} {1983})}\BibitemShut {NoStop}%
\bibitem [{\citenamefont {{Isenberg}}\ and\ \citenamefont
  {{Hollweg}}(1983)}]{Isenberg:1983}%
  \BibitemOpen
  \bibfield  {author} {\bibinfo {author} {\bibfnamefont {P.~A.}\ \bibnamefont
  {{Isenberg}}}\ and\ \bibinfo {author} {\bibfnamefont {J.~V.}\ \bibnamefont
  {{Hollweg}}},\ }\bibfield  {title} {\enquote {\bibinfo {title} {{On the
  preferential acceleration and heating of solar wind heavy ions}},}\ }\href
  {\doibase 10.1029/JA088iA05p03923} {\bibfield  {journal} {\bibinfo  {journal}
  {J.~Geophys.~Res.}\ }\textbf {\bibinfo {volume} {88}},\ \bibinfo {pages}
  {3923--3935} (\bibinfo {year} {1983})}\BibitemShut {NoStop}%
\bibitem [{\citenamefont {{Goldstein}}, \citenamefont {{Roberts}},\ and\
  \citenamefont {{Fitch}}(1994)}]{Goldstein:1994}%
  \BibitemOpen
  \bibfield  {author} {\bibinfo {author} {\bibfnamefont {M.~L.}\ \bibnamefont
  {{Goldstein}}}, \bibinfo {author} {\bibfnamefont {D.~A.}\ \bibnamefont
  {{Roberts}}}, \ and\ \bibinfo {author} {\bibfnamefont {C.~A.}\ \bibnamefont
  {{Fitch}}},\ }\bibfield  {title} {\enquote {\bibinfo {title} {{Properties of
  the fluctuating magnetic helicity in the inertial and dissipation ranges of
  solar wind turbulence}},}\ }\href {\doibase 10.1029/94JA00789} {\bibfield
  {journal} {\bibinfo  {journal} {J.~Geophys.~Res.}\ }\textbf {\bibinfo
  {volume} {99}},\ \bibinfo {pages} {11519--11538} (\bibinfo {year}
  {1994})}\BibitemShut {NoStop}%
\bibitem [{\citenamefont {Leamon}\ \emph
  {et~al.}(1998{\natexlab{a}})\citenamefont {Leamon}, \citenamefont {Smith},
  \citenamefont {Ness}, \citenamefont {Matthaeus},\ and\ \citenamefont
  {Wong}}]{Leamon:1998a}%
  \BibitemOpen
  \bibfield  {author} {\bibinfo {author} {\bibfnamefont {R.~J.}\ \bibnamefont
  {Leamon}}, \bibinfo {author} {\bibfnamefont {C.~W.}\ \bibnamefont {Smith}},
  \bibinfo {author} {\bibfnamefont {N.~F.}\ \bibnamefont {Ness}}, \bibinfo
  {author} {\bibfnamefont {W.~H.}\ \bibnamefont {Matthaeus}}, \ and\ \bibinfo
  {author} {\bibfnamefont {H.~K.}\ \bibnamefont {Wong}},\ }\bibfield  {title}
  {\enquote {\bibinfo {title} {Observational constraints on the dynamics of the
  interplanetary magnetic field dissipation range},}\ }\href@noop {} {\bibfield
   {journal} {\bibinfo  {journal} {J.~Geophys.~Res.}\ }\textbf {\bibinfo
  {volume} {103}},\ \bibinfo {pages} {4775--4787} (\bibinfo {year}
  {1998}{\natexlab{a}})}\BibitemShut {NoStop}%
\bibitem [{\citenamefont {Leamon}\ \emph
  {et~al.}(1998{\natexlab{b}})\citenamefont {Leamon}, \citenamefont
  {Matthaeus}, \citenamefont {Smith},\ and\ \citenamefont
  {Wong}}]{Leamon:1998b}%
  \BibitemOpen
  \bibfield  {author} {\bibinfo {author} {\bibfnamefont {R.~J.}\ \bibnamefont
  {Leamon}}, \bibinfo {author} {\bibfnamefont {W.~H.}\ \bibnamefont
  {Matthaeus}}, \bibinfo {author} {\bibfnamefont {C.~W.}\ \bibnamefont
  {Smith}}, \ and\ \bibinfo {author} {\bibfnamefont {H.~K.}\ \bibnamefont
  {Wong}},\ }\bibfield  {title} {\enquote {\bibinfo {title} {Contribution of
  cyclotron-resonant damping to kinetic dissipation of interplanetary
  turbulence},}\ }\href@noop {} {\bibfield  {journal} {\bibinfo  {journal}
  {Astrophys.~J.}\ }\textbf {\bibinfo {volume} {507}},\ \bibinfo {pages}
  {L181--L184} (\bibinfo {year} {1998}{\natexlab{b}})}\BibitemShut {NoStop}%
\bibitem [{\citenamefont {{Quataert}}(1998)}]{Quataert:1998}%
  \BibitemOpen
  \bibfield  {author} {\bibinfo {author} {\bibfnamefont {E.}~\bibnamefont
  {{Quataert}}},\ }\bibfield  {title} {\enquote {\bibinfo {title} {{Particle
  Heating by Alfv\'enic Turbulence in Hot Accretion Flows}},}\ }\href {\doibase
  10.1086/305770} {\bibfield  {journal} {\bibinfo  {journal} {Astrophys.~J.}\
  }\textbf {\bibinfo {volume} {500}},\ \bibinfo {pages} {978--991} (\bibinfo
  {year} {1998})},\ \Eprint {http://arxiv.org/abs/astro-ph/9710127}
  {astro-ph/9710127} \BibitemShut {NoStop}%
\bibitem [{\citenamefont {{Gary}}(1999)}]{Gary:1999a}%
  \BibitemOpen
  \bibfield  {author} {\bibinfo {author} {\bibfnamefont {S.~P.}\ \bibnamefont
  {{Gary}}},\ }\bibfield  {title} {\enquote {\bibinfo {title} {{Collisionless
  dissipation wavenumber: Linear theory}},}\ }\href {\doibase
  10.1029/1998JA900161} {\bibfield  {journal} {\bibinfo  {journal}
  {J.~Geophys.~Res.}\ }\textbf {\bibinfo {volume} {104}},\ \bibinfo {pages}
  {6759--6762} (\bibinfo {year} {1999})}\BibitemShut {NoStop}%
\bibitem [{\citenamefont {Leamon}\ \emph {et~al.}(1999)\citenamefont {Leamon},
  \citenamefont {Smith}, \citenamefont {Ness},\ and\ \citenamefont
  {Wong}}]{Leamon:1999}%
  \BibitemOpen
  \bibfield  {author} {\bibinfo {author} {\bibfnamefont {R.~J.}\ \bibnamefont
  {Leamon}}, \bibinfo {author} {\bibfnamefont {C.~W.}\ \bibnamefont {Smith}},
  \bibinfo {author} {\bibfnamefont {N.~F.}\ \bibnamefont {Ness}}, \ and\
  \bibinfo {author} {\bibfnamefont {H.~K.}\ \bibnamefont {Wong}},\ }\bibfield
  {title} {\enquote {\bibinfo {title} {Dissipation range dynamics: Kinetic
  alfv\'en waves and the importance of $\beta_e$},}\ }\href@noop {} {\bibfield
  {journal} {\bibinfo  {journal} {J.~Geophys.~Res.}\ }\textbf {\bibinfo
  {volume} {104}},\ \bibinfo {pages} {22331--22344} (\bibinfo {year}
  {1999})}\BibitemShut {NoStop}%
\bibitem [{\citenamefont {{Quataert}}\ and\ \citenamefont
  {{Gruzinov}}(1999)}]{Quataert:1999}%
  \BibitemOpen
  \bibfield  {author} {\bibinfo {author} {\bibfnamefont {E.}~\bibnamefont
  {{Quataert}}}\ and\ \bibinfo {author} {\bibfnamefont {A.}~\bibnamefont
  {{Gruzinov}}},\ }\bibfield  {title} {\enquote {\bibinfo {title} {{Turbulence
  and Particle Heating in Advection-dominated Accretion Flows}},}\ }\href
  {\doibase 10.1086/307423} {\bibfield  {journal} {\bibinfo  {journal}
  {Astrophys.~J.}\ }\textbf {\bibinfo {volume} {520}},\ \bibinfo {pages}
  {248--255} (\bibinfo {year} {1999})},\ \Eprint
  {http://arxiv.org/abs/astro-ph/9803112} {astro-ph/9803112} \BibitemShut
  {NoStop}%
\bibitem [{\citenamefont {{Leamon}}\ \emph {et~al.}(2000)\citenamefont
  {{Leamon}}, \citenamefont {{Matthaeus}}, \citenamefont {{Smith}},
  \citenamefont {{Zank}}, \citenamefont {{Mullan}},\ and\ \citenamefont
  {{Oughton}}}]{Leamon:2000}%
  \BibitemOpen
  \bibfield  {author} {\bibinfo {author} {\bibfnamefont {R.~J.}\ \bibnamefont
  {{Leamon}}}, \bibinfo {author} {\bibfnamefont {W.~H.}\ \bibnamefont
  {{Matthaeus}}}, \bibinfo {author} {\bibfnamefont {C.~W.}\ \bibnamefont
  {{Smith}}}, \bibinfo {author} {\bibfnamefont {G.~P.}\ \bibnamefont {{Zank}}},
  \bibinfo {author} {\bibfnamefont {D.~J.}\ \bibnamefont {{Mullan}}}, \ and\
  \bibinfo {author} {\bibfnamefont {S.}~\bibnamefont {{Oughton}}},\ }\bibfield
  {title} {\enquote {\bibinfo {title} {{MHD-driven Kinetic Dissipation in the
  Solar Wind and Corona}},}\ }\href {\doibase 10.1086/309059} {\bibfield
  {journal} {\bibinfo  {journal} {Astrophys.~J.}\ }\textbf {\bibinfo {volume}
  {537}},\ \bibinfo {pages} {1054--1062} (\bibinfo {year} {2000})}\BibitemShut
  {NoStop}%
\bibitem [{\citenamefont {{Isenberg}}, \citenamefont {{Lee}},\ and\
  \citenamefont {{Hollweg}}(2001)}]{Isenberg:2001}%
  \BibitemOpen
  \bibfield  {author} {\bibinfo {author} {\bibfnamefont {P.~A.}\ \bibnamefont
  {{Isenberg}}}, \bibinfo {author} {\bibfnamefont {M.~A.}\ \bibnamefont
  {{Lee}}}, \ and\ \bibinfo {author} {\bibfnamefont {J.~V.}\ \bibnamefont
  {{Hollweg}}},\ }\bibfield  {title} {\enquote {\bibinfo {title} {{The kinetic
  shell model of coronal heating and acceleration by ion cyclotron waves: 1.
  Outward propagating waves}},}\ }\href {\doibase 10.1029/2000JA000099}
  {\bibfield  {journal} {\bibinfo  {journal} {J.~Geophys.~Res.}\ }\textbf
  {\bibinfo {volume} {106}},\ \bibinfo {pages} {5649--5660} (\bibinfo {year}
  {2001})}\BibitemShut {NoStop}%
\bibitem [{\citenamefont {{Hollweg}}\ and\ \citenamefont
  {{Isenberg}}(2002)}]{Hollweg:2002}%
  \BibitemOpen
  \bibfield  {author} {\bibinfo {author} {\bibfnamefont {J.~V.}\ \bibnamefont
  {{Hollweg}}}\ and\ \bibinfo {author} {\bibfnamefont {P.~A.}\ \bibnamefont
  {{Isenberg}}},\ }\bibfield  {title} {\enquote {\bibinfo {title} {{Generation
  of the fast solar wind: A review with emphasis on the resonant cyclotron
  interaction}},}\ }\href {\doibase 10.1029/2001JA000270} {\bibfield  {journal}
  {\bibinfo  {journal} {J.~Geophys.~Res.}\ }\textbf {\bibinfo {volume} {107}},\
  \bibinfo {eid} {1147} (\bibinfo {year} {2002})}\BibitemShut {NoStop}%
\bibitem [{\citenamefont {{Howes}}\ \emph {et~al.}(2008)\citenamefont
  {{Howes}}, \citenamefont {{Cowley}}, \citenamefont {{Dorland}}, \citenamefont
  {{Hammett}}, \citenamefont {{Quataert}},\ and\ \citenamefont
  {{Schekochihin}}}]{Howes:2008b}%
  \BibitemOpen
  \bibfield  {author} {\bibinfo {author} {\bibfnamefont {G.~G.}\ \bibnamefont
  {{Howes}}}, \bibinfo {author} {\bibfnamefont {S.~C.}\ \bibnamefont
  {{Cowley}}}, \bibinfo {author} {\bibfnamefont {W.}~\bibnamefont {{Dorland}}},
  \bibinfo {author} {\bibfnamefont {G.~W.}\ \bibnamefont {{Hammett}}}, \bibinfo
  {author} {\bibfnamefont {E.}~\bibnamefont {{Quataert}}}, \ and\ \bibinfo
  {author} {\bibfnamefont {A.~A.}\ \bibnamefont {{Schekochihin}}},\ }\bibfield
  {title} {\enquote {\bibinfo {title} {{A model of turbulence in magnetized
  plasmas: Implications for the dissipation range in the solar wind}},}\ }\href
  {\doibase 10.1029/2007JA012665} {\bibfield  {journal} {\bibinfo  {journal}
  {J.~Geophys.~Res.}\ }\textbf {\bibinfo {volume} {113}},\ \bibinfo {pages}
  {A05103} (\bibinfo {year} {2008})},\ \Eprint
  {http://arxiv.org/abs/arXiv:0707.3147} {arXiv:0707.3147} \BibitemShut
  {NoStop}%
\bibitem [{\citenamefont {{Schekochihin}}\ \emph {et~al.}(2009)\citenamefont
  {{Schekochihin}}, \citenamefont {{Cowley}}, \citenamefont {{Dorland}},
  \citenamefont {{Hammett}}, \citenamefont {{Howes}}, \citenamefont
  {{Quataert}},\ and\ \citenamefont {{Tatsuno}}}]{Schekochihin:2009}%
  \BibitemOpen
  \bibfield  {author} {\bibinfo {author} {\bibfnamefont {A.~A.}\ \bibnamefont
  {{Schekochihin}}}, \bibinfo {author} {\bibfnamefont {S.~C.}\ \bibnamefont
  {{Cowley}}}, \bibinfo {author} {\bibfnamefont {W.}~\bibnamefont {{Dorland}}},
  \bibinfo {author} {\bibfnamefont {G.~W.}\ \bibnamefont {{Hammett}}}, \bibinfo
  {author} {\bibfnamefont {G.~G.}\ \bibnamefont {{Howes}}}, \bibinfo {author}
  {\bibfnamefont {E.}~\bibnamefont {{Quataert}}}, \ and\ \bibinfo {author}
  {\bibfnamefont {T.}~\bibnamefont {{Tatsuno}}},\ }\bibfield  {title} {\enquote
  {\bibinfo {title} {{Astrophysical Gyrokinetics: Kinetic and Fluid Turbulent
  Cascades in Magnetized Weakly Collisional Plasmas}},}\ }\href {\doibase
  10.1088/0067-0049/182/1/310} {\bibfield  {journal} {\bibinfo  {journal}
  {Astrophys.~J.~Supp.}\ }\textbf {\bibinfo {volume} {182}},\ \bibinfo {pages}
  {310--377} (\bibinfo {year} {2009})}\BibitemShut {NoStop}%
\bibitem [{\citenamefont {{TenBarge}}\ and\ \citenamefont
  {{Howes}}(2013)}]{TenBarge:2013a}%
  \BibitemOpen
  \bibfield  {author} {\bibinfo {author} {\bibfnamefont {J.~M.}\ \bibnamefont
  {{TenBarge}}}\ and\ \bibinfo {author} {\bibfnamefont {G.~G.}\ \bibnamefont
  {{Howes}}},\ }\bibfield  {title} {\enquote {\bibinfo {title} {{Current Sheets
  and Collisionless Damping in Kinetic Plasma Turbulence}},}\ }\href {\doibase
  10.1088/2041-8205/771/2/L27} {\bibfield  {journal} {\bibinfo  {journal}
  {Astrophys.~J.~Lett.}\ }\textbf {\bibinfo {volume} {771}},\ \bibinfo {eid}
  {L27} (\bibinfo {year} {2013})},\ \Eprint {http://arxiv.org/abs/1304.2958}
  {arXiv:1304.2958 [physics.plasm-ph]} \BibitemShut {NoStop}%
\bibitem [{\citenamefont {Howes}(2015)}]{Howes:2015b}%
  \BibitemOpen
  \bibfield  {author} {\bibinfo {author} {\bibfnamefont {G.~G.}\ \bibnamefont
  {Howes}},\ }\bibfield  {title} {\enquote {\bibinfo {title} {A dynamical model
  of plasma turbulence in the solar wind},}\ }\href {\doibase
  10.1098/rsta.2014.0145} {\bibfield  {journal} {\bibinfo  {journal}
  {Philosophical Transactions of the Royal Society of London A: Mathematical,
  Physical and Engineering Sciences}\ }\textbf {\bibinfo {volume} {373}},\
  \bibinfo {pages} {20140145} (\bibinfo {year} {2015})}\BibitemShut {NoStop}%
\bibitem [{\citenamefont {{Li}}\ \emph {et~al.}(2016)\citenamefont {{Li}},
  \citenamefont {{Howes}}, \citenamefont {{Klein}},\ and\ \citenamefont
  {{TenBarge}}}]{TCLi:2016}%
  \BibitemOpen
  \bibfield  {author} {\bibinfo {author} {\bibfnamefont {T.~C.}\ \bibnamefont
  {{Li}}}, \bibinfo {author} {\bibfnamefont {G.~G.}\ \bibnamefont {{Howes}}},
  \bibinfo {author} {\bibfnamefont {K.~G.}\ \bibnamefont {{Klein}}}, \ and\
  \bibinfo {author} {\bibfnamefont {J.~M.}\ \bibnamefont {{TenBarge}}},\
  }\bibfield  {title} {\enquote {\bibinfo {title} {{Energy Dissipation and
  Landau Damping in Two- and Three-dimensional Plasma Turbulence}},}\ }\href
  {\doibase 10.3847/2041-8205/832/2/L24} {\bibfield  {journal} {\bibinfo
  {journal} {Astrophys.~J.~Lett.}\ }\textbf {\bibinfo {volume} {832}},\
  \bibinfo {eid} {L24} (\bibinfo {year} {2016})},\ \Eprint
  {http://arxiv.org/abs/1510.02842} {arXiv:1510.02842 [astro-ph.SR]}
  \BibitemShut {NoStop}%
\bibitem [{\citenamefont {{Howes}}, \citenamefont {{McCubbin}},\ and\
  \citenamefont {{Klein}}(2018{\natexlab{a}})}]{Howes:2018a}%
  \BibitemOpen
  \bibfield  {author} {\bibinfo {author} {\bibfnamefont {G.~G.}\ \bibnamefont
  {{Howes}}}, \bibinfo {author} {\bibfnamefont {A.~J.}\ \bibnamefont
  {{McCubbin}}}, \ and\ \bibinfo {author} {\bibfnamefont {K.~G.}\ \bibnamefont
  {{Klein}}},\ }\bibfield  {title} {\enquote {\bibinfo {title} {{Spatial
  Localization of Particle Enegization in Current Sheets Produced by Alfv\'en
  Wave Collisions}},}\ }\href@noop {} {\bibfield  {journal} {\bibinfo
  {journal} {J.~Plasma Phys.}\ }\textbf {\bibinfo {volume} {84}},\ \bibinfo
  {pages} {905840105} (\bibinfo {year} {2018}{\natexlab{a}})}\BibitemShut
  {NoStop}%
\bibitem [{\citenamefont {{McChesney}}, \citenamefont {{Bellan}},\ and\
  \citenamefont {{Stern}}(1991)}]{McChesney:1991}%
  \BibitemOpen
  \bibfield  {author} {\bibinfo {author} {\bibfnamefont {J.~M.}\ \bibnamefont
  {{McChesney}}}, \bibinfo {author} {\bibfnamefont {P.~M.}\ \bibnamefont
  {{Bellan}}}, \ and\ \bibinfo {author} {\bibfnamefont {R.~A.}\ \bibnamefont
  {{Stern}}},\ }\bibfield  {title} {\enquote {\bibinfo {title} {{Observation of
  fast stochastic ion heating by drift waves}},}\ }\href {\doibase
  10.1063/1.859768} {\bibfield  {journal} {\bibinfo  {journal} {Phys.~Fluids
  B}\ }\textbf {\bibinfo {volume} {3}},\ \bibinfo {pages} {3363--3378}
  (\bibinfo {year} {1991})}\BibitemShut {NoStop}%
\bibitem [{\citenamefont {{Johnson}}\ and\ \citenamefont
  {{Cheng}}(2001)}]{Johnson:2001}%
  \BibitemOpen
  \bibfield  {author} {\bibinfo {author} {\bibfnamefont {J.~R.}\ \bibnamefont
  {{Johnson}}}\ and\ \bibinfo {author} {\bibfnamefont {C.~Z.}\ \bibnamefont
  {{Cheng}}},\ }\bibfield  {title} {\enquote {\bibinfo {title} {{Stochastic ion
  heating at the magnetopause due to kinetic Alfv{\'e}n waves}},}\ }\href
  {\doibase 10.1029/2001GL013509} {\bibfield  {journal} {\bibinfo  {journal}
  {Geophys.~Res.~Lett.}\ }\textbf {\bibinfo {volume} {28}},\ \bibinfo {pages}
  {4421--4424} (\bibinfo {year} {2001})}\BibitemShut {NoStop}%
\bibitem [{\citenamefont {{Chen}}, \citenamefont {{Lin}},\ and\ \citenamefont
  {{White}}(2001)}]{Chen:2001}%
  \BibitemOpen
  \bibfield  {author} {\bibinfo {author} {\bibfnamefont {L.}~\bibnamefont
  {{Chen}}}, \bibinfo {author} {\bibfnamefont {Z.}~\bibnamefont {{Lin}}}, \
  and\ \bibinfo {author} {\bibfnamefont {R.}~\bibnamefont {{White}}},\
  }\bibfield  {title} {\enquote {\bibinfo {title} {{On resonant heating below
  the cyclotron frequency}},}\ }\href {\doibase 10.1063/1.1406939} {\bibfield
  {journal} {\bibinfo  {journal} {Phys.~Plasmas}\ }\textbf {\bibinfo {volume}
  {8}},\ \bibinfo {pages} {4713--4716} (\bibinfo {year} {2001})}\BibitemShut
  {NoStop}%
\bibitem [{\citenamefont {{White}}, \citenamefont {{Chen}},\ and\ \citenamefont
  {{Lin}}(2002)}]{White:2002}%
  \BibitemOpen
  \bibfield  {author} {\bibinfo {author} {\bibfnamefont {R.}~\bibnamefont
  {{White}}}, \bibinfo {author} {\bibfnamefont {L.}~\bibnamefont {{Chen}}}, \
  and\ \bibinfo {author} {\bibfnamefont {Z.}~\bibnamefont {{Lin}}},\ }\bibfield
   {title} {\enquote {\bibinfo {title} {{Resonant plasma heating below the
  cyclotron frequency}},}\ }\href {\doibase 10.1063/1.1445180} {\bibfield
  {journal} {\bibinfo  {journal} {Phys.~Plasmas}\ }\textbf {\bibinfo {volume}
  {9}},\ \bibinfo {pages} {1890--1897} (\bibinfo {year} {2002})}\BibitemShut
  {NoStop}%
\bibitem [{\citenamefont {{Voitenko}}\ and\ \citenamefont
  {{Goossens}}(2004)}]{Voitenko:2004}%
  \BibitemOpen
  \bibfield  {author} {\bibinfo {author} {\bibfnamefont {Y.}~\bibnamefont
  {{Voitenko}}}\ and\ \bibinfo {author} {\bibfnamefont {M.}~\bibnamefont
  {{Goossens}}},\ }\bibfield  {title} {\enquote {\bibinfo {title} {{Excitation
  of kinetic Alfv{\'e}n turbulence by MHD waves and energization of space
  plasmas}},}\ }\href@noop {} {\bibfield  {journal} {\bibinfo  {journal}
  {Nonlin.~Proc.~Geophys.}\ }\textbf {\bibinfo {volume} {11}},\ \bibinfo
  {pages} {535--543} (\bibinfo {year} {2004})}\BibitemShut {NoStop}%
\bibitem [{\citenamefont {{Bourouaine}}, \citenamefont {{Marsch}},\ and\
  \citenamefont {{Vocks}}(2008)}]{Bourouaine:2008}%
  \BibitemOpen
  \bibfield  {author} {\bibinfo {author} {\bibfnamefont {S.}~\bibnamefont
  {{Bourouaine}}}, \bibinfo {author} {\bibfnamefont {E.}~\bibnamefont
  {{Marsch}}}, \ and\ \bibinfo {author} {\bibfnamefont {C.}~\bibnamefont
  {{Vocks}}},\ }\bibfield  {title} {\enquote {\bibinfo {title} {{On the
  Efficiency of Nonresonant Ion Heating by Coronal Alfv{\'e}n Waves}},}\ }\href
  {\doibase 10.1086/592243} {\bibfield  {journal} {\bibinfo  {journal}
  {Astrophys.~J.~Lett.}\ }\textbf {\bibinfo {volume} {684}},\ \bibinfo {pages}
  {L119--L122} (\bibinfo {year} {2008})}\BibitemShut {NoStop}%
\bibitem [{\citenamefont {{Chandran}}\ \emph {et~al.}(2010)\citenamefont
  {{Chandran}}, \citenamefont {{Li}}, \citenamefont {{Rogers}}, \citenamefont
  {{Quataert}},\ and\ \citenamefont {{Germaschewski}}}]{Chandran:2010a}%
  \BibitemOpen
  \bibfield  {author} {\bibinfo {author} {\bibfnamefont {B.~D.~G.}\
  \bibnamefont {{Chandran}}}, \bibinfo {author} {\bibfnamefont
  {B.}~\bibnamefont {{Li}}}, \bibinfo {author} {\bibfnamefont {B.~N.}\
  \bibnamefont {{Rogers}}}, \bibinfo {author} {\bibfnamefont {E.}~\bibnamefont
  {{Quataert}}}, \ and\ \bibinfo {author} {\bibfnamefont {K.}~\bibnamefont
  {{Germaschewski}}},\ }\bibfield  {title} {\enquote {\bibinfo {title}
  {{Perpendicular Ion Heating by Low-frequency Alfv{\'e}n-wave Turbulence in
  the Solar Wind}},}\ }\href {\doibase 10.1088/0004-637X/720/1/503} {\bibfield
  {journal} {\bibinfo  {journal} {Astrophys.~J.}\ }\textbf {\bibinfo {volume}
  {720}},\ \bibinfo {pages} {503--515} (\bibinfo {year} {2010})}\BibitemShut
  {NoStop}%
\bibitem [{\citenamefont {{Chandran}}\ \emph {et~al.}(2011)\citenamefont
  {{Chandran}}, \citenamefont {{Dennis}}, \citenamefont {{Quataert}},\ and\
  \citenamefont {{Bale}}}]{Chandran:2011}%
  \BibitemOpen
  \bibfield  {author} {\bibinfo {author} {\bibfnamefont {B.~D.~G.}\
  \bibnamefont {{Chandran}}}, \bibinfo {author} {\bibfnamefont {T.~J.}\
  \bibnamefont {{Dennis}}}, \bibinfo {author} {\bibfnamefont {E.}~\bibnamefont
  {{Quataert}}}, \ and\ \bibinfo {author} {\bibfnamefont {S.~D.}\ \bibnamefont
  {{Bale}}},\ }\bibfield  {title} {\enquote {\bibinfo {title} {{Incorporating
  Kinetic Physics into a Two-fluid Solar-wind Model with Temperature Anisotropy
  and Low-frequency Alfv{\'e}n-wave Turbulence}},}\ }\href {\doibase
  10.1088/0004-637X/743/2/197} {\bibfield  {journal} {\bibinfo  {journal}
  {Astrophys.~J.}\ }\textbf {\bibinfo {volume} {743}},\ \bibinfo {eid} {197}
  (\bibinfo {year} {2011})},\ \Eprint {http://arxiv.org/abs/1110.3029}
  {arXiv:1110.3029 [astro-ph.SR]} \BibitemShut {NoStop}%
\bibitem [{\citenamefont {{Bourouaine}}\ and\ \citenamefont
  {{Chandran}}(2013)}]{Bourouaine:2013}%
  \BibitemOpen
  \bibfield  {author} {\bibinfo {author} {\bibfnamefont {S.}~\bibnamefont
  {{Bourouaine}}}\ and\ \bibinfo {author} {\bibfnamefont {B.~D.~G.}\
  \bibnamefont {{Chandran}}},\ }\bibfield  {title} {\enquote {\bibinfo {title}
  {{Observational Test of Stochastic Heating in Low-{$\beta$} Fast-solar-wind
  Streams}},}\ }\href {\doibase 10.1088/0004-637X/774/2/96} {\bibfield
  {journal} {\bibinfo  {journal} {Astrophys.~J.}\ }\textbf {\bibinfo {volume}
  {774}},\ \bibinfo {eid} {96} (\bibinfo {year} {2013})},\ \Eprint
  {http://arxiv.org/abs/1307.3789} {arXiv:1307.3789 [astro-ph.SR]} \BibitemShut
  {NoStop}%
\bibitem [{\citenamefont {{Ambrosiano}}\ \emph {et~al.}(1988)\citenamefont
  {{Ambrosiano}}, \citenamefont {{Matthaeus}}, \citenamefont {{Goldstein}},\
  and\ \citenamefont {{Plante}}}]{Ambrosiano:1988}%
  \BibitemOpen
  \bibfield  {author} {\bibinfo {author} {\bibfnamefont {J.}~\bibnamefont
  {{Ambrosiano}}}, \bibinfo {author} {\bibfnamefont {W.~H.}\ \bibnamefont
  {{Matthaeus}}}, \bibinfo {author} {\bibfnamefont {M.~L.}\ \bibnamefont
  {{Goldstein}}}, \ and\ \bibinfo {author} {\bibfnamefont {D.}~\bibnamefont
  {{Plante}}},\ }\bibfield  {title} {\enquote {\bibinfo {title} {{Test particle
  acceleration in turbulent reconnecting magnetic fields}},}\ }\href {\doibase
  10.1029/JA093iA12p14383} {\bibfield  {journal} {\bibinfo  {journal}
  {J.~Geophys.~Res.}\ }\textbf {\bibinfo {volume} {93}},\ \bibinfo {pages}
  {14383--14400} (\bibinfo {year} {1988})}\BibitemShut {NoStop}%
\bibitem [{\citenamefont {{Dmitruk}}, \citenamefont {{Matthaeus}},\ and\
  \citenamefont {{Seenu}}(2004)}]{Dmitruk:2004}%
  \BibitemOpen
  \bibfield  {author} {\bibinfo {author} {\bibfnamefont {P.}~\bibnamefont
  {{Dmitruk}}}, \bibinfo {author} {\bibfnamefont {W.~H.}\ \bibnamefont
  {{Matthaeus}}}, \ and\ \bibinfo {author} {\bibfnamefont {N.}~\bibnamefont
  {{Seenu}}},\ }\bibfield  {title} {\enquote {\bibinfo {title} {{Test Particle
  Energization by Current Sheets and Nonuniform Fields in Magnetohydrodynamic
  Turbulence}},}\ }\href {\doibase 10.1086/425301} {\bibfield  {journal}
  {\bibinfo  {journal} {Astrophys.~J.}\ }\textbf {\bibinfo {volume} {617}},\
  \bibinfo {pages} {667--679} (\bibinfo {year} {2004})}\BibitemShut {NoStop}%
\bibitem [{\citenamefont {{Markovskii}}\ and\ \citenamefont
  {{Vasquez}}(2011)}]{Markovskii:2011}%
  \BibitemOpen
  \bibfield  {author} {\bibinfo {author} {\bibfnamefont {S.~A.}\ \bibnamefont
  {{Markovskii}}}\ and\ \bibinfo {author} {\bibfnamefont {B.~J.}\ \bibnamefont
  {{Vasquez}}},\ }\bibfield  {title} {\enquote {\bibinfo {title} {{A
  Short-timescale Channel of Dissipation of the Strong Solar Wind
  Turbulence}},}\ }\href {\doibase 10.1088/0004-637X/739/1/22} {\bibfield
  {journal} {\bibinfo  {journal} {Astrophys.~J.}\ }\textbf {\bibinfo {volume}
  {739}},\ \bibinfo {eid} {22} (\bibinfo {year} {2011})}\BibitemShut {NoStop}%
\bibitem [{\citenamefont {{Matthaeus}}\ and\ \citenamefont
  {{Velli}}(2011)}]{Matthaeus:2011}%
  \BibitemOpen
  \bibfield  {author} {\bibinfo {author} {\bibfnamefont {W.~H.}\ \bibnamefont
  {{Matthaeus}}}\ and\ \bibinfo {author} {\bibfnamefont {M.}~\bibnamefont
  {{Velli}}},\ }\bibfield  {title} {\enquote {\bibinfo {title} {{Who Needs
  Turbulence?. A Review of Turbulence Effects in the Heliosphere and on the
  Fundamental Process of Reconnection}},}\ }\href {\doibase
  10.1007/s11214-011-9793-9} {\bibfield  {journal} {\bibinfo  {journal} {Space
  Sci.~Rev.}\ }\textbf {\bibinfo {volume} {160}},\ \bibinfo {pages} {145--168}
  (\bibinfo {year} {2011})}\BibitemShut {NoStop}%
\bibitem [{\citenamefont {{Osman}}\ \emph {et~al.}(2011)\citenamefont
  {{Osman}}, \citenamefont {{Matthaeus}}, \citenamefont {{Greco}},\ and\
  \citenamefont {{Servidio}}}]{Osman:2011}%
  \BibitemOpen
  \bibfield  {author} {\bibinfo {author} {\bibfnamefont {K.~T.}\ \bibnamefont
  {{Osman}}}, \bibinfo {author} {\bibfnamefont {W.~H.}\ \bibnamefont
  {{Matthaeus}}}, \bibinfo {author} {\bibfnamefont {A.}~\bibnamefont
  {{Greco}}}, \ and\ \bibinfo {author} {\bibfnamefont {S.}~\bibnamefont
  {{Servidio}}},\ }\bibfield  {title} {\enquote {\bibinfo {title} {{Evidence
  for Inhomogeneous Heating in the Solar Wind}},}\ }\href {\doibase
  10.1088/2041-8205/727/1/L11} {\bibfield  {journal} {\bibinfo  {journal}
  {Astrophys.~J.~Lett.}\ }\textbf {\bibinfo {volume} {727}},\ \bibinfo {pages}
  {L11} (\bibinfo {year} {2011})}\BibitemShut {NoStop}%
\bibitem [{\citenamefont {{Servidio}}\ \emph {et~al.}(2011)\citenamefont
  {{Servidio}}, \citenamefont {{Greco}}, \citenamefont {{Matthaeus}},
  \citenamefont {{Osman}},\ and\ \citenamefont {{Dmitruk}}}]{Servidio:2011a}%
  \BibitemOpen
  \bibfield  {author} {\bibinfo {author} {\bibfnamefont {S.}~\bibnamefont
  {{Servidio}}}, \bibinfo {author} {\bibfnamefont {A.}~\bibnamefont {{Greco}}},
  \bibinfo {author} {\bibfnamefont {W.~H.}\ \bibnamefont {{Matthaeus}}},
  \bibinfo {author} {\bibfnamefont {K.~T.}\ \bibnamefont {{Osman}}}, \ and\
  \bibinfo {author} {\bibfnamefont {P.}~\bibnamefont {{Dmitruk}}},\ }\bibfield
  {title} {\enquote {\bibinfo {title} {{Statistical association of
  discontinuities and reconnection in magnetohydrodynamic turbulence}},}\
  }\href {\doibase 10.1029/2011JA016569} {\bibfield  {journal} {\bibinfo
  {journal} {J.~Geophys.~Res.}\ }\textbf {\bibinfo {volume} {116}},\ \bibinfo
  {eid} {A09102} (\bibinfo {year} {2011})}\BibitemShut {NoStop}%
\bibitem [{\citenamefont {{Osman}}\ \emph
  {et~al.}(2012{\natexlab{a}})\citenamefont {{Osman}}, \citenamefont
  {{Matthaeus}}, \citenamefont {{Wan}},\ and\ \citenamefont
  {{Rappazzo}}}]{Osman:2012a}%
  \BibitemOpen
  \bibfield  {author} {\bibinfo {author} {\bibfnamefont {K.~T.}\ \bibnamefont
  {{Osman}}}, \bibinfo {author} {\bibfnamefont {W.~H.}\ \bibnamefont
  {{Matthaeus}}}, \bibinfo {author} {\bibfnamefont {M.}~\bibnamefont {{Wan}}},
  \ and\ \bibinfo {author} {\bibfnamefont {A.~F.}\ \bibnamefont {{Rappazzo}}},\
  }\bibfield  {title} {\enquote {\bibinfo {title} {{Intermittency and Local
  Heating in the Solar Wind}},}\ }\href {\doibase
  10.1103/PhysRevLett.108.261102} {\bibfield  {journal} {\bibinfo  {journal}
  {Phys.~Rev.~Lett.}\ }\textbf {\bibinfo {volume} {108}},\ \bibinfo {eid}
  {261102} (\bibinfo {year} {2012}{\natexlab{a}})},\ \Eprint
  {http://arxiv.org/abs/1111.6921} {arXiv:1111.6921 [physics.space-ph]}
  \BibitemShut {NoStop}%
\bibitem [{\citenamefont {{Osman}}\ \emph
  {et~al.}(2012{\natexlab{b}})\citenamefont {{Osman}}, \citenamefont
  {{Matthaeus}}, \citenamefont {{Hnat}},\ and\ \citenamefont
  {{Chapman}}}]{Osman:2012b}%
  \BibitemOpen
  \bibfield  {author} {\bibinfo {author} {\bibfnamefont {K.~T.}\ \bibnamefont
  {{Osman}}}, \bibinfo {author} {\bibfnamefont {W.~H.}\ \bibnamefont
  {{Matthaeus}}}, \bibinfo {author} {\bibfnamefont {B.}~\bibnamefont {{Hnat}}},
  \ and\ \bibinfo {author} {\bibfnamefont {S.~C.}\ \bibnamefont {{Chapman}}},\
  }\bibfield  {title} {\enquote {\bibinfo {title} {{Kinetic Signatures and
  Intermittent Turbulence in the Solar Wind Plasma}},}\ }\href {\doibase
  10.1103/PhysRevLett.108.261103} {\bibfield  {journal} {\bibinfo  {journal}
  {Phys.~Rev.~Lett.}\ }\textbf {\bibinfo {volume} {108}},\ \bibinfo {eid}
  {261103} (\bibinfo {year} {2012}{\natexlab{b}})},\ \Eprint
  {http://arxiv.org/abs/1203.6596} {arXiv:1203.6596 [physics.space-ph]}
  \BibitemShut {NoStop}%
\bibitem [{\citenamefont {{Wan}}\ \emph {et~al.}(2012)\citenamefont {{Wan}},
  \citenamefont {{Matthaeus}}, \citenamefont {{Karimabadi}}, \citenamefont
  {{Roytershteyn}}, \citenamefont {{Shay}}, \citenamefont {{Wu}}, \citenamefont
  {{Daughton}}, \citenamefont {{Loring}},\ and\ \citenamefont
  {{Chapman}}}]{Wan:2012}%
  \BibitemOpen
  \bibfield  {author} {\bibinfo {author} {\bibfnamefont {M.}~\bibnamefont
  {{Wan}}}, \bibinfo {author} {\bibfnamefont {W.~H.}\ \bibnamefont
  {{Matthaeus}}}, \bibinfo {author} {\bibfnamefont {H.}~\bibnamefont
  {{Karimabadi}}}, \bibinfo {author} {\bibfnamefont {V.}~\bibnamefont
  {{Roytershteyn}}}, \bibinfo {author} {\bibfnamefont {M.}~\bibnamefont
  {{Shay}}}, \bibinfo {author} {\bibfnamefont {P.}~\bibnamefont {{Wu}}},
  \bibinfo {author} {\bibfnamefont {W.}~\bibnamefont {{Daughton}}}, \bibinfo
  {author} {\bibfnamefont {B.}~\bibnamefont {{Loring}}}, \ and\ \bibinfo
  {author} {\bibfnamefont {S.~C.}\ \bibnamefont {{Chapman}}},\ }\bibfield
  {title} {\enquote {\bibinfo {title} {{Intermittent Dissipation at Kinetic
  Scales in Collisionless Plasma Turbulence}},}\ }\href {\doibase
  10.1103/PhysRevLett.109.195001} {\bibfield  {journal} {\bibinfo  {journal}
  {Phys.~Rev.~Lett.}\ }\textbf {\bibinfo {volume} {109}},\ \bibinfo {eid}
  {195001} (\bibinfo {year} {2012})}\BibitemShut {NoStop}%
\bibitem [{\citenamefont {{Karimabadi}}\ \emph {et~al.}(2013)\citenamefont
  {{Karimabadi}}, \citenamefont {{Roytershteyn}}, \citenamefont {{Wan}},
  \citenamefont {{Matthaeus}}, \citenamefont {{Daughton}}, \citenamefont
  {{Wu}}, \citenamefont {{Shay}}, \citenamefont {{Loring}}, \citenamefont
  {{Borovsky}}, \citenamefont {{Leonardis}}, \citenamefont {{Chapman}},\ and\
  \citenamefont {{Nakamura}}}]{Karimabadi:2013}%
  \BibitemOpen
  \bibfield  {author} {\bibinfo {author} {\bibfnamefont {H.}~\bibnamefont
  {{Karimabadi}}}, \bibinfo {author} {\bibfnamefont {V.}~\bibnamefont
  {{Roytershteyn}}}, \bibinfo {author} {\bibfnamefont {M.}~\bibnamefont
  {{Wan}}}, \bibinfo {author} {\bibfnamefont {W.~H.}\ \bibnamefont
  {{Matthaeus}}}, \bibinfo {author} {\bibfnamefont {W.}~\bibnamefont
  {{Daughton}}}, \bibinfo {author} {\bibfnamefont {P.}~\bibnamefont {{Wu}}},
  \bibinfo {author} {\bibfnamefont {M.}~\bibnamefont {{Shay}}}, \bibinfo
  {author} {\bibfnamefont {B.}~\bibnamefont {{Loring}}}, \bibinfo {author}
  {\bibfnamefont {J.}~\bibnamefont {{Borovsky}}}, \bibinfo {author}
  {\bibfnamefont {E.}~\bibnamefont {{Leonardis}}}, \bibinfo {author}
  {\bibfnamefont {S.~C.}\ \bibnamefont {{Chapman}}}, \ and\ \bibinfo {author}
  {\bibfnamefont {T.~K.~M.}\ \bibnamefont {{Nakamura}}},\ }\bibfield  {title}
  {\enquote {\bibinfo {title} {{Coherent structures, intermittent turbulence,
  and dissipation in high-temperature plasmas}},}\ }\href {\doibase
  10.1063/1.4773205} {\bibfield  {journal} {\bibinfo  {journal}
  {Phys.~Plasmas}\ }\textbf {\bibinfo {volume} {20}},\ \bibinfo {pages}
  {012303} (\bibinfo {year} {2013})}\BibitemShut {NoStop}%
\bibitem [{\citenamefont {{Zhdankin}}\ \emph {et~al.}(2013)\citenamefont
  {{Zhdankin}}, \citenamefont {{Uzdensky}}, \citenamefont {{Perez}},\ and\
  \citenamefont {{Boldyrev}}}]{Zhdankin:2013}%
  \BibitemOpen
  \bibfield  {author} {\bibinfo {author} {\bibfnamefont {V.}~\bibnamefont
  {{Zhdankin}}}, \bibinfo {author} {\bibfnamefont {D.~A.}\ \bibnamefont
  {{Uzdensky}}}, \bibinfo {author} {\bibfnamefont {J.~C.}\ \bibnamefont
  {{Perez}}}, \ and\ \bibinfo {author} {\bibfnamefont {S.}~\bibnamefont
  {{Boldyrev}}},\ }\bibfield  {title} {\enquote {\bibinfo {title} {{Statistical
  Analysis of Current Sheets in Three-dimensional Magnetohydrodynamic
  Turbulence}},}\ }\href {\doibase 10.1088/0004-637X/771/2/124} {\bibfield
  {journal} {\bibinfo  {journal} {Astrophys.~J.}\ }\textbf {\bibinfo {volume}
  {771}},\ \bibinfo {eid} {124} (\bibinfo {year} {2013})},\ \Eprint
  {http://arxiv.org/abs/1302.1460} {arXiv:1302.1460 [astro-ph.HE]} \BibitemShut
  {NoStop}%
\bibitem [{\citenamefont {{Dalena}}\ \emph {et~al.}(2014)\citenamefont
  {{Dalena}}, \citenamefont {{Rappazzo}}, \citenamefont {{Dmitruk}},
  \citenamefont {{Greco}},\ and\ \citenamefont {{Matthaeus}}}]{Dalena:2014}%
  \BibitemOpen
  \bibfield  {author} {\bibinfo {author} {\bibfnamefont {S.}~\bibnamefont
  {{Dalena}}}, \bibinfo {author} {\bibfnamefont {A.~F.}\ \bibnamefont
  {{Rappazzo}}}, \bibinfo {author} {\bibfnamefont {P.}~\bibnamefont
  {{Dmitruk}}}, \bibinfo {author} {\bibfnamefont {A.}~\bibnamefont {{Greco}}},
  \ and\ \bibinfo {author} {\bibfnamefont {W.~H.}\ \bibnamefont
  {{Matthaeus}}},\ }\bibfield  {title} {\enquote {\bibinfo {title}
  {{Test-particle Acceleration in a Hierarchical Three-dimensional Turbulence
  Model}},}\ }\href {\doibase 10.1088/0004-637X/783/2/143} {\bibfield
  {journal} {\bibinfo  {journal} {Astrophys.~J.}\ }\textbf {\bibinfo {volume}
  {783}},\ \bibinfo {eid} {143} (\bibinfo {year} {2014})},\ \Eprint
  {http://arxiv.org/abs/1402.3745} {arXiv:1402.3745 [astro-ph.SR]} \BibitemShut
  {NoStop}%
\bibitem [{\citenamefont {{Osman}}\ \emph
  {et~al.}(2014{\natexlab{a}})\citenamefont {{Osman}}, \citenamefont
  {{Kiyani}}, \citenamefont {{Chapman}},\ and\ \citenamefont
  {{Hnat}}}]{Osman:2014a}%
  \BibitemOpen
  \bibfield  {author} {\bibinfo {author} {\bibfnamefont {K.~T.}\ \bibnamefont
  {{Osman}}}, \bibinfo {author} {\bibfnamefont {K.~H.}\ \bibnamefont
  {{Kiyani}}}, \bibinfo {author} {\bibfnamefont {S.~C.}\ \bibnamefont
  {{Chapman}}}, \ and\ \bibinfo {author} {\bibfnamefont {B.}~\bibnamefont
  {{Hnat}}},\ }\bibfield  {title} {\enquote {\bibinfo {title} {{Anisotropic
  Intermittency of Magnetohydrodynamic Turbulence}},}\ }\href {\doibase
  10.1088/2041-8205/783/2/L27} {\bibfield  {journal} {\bibinfo  {journal}
  {Astrophys.~J.~Lett.}\ }\textbf {\bibinfo {volume} {783}},\ \bibinfo {eid}
  {L27} (\bibinfo {year} {2014}{\natexlab{a}})},\ \Eprint
  {http://arxiv.org/abs/1311.5938} {arXiv:1311.5938 [physics.space-ph]}
  \BibitemShut {NoStop}%
\bibitem [{\citenamefont {{Osman}}\ \emph
  {et~al.}(2014{\natexlab{b}})\citenamefont {{Osman}}, \citenamefont
  {{Matthaeus}}, \citenamefont {{Gosling}}, \citenamefont {{Greco}},
  \citenamefont {{Servidio}}, \citenamefont {{Hnat}}, \citenamefont
  {{Chapman}},\ and\ \citenamefont {{Phan}}}]{Osman:2014b}%
  \BibitemOpen
  \bibfield  {author} {\bibinfo {author} {\bibfnamefont {K.~T.}\ \bibnamefont
  {{Osman}}}, \bibinfo {author} {\bibfnamefont {W.~H.}\ \bibnamefont
  {{Matthaeus}}}, \bibinfo {author} {\bibfnamefont {J.~T.}\ \bibnamefont
  {{Gosling}}}, \bibinfo {author} {\bibfnamefont {A.}~\bibnamefont {{Greco}}},
  \bibinfo {author} {\bibfnamefont {S.}~\bibnamefont {{Servidio}}}, \bibinfo
  {author} {\bibfnamefont {B.}~\bibnamefont {{Hnat}}}, \bibinfo {author}
  {\bibfnamefont {S.~C.}\ \bibnamefont {{Chapman}}}, \ and\ \bibinfo {author}
  {\bibfnamefont {T.~D.}\ \bibnamefont {{Phan}}},\ }\bibfield  {title}
  {\enquote {\bibinfo {title} {{Magnetic Reconnection and Intermittent
  Turbulence in the Solar Wind}},}\ }\href {\doibase
  10.1103/PhysRevLett.112.215002} {\bibfield  {journal} {\bibinfo  {journal}
  {Phys.~Rev.~Lett.}\ }\textbf {\bibinfo {volume} {112}},\ \bibinfo {eid}
  {215002} (\bibinfo {year} {2014}{\natexlab{b}})},\ \Eprint
  {http://arxiv.org/abs/1403.4590} {arXiv:1403.4590 [physics.space-ph]}
  \BibitemShut {NoStop}%
\bibitem [{\citenamefont {{Zhdankin}}, \citenamefont {{Uzdensky}},\ and\
  \citenamefont {{Boldyrev}}(2015{\natexlab{a}})}]{Zhdankin:2015a}%
  \BibitemOpen
  \bibfield  {author} {\bibinfo {author} {\bibfnamefont {V.}~\bibnamefont
  {{Zhdankin}}}, \bibinfo {author} {\bibfnamefont {D.~A.}\ \bibnamefont
  {{Uzdensky}}}, \ and\ \bibinfo {author} {\bibfnamefont {S.}~\bibnamefont
  {{Boldyrev}}},\ }\bibfield  {title} {\enquote {\bibinfo {title} {{Temporal
  Intermittency of Energy Dissipation in Magnetohydrodynamic Turbulence}},}\
  }\href {\doibase 10.1103/PhysRevLett.114.065002} {\bibfield  {journal}
  {\bibinfo  {journal} {Phys.~Rev.~Lett.}\ }\textbf {\bibinfo {volume} {114}},\
  \bibinfo {eid} {065002} (\bibinfo {year} {2015}{\natexlab{a}})},\ \Eprint
  {http://arxiv.org/abs/1501.01664} {arXiv:1501.01664 [astro-ph.HE]}
  \BibitemShut {NoStop}%
\bibitem [{\citenamefont {{Zhdankin}}, \citenamefont {{Uzdensky}},\ and\
  \citenamefont {{Boldyrev}}(2015{\natexlab{b}})}]{Zhdankin:2015b}%
  \BibitemOpen
  \bibfield  {author} {\bibinfo {author} {\bibfnamefont {V.}~\bibnamefont
  {{Zhdankin}}}, \bibinfo {author} {\bibfnamefont {D.~A.}\ \bibnamefont
  {{Uzdensky}}}, \ and\ \bibinfo {author} {\bibfnamefont {S.}~\bibnamefont
  {{Boldyrev}}},\ }\bibfield  {title} {\enquote {\bibinfo {title} {{Temporal
  Analysis of Dissipative Structures in Magnetohydrodynamic Turbulence}},}\
  }\href {\doibase 10.1088/0004-637X/811/1/6} {\bibfield  {journal} {\bibinfo
  {journal} {Astrophys.~J.}\ }\textbf {\bibinfo {volume} {811}},\ \bibinfo
  {eid} {6} (\bibinfo {year} {2015}{\natexlab{b}})},\ \Eprint
  {http://arxiv.org/abs/1506.08356} {arXiv:1506.08356 [astro-ph.HE]}
  \BibitemShut {NoStop}%
\bibitem [{\citenamefont {{Cranmer}}\ \emph {et~al.}(2009)\citenamefont
  {{Cranmer}}, \citenamefont {{Matthaeus}}, \citenamefont {{Breech}},\ and\
  \citenamefont {{Kasper}}}]{Cranmer:2009a}%
  \BibitemOpen
  \bibfield  {author} {\bibinfo {author} {\bibfnamefont {S.~R.}\ \bibnamefont
  {{Cranmer}}}, \bibinfo {author} {\bibfnamefont {W.~H.}\ \bibnamefont
  {{Matthaeus}}}, \bibinfo {author} {\bibfnamefont {B.~A.}\ \bibnamefont
  {{Breech}}}, \ and\ \bibinfo {author} {\bibfnamefont {J.~C.}\ \bibnamefont
  {{Kasper}}},\ }\bibfield  {title} {\enquote {\bibinfo {title} {{Empirical
  Constraints on Proton and Electron Heating in the Fast Solar Wind}},}\ }\href
  {\doibase 10.1088/0004-637X/702/2/1604} {\bibfield  {journal} {\bibinfo
  {journal} {Astrophys.~J.}\ }\textbf {\bibinfo {volume} {702}},\ \bibinfo
  {pages} {1604--1614} (\bibinfo {year} {2009})},\ \Eprint
  {http://arxiv.org/abs/0907.2650} {arXiv:0907.2650 [astro-ph.SR]} \BibitemShut
  {NoStop}%
\bibitem [{\citenamefont {{Howes}}(2011)}]{Howes:2011c}%
  \BibitemOpen
  \bibfield  {author} {\bibinfo {author} {\bibfnamefont {G.~G.}\ \bibnamefont
  {{Howes}}},\ }\bibfield  {title} {\enquote {\bibinfo {title} {{Prediction of
  the Proton-to-total Turbulent Heating in the Solar Wind}},}\ }\href {\doibase
  10.1088/0004-637X/738/1/40} {\bibfield  {journal} {\bibinfo  {journal}
  {Astrophys.~J.}\ }\textbf {\bibinfo {volume} {738}},\ \bibinfo {eid} {40}
  (\bibinfo {year} {2011})},\ \Eprint {http://arxiv.org/abs/1106.4328}
  {arXiv:1106.4328} \BibitemShut {NoStop}%
\bibitem [{\citenamefont {{Fox}}\ \emph {et~al.}(2016)\citenamefont {{Fox}},
  \citenamefont {{Velli}}, \citenamefont {{Bale}}, \citenamefont {{Decker}},
  \citenamefont {{Driesman}}, \citenamefont {{Howard}}, \citenamefont
  {{Kasper}}, \citenamefont {{Kinnison}}, \citenamefont {{Kusterer}},
  \citenamefont {{Lario}}, \citenamefont {{Lockwood}}, \citenamefont
  {{McComas}}, \citenamefont {{Raouafi}},\ and\ \citenamefont
  {{Szabo}}}]{Fox:2016}%
  \BibitemOpen
  \bibfield  {author} {\bibinfo {author} {\bibfnamefont {N.~J.}\ \bibnamefont
  {{Fox}}}, \bibinfo {author} {\bibfnamefont {M.~C.}\ \bibnamefont {{Velli}}},
  \bibinfo {author} {\bibfnamefont {S.~D.}\ \bibnamefont {{Bale}}}, \bibinfo
  {author} {\bibfnamefont {R.}~\bibnamefont {{Decker}}}, \bibinfo {author}
  {\bibfnamefont {A.}~\bibnamefont {{Driesman}}}, \bibinfo {author}
  {\bibfnamefont {R.~A.}\ \bibnamefont {{Howard}}}, \bibinfo {author}
  {\bibfnamefont {J.~C.}\ \bibnamefont {{Kasper}}}, \bibinfo {author}
  {\bibfnamefont {J.}~\bibnamefont {{Kinnison}}}, \bibinfo {author}
  {\bibfnamefont {M.}~\bibnamefont {{Kusterer}}}, \bibinfo {author}
  {\bibfnamefont {D.}~\bibnamefont {{Lario}}}, \bibinfo {author} {\bibfnamefont
  {M.~K.}\ \bibnamefont {{Lockwood}}}, \bibinfo {author} {\bibfnamefont
  {D.~J.}\ \bibnamefont {{McComas}}}, \bibinfo {author} {\bibfnamefont {N.~E.}\
  \bibnamefont {{Raouafi}}}, \ and\ \bibinfo {author} {\bibfnamefont
  {A.}~\bibnamefont {{Szabo}}},\ }\bibfield  {title} {\enquote {\bibinfo
  {title} {{The Solar Probe Plus Mission: Humanity's First Visit to Our
  Star}},}\ }\href {\doibase 10.1007/s11214-015-0211-6} {\bibfield  {journal}
  {\bibinfo  {journal} {Space Sci.~Rev.}\ }\textbf {\bibinfo {volume} {204}},\
  \bibinfo {pages} {7--48} (\bibinfo {year} {2016})}\BibitemShut {NoStop}%
\bibitem [{\citenamefont {{Similon}}\ and\ \citenamefont
  {{Sudan}}(1989)}]{Similon:1989}%
  \BibitemOpen
  \bibfield  {author} {\bibinfo {author} {\bibfnamefont {P.~L.}\ \bibnamefont
  {{Similon}}}\ and\ \bibinfo {author} {\bibfnamefont {R.~N.}\ \bibnamefont
  {{Sudan}}},\ }\bibfield  {title} {\enquote {\bibinfo {title} {{Energy
  dissipation of Alfven wave packets deformed by irregular magnetic fields in
  solar-coronal arches}},}\ }\href {\doibase 10.1086/167023} {\bibfield
  {journal} {\bibinfo  {journal} {Astrophys.~J.}\ }\textbf {\bibinfo {volume}
  {336}},\ \bibinfo {pages} {442--453} (\bibinfo {year} {1989})}\BibitemShut
  {NoStop}%
\bibitem [{\citenamefont {Maron}\ and\ \citenamefont
  {Goldreich}(2001)}]{Maron:2001}%
  \BibitemOpen
  \bibfield  {author} {\bibinfo {author} {\bibfnamefont {J.}~\bibnamefont
  {Maron}}\ and\ \bibinfo {author} {\bibfnamefont {P.}~\bibnamefont
  {Goldreich}},\ }\bibfield  {title} {\enquote {\bibinfo {title} {Simulations
  of incompressible magnetohydrodynamic turbulence},}\ }\href@noop {}
  {\bibfield  {journal} {\bibinfo  {journal} {Astrophys.~J.}\ }\textbf
  {\bibinfo {volume} {554}},\ \bibinfo {pages} {1175--1196} (\bibinfo {year}
  {2001})}\BibitemShut {NoStop}%
\bibitem [{\citenamefont {{Howes}}\ and\ \citenamefont
  {{Bourouaine}}(2017)}]{Howes:2017b}%
  \BibitemOpen
  \bibfield  {author} {\bibinfo {author} {\bibfnamefont {G.~G.}\ \bibnamefont
  {{Howes}}}\ and\ \bibinfo {author} {\bibfnamefont {S.}~\bibnamefont
  {{Bourouaine}}},\ }\bibfield  {title} {\enquote {\bibinfo {title} {{The
  Development of Magnetic Field Line Wander by Plasma Turbulence}},}\ }\href
  {\doibase 10.1017/S0022377817000617} {\bibfield  {journal} {\bibinfo
  {journal} {J.~Plasma Phys.}\ }\textbf {\bibinfo {volume} {83}},\ \bibinfo
  {pages} {905830408} (\bibinfo {year} {2017})}\BibitemShut {NoStop}%
\bibitem [{\citenamefont {{Bourouaine}}\ and\ \citenamefont
  {{Howes}}(2017)}]{Bourouaine:2017}%
  \BibitemOpen
  \bibfield  {author} {\bibinfo {author} {\bibfnamefont {S.}~\bibnamefont
  {{Bourouaine}}}\ and\ \bibinfo {author} {\bibfnamefont {G.~G.}\ \bibnamefont
  {{Howes}}},\ }\bibfield  {title} {\enquote {\bibinfo {title} {{The
  development of magnetic field line wander in gyrokinetic plasma turbulence:
  dependence on amplitude of turbulence}},}\ }\href {\doibase
  10.1017/S0022377817000319} {\bibfield  {journal} {\bibinfo  {journal}
  {Journal of Plasma Physics}\ }\textbf {\bibinfo {volume} {83}},\ \bibinfo
  {eid} {905830301} (\bibinfo {year} {2017})}\BibitemShut {NoStop}%
\bibitem [{\citenamefont {{Parker}}(1965)}]{Parker:1965}%
  \BibitemOpen
  \bibfield  {author} {\bibinfo {author} {\bibfnamefont {E.~N.}\ \bibnamefont
  {{Parker}}},\ }\bibfield  {title} {\enquote {\bibinfo {title} {{The passage
  of energetic charged particles through interplanetary space}},}\ }\href
  {\doibase 10.1016/0032-0633(65)90131-5} {\bibfield  {journal} {\bibinfo
  {journal} {Planet. Space Sci.}\ }\textbf {\bibinfo {volume} {13}},\ \bibinfo
  {pages} {9--49} (\bibinfo {year} {1965})}\BibitemShut {NoStop}%
\bibitem [{\citenamefont {{Jokipii}}(1966)}]{Jokipii:1966}%
  \BibitemOpen
  \bibfield  {author} {\bibinfo {author} {\bibfnamefont {J.~R.}\ \bibnamefont
  {{Jokipii}}},\ }\bibfield  {title} {\enquote {\bibinfo {title} {{Cosmic-Ray
  Propagation. I. Charged Particles in a Random Magnetic Field}},}\ }\href@noop
  {} {\bibfield  {journal} {\bibinfo  {journal} {Astrophys.~J.}\ }\textbf
  {\bibinfo {volume} {146}},\ \bibinfo {pages} {480} (\bibinfo {year}
  {1966})}\BibitemShut {NoStop}%
\bibitem [{\citenamefont {{Reames}}(2013)}]{Reames:2013}%
  \BibitemOpen
  \bibfield  {author} {\bibinfo {author} {\bibfnamefont {D.~V.}\ \bibnamefont
  {{Reames}}},\ }\bibfield  {title} {\enquote {\bibinfo {title} {{The Two
  Sources of Solar Energetic Particles}},}\ }\href {\doibase
  10.1007/s11214-013-9958-9} {\bibfield  {journal} {\bibinfo  {journal} {Space
  Sci.~Rev.}\ }\textbf {\bibinfo {volume} {175}},\ \bibinfo {pages} {53--92}
  (\bibinfo {year} {2013})},\ \Eprint {http://arxiv.org/abs/1306.3608}
  {arXiv:1306.3608 [astro-ph.SR]} \BibitemShut {NoStop}%
\bibitem [{\citenamefont {{Cummings}}\ and\ \citenamefont
  {{Stone}}(1996)}]{Cummings:1996}%
  \BibitemOpen
  \bibfield  {author} {\bibinfo {author} {\bibfnamefont {A.~C.}\ \bibnamefont
  {{Cummings}}}\ and\ \bibinfo {author} {\bibfnamefont {E.~C.}\ \bibnamefont
  {{Stone}}},\ }\bibfield  {title} {\enquote {\bibinfo {title} {{Composition of
  Anomalous Cosmic Rays and Implications for the Heliosphere}},}\ }\href
  {\doibase 10.1007/BF00170798} {\bibfield  {journal} {\bibinfo  {journal}
  {Space Sci.~Rev.}\ }\textbf {\bibinfo {volume} {78}},\ \bibinfo {pages}
  {117--128} (\bibinfo {year} {1996})}\BibitemShut {NoStop}%
\bibitem [{\citenamefont {{Cummings}}\ and\ \citenamefont
  {{Stone}}(2007)}]{Cummings:2007}%
  \BibitemOpen
  \bibfield  {author} {\bibinfo {author} {\bibfnamefont {A.~C.}\ \bibnamefont
  {{Cummings}}}\ and\ \bibinfo {author} {\bibfnamefont {E.~C.}\ \bibnamefont
  {{Stone}}},\ }\bibfield  {title} {\enquote {\bibinfo {title} {{Composition of
  Anomalous Cosmic Rays}},}\ }\href {\doibase 10.1007/s11214-007-9161-y}
  {\bibfield  {journal} {\bibinfo  {journal} {ssr}\ }\textbf {\bibinfo {volume}
  {130}},\ \bibinfo {pages} {389--399} (\bibinfo {year} {2007})}\BibitemShut
  {NoStop}%
\bibitem [{\citenamefont {{Jokipii}}, \citenamefont {{Levy}},\ and\
  \citenamefont {{Hubbard}}(1977)}]{Jokipii:1977}%
  \BibitemOpen
  \bibfield  {author} {\bibinfo {author} {\bibfnamefont {J.~R.}\ \bibnamefont
  {{Jokipii}}}, \bibinfo {author} {\bibfnamefont {E.~H.}\ \bibnamefont
  {{Levy}}}, \ and\ \bibinfo {author} {\bibfnamefont {W.~B.}\ \bibnamefont
  {{Hubbard}}},\ }\bibfield  {title} {\enquote {\bibinfo {title} {{Effects of
  particle drift on cosmic-ray transport. I - General properties, application
  to solar modulation}},}\ }\href {\doibase 10.1086/155218} {\bibfield
  {journal} {\bibinfo  {journal} {Astrophys.~J.}\ }\textbf {\bibinfo {volume}
  {213}},\ \bibinfo {pages} {861--868} (\bibinfo {year} {1977})}\BibitemShut
  {NoStop}%
\bibitem [{\citenamefont {{Kota}}\ and\ \citenamefont
  {{Jokipii}}(1983)}]{Kota:1983}%
  \BibitemOpen
  \bibfield  {author} {\bibinfo {author} {\bibfnamefont {J.}~\bibnamefont
  {{Kota}}}\ and\ \bibinfo {author} {\bibfnamefont {J.~R.}\ \bibnamefont
  {{Jokipii}}},\ }\bibfield  {title} {\enquote {\bibinfo {title} {{Effects of
  drift on the transport of cosmic rays. VI - A three-dimensional model
  including diffusion}},}\ }\href {\doibase 10.1086/160701} {\bibfield
  {journal} {\bibinfo  {journal} {Astrophys.~J.}\ }\textbf {\bibinfo {volume}
  {265}},\ \bibinfo {pages} {573--581} (\bibinfo {year} {1983})}\BibitemShut
  {NoStop}%
\bibitem [{\citenamefont {{Basu}}\ \emph {et~al.}(2002)\citenamefont {{Basu}},
  \citenamefont {{Groves}}, \citenamefont {{Basu}},\ and\ \citenamefont
  {{Sultan}}}]{Basu:2002}%
  \BibitemOpen
  \bibfield  {author} {\bibinfo {author} {\bibfnamefont {S.}~\bibnamefont
  {{Basu}}}, \bibinfo {author} {\bibfnamefont {K.~M.}\ \bibnamefont
  {{Groves}}}, \bibinfo {author} {\bibfnamefont {S.}~\bibnamefont {{Basu}}}, \
  and\ \bibinfo {author} {\bibfnamefont {P.~J.}\ \bibnamefont {{Sultan}}},\
  }\bibfield  {title} {\enquote {\bibinfo {title} {{Specification and
  forecasting of scintillations in communication/navigation links: current
  status and future plans}},}\ }\href {\doibase 10.1016/S1364-6826(02)00124-4}
  {\bibfield  {journal} {\bibinfo  {journal} {Journal of Atmospheric and
  Solar-Terrestrial Physics}\ }\textbf {\bibinfo {volume} {64}},\ \bibinfo
  {pages} {1745--1754} (\bibinfo {year} {2002})}\BibitemShut {NoStop}%
\bibitem [{\citenamefont {{Bell}}(1978)}]{Bell:1978}%
  \BibitemOpen
  \bibfield  {author} {\bibinfo {author} {\bibfnamefont {A.~R.}\ \bibnamefont
  {{Bell}}},\ }\bibfield  {title} {\enquote {\bibinfo {title} {{The
  acceleration of cosmic rays in shock fronts. I}},}\ }\href {\doibase
  10.1093/mnras/182.2.147} {\bibfield  {journal} {\bibinfo  {journal}
  {Mon.~Not.~Roy.~Astron.~Soc.}\ }\textbf {\bibinfo {volume} {182}},\ \bibinfo
  {pages} {147--156} (\bibinfo {year} {1978})}\BibitemShut {NoStop}%
\bibitem [{\citenamefont {{Malkov}}\ and\ \citenamefont
  {{Drury}}(2001)}]{Malkov:2001}%
  \BibitemOpen
  \bibfield  {author} {\bibinfo {author} {\bibfnamefont {M.~A.}\ \bibnamefont
  {{Malkov}}}\ and\ \bibinfo {author} {\bibfnamefont {L.~O.}\ \bibnamefont
  {{Drury}}},\ }\bibfield  {title} {\enquote {\bibinfo {title} {{Nonlinear
  theory of diffusive acceleration of particles by shock waves}},}\ }\href
  {\doibase 10.1088/0034-4885/64/4/201} {\bibfield  {journal} {\bibinfo
  {journal} {Reports on Progress in Physics}\ }\textbf {\bibinfo {volume}
  {64}},\ \bibinfo {pages} {429--481} (\bibinfo {year} {2001})}\BibitemShut
  {NoStop}%
\bibitem [{\citenamefont {{Howes}}\ and\ \citenamefont
  {{Nielson}}(2013)}]{Howes:2013a}%
  \BibitemOpen
  \bibfield  {author} {\bibinfo {author} {\bibfnamefont {G.~G.}\ \bibnamefont
  {{Howes}}}\ and\ \bibinfo {author} {\bibfnamefont {K.~D.}\ \bibnamefont
  {{Nielson}}},\ }\bibfield  {title} {\enquote {\bibinfo {title} {{Alfv{\'e}n
  wave collisions, the fundamental building block of plasma turbulence. I.
  Asymptotic solution}},}\ }\href {\doibase 10.1063/1.4812805} {\bibfield
  {journal} {\bibinfo  {journal} {Phys.~Plasmas}\ }\textbf {\bibinfo {volume}
  {20}},\ \bibinfo {pages} {072302} (\bibinfo {year} {2013})},\ \Eprint
  {http://arxiv.org/abs/1306.1455} {arXiv:1306.1455 [astro-ph.SR]} \BibitemShut
  {NoStop}%
\bibitem [{\citenamefont {Iroshnikov}(1963)}]{Iroshnikov:1963}%
  \BibitemOpen
  \bibfield  {author} {\bibinfo {author} {\bibfnamefont {R.~S.}\ \bibnamefont
  {Iroshnikov}},\ }\bibfield  {title} {\enquote {\bibinfo {title} {The
  turbulence of a conducting fluid in a strong magnetic field},}\ }\href@noop
  {} {\bibfield  {journal} {\bibinfo  {journal} {Astron. Zh.}\ }\textbf
  {\bibinfo {volume} {40}},\ \bibinfo {pages} {742} (\bibinfo {year} {1963})},\
  \bibinfo {note} {{English} Translation: Sov. Astron., 7 566
  (1964)}\BibitemShut {NoStop}%
\bibitem [{\citenamefont {Kraichnan}(1965)}]{Kraichnan:1965}%
  \BibitemOpen
  \bibfield  {author} {\bibinfo {author} {\bibfnamefont {R.~H.}\ \bibnamefont
  {Kraichnan}},\ }\bibfield  {title} {\enquote {\bibinfo {title} {Inertial
  range spectrum of hyromagnetic turbulence},}\ }\href@noop {} {\bibfield
  {journal} {\bibinfo  {journal} {Phys.~Fluids}\ }\textbf {\bibinfo {volume}
  {8}},\ \bibinfo {pages} {1385--1387} (\bibinfo {year} {1965})}\BibitemShut
  {NoStop}%
\bibitem [{\citenamefont {{Nielson}}, \citenamefont {{Howes}},\ and\
  \citenamefont {{Dorland}}(2013)}]{Nielson:2013a}%
  \BibitemOpen
  \bibfield  {author} {\bibinfo {author} {\bibfnamefont {K.~D.}\ \bibnamefont
  {{Nielson}}}, \bibinfo {author} {\bibfnamefont {G.~G.}\ \bibnamefont
  {{Howes}}}, \ and\ \bibinfo {author} {\bibfnamefont {W.}~\bibnamefont
  {{Dorland}}},\ }\bibfield  {title} {\enquote {\bibinfo {title} {{Alfv{\'e}n
  wave collisions, the fundamental building block of plasma turbulence. II.
  Numerical solution}},}\ }\href {\doibase 10.1063/1.4812807} {\bibfield
  {journal} {\bibinfo  {journal} {Physics of Plasmas}\ }\textbf {\bibinfo
  {volume} {20}},\ \bibinfo {pages} {072303} (\bibinfo {year} {2013})},\
  \Eprint {http://arxiv.org/abs/1306.1456} {arXiv:1306.1456 [astro-ph.SR]}
  \BibitemShut {NoStop}%
\bibitem [{\citenamefont {{Howes}}\ \emph {et~al.}(2012)\citenamefont
  {{Howes}}, \citenamefont {{Drake}}, \citenamefont {{Nielson}}, \citenamefont
  {{Carter}}, \citenamefont {{Kletzing}},\ and\ \citenamefont
  {{Skiff}}}]{Howes:2012b}%
  \BibitemOpen
  \bibfield  {author} {\bibinfo {author} {\bibfnamefont {G.~G.}\ \bibnamefont
  {{Howes}}}, \bibinfo {author} {\bibfnamefont {D.~J.}\ \bibnamefont
  {{Drake}}}, \bibinfo {author} {\bibfnamefont {K.~D.}\ \bibnamefont
  {{Nielson}}}, \bibinfo {author} {\bibfnamefont {T.~A.}\ \bibnamefont
  {{Carter}}}, \bibinfo {author} {\bibfnamefont {C.~A.}\ \bibnamefont
  {{Kletzing}}}, \ and\ \bibinfo {author} {\bibfnamefont {F.}~\bibnamefont
  {{Skiff}}},\ }\bibfield  {title} {\enquote {\bibinfo {title} {{Toward
  Astrophysical Turbulence in the Laboratory}},}\ }\href {\doibase
  10.1103/PhysRevLett.109.255001} {\bibfield  {journal} {\bibinfo  {journal}
  {Phys.~Rev.~Lett.}\ }\textbf {\bibinfo {volume} {109}},\ \bibinfo {eid}
  {255001} (\bibinfo {year} {2012})},\ \Eprint {http://arxiv.org/abs/1210.4568}
  {arXiv:1210.4568 [physics.plasm-ph]} \BibitemShut {NoStop}%
\bibitem [{\citenamefont {{Howes}}\ \emph {et~al.}(2013)\citenamefont
  {{Howes}}, \citenamefont {{Nielson}}, \citenamefont {{Drake}}, \citenamefont
  {{Schroeder}}, \citenamefont {{Skiff}}, \citenamefont {{Kletzing}},\ and\
  \citenamefont {{Carter}}}]{Howes:2013b}%
  \BibitemOpen
  \bibfield  {author} {\bibinfo {author} {\bibfnamefont {G.~G.}\ \bibnamefont
  {{Howes}}}, \bibinfo {author} {\bibfnamefont {K.~D.}\ \bibnamefont
  {{Nielson}}}, \bibinfo {author} {\bibfnamefont {D.~J.}\ \bibnamefont
  {{Drake}}}, \bibinfo {author} {\bibfnamefont {J.~W.~R.}\ \bibnamefont
  {{Schroeder}}}, \bibinfo {author} {\bibfnamefont {F.}~\bibnamefont
  {{Skiff}}}, \bibinfo {author} {\bibfnamefont {C.~A.}\ \bibnamefont
  {{Kletzing}}}, \ and\ \bibinfo {author} {\bibfnamefont {T.~A.}\ \bibnamefont
  {{Carter}}},\ }\bibfield  {title} {\enquote {\bibinfo {title} {{Alfv{\'e}n
  wave collisions, the fundamental building block of plasma turbulence. III.
  Theory for experimental design}},}\ }\href {\doibase 10.1063/1.4812808}
  {\bibfield  {journal} {\bibinfo  {journal} {Physics of Plasmas}\ }\textbf
  {\bibinfo {volume} {20}},\ \bibinfo {pages} {072304} (\bibinfo {year}
  {2013})},\ \Eprint {http://arxiv.org/abs/1306.1460} {arXiv:1306.1460
  [astro-ph.SR]} \BibitemShut {NoStop}%
\bibitem [{\citenamefont {{Drake}}\ \emph {et~al.}(2013)\citenamefont
  {{Drake}}, \citenamefont {{Schroeder}}, \citenamefont {{Howes}},
  \citenamefont {{Kletzing}}, \citenamefont {{Skiff}}, \citenamefont
  {{Carter}},\ and\ \citenamefont {{Auerbach}}}]{Drake:2013}%
  \BibitemOpen
  \bibfield  {author} {\bibinfo {author} {\bibfnamefont {D.~J.}\ \bibnamefont
  {{Drake}}}, \bibinfo {author} {\bibfnamefont {J.~W.~R.}\ \bibnamefont
  {{Schroeder}}}, \bibinfo {author} {\bibfnamefont {G.~G.}\ \bibnamefont
  {{Howes}}}, \bibinfo {author} {\bibfnamefont {C.~A.}\ \bibnamefont
  {{Kletzing}}}, \bibinfo {author} {\bibfnamefont {F.}~\bibnamefont {{Skiff}}},
  \bibinfo {author} {\bibfnamefont {T.~A.}\ \bibnamefont {{Carter}}}, \ and\
  \bibinfo {author} {\bibfnamefont {D.~W.}\ \bibnamefont {{Auerbach}}},\
  }\bibfield  {title} {\enquote {\bibinfo {title} {{Alfv{\'e}n wave collisions,
  the fundamental building block of plasma turbulence. IV. Laboratory
  experiment}},}\ }\href {\doibase 10.1063/1.4813242} {\bibfield  {journal}
  {\bibinfo  {journal} {Physics of Plasmas}\ }\textbf {\bibinfo {volume}
  {20}},\ \bibinfo {pages} {072901} (\bibinfo {year} {2013})},\ \Eprint
  {http://arxiv.org/abs/1306.1130} {arXiv:1306.1130 [astro-ph.SR]} \BibitemShut
  {NoStop}%
\bibitem [{\citenamefont {{Drake}}\ \emph {et~al.}(2014)\citenamefont
  {{Drake}}, \citenamefont {{Schroeder}}, \citenamefont {{Shanken}},
  \citenamefont {{Howes}}, \citenamefont {{Skiff}}, \citenamefont {{Kletzing}},
  \citenamefont {{Carter}},\ and\ \citenamefont {{Dorfman}}}]{Drake:2014}%
  \BibitemOpen
  \bibfield  {author} {\bibinfo {author} {\bibfnamefont {D.~J.}\ \bibnamefont
  {{Drake}}}, \bibinfo {author} {\bibfnamefont {J.~W.~R.}\ \bibnamefont
  {{Schroeder}}}, \bibinfo {author} {\bibfnamefont {B.~C.}\ \bibnamefont
  {{Shanken}}}, \bibinfo {author} {\bibfnamefont {G.~G.}\ \bibnamefont
  {{Howes}}}, \bibinfo {author} {\bibfnamefont {F.}~\bibnamefont {{Skiff}}},
  \bibinfo {author} {\bibfnamefont {C.~A.}\ \bibnamefont {{Kletzing}}},
  \bibinfo {author} {\bibfnamefont {T.~A.}\ \bibnamefont {{Carter}}}, \ and\
  \bibinfo {author} {\bibfnamefont {S.}~\bibnamefont {{Dorfman}}},\ }\bibfield
  {title} {\enquote {\bibinfo {title} {{Analysis of Magnetic Fields in Inertial
  Alfv{\'e}n Wave Collisions}},}\ }\href {\doibase 10.1109/TPS.2014.2321996}
  {\bibfield  {journal} {\bibinfo  {journal} {IEEE Trans. Plasma Sci.}\
  }\textbf {\bibinfo {volume} {42}},\ \bibinfo {pages} {2534--2535} (\bibinfo
  {year} {2014})},\ \Eprint {http://arxiv.org/abs/1406.3357} {arXiv:1406.3357
  [astro-ph.SR]} \BibitemShut {NoStop}%
\bibitem [{\citenamefont {{Drake}}\ \emph {et~al.}(2016)\citenamefont
  {{Drake}}, \citenamefont {{Howes}}, \citenamefont {{Rhudy}}, \citenamefont
  {{Terry}}, \citenamefont {{Carter}}, \citenamefont {{Kletzing}},
  \citenamefont {{Schroeder}},\ and\ \citenamefont {{Skiff}}}]{Drake:2016}%
  \BibitemOpen
  \bibfield  {author} {\bibinfo {author} {\bibfnamefont {D.~J.}\ \bibnamefont
  {{Drake}}}, \bibinfo {author} {\bibfnamefont {G.~G.}\ \bibnamefont
  {{Howes}}}, \bibinfo {author} {\bibfnamefont {J.~D.}\ \bibnamefont
  {{Rhudy}}}, \bibinfo {author} {\bibfnamefont {S.~K.}\ \bibnamefont
  {{Terry}}}, \bibinfo {author} {\bibfnamefont {T.~A.}\ \bibnamefont
  {{Carter}}}, \bibinfo {author} {\bibfnamefont {C.~A.}\ \bibnamefont
  {{Kletzing}}}, \bibinfo {author} {\bibfnamefont {J.~W.~R.}\ \bibnamefont
  {{Schroeder}}}, \ and\ \bibinfo {author} {\bibfnamefont {F.}~\bibnamefont
  {{Skiff}}},\ }\bibfield  {title} {\enquote {\bibinfo {title} {{Measurements
  of the nonlinear beat wave produced by the interaction of counterpropagating
  Alfven waves}},}\ }\href {\doibase 10.1063/1.4941977} {\bibfield  {journal}
  {\bibinfo  {journal} {Phys.~Plasmas}\ }\textbf {\bibinfo {volume} {23}},\
  \bibinfo {eid} {022305} (\bibinfo {year} {2016})}\BibitemShut {NoStop}%
\bibitem [{\citenamefont {{Howes}}(2016)}]{Howes:2016b}%
  \BibitemOpen
  \bibfield  {author} {\bibinfo {author} {\bibfnamefont {G.~G.}\ \bibnamefont
  {{Howes}}},\ }\bibfield  {title} {\enquote {\bibinfo {title} {{The Dynamical
  Generation of Current Sheets in Astrophysical Plasma Turbulence}},}\ }\href
  {\doibase 10.3847/2041-8205/827/2/L28} {\bibfield  {journal} {\bibinfo
  {journal} {Astrophys.~J.~Lett.}\ }\textbf {\bibinfo {volume} {82}},\ \bibinfo
  {eid} {L28} (\bibinfo {year} {2016})},\ \Eprint
  {http://arxiv.org/abs/1607.07465} {arXiv:1607.07465 [astro-ph.SR]}
  \BibitemShut {NoStop}%
\bibitem [{\citenamefont {{Verniero}}, \citenamefont {{Howes}},\ and\
  \citenamefont {{Klein}}(2018)}]{Verniero:2018a}%
  \BibitemOpen
  \bibfield  {author} {\bibinfo {author} {\bibfnamefont {J.~L.}\ \bibnamefont
  {{Verniero}}}, \bibinfo {author} {\bibfnamefont {G.~G.}\ \bibnamefont
  {{Howes}}}, \ and\ \bibinfo {author} {\bibfnamefont {K.~G.}\ \bibnamefont
  {{Klein}}},\ }\bibfield  {title} {\enquote {\bibinfo {title} {{Nonlinear
  energy transfer and current sheet development in localized Alfv{\'e}n
  wavepacket collisions in the strong turbulence limit}},}\ }\href {\doibase
  10.1017/S0022377817001003} {\bibfield  {journal} {\bibinfo  {journal}
  {Journal of Plasma Physics}\ }\textbf {\bibinfo {volume} {84}},\ \bibinfo
  {eid} {905840103} (\bibinfo {year} {2018})},\ \Eprint
  {http://arxiv.org/abs/1705.07046} {arXiv:1705.07046 [physics.plasm-ph]}
  \BibitemShut {NoStop}%
\bibitem [{\citenamefont {{Verniero}}\ and\ \citenamefont
  {{Howes}}(2018)}]{Verniero:2018b}%
  \BibitemOpen
  \bibfield  {author} {\bibinfo {author} {\bibfnamefont {J.~L.}\ \bibnamefont
  {{Verniero}}}\ and\ \bibinfo {author} {\bibfnamefont {G.~G.}\ \bibnamefont
  {{Howes}}},\ }\bibfield  {title} {\enquote {\bibinfo {title} {{The Alfv\'enic
  nature of energy transfer mediation in localized, strongly nonlinear \Alfven
  wavepacket collisions}},}\ }\href@noop {} {\bibfield  {journal} {\bibinfo
  {journal} {J.~Plasma Phys.}\ } (\bibinfo {year} {2018})},\ \bibinfo {note}
  {accepted}\BibitemShut {NoStop}%
\bibitem [{\citenamefont {{Kohl}}\ \emph {et~al.}(1998)\citenamefont {{Kohl}},
  \citenamefont {{Noci}}, \citenamefont {{Antonucci}}, \citenamefont
  {{Tondello}}, \citenamefont {{Huber}}, \citenamefont {{Cranmer}},
  \citenamefont {{Strachan}}, \citenamefont {{Panasyuk}}, \citenamefont
  {{Gardner}}, \citenamefont {{Romoli}}, \citenamefont {{Fineschi}},
  \citenamefont {{Dobrzycka}}, \citenamefont {{Raymond}}, \citenamefont
  {{Nicolosi}}, \citenamefont {{Siegmund}}, \citenamefont {{Spadaro}},
  \citenamefont {{Benna}}, \citenamefont {{Ciaravella}}, \citenamefont
  {{Giordano}}, \citenamefont {{Habbal}}, \citenamefont {{Karovska}},
  \citenamefont {{Li}}, \citenamefont {{Martin}}, \citenamefont {{Michels}},
  \citenamefont {{Modigliani}}, \citenamefont {{Naletto}}, \citenamefont
  {{O'Neal}}, \citenamefont {{Pernechele}}, \citenamefont {{Poletto}},
  \citenamefont {{Smith}},\ and\ \citenamefont {{Suleiman}}}]{Kohl:1998}%
  \BibitemOpen
  \bibfield  {author} {\bibinfo {author} {\bibfnamefont {J.~L.}\ \bibnamefont
  {{Kohl}}}, \bibinfo {author} {\bibfnamefont {G.}~\bibnamefont {{Noci}}},
  \bibinfo {author} {\bibfnamefont {E.}~\bibnamefont {{Antonucci}}}, \bibinfo
  {author} {\bibfnamefont {G.}~\bibnamefont {{Tondello}}}, \bibinfo {author}
  {\bibfnamefont {M.~C.~E.}\ \bibnamefont {{Huber}}}, \bibinfo {author}
  {\bibfnamefont {S.~R.}\ \bibnamefont {{Cranmer}}}, \bibinfo {author}
  {\bibfnamefont {L.}~\bibnamefont {{Strachan}}}, \bibinfo {author}
  {\bibfnamefont {A.~V.}\ \bibnamefont {{Panasyuk}}}, \bibinfo {author}
  {\bibfnamefont {L.~D.}\ \bibnamefont {{Gardner}}}, \bibinfo {author}
  {\bibfnamefont {M.}~\bibnamefont {{Romoli}}}, \bibinfo {author}
  {\bibfnamefont {S.}~\bibnamefont {{Fineschi}}}, \bibinfo {author}
  {\bibfnamefont {D.}~\bibnamefont {{Dobrzycka}}}, \bibinfo {author}
  {\bibfnamefont {J.~C.}\ \bibnamefont {{Raymond}}}, \bibinfo {author}
  {\bibfnamefont {P.}~\bibnamefont {{Nicolosi}}}, \bibinfo {author}
  {\bibfnamefont {O.~H.~W.}\ \bibnamefont {{Siegmund}}}, \bibinfo {author}
  {\bibfnamefont {D.}~\bibnamefont {{Spadaro}}}, \bibinfo {author}
  {\bibfnamefont {C.}~\bibnamefont {{Benna}}}, \bibinfo {author} {\bibfnamefont
  {A.}~\bibnamefont {{Ciaravella}}}, \bibinfo {author} {\bibfnamefont
  {S.}~\bibnamefont {{Giordano}}}, \bibinfo {author} {\bibfnamefont {S.~R.}\
  \bibnamefont {{Habbal}}}, \bibinfo {author} {\bibfnamefont {M.}~\bibnamefont
  {{Karovska}}}, \bibinfo {author} {\bibfnamefont {X.}~\bibnamefont {{Li}}},
  \bibinfo {author} {\bibfnamefont {R.}~\bibnamefont {{Martin}}}, \bibinfo
  {author} {\bibfnamefont {J.~G.}\ \bibnamefont {{Michels}}}, \bibinfo {author}
  {\bibfnamefont {A.}~\bibnamefont {{Modigliani}}}, \bibinfo {author}
  {\bibfnamefont {G.}~\bibnamefont {{Naletto}}}, \bibinfo {author}
  {\bibfnamefont {R.~H.}\ \bibnamefont {{O'Neal}}}, \bibinfo {author}
  {\bibfnamefont {C.}~\bibnamefont {{Pernechele}}}, \bibinfo {author}
  {\bibfnamefont {G.}~\bibnamefont {{Poletto}}}, \bibinfo {author}
  {\bibfnamefont {P.~L.}\ \bibnamefont {{Smith}}}, \ and\ \bibinfo {author}
  {\bibfnamefont {R.~M.}\ \bibnamefont {{Suleiman}}},\ }\bibfield  {title}
  {\enquote {\bibinfo {title} {{UVCS/SOHO Empirical Determinations of
  Anisotropic Velocity Distributions in the Solar Corona}},}\ }\href {\doibase
  10.1086/311434} {\bibfield  {journal} {\bibinfo  {journal}
  {Astrophys.~J.~Lett.}\ }\textbf {\bibinfo {volume} {501}},\ \bibinfo {pages}
  {L127+} (\bibinfo {year} {1998})}\BibitemShut {NoStop}%
\bibitem [{\citenamefont {{Ren}}\ \emph {et~al.}(2011)\citenamefont {{Ren}},
  \citenamefont {{Almagri}}, \citenamefont {{Fiksel}}, \citenamefont
  {{Prager}}, \citenamefont {{Sarff}},\ and\ \citenamefont
  {{Terry}}}]{Ren:2011}%
  \BibitemOpen
  \bibfield  {author} {\bibinfo {author} {\bibfnamefont {Y.}~\bibnamefont
  {{Ren}}}, \bibinfo {author} {\bibfnamefont {A.~F.}\ \bibnamefont
  {{Almagri}}}, \bibinfo {author} {\bibfnamefont {G.}~\bibnamefont {{Fiksel}}},
  \bibinfo {author} {\bibfnamefont {S.~C.}\ \bibnamefont {{Prager}}}, \bibinfo
  {author} {\bibfnamefont {J.~S.}\ \bibnamefont {{Sarff}}}, \ and\ \bibinfo
  {author} {\bibfnamefont {P.~W.}\ \bibnamefont {{Terry}}},\ }\bibfield
  {title} {\enquote {\bibinfo {title} {{Experimental Observation of Anisotropic
  Magnetic Turbulence in a Reversed Field Pinch Plasma}},}\ }\href {\doibase
  10.1103/PhysRevLett.107.195002} {\bibfield  {journal} {\bibinfo  {journal}
  {Phys.~Rev.~Lett.}\ }\textbf {\bibinfo {volume} {107}},\ \bibinfo {eid}
  {195002} (\bibinfo {year} {2011})}\BibitemShut {NoStop}%
\bibitem [{\citenamefont {{Reardon}}\ \emph {et~al.}(2001)\citenamefont
  {{Reardon}}, \citenamefont {{Fiksel}}, \citenamefont {{Forest}},
  \citenamefont {{Abdrashitov}}, \citenamefont {{Davydenko}}, \citenamefont
  {{Ivanov}}, \citenamefont {{Korepanov}}, \citenamefont {{Murachtin}},\ and\
  \citenamefont {{Shulzhenko}}}]{Reardon:2001}%
  \BibitemOpen
  \bibfield  {author} {\bibinfo {author} {\bibfnamefont {J.~C.}\ \bibnamefont
  {{Reardon}}}, \bibinfo {author} {\bibfnamefont {G.}~\bibnamefont {{Fiksel}}},
  \bibinfo {author} {\bibfnamefont {C.~B.}\ \bibnamefont {{Forest}}}, \bibinfo
  {author} {\bibfnamefont {A.~F.}\ \bibnamefont {{Abdrashitov}}}, \bibinfo
  {author} {\bibfnamefont {V.~I.}\ \bibnamefont {{Davydenko}}}, \bibinfo
  {author} {\bibfnamefont {A.~A.}\ \bibnamefont {{Ivanov}}}, \bibinfo {author}
  {\bibfnamefont {S.~A.}\ \bibnamefont {{Korepanov}}}, \bibinfo {author}
  {\bibfnamefont {S.~V.}\ \bibnamefont {{Murachtin}}}, \ and\ \bibinfo {author}
  {\bibfnamefont {G.~I.}\ \bibnamefont {{Shulzhenko}}},\ }\bibfield  {title}
  {\enquote {\bibinfo {title} {{Rutherford scattering diagnostic for the
  Madison symmetric torus reversed-field pinch}},}\ }\href {\doibase
  10.1063/1.1321737} {\bibfield  {journal} {\bibinfo  {journal} {Rev. Sci.
  Instrum.}\ }\textbf {\bibinfo {volume} {72}},\ \bibinfo {pages} {598--601}
  (\bibinfo {year} {2001})}\BibitemShut {NoStop}%
\bibitem [{\citenamefont {{Fiksel}}\ \emph {et~al.}(2009)\citenamefont
  {{Fiksel}}, \citenamefont {{Almagri}}, \citenamefont {{Chapman}},
  \citenamefont {{Mirnov}}, \citenamefont {{Ren}}, \citenamefont {{Sarff}},\
  and\ \citenamefont {{Terry}}}]{Fiksel:2009}%
  \BibitemOpen
  \bibfield  {author} {\bibinfo {author} {\bibfnamefont {G.}~\bibnamefont
  {{Fiksel}}}, \bibinfo {author} {\bibfnamefont {A.~F.}\ \bibnamefont
  {{Almagri}}}, \bibinfo {author} {\bibfnamefont {B.~E.}\ \bibnamefont
  {{Chapman}}}, \bibinfo {author} {\bibfnamefont {V.~V.}\ \bibnamefont
  {{Mirnov}}}, \bibinfo {author} {\bibfnamefont {Y.}~\bibnamefont {{Ren}}},
  \bibinfo {author} {\bibfnamefont {J.~S.}\ \bibnamefont {{Sarff}}}, \ and\
  \bibinfo {author} {\bibfnamefont {P.~W.}\ \bibnamefont {{Terry}}},\
  }\bibfield  {title} {\enquote {\bibinfo {title} {{Mass-Dependent Ion Heating
  during Magnetic Reconnection in a Laboratory Plasma}},}\ }\href {\doibase
  10.1103/PhysRevLett.103.145002} {\bibfield  {journal} {\bibinfo  {journal}
  {Physical Review Letters}\ }\textbf {\bibinfo {volume} {103}},\ \bibinfo
  {eid} {145002} (\bibinfo {year} {2009})}\BibitemShut {NoStop}%
\bibitem [{\citenamefont {{Schaffner}}, \citenamefont {{Wan}},\ and\
  \citenamefont {{Brown}}(2014)}]{Schaffner:2014}%
  \BibitemOpen
  \bibfield  {author} {\bibinfo {author} {\bibfnamefont {D.~A.}\ \bibnamefont
  {{Schaffner}}}, \bibinfo {author} {\bibfnamefont {A.}~\bibnamefont {{Wan}}},
  \ and\ \bibinfo {author} {\bibfnamefont {M.~R.}\ \bibnamefont {{Brown}}},\
  }\bibfield  {title} {\enquote {\bibinfo {title} {{Observation of Turbulent
  Intermittency Scaling with Magnetic Helicity in an MHD Plasma Wind
  Tunnel}},}\ }\href {\doibase 10.1103/PhysRevLett.112.165001} {\bibfield
  {journal} {\bibinfo  {journal} {Phys.~Rev.~Lett.}\ }\textbf {\bibinfo
  {volume} {112}},\ \bibinfo {eid} {165001} (\bibinfo {year}
  {2014})}\BibitemShut {NoStop}%
\bibitem [{\citenamefont {{Weck}}\ \emph {et~al.}(2015)\citenamefont {{Weck}},
  \citenamefont {{Schaffner}}, \citenamefont {{Brown}},\ and\ \citenamefont
  {{Wicks}}}]{Weck:2015}%
  \BibitemOpen
  \bibfield  {author} {\bibinfo {author} {\bibfnamefont {P.~J.}\ \bibnamefont
  {{Weck}}}, \bibinfo {author} {\bibfnamefont {D.~A.}\ \bibnamefont
  {{Schaffner}}}, \bibinfo {author} {\bibfnamefont {M.~R.}\ \bibnamefont
  {{Brown}}}, \ and\ \bibinfo {author} {\bibfnamefont {R.~T.}\ \bibnamefont
  {{Wicks}}},\ }\bibfield  {title} {\enquote {\bibinfo {title} {{Permutation
  entropy and statistical complexity analysis of turbulence in laboratory
  plasmas and the solar wind}},}\ }\href {\doibase 10.1103/PhysRevE.91.023101}
  {\bibfield  {journal} {\bibinfo  {journal} {Phys.~Rev.~E}\ }\textbf {\bibinfo
  {volume} {91}},\ \bibinfo {eid} {023101} (\bibinfo {year} {2015})},\ \Eprint
  {http://arxiv.org/abs/1409.5455} {arXiv:1409.5455 [physics.plasm-ph]}
  \BibitemShut {NoStop}%
\bibitem [{\citenamefont {{Cooper}}\ \emph {et~al.}(2014)\citenamefont
  {{Cooper}}, \citenamefont {{Wallace}}, \citenamefont {{Brookhart}},
  \citenamefont {{Clark}}, \citenamefont {{Collins}}, \citenamefont {{Ding}},
  \citenamefont {{Flanagan}}, \citenamefont {{Khalzov}}, \citenamefont {{Li}},
  \citenamefont {{Milhone}}, \citenamefont {{Nornberg}}, \citenamefont
  {{Nonn}}, \citenamefont {{Weisberg}}, \citenamefont {{Whyte}}, \citenamefont
  {{Zweibel}},\ and\ \citenamefont {{Forest}}}]{Cooper:2014}%
  \BibitemOpen
  \bibfield  {author} {\bibinfo {author} {\bibfnamefont {C.~M.}\ \bibnamefont
  {{Cooper}}}, \bibinfo {author} {\bibfnamefont {J.}~\bibnamefont {{Wallace}}},
  \bibinfo {author} {\bibfnamefont {M.}~\bibnamefont {{Brookhart}}}, \bibinfo
  {author} {\bibfnamefont {M.}~\bibnamefont {{Clark}}}, \bibinfo {author}
  {\bibfnamefont {C.}~\bibnamefont {{Collins}}}, \bibinfo {author}
  {\bibfnamefont {W.~X.}\ \bibnamefont {{Ding}}}, \bibinfo {author}
  {\bibfnamefont {K.}~\bibnamefont {{Flanagan}}}, \bibinfo {author}
  {\bibfnamefont {I.}~\bibnamefont {{Khalzov}}}, \bibinfo {author}
  {\bibfnamefont {Y.}~\bibnamefont {{Li}}}, \bibinfo {author} {\bibfnamefont
  {J.}~\bibnamefont {{Milhone}}}, \bibinfo {author} {\bibfnamefont
  {M.}~\bibnamefont {{Nornberg}}}, \bibinfo {author} {\bibfnamefont
  {P.}~\bibnamefont {{Nonn}}}, \bibinfo {author} {\bibfnamefont
  {D.}~\bibnamefont {{Weisberg}}}, \bibinfo {author} {\bibfnamefont {D.~G.}\
  \bibnamefont {{Whyte}}}, \bibinfo {author} {\bibfnamefont {E.}~\bibnamefont
  {{Zweibel}}}, \ and\ \bibinfo {author} {\bibfnamefont {C.~B.}\ \bibnamefont
  {{Forest}}},\ }\bibfield  {title} {\enquote {\bibinfo {title} {{The Madison
  plasma dynamo experiment: A facility for studying laboratory plasma
  astrophysics}},}\ }\href {\doibase 10.1063/1.4861609} {\bibfield  {journal}
  {\bibinfo  {journal} {Phys.~Plasmas}\ }\textbf {\bibinfo {volume} {21}},\
  \bibinfo {eid} {013505} (\bibinfo {year} {2014})},\ \Eprint
  {http://arxiv.org/abs/1310.8637} {arXiv:1310.8637 [physics.plasm-ph]}
  \BibitemShut {NoStop}%
\bibitem [{\citenamefont {{Forest}}\ \emph {et~al.}(2015)\citenamefont
  {{Forest}}, \citenamefont {{Flanagan}}, \citenamefont {{Brookhart}},
  \citenamefont {{Clark}}, \citenamefont {{Cooper}}, \citenamefont
  {{D{\'e}sangles}}, \citenamefont {{Egedal}}, \citenamefont {{Endrizzi}},
  \citenamefont {{Khalzov}}, \citenamefont {{Li}}, \citenamefont {{Miesch}},
  \citenamefont {{Milhone}}, \citenamefont {{Nornberg}}, \citenamefont
  {{Olson}}, \citenamefont {{Peterson}}, \citenamefont {{Roesler}},
  \citenamefont {{Schekochihin}}, \citenamefont {{Schmitz}}, \citenamefont
  {{Siller}}, \citenamefont {{Spitkovsky}}, \citenamefont {{Stemo}},
  \citenamefont {{Wallace}}, \citenamefont {{Weisberg}},\ and\ \citenamefont
  {{Zweibel}}}]{Forest:2015}%
  \BibitemOpen
  \bibfield  {author} {\bibinfo {author} {\bibfnamefont {C.~B.}\ \bibnamefont
  {{Forest}}}, \bibinfo {author} {\bibfnamefont {K.}~\bibnamefont
  {{Flanagan}}}, \bibinfo {author} {\bibfnamefont {M.}~\bibnamefont
  {{Brookhart}}}, \bibinfo {author} {\bibfnamefont {M.}~\bibnamefont
  {{Clark}}}, \bibinfo {author} {\bibfnamefont {C.~M.}\ \bibnamefont
  {{Cooper}}}, \bibinfo {author} {\bibfnamefont {V.}~\bibnamefont
  {{D{\'e}sangles}}}, \bibinfo {author} {\bibfnamefont {J.}~\bibnamefont
  {{Egedal}}}, \bibinfo {author} {\bibfnamefont {D.}~\bibnamefont
  {{Endrizzi}}}, \bibinfo {author} {\bibfnamefont {I.~V.}\ \bibnamefont
  {{Khalzov}}}, \bibinfo {author} {\bibfnamefont {H.}~\bibnamefont {{Li}}},
  \bibinfo {author} {\bibfnamefont {M.}~\bibnamefont {{Miesch}}}, \bibinfo
  {author} {\bibfnamefont {J.}~\bibnamefont {{Milhone}}}, \bibinfo {author}
  {\bibfnamefont {M.}~\bibnamefont {{Nornberg}}}, \bibinfo {author}
  {\bibfnamefont {J.}~\bibnamefont {{Olson}}}, \bibinfo {author} {\bibfnamefont
  {E.}~\bibnamefont {{Peterson}}}, \bibinfo {author} {\bibfnamefont
  {F.}~\bibnamefont {{Roesler}}}, \bibinfo {author} {\bibfnamefont
  {A.}~\bibnamefont {{Schekochihin}}}, \bibinfo {author} {\bibfnamefont
  {O.}~\bibnamefont {{Schmitz}}}, \bibinfo {author} {\bibfnamefont
  {R.}~\bibnamefont {{Siller}}}, \bibinfo {author} {\bibfnamefont
  {A.}~\bibnamefont {{Spitkovsky}}}, \bibinfo {author} {\bibfnamefont
  {A.}~\bibnamefont {{Stemo}}}, \bibinfo {author} {\bibfnamefont
  {J.}~\bibnamefont {{Wallace}}}, \bibinfo {author} {\bibfnamefont
  {D.}~\bibnamefont {{Weisberg}}}, \ and\ \bibinfo {author} {\bibfnamefont
  {E.}~\bibnamefont {{Zweibel}}},\ }\bibfield  {title} {\enquote {\bibinfo
  {title} {{The Wisconsin Plasma Astrophysics Laboratory}},}\ }\href {\doibase
  10.1017/S0022377815000975} {\bibfield  {journal} {\bibinfo  {journal}
  {J.~Plasma Phys.}\ }\textbf {\bibinfo {volume} {81}},\ \bibinfo {eid}
  {345810501} (\bibinfo {year} {2015})},\ \Eprint
  {http://arxiv.org/abs/1506.07195} {arXiv:1506.07195 [physics.plasm-ph]}
  \BibitemShut {NoStop}%
\bibitem [{\citenamefont {{Hsu}}\ \emph {et~al.}(2015)\citenamefont {{Hsu}},
  \citenamefont {{Moser}}, \citenamefont {{Merritt}}, \citenamefont {{Adams}},
  \citenamefont {{Dunn}}, \citenamefont {{Brockington}}, \citenamefont
  {{Case}}, \citenamefont {{Gilmore}}, \citenamefont {{Lynn}}, \citenamefont
  {{Messer}},\ and\ \citenamefont {{Witherspoon}}}]{Hsu:2015}%
  \BibitemOpen
  \bibfield  {author} {\bibinfo {author} {\bibfnamefont {S.~C.}\ \bibnamefont
  {{Hsu}}}, \bibinfo {author} {\bibfnamefont {A.~L.}\ \bibnamefont {{Moser}}},
  \bibinfo {author} {\bibfnamefont {E.~C.}\ \bibnamefont {{Merritt}}}, \bibinfo
  {author} {\bibfnamefont {C.~S.}\ \bibnamefont {{Adams}}}, \bibinfo {author}
  {\bibfnamefont {J.~P.}\ \bibnamefont {{Dunn}}}, \bibinfo {author}
  {\bibfnamefont {S.}~\bibnamefont {{Brockington}}}, \bibinfo {author}
  {\bibfnamefont {A.}~\bibnamefont {{Case}}}, \bibinfo {author} {\bibfnamefont
  {M.}~\bibnamefont {{Gilmore}}}, \bibinfo {author} {\bibfnamefont {A.~G.}\
  \bibnamefont {{Lynn}}}, \bibinfo {author} {\bibfnamefont {S.~J.}\
  \bibnamefont {{Messer}}}, \ and\ \bibinfo {author} {\bibfnamefont {F.~D.}\
  \bibnamefont {{Witherspoon}}},\ }\bibfield  {title} {\enquote {\bibinfo
  {title} {{Laboratory plasma physics experiments using merging supersonic
  plasma jets}},}\ }\href {\doibase 10.1017/S0022377814001184} {\bibfield
  {journal} {\bibinfo  {journal} {J.~Plasma Phys.}\ }\textbf {\bibinfo {volume}
  {81}},\ \bibinfo {eid} {345810201} (\bibinfo {year} {2015})},\ \Eprint
  {http://arxiv.org/abs/1408.0323} {arXiv:1408.0323 [physics.plasm-ph]}
  \BibitemShut {NoStop}%
\bibitem [{\citenamefont {Vasyliunas}(1975)}]{Vasyliunas:1975}%
  \BibitemOpen
  \bibfield  {author} {\bibinfo {author} {\bibfnamefont {V.}~\bibnamefont
  {Vasyliunas}},\ }\bibfield  {title} {\enquote {\bibinfo {title} {Theoretical
  models of field line merging, i.}}\ }\href@noop {} {\bibfield  {journal}
  {\bibinfo  {journal} {Rev. Geophys. Space Phys.}\ }\textbf {\bibinfo {volume}
  {13}},\ \bibinfo {pages} {303} (\bibinfo {year} {1975})}\BibitemShut
  {NoStop}%
\bibitem [{\citenamefont {Biskamp}(2000)}]{Biskamp:2000}%
  \BibitemOpen
  \bibfield  {author} {\bibinfo {author} {\bibfnamefont {D.}~\bibnamefont
  {Biskamp}},\ }\href@noop {} {\emph {\bibinfo {title} {Magnetic reconnection
  in plasmas}}}\ (\bibinfo  {publisher} {Cambridge University Press},\ \bibinfo
  {address} {Cambridge},\ \bibinfo {year} {2000})\BibitemShut {NoStop}%
\bibitem [{\citenamefont {Priest}\ and\ \citenamefont
  {Forbes}(2000)}]{Priest:2000}%
  \BibitemOpen
  \bibfield  {author} {\bibinfo {author} {\bibfnamefont {E.}~\bibnamefont
  {Priest}}\ and\ \bibinfo {author} {\bibfnamefont {T.}~\bibnamefont
  {Forbes}},\ }\href@noop {} {\emph {\bibinfo {title} {Magnetic reconnection --
  MHD theory and applications}}}\ (\bibinfo  {publisher} {Cambridge University
  Press},\ \bibinfo {address} {Cambridge, UK},\ \bibinfo {year}
  {2000})\BibitemShut {NoStop}%
\bibitem [{\citenamefont {Zweibel}\ and\ \citenamefont
  {Yamada}(2009)}]{Zweibel:2009}%
  \BibitemOpen
  \bibfield  {author} {\bibinfo {author} {\bibfnamefont {E.}~\bibnamefont
  {Zweibel}}\ and\ \bibinfo {author} {\bibfnamefont {M.}~\bibnamefont
  {Yamada}},\ }\bibfield  {title} {\enquote {\bibinfo {title} {Magnetic
  reconnection in astrophysical and laboratory plasmas},}\ }\href@noop {}
  {\bibfield  {journal} {\bibinfo  {journal} {Annu. Rev. Astron. Astrophys.}\
  }\textbf {\bibinfo {volume} {47}},\ \bibinfo {pages} {291} (\bibinfo {year}
  {2009})}\BibitemShut {NoStop}%
\bibitem [{\citenamefont {Yamada}, \citenamefont {Kulsrud},\ and\ \citenamefont
  {Ji}(2010)}]{Yamada:2010}%
  \BibitemOpen
  \bibfield  {author} {\bibinfo {author} {\bibfnamefont {M.}~\bibnamefont
  {Yamada}}, \bibinfo {author} {\bibfnamefont {R.}~\bibnamefont {Kulsrud}}, \
  and\ \bibinfo {author} {\bibfnamefont {H.}~\bibnamefont {Ji}},\ }\bibfield
  {title} {\enquote {\bibinfo {title} {Magnetic reconnection},}\ }\href@noop {}
  {\bibfield  {journal} {\bibinfo  {journal} {Rev. Mod. Phys.}\ }\textbf
  {\bibinfo {volume} {82}},\ \bibinfo {pages} {603} (\bibinfo {year}
  {2010})}\BibitemShut {NoStop}%
\bibitem [{\citenamefont {{Zweibel}}\ and\ \citenamefont
  {{Yamada}}(2016)}]{Zweibel:2016}%
  \BibitemOpen
  \bibfield  {author} {\bibinfo {author} {\bibfnamefont {E.~G.}\ \bibnamefont
  {{Zweibel}}}\ and\ \bibinfo {author} {\bibfnamefont {M.}~\bibnamefont
  {{Yamada}}},\ }\bibfield  {title} {\enquote {\bibinfo {title} {{Perspectives
  on magnetic reconnection}},}\ }\href@noop {} {\bibfield  {journal} {\bibinfo
  {journal} {Proceedings of the Royal Society of London Series A}\ }\textbf
  {\bibinfo {volume} {472}},\ \bibinfo {pages} {20160479} (\bibinfo {year}
  {2016})}\BibitemShut {NoStop}%
\bibitem [{\citenamefont {{Masuda}}\ \emph {et~al.}(1994)\citenamefont
  {{Masuda}}, \citenamefont {{Kosugi}}, \citenamefont {{Hara}}, \citenamefont
  {{Tsuneta}},\ and\ \citenamefont {{Ogawara}}}]{Masuda:1994}%
  \BibitemOpen
  \bibfield  {author} {\bibinfo {author} {\bibfnamefont {S.}~\bibnamefont
  {{Masuda}}}, \bibinfo {author} {\bibfnamefont {T.}~\bibnamefont {{Kosugi}}},
  \bibinfo {author} {\bibfnamefont {H.}~\bibnamefont {{Hara}}}, \bibinfo
  {author} {\bibfnamefont {S.}~\bibnamefont {{Tsuneta}}}, \ and\ \bibinfo
  {author} {\bibfnamefont {Y.}~\bibnamefont {{Ogawara}}},\ }\bibfield  {title}
  {\enquote {\bibinfo {title} {{A loop-top hard X-ray source in a compact solar
  flare as evidence for magnetic reconnection}},}\ }\href {\doibase
  10.1038/371495a0} {\bibfield  {journal} {\bibinfo  {journal} {Nature}\
  }\textbf {\bibinfo {volume} {371}},\ \bibinfo {pages} {495--497} (\bibinfo
  {year} {1994})}\BibitemShut {NoStop}%
\bibitem [{\citenamefont {{Antiochos}}, \citenamefont {{DeVore}},\ and\
  \citenamefont {{Klimchuk}}(1999)}]{Antiochos:1999}%
  \BibitemOpen
  \bibfield  {author} {\bibinfo {author} {\bibfnamefont {S.~K.}\ \bibnamefont
  {{Antiochos}}}, \bibinfo {author} {\bibfnamefont {C.~R.}\ \bibnamefont
  {{DeVore}}}, \ and\ \bibinfo {author} {\bibfnamefont {J.~A.}\ \bibnamefont
  {{Klimchuk}}},\ }\bibfield  {title} {\enquote {\bibinfo {title} {{A Model for
  Solar Coronal Mass Ejections}},}\ }\href {\doibase 10.1086/306563} {\bibfield
   {journal} {\bibinfo  {journal} {Astrophys.~J.}\ }\textbf {\bibinfo {volume}
  {510}},\ \bibinfo {pages} {485--493} (\bibinfo {year} {1999})},\ \Eprint
  {http://arxiv.org/abs/astro-ph/9807220} {astro-ph/9807220} \BibitemShut
  {NoStop}%
\bibitem [{\citenamefont {{Forbes}}\ \emph {et~al.}(2006)\citenamefont
  {{Forbes}}, \citenamefont {{Linker}}, \citenamefont {{Chen}}, \citenamefont
  {{Cid}}, \citenamefont {{K{\'o}ta}}, \citenamefont {{Lee}}, \citenamefont
  {{Mann}}, \citenamefont {{Miki{\'c}}}, \citenamefont {{Potgieter}},
  \citenamefont {{Schmidt}}, \citenamefont {{Siscoe}}, \citenamefont
  {{Vainio}}, \citenamefont {{Antiochos}},\ and\ \citenamefont
  {{Riley}}}]{Forbes:2006}%
  \BibitemOpen
  \bibfield  {author} {\bibinfo {author} {\bibfnamefont {T.~G.}\ \bibnamefont
  {{Forbes}}}, \bibinfo {author} {\bibfnamefont {J.~A.}\ \bibnamefont
  {{Linker}}}, \bibinfo {author} {\bibfnamefont {J.}~\bibnamefont {{Chen}}},
  \bibinfo {author} {\bibfnamefont {C.}~\bibnamefont {{Cid}}}, \bibinfo
  {author} {\bibfnamefont {J.}~\bibnamefont {{K{\'o}ta}}}, \bibinfo {author}
  {\bibfnamefont {M.~A.}\ \bibnamefont {{Lee}}}, \bibinfo {author}
  {\bibfnamefont {G.}~\bibnamefont {{Mann}}}, \bibinfo {author} {\bibfnamefont
  {Z.}~\bibnamefont {{Miki{\'c}}}}, \bibinfo {author} {\bibfnamefont {M.~S.}\
  \bibnamefont {{Potgieter}}}, \bibinfo {author} {\bibfnamefont {J.~M.}\
  \bibnamefont {{Schmidt}}}, \bibinfo {author} {\bibfnamefont {G.~L.}\
  \bibnamefont {{Siscoe}}}, \bibinfo {author} {\bibfnamefont {R.}~\bibnamefont
  {{Vainio}}}, \bibinfo {author} {\bibfnamefont {S.~K.}\ \bibnamefont
  {{Antiochos}}}, \ and\ \bibinfo {author} {\bibfnamefont {P.}~\bibnamefont
  {{Riley}}},\ }\bibfield  {title} {\enquote {\bibinfo {title} {{CME Theory and
  Models}},}\ }\href {\doibase 10.1007/s11214-006-9019-8} {\bibfield  {journal}
  {\bibinfo  {journal} {Space Sci.~Rev.}\ }\textbf {\bibinfo {volume} {123}},\
  \bibinfo {pages} {251--302} (\bibinfo {year} {2006})}\BibitemShut {NoStop}%
\bibitem [{\citenamefont {{Birn}}\ \emph {et~al.}(2001)\citenamefont {{Birn}},
  \citenamefont {{Drake}}, \citenamefont {{Shay}}, \citenamefont {{Rogers}},
  \citenamefont {{Denton}}, \citenamefont {{Hesse}}, \citenamefont
  {{Kuznetsova}}, \citenamefont {{Ma}}, \citenamefont {{Bhattacharjee}},
  \citenamefont {{Otto}},\ and\ \citenamefont {{Pritchett}}}]{Birn:2001}%
  \BibitemOpen
  \bibfield  {author} {\bibinfo {author} {\bibfnamefont {J.}~\bibnamefont
  {{Birn}}}, \bibinfo {author} {\bibfnamefont {J.~F.}\ \bibnamefont {{Drake}}},
  \bibinfo {author} {\bibfnamefont {M.~A.}\ \bibnamefont {{Shay}}}, \bibinfo
  {author} {\bibfnamefont {B.~N.}\ \bibnamefont {{Rogers}}}, \bibinfo {author}
  {\bibfnamefont {R.~E.}\ \bibnamefont {{Denton}}}, \bibinfo {author}
  {\bibfnamefont {M.}~\bibnamefont {{Hesse}}}, \bibinfo {author} {\bibfnamefont
  {M.}~\bibnamefont {{Kuznetsova}}}, \bibinfo {author} {\bibfnamefont {Z.~W.}\
  \bibnamefont {{Ma}}}, \bibinfo {author} {\bibfnamefont {A.}~\bibnamefont
  {{Bhattacharjee}}}, \bibinfo {author} {\bibfnamefont {A.}~\bibnamefont
  {{Otto}}}, \ and\ \bibinfo {author} {\bibfnamefont {P.~L.}\ \bibnamefont
  {{Pritchett}}},\ }\bibfield  {title} {\enquote {\bibinfo {title} {{Geospace
  Environmental Modeling (GEM) magnetic reconnection challenge}},}\ }\href
  {\doibase 10.1029/1999JA900449} {\bibfield  {journal} {\bibinfo  {journal}
  {J.~Geophys.~Res.}\ }\textbf {\bibinfo {volume} {106}},\ \bibinfo {pages}
  {3715--3720} (\bibinfo {year} {2001})}\BibitemShut {NoStop}%
\bibitem [{\citenamefont {{Shay}}\ \emph {et~al.}(2001)\citenamefont {{Shay}},
  \citenamefont {{Drake}}, \citenamefont {{Rogers}},\ and\ \citenamefont
  {{Denton}}}]{Shay:2001}%
  \BibitemOpen
  \bibfield  {author} {\bibinfo {author} {\bibfnamefont {M.~A.}\ \bibnamefont
  {{Shay}}}, \bibinfo {author} {\bibfnamefont {J.~F.}\ \bibnamefont {{Drake}}},
  \bibinfo {author} {\bibfnamefont {B.~N.}\ \bibnamefont {{Rogers}}}, \ and\
  \bibinfo {author} {\bibfnamefont {R.~E.}\ \bibnamefont {{Denton}}},\
  }\bibfield  {title} {\enquote {\bibinfo {title} {{Alfv{\'e}nic collisionless
  magnetic reconnection and the Hall term}},}\ }\href {\doibase
  10.1029/1999JA001007} {\bibfield  {journal} {\bibinfo  {journal}
  {J.~Geophys.~Res.}\ }\textbf {\bibinfo {volume} {106}},\ \bibinfo {pages}
  {3759--3772} (\bibinfo {year} {2001})}\BibitemShut {NoStop}%
\bibitem [{\citenamefont {{Ricci}}\ \emph {et~al.}(2004)\citenamefont
  {{Ricci}}, \citenamefont {{Brackbill}}, \citenamefont {{Daughton}},\ and\
  \citenamefont {{Lapenta}}}]{Ricci:2004}%
  \BibitemOpen
  \bibfield  {author} {\bibinfo {author} {\bibfnamefont {P.}~\bibnamefont
  {{Ricci}}}, \bibinfo {author} {\bibfnamefont {J.~U.}\ \bibnamefont
  {{Brackbill}}}, \bibinfo {author} {\bibfnamefont {W.}~\bibnamefont
  {{Daughton}}}, \ and\ \bibinfo {author} {\bibfnamefont {G.}~\bibnamefont
  {{Lapenta}}},\ }\bibfield  {title} {\enquote {\bibinfo {title}
  {{Collisionless magnetic reconnection in the presence of a guide field}},}\
  }\href {\doibase 10.1063/1.1768552} {\bibfield  {journal} {\bibinfo
  {journal} {Phys.~Plasmas}\ }\textbf {\bibinfo {volume} {11}},\ \bibinfo
  {pages} {4102--4114} (\bibinfo {year} {2004})},\ \Eprint
  {http://arxiv.org/abs/astro-ph/0304224} {astro-ph/0304224} \BibitemShut
  {NoStop}%
\bibitem [{\citenamefont {{Drake}}, \citenamefont {{Shay}},\ and\ \citenamefont
  {{Swisdak}}(2008)}]{Drake:2008}%
  \BibitemOpen
  \bibfield  {author} {\bibinfo {author} {\bibfnamefont {J.~F.}\ \bibnamefont
  {{Drake}}}, \bibinfo {author} {\bibfnamefont {M.~A.}\ \bibnamefont {{Shay}}},
  \ and\ \bibinfo {author} {\bibfnamefont {M.}~\bibnamefont {{Swisdak}}},\
  }\bibfield  {title} {\enquote {\bibinfo {title} {{The Hall fields and fast
  magnetic reconnection}},}\ }\href {\doibase 10.1063/1.2901194} {\bibfield
  {journal} {\bibinfo  {journal} {Phys.~Plasmas}\ }\textbf {\bibinfo {volume}
  {15}},\ \bibinfo {eid} {042306} (\bibinfo {year} {2008})}\BibitemShut
  {NoStop}%
\bibitem [{\citenamefont {{Pritchett}}(2008)}]{Pritchett:2008}%
  \BibitemOpen
  \bibfield  {author} {\bibinfo {author} {\bibfnamefont {P.~L.}\ \bibnamefont
  {{Pritchett}}},\ }\bibfield  {title} {\enquote {\bibinfo {title}
  {{Collisionless magnetic reconnection in an asymmetric current sheet}},}\
  }\href {\doibase 10.1029/2007JA012930} {\bibfield  {journal} {\bibinfo
  {journal} {J.~Geophys.~Res.}\ }\textbf {\bibinfo {volume} {113}},\ \bibinfo
  {eid} {A06210} (\bibinfo {year} {2008})}\BibitemShut {NoStop}%
\bibitem [{\citenamefont {{Dungey}}(1961)}]{Dungey:1961}%
  \BibitemOpen
  \bibfield  {author} {\bibinfo {author} {\bibfnamefont {J.~W.}\ \bibnamefont
  {{Dungey}}},\ }\bibfield  {title} {\enquote {\bibinfo {title}
  {{Interplanetary Magnetic Field and the Auroral Zones}},}\ }\href {\doibase
  10.1103/PhysRevLett.6.47} {\bibfield  {journal} {\bibinfo  {journal}
  {Physical Review Letters}\ }\textbf {\bibinfo {volume} {6}},\ \bibinfo
  {pages} {47--48} (\bibinfo {year} {1961})}\BibitemShut {NoStop}%
\bibitem [{\citenamefont {{Crooker}}(1979)}]{Crooker:1979}%
  \BibitemOpen
  \bibfield  {author} {\bibinfo {author} {\bibfnamefont {N.~U.}\ \bibnamefont
  {{Crooker}}},\ }\bibfield  {title} {\enquote {\bibinfo {title} {{Dayside
  merging and cusp geometry}},}\ }\href {\doibase 10.1029/JA084iA03p00951}
  {\bibfield  {journal} {\bibinfo  {journal} {J.~Geophys.~Res.}\ }\textbf
  {\bibinfo {volume} {84}},\ \bibinfo {pages} {951--959} (\bibinfo {year}
  {1979})}\BibitemShut {NoStop}%
\bibitem [{\citenamefont {{Angelopoulos}}\ \emph {et~al.}(2008)\citenamefont
  {{Angelopoulos}}, \citenamefont {{McFadden}}, \citenamefont {{Larson}},
  \citenamefont {{Carlson}}, \citenamefont {{Mende}}, \citenamefont {{Frey}},
  \citenamefont {{Phan}}, \citenamefont {{Sibeck}}, \citenamefont
  {{Glassmeier}}, \citenamefont {{Auster}}, \citenamefont {{Donovan}},
  \citenamefont {{Mann}}, \citenamefont {{Rae}}, \citenamefont {{Russell}},
  \citenamefont {{Runov}}, \citenamefont {{Zhou}},\ and\ \citenamefont
  {{Kepko}}}]{Angelopoulos:2008}%
  \BibitemOpen
  \bibfield  {author} {\bibinfo {author} {\bibfnamefont {V.}~\bibnamefont
  {{Angelopoulos}}}, \bibinfo {author} {\bibfnamefont {J.~P.}\ \bibnamefont
  {{McFadden}}}, \bibinfo {author} {\bibfnamefont {D.}~\bibnamefont
  {{Larson}}}, \bibinfo {author} {\bibfnamefont {C.~W.}\ \bibnamefont
  {{Carlson}}}, \bibinfo {author} {\bibfnamefont {S.~B.}\ \bibnamefont
  {{Mende}}}, \bibinfo {author} {\bibfnamefont {H.}~\bibnamefont {{Frey}}},
  \bibinfo {author} {\bibfnamefont {T.}~\bibnamefont {{Phan}}}, \bibinfo
  {author} {\bibfnamefont {D.~G.}\ \bibnamefont {{Sibeck}}}, \bibinfo {author}
  {\bibfnamefont {K.-H.}\ \bibnamefont {{Glassmeier}}}, \bibinfo {author}
  {\bibfnamefont {U.}~\bibnamefont {{Auster}}}, \bibinfo {author}
  {\bibfnamefont {E.}~\bibnamefont {{Donovan}}}, \bibinfo {author}
  {\bibfnamefont {I.~R.}\ \bibnamefont {{Mann}}}, \bibinfo {author}
  {\bibfnamefont {I.~J.}\ \bibnamefont {{Rae}}}, \bibinfo {author}
  {\bibfnamefont {C.~T.}\ \bibnamefont {{Russell}}}, \bibinfo {author}
  {\bibfnamefont {A.}~\bibnamefont {{Runov}}}, \bibinfo {author} {\bibfnamefont
  {X.-Z.}\ \bibnamefont {{Zhou}}}, \ and\ \bibinfo {author} {\bibfnamefont
  {L.}~\bibnamefont {{Kepko}}},\ }\bibfield  {title} {\enquote {\bibinfo
  {title} {{Tail Reconnection Triggering Substorm Onset}},}\ }\href {\doibase
  10.1126/science.1160495} {\bibfield  {journal} {\bibinfo  {journal}
  {Science}\ }\textbf {\bibinfo {volume} {321}},\ \bibinfo {pages} {931}
  (\bibinfo {year} {2008})}\BibitemShut {NoStop}%
\bibitem [{\citenamefont {{Stasiewicz}}\ \emph {et~al.}(2000)\citenamefont
  {{Stasiewicz}}, \citenamefont {{Bellan}}, \citenamefont {{Chaston}},
  \citenamefont {{Kletzing}}, \citenamefont {{Lysak}}, \citenamefont {{Maggs}},
  \citenamefont {{Pokhotelov}}, \citenamefont {{Seyler}}, \citenamefont
  {{Shukla}}, \citenamefont {{Stenflo}}, \citenamefont {{Streltsov}},\ and\
  \citenamefont {{Wahlund}}}]{Stasiewicz:2000}%
  \BibitemOpen
  \bibfield  {author} {\bibinfo {author} {\bibfnamefont {K.}~\bibnamefont
  {{Stasiewicz}}}, \bibinfo {author} {\bibfnamefont {P.}~\bibnamefont
  {{Bellan}}}, \bibinfo {author} {\bibfnamefont {C.}~\bibnamefont {{Chaston}}},
  \bibinfo {author} {\bibfnamefont {C.}~\bibnamefont {{Kletzing}}}, \bibinfo
  {author} {\bibfnamefont {R.}~\bibnamefont {{Lysak}}}, \bibinfo {author}
  {\bibfnamefont {J.}~\bibnamefont {{Maggs}}}, \bibinfo {author} {\bibfnamefont
  {O.}~\bibnamefont {{Pokhotelov}}}, \bibinfo {author} {\bibfnamefont
  {C.}~\bibnamefont {{Seyler}}}, \bibinfo {author} {\bibfnamefont
  {P.}~\bibnamefont {{Shukla}}}, \bibinfo {author} {\bibfnamefont
  {L.}~\bibnamefont {{Stenflo}}}, \bibinfo {author} {\bibfnamefont
  {A.}~\bibnamefont {{Streltsov}}}, \ and\ \bibinfo {author} {\bibfnamefont
  {J.-E.}\ \bibnamefont {{Wahlund}}},\ }\bibfield  {title} {\enquote {\bibinfo
  {title} {{Small Scale Alfv{\'e}nic Structure in the Aurora}},}\ }\href@noop
  {} {\bibfield  {journal} {\bibinfo  {journal} {Space Sci.~Rev.}\ }\textbf
  {\bibinfo {volume} {92}},\ \bibinfo {pages} {423--533} (\bibinfo {year}
  {2000})}\BibitemShut {NoStop}%
\bibitem [{\citenamefont {{Zweibel}}\ \emph {et~al.}(2011)\citenamefont
  {{Zweibel}}, \citenamefont {{Lawrence}}, \citenamefont {{Yoo}}, \citenamefont
  {{Ji}}, \citenamefont {{Yamada}},\ and\ \citenamefont
  {{Malyshkin}}}]{Zweibel:2011}%
  \BibitemOpen
  \bibfield  {author} {\bibinfo {author} {\bibfnamefont {E.~G.}\ \bibnamefont
  {{Zweibel}}}, \bibinfo {author} {\bibfnamefont {E.}~\bibnamefont
  {{Lawrence}}}, \bibinfo {author} {\bibfnamefont {J.}~\bibnamefont {{Yoo}}},
  \bibinfo {author} {\bibfnamefont {H.}~\bibnamefont {{Ji}}}, \bibinfo {author}
  {\bibfnamefont {M.}~\bibnamefont {{Yamada}}}, \ and\ \bibinfo {author}
  {\bibfnamefont {L.~M.}\ \bibnamefont {{Malyshkin}}},\ }\bibfield  {title}
  {\enquote {\bibinfo {title} {{Magnetic reconnection in partially ionized
  plasmas}},}\ }\href {\doibase 10.1063/1.3656960} {\bibfield  {journal}
  {\bibinfo  {journal} {Physics of Plasmas}\ }\textbf {\bibinfo {volume}
  {18}},\ \bibinfo {eid} {111211} (\bibinfo {year} {2011})}\BibitemShut
  {NoStop}%
\bibitem [{\citenamefont {{Brown}}(1999)}]{Brown:1999}%
  \BibitemOpen
  \bibfield  {author} {\bibinfo {author} {\bibfnamefont {M.~R.}\ \bibnamefont
  {{Brown}}},\ }\bibfield  {title} {\enquote {\bibinfo {title} {{Experimental
  studies of magnetic reconnection}},}\ }\href {\doibase 10.1063/1.873430}
  {\bibfield  {journal} {\bibinfo  {journal} {Physics of Plasmas}\ }\textbf
  {\bibinfo {volume} {6}},\ \bibinfo {pages} {1717--1724} (\bibinfo {year}
  {1999})}\BibitemShut {NoStop}%
\bibitem [{\citenamefont {{Egedal}}\ \emph {et~al.}(2000)\citenamefont
  {{Egedal}}, \citenamefont {{Fasoli}}, \citenamefont {{Porkolab}},\ and\
  \citenamefont {{Tarkowski}}}]{Egedal:2000}%
  \BibitemOpen
  \bibfield  {author} {\bibinfo {author} {\bibfnamefont {J.}~\bibnamefont
  {{Egedal}}}, \bibinfo {author} {\bibfnamefont {A.}~\bibnamefont {{Fasoli}}},
  \bibinfo {author} {\bibfnamefont {M.}~\bibnamefont {{Porkolab}}}, \ and\
  \bibinfo {author} {\bibfnamefont {D.}~\bibnamefont {{Tarkowski}}},\
  }\bibfield  {title} {\enquote {\bibinfo {title} {{Plasma generation and
  confinement in a toroidal magnetic cusp}},}\ }\href {\doibase
  10.1063/1.1287340} {\bibfield  {journal} {\bibinfo  {journal} {Rev. Sci.
  Instrum.}\ }\textbf {\bibinfo {volume} {71}},\ \bibinfo {pages} {3351--3361}
  (\bibinfo {year} {2000})}\BibitemShut {NoStop}%
\bibitem [{\citenamefont {{Nilson}}\ \emph {et~al.}(2006)\citenamefont
  {{Nilson}}, \citenamefont {{Willingale}}, \citenamefont {{Kaluza}},
  \citenamefont {{Kamperidis}}, \citenamefont {{Minardi}}, \citenamefont
  {{Wei}}, \citenamefont {{Fernandes}}, \citenamefont {{Notley}}, \citenamefont
  {{Bandyopadhyay}}, \citenamefont {{Sherlock}}, \citenamefont {{Kingham}},
  \citenamefont {{Tatarakis}}, \citenamefont {{Najmudin}}, \citenamefont
  {{Rozmus}}, \citenamefont {{Evans}}, \citenamefont {{Haines}}, \citenamefont
  {{Dangor}},\ and\ \citenamefont {{Krushelnick}}}]{Nilson:2006}%
  \BibitemOpen
  \bibfield  {author} {\bibinfo {author} {\bibfnamefont {P.~M.}\ \bibnamefont
  {{Nilson}}}, \bibinfo {author} {\bibfnamefont {L.}~\bibnamefont
  {{Willingale}}}, \bibinfo {author} {\bibfnamefont {M.~C.}\ \bibnamefont
  {{Kaluza}}}, \bibinfo {author} {\bibfnamefont {C.}~\bibnamefont
  {{Kamperidis}}}, \bibinfo {author} {\bibfnamefont {S.}~\bibnamefont
  {{Minardi}}}, \bibinfo {author} {\bibfnamefont {M.~S.}\ \bibnamefont
  {{Wei}}}, \bibinfo {author} {\bibfnamefont {P.}~\bibnamefont {{Fernandes}}},
  \bibinfo {author} {\bibfnamefont {M.}~\bibnamefont {{Notley}}}, \bibinfo
  {author} {\bibfnamefont {S.}~\bibnamefont {{Bandyopadhyay}}}, \bibinfo
  {author} {\bibfnamefont {M.}~\bibnamefont {{Sherlock}}}, \bibinfo {author}
  {\bibfnamefont {R.~J.}\ \bibnamefont {{Kingham}}}, \bibinfo {author}
  {\bibfnamefont {M.}~\bibnamefont {{Tatarakis}}}, \bibinfo {author}
  {\bibfnamefont {Z.}~\bibnamefont {{Najmudin}}}, \bibinfo {author}
  {\bibfnamefont {W.}~\bibnamefont {{Rozmus}}}, \bibinfo {author}
  {\bibfnamefont {R.~G.}\ \bibnamefont {{Evans}}}, \bibinfo {author}
  {\bibfnamefont {M.~G.}\ \bibnamefont {{Haines}}}, \bibinfo {author}
  {\bibfnamefont {A.~E.}\ \bibnamefont {{Dangor}}}, \ and\ \bibinfo {author}
  {\bibfnamefont {K.}~\bibnamefont {{Krushelnick}}},\ }\bibfield  {title}
  {\enquote {\bibinfo {title} {{Magnetic Reconnection and Plasma Dynamics in
  Two-Beam Laser-Solid Interactions}},}\ }\href {\doibase
  10.1103/PhysRevLett.97.255001} {\bibfield  {journal} {\bibinfo  {journal}
  {Phys.~Rev.~Lett.}\ }\textbf {\bibinfo {volume} {97}},\ \bibinfo {eid}
  {255001} (\bibinfo {year} {2006})}\BibitemShut {NoStop}%
\bibitem [{\citenamefont {{Fiksel}}\ \emph {et~al.}(2014)\citenamefont
  {{Fiksel}}, \citenamefont {{Fox}}, \citenamefont {{Bhattacharjee}},
  \citenamefont {{Barnak}}, \citenamefont {{Chang}}, \citenamefont
  {{Germaschewski}}, \citenamefont {{Hu}},\ and\ \citenamefont
  {{Nilson}}}]{Fiksel:2014}%
  \BibitemOpen
  \bibfield  {author} {\bibinfo {author} {\bibfnamefont {G.}~\bibnamefont
  {{Fiksel}}}, \bibinfo {author} {\bibfnamefont {W.}~\bibnamefont {{Fox}}},
  \bibinfo {author} {\bibfnamefont {A.}~\bibnamefont {{Bhattacharjee}}},
  \bibinfo {author} {\bibfnamefont {D.~H.}\ \bibnamefont {{Barnak}}}, \bibinfo
  {author} {\bibfnamefont {P.-Y.}\ \bibnamefont {{Chang}}}, \bibinfo {author}
  {\bibfnamefont {K.}~\bibnamefont {{Germaschewski}}}, \bibinfo {author}
  {\bibfnamefont {S.~X.}\ \bibnamefont {{Hu}}}, \ and\ \bibinfo {author}
  {\bibfnamefont {P.~M.}\ \bibnamefont {{Nilson}}},\ }\bibfield  {title}
  {\enquote {\bibinfo {title} {{Magnetic Reconnection between Colliding
  Magnetized Laser-Produced Plasma Plumes}},}\ }\href {\doibase
  10.1103/PhysRevLett.113.105003} {\bibfield  {journal} {\bibinfo  {journal}
  {Physical Review Letters}\ }\textbf {\bibinfo {volume} {113}},\ \bibinfo
  {eid} {105003} (\bibinfo {year} {2014})}\BibitemShut {NoStop}%
\bibitem [{\citenamefont {Ji}\ \emph {et~al.}(1998)\citenamefont {Ji},
  \citenamefont {Yamada}, \citenamefont {Hsu},\ and\ \citenamefont
  {Kulsrud}}]{Ji:1998}%
  \BibitemOpen
  \bibfield  {author} {\bibinfo {author} {\bibfnamefont {H.}~\bibnamefont
  {Ji}}, \bibinfo {author} {\bibfnamefont {M.}~\bibnamefont {Yamada}}, \bibinfo
  {author} {\bibfnamefont {S.}~\bibnamefont {Hsu}}, \ and\ \bibinfo {author}
  {\bibfnamefont {R.}~\bibnamefont {Kulsrud}},\ }\bibfield  {title} {\enquote
  {\bibinfo {title} {Experimental test of the sweet-parker model of magnetic
  reconnection},}\ }\href@noop {} {\bibfield  {journal} {\bibinfo  {journal}
  {Phys.~Rev.~Lett.}\ }\textbf {\bibinfo {volume} {80}},\ \bibinfo {pages}
  {3256} (\bibinfo {year} {1998})}\BibitemShut {NoStop}%
\bibitem [{\citenamefont {Kuritsyn}\ \emph {et~al.}(2007)\citenamefont
  {Kuritsyn}, \citenamefont {Ji}, \citenamefont {Gerhardt}, \citenamefont
  {Ren},\ and\ \citenamefont {Yamada}}]{Kuritsyn:2007}%
  \BibitemOpen
  \bibfield  {author} {\bibinfo {author} {\bibfnamefont {A.}~\bibnamefont
  {Kuritsyn}}, \bibinfo {author} {\bibfnamefont {H.}~\bibnamefont {Ji}},
  \bibinfo {author} {\bibfnamefont {S.}~\bibnamefont {Gerhardt}}, \bibinfo
  {author} {\bibfnamefont {Y.}~\bibnamefont {Ren}}, \ and\ \bibinfo {author}
  {\bibfnamefont {M.}~\bibnamefont {Yamada}},\ }\bibfield  {title} {\enquote
  {\bibinfo {title} {Effects of global boundary and local collisionality on
  magnetic reconnection in a laboratory plasma},}\ }\href@noop {} {\bibfield
  {journal} {\bibinfo  {journal} {Geophys. Res. Lett.}\ }\textbf {\bibinfo
  {volume} {34}},\ \bibinfo {pages} {L16106} (\bibinfo {year}
  {2007})}\BibitemShut {NoStop}%
\bibitem [{\citenamefont {Jara-Almonte}\ \emph {et~al.}(2016)\citenamefont
  {Jara-Almonte}, \citenamefont {Ji}, \citenamefont {Yamada}, \citenamefont
  {Yoo},\ and\ \citenamefont {Fox}}]{Jara-Almonte:2016}%
  \BibitemOpen
  \bibfield  {author} {\bibinfo {author} {\bibfnamefont {J.}~\bibnamefont
  {Jara-Almonte}}, \bibinfo {author} {\bibfnamefont {H.}~\bibnamefont {Ji}},
  \bibinfo {author} {\bibfnamefont {M.}~\bibnamefont {Yamada}}, \bibinfo
  {author} {\bibfnamefont {J.}~\bibnamefont {Yoo}}, \ and\ \bibinfo {author}
  {\bibfnamefont {W.}~\bibnamefont {Fox}},\ }\bibfield  {title} {\enquote
  {\bibinfo {title} {{Laboratory Observation of Resistive Electron Tearing in a
  Two-Fluid Reconnecting Current Sheet}},}\ }\href@noop {} {\bibfield
  {journal} {\bibinfo  {journal} {Physical Review Letters}\ }\textbf {\bibinfo
  {volume} {117}},\ \bibinfo {pages} {095001} (\bibinfo {year}
  {2016})}\BibitemShut {NoStop}%
\bibitem [{\citenamefont {{Ren}}\ \emph {et~al.}(2005)\citenamefont {{Ren}},
  \citenamefont {{Yamada}}, \citenamefont {{Gerhardt}}, \citenamefont {{Ji}},
  \citenamefont {{Kulsrud}},\ and\ \citenamefont {{Kuritsyn}}}]{Ren:2005}%
  \BibitemOpen
  \bibfield  {author} {\bibinfo {author} {\bibfnamefont {Y.}~\bibnamefont
  {{Ren}}}, \bibinfo {author} {\bibfnamefont {M.}~\bibnamefont {{Yamada}}},
  \bibinfo {author} {\bibfnamefont {S.}~\bibnamefont {{Gerhardt}}}, \bibinfo
  {author} {\bibfnamefont {H.}~\bibnamefont {{Ji}}}, \bibinfo {author}
  {\bibfnamefont {R.}~\bibnamefont {{Kulsrud}}}, \ and\ \bibinfo {author}
  {\bibfnamefont {A.}~\bibnamefont {{Kuritsyn}}},\ }\bibfield  {title}
  {\enquote {\bibinfo {title} {{Experimental Verification of the Hall Effect
  during Magnetic Reconnection in a Laboratory Plasma}},}\ }\href@noop {}
  {\bibfield  {journal} {\bibinfo  {journal} {Phys.~Rev.~Lett.}\ }\textbf
  {\bibinfo {volume} {95}},\ \bibinfo {pages} {055003} (\bibinfo {year}
  {2005})}\BibitemShut {NoStop}%
\bibitem [{\citenamefont {{Yamada}}\ \emph {et~al.}(2006)\citenamefont
  {{Yamada}}, \citenamefont {{Ren}}, \citenamefont {{Ji}}, \citenamefont
  {{Breslau}}, \citenamefont {{Gerhardt}}, \citenamefont {{Kulsrud}},\ and\
  \citenamefont {{Kuritsyn}}}]{Yamada:2006}%
  \BibitemOpen
  \bibfield  {author} {\bibinfo {author} {\bibfnamefont {M.}~\bibnamefont
  {{Yamada}}}, \bibinfo {author} {\bibfnamefont {Y.}~\bibnamefont {{Ren}}},
  \bibinfo {author} {\bibfnamefont {H.}~\bibnamefont {{Ji}}}, \bibinfo {author}
  {\bibfnamefont {J.}~\bibnamefont {{Breslau}}}, \bibinfo {author}
  {\bibfnamefont {S.}~\bibnamefont {{Gerhardt}}}, \bibinfo {author}
  {\bibfnamefont {R.}~\bibnamefont {{Kulsrud}}}, \ and\ \bibinfo {author}
  {\bibfnamefont {A.}~\bibnamefont {{Kuritsyn}}},\ }\bibfield  {title}
  {\enquote {\bibinfo {title} {{Experimental study of two-fluid effects on
  magnetic reconnection in a laboratory plasma with variable
  collisionality}},}\ }\href@noop {} {\bibfield  {journal} {\bibinfo  {journal}
  {Phys. Plasmas}\ }\textbf {\bibinfo {volume} {13}},\ \bibinfo {pages}
  {052119} (\bibinfo {year} {2006})}\BibitemShut {NoStop}%
\bibitem [{\citenamefont {{Ji}}\ \emph {et~al.}(2008)\citenamefont {{Ji}},
  \citenamefont {{Ren}}, \citenamefont {{Yamada}}, \citenamefont {{Dorfman}},
  \citenamefont {{Daughton}},\ and\ \citenamefont {{Gerhardt}}}]{Ji:2008}%
  \BibitemOpen
  \bibfield  {author} {\bibinfo {author} {\bibfnamefont {H.}~\bibnamefont
  {{Ji}}}, \bibinfo {author} {\bibfnamefont {Y.}~\bibnamefont {{Ren}}},
  \bibinfo {author} {\bibfnamefont {M.}~\bibnamefont {{Yamada}}}, \bibinfo
  {author} {\bibfnamefont {S.}~\bibnamefont {{Dorfman}}}, \bibinfo {author}
  {\bibfnamefont {W.}~\bibnamefont {{Daughton}}}, \ and\ \bibinfo {author}
  {\bibfnamefont {S.~P.}\ \bibnamefont {{Gerhardt}}},\ }\bibfield  {title}
  {\enquote {\bibinfo {title} {{New insights into dissipation in the electron
  layer during magnetic reconnection}},}\ }\href@noop {} {\bibfield  {journal}
  {\bibinfo  {journal} {Geophys.~Res.~Lett.}\ }\textbf {\bibinfo {volume}
  {35}},\ \bibinfo {pages} {L13106} (\bibinfo {year} {2008})}\BibitemShut
  {NoStop}%
\bibitem [{\citenamefont {{Ren}}\ \emph {et~al.}(2008)\citenamefont {{Ren}},
  \citenamefont {{Yamada}}, \citenamefont {Ji}, \citenamefont {{Gerhardt}},\
  and\ \citenamefont {{Kulsrud}}}]{Ren:2008}%
  \BibitemOpen
  \bibfield  {author} {\bibinfo {author} {\bibfnamefont {Y.}~\bibnamefont
  {{Ren}}}, \bibinfo {author} {\bibfnamefont {M.}~\bibnamefont {{Yamada}}},
  \bibinfo {author} {\bibfnamefont {H.}~\bibnamefont {Ji}}, \bibinfo {author}
  {\bibfnamefont {S.}~\bibnamefont {{Gerhardt}}}, \ and\ \bibinfo {author}
  {\bibfnamefont {R.}~\bibnamefont {{Kulsrud}}},\ }\bibfield  {title} {\enquote
  {\bibinfo {title} {{Identification of the Electron Diffusion Region during
  Magnetic Reconnection in a Laboratory Plasma}},}\ }\href@noop {} {\bibfield
  {journal} {\bibinfo  {journal} {Phys. Rev. Lett.}\ }\textbf {\bibinfo
  {volume} {101}},\ \bibinfo {pages} {085003} (\bibinfo {year}
  {2008})}\BibitemShut {NoStop}%
\bibitem [{\citenamefont {Yoo}\ \emph {et~al.}(2013)\citenamefont {Yoo},
  \citenamefont {Yamada}, \citenamefont {Ji},\ and\ \citenamefont
  {Myers}}]{Yoo:2013}%
  \BibitemOpen
  \bibfield  {author} {\bibinfo {author} {\bibfnamefont {J.}~\bibnamefont
  {Yoo}}, \bibinfo {author} {\bibfnamefont {M.}~\bibnamefont {Yamada}},
  \bibinfo {author} {\bibfnamefont {H.}~\bibnamefont {Ji}}, \ and\ \bibinfo
  {author} {\bibfnamefont {C.}~\bibnamefont {Myers}},\ }\bibfield  {title}
  {\enquote {\bibinfo {title} {Observation of ion acceleration and heating
  during collisionless magnetic reconnection in a laboratory plasma},}\
  }\href@noop {} {\bibfield  {journal} {\bibinfo  {journal} {Phys.~Rev.~Lett.}\
  }\textbf {\bibinfo {volume} {110}},\ \bibinfo {pages} {215007} (\bibinfo
  {year} {2013})}\BibitemShut {NoStop}%
\bibitem [{\citenamefont {{Oz}}\ \emph {et~al.}(2011)\citenamefont {{Oz}},
  \citenamefont {{Myers}}, \citenamefont {{Yamada}}, \citenamefont {{Ji}},
  \citenamefont {{Kulsrud}},\ and\ \citenamefont {{Xie}}}]{Oz:2011}%
  \BibitemOpen
  \bibfield  {author} {\bibinfo {author} {\bibfnamefont {E.}~\bibnamefont
  {{Oz}}}, \bibinfo {author} {\bibfnamefont {C.~E.}\ \bibnamefont {{Myers}}},
  \bibinfo {author} {\bibfnamefont {M.}~\bibnamefont {{Yamada}}}, \bibinfo
  {author} {\bibfnamefont {H.}~\bibnamefont {{Ji}}}, \bibinfo {author}
  {\bibfnamefont {R.~M.}\ \bibnamefont {{Kulsrud}}}, \ and\ \bibinfo {author}
  {\bibfnamefont {J.}~\bibnamefont {{Xie}}},\ }\bibfield  {title} {\enquote
  {\bibinfo {title} {{Experimental verification of the Kruskal-Shafranov
  stability limit in line-tied partial-toroidal plasmas}},}\ }\href@noop {}
  {\bibfield  {journal} {\bibinfo  {journal} {Phys.~Plasmas}\ }\textbf
  {\bibinfo {volume} {18}},\ \bibinfo {pages} {102107} (\bibinfo {year}
  {2011})}\BibitemShut {NoStop}%
\bibitem [{\citenamefont {Myers}\ \emph {et~al.}(2015)\citenamefont {Myers},
  \citenamefont {Yamada}, \citenamefont {Ji}, \citenamefont {Yoo},
  \citenamefont {Fox}, \citenamefont {Jara-Almonte}, \citenamefont {Savcheva},\
  and\ \citenamefont {DeLuca}}]{Myers:2015}%
  \BibitemOpen
  \bibfield  {author} {\bibinfo {author} {\bibfnamefont {C.~E.}\ \bibnamefont
  {Myers}}, \bibinfo {author} {\bibfnamefont {M.}~\bibnamefont {Yamada}},
  \bibinfo {author} {\bibfnamefont {H.}~\bibnamefont {Ji}}, \bibinfo {author}
  {\bibfnamefont {J.}~\bibnamefont {Yoo}}, \bibinfo {author} {\bibfnamefont
  {W.}~\bibnamefont {Fox}}, \bibinfo {author} {\bibfnamefont {J.}~\bibnamefont
  {Jara-Almonte}}, \bibinfo {author} {\bibfnamefont {A.}~\bibnamefont
  {Savcheva}}, \ and\ \bibinfo {author} {\bibfnamefont {E.~E.}\ \bibnamefont
  {DeLuca}},\ }\bibfield  {title} {\enquote {\bibinfo {title} {{A dynamic
  magnetic tension force as the cause of failed solar eruptions}},}\
  }\href@noop {} {\bibfield  {journal} {\bibinfo  {journal} {Nature}\ }\textbf
  {\bibinfo {volume} {528}},\ \bibinfo {pages} {526--529} (\bibinfo {year}
  {2015})}\BibitemShut {NoStop}%
\bibitem [{\citenamefont {Tharp}\ \emph {et~al.}(2012)\citenamefont {Tharp},
  \citenamefont {Yamada}, \citenamefont {Ji}, \citenamefont {Lawrence},
  \citenamefont {Dorfman}, \citenamefont {Myers},\ and\ \citenamefont
  {Yoo}}]{Tharp:2012}%
  \BibitemOpen
  \bibfield  {author} {\bibinfo {author} {\bibfnamefont {T.}~\bibnamefont
  {Tharp}}, \bibinfo {author} {\bibfnamefont {M.}~\bibnamefont {Yamada}},
  \bibinfo {author} {\bibfnamefont {H.}~\bibnamefont {Ji}}, \bibinfo {author}
  {\bibfnamefont {E.}~\bibnamefont {Lawrence}}, \bibinfo {author}
  {\bibfnamefont {S.}~\bibnamefont {Dorfman}}, \bibinfo {author} {\bibfnamefont
  {C.}~\bibnamefont {Myers}}, \ and\ \bibinfo {author} {\bibfnamefont
  {J.}~\bibnamefont {Yoo}},\ }\bibfield  {title} {\enquote {\bibinfo {title}
  {Quantitative study of guide-field effects on hall reconnection in a
  laboratory plasma},}\ }\href@noop {} {\bibfield  {journal} {\bibinfo
  {journal} {Phys.~Rev.~Lett.}\ }\textbf {\bibinfo {volume} {109}},\ \bibinfo
  {pages} {169002} (\bibinfo {year} {2012})}\BibitemShut {NoStop}%
\bibitem [{\citenamefont {Fox}\ \emph {et~al.}(2017)\citenamefont {Fox},
  \citenamefont {Sciortino}, \citenamefont {von Stechow}, \citenamefont
  {Jara-Almonte}, \citenamefont {Yoo}, \citenamefont {Ji},\ and\ \citenamefont
  {Yamada}}]{Fox:2017}%
  \BibitemOpen
  \bibfield  {author} {\bibinfo {author} {\bibfnamefont {W.}~\bibnamefont
  {Fox}}, \bibinfo {author} {\bibfnamefont {F.}~\bibnamefont {Sciortino}},
  \bibinfo {author} {\bibfnamefont {A.}~\bibnamefont {von Stechow}}, \bibinfo
  {author} {\bibfnamefont {J.}~\bibnamefont {Jara-Almonte}}, \bibinfo {author}
  {\bibfnamefont {J.}~\bibnamefont {Yoo}}, \bibinfo {author} {\bibfnamefont
  {H.}~\bibnamefont {Ji}}, \ and\ \bibinfo {author} {\bibfnamefont
  {M.}~\bibnamefont {Yamada}},\ }\bibfield  {title} {\enquote {\bibinfo {title}
  {Experimental verification of the role of electron pressure in fast magnetic
  reconnection with a guide field},}\ }\href@noop {} {\bibfield  {journal}
  {\bibinfo  {journal} {Phys.~Rev.~Lett.}\ }\textbf {\bibinfo {volume} {118}},\
  \bibinfo {pages} {125002} (\bibinfo {year} {2017})}\BibitemShut {NoStop}%
\bibitem [{\citenamefont {Carter}\ \emph {et~al.}(2002)\citenamefont {Carter},
  \citenamefont {Ji}, \citenamefont {Trintchouk}, \citenamefont {Yamada},\ and\
  \citenamefont {Kulsrud}}]{Carter:2002a}%
  \BibitemOpen
  \bibfield  {author} {\bibinfo {author} {\bibfnamefont {T.}~\bibnamefont
  {Carter}}, \bibinfo {author} {\bibfnamefont {H.}~\bibnamefont {Ji}}, \bibinfo
  {author} {\bibfnamefont {F.}~\bibnamefont {Trintchouk}}, \bibinfo {author}
  {\bibfnamefont {M.}~\bibnamefont {Yamada}}, \ and\ \bibinfo {author}
  {\bibfnamefont {R.}~\bibnamefont {Kulsrud}},\ }\bibfield  {title} {\enquote
  {\bibinfo {title} {Measurement of lower-hybrid drift turbulence in a
  reconnecting current sheet},}\ }\href@noop {} {\bibfield  {journal} {\bibinfo
   {journal} {Phys.~Rev.~Lett.}\ }\textbf {\bibinfo {volume} {88}},\ \bibinfo
  {pages} {015001} (\bibinfo {year} {2002})}\BibitemShut {NoStop}%
\bibitem [{\citenamefont {Ji}\ \emph {et~al.}(2004)\citenamefont {Ji},
  \citenamefont {Terry}, \citenamefont {Yamada}, \citenamefont {Kulsrud},
  \citenamefont {Kuritsyn},\ and\ \citenamefont {Ren}}]{Ji:2004}%
  \BibitemOpen
  \bibfield  {author} {\bibinfo {author} {\bibfnamefont {H.}~\bibnamefont
  {Ji}}, \bibinfo {author} {\bibfnamefont {S.}~\bibnamefont {Terry}}, \bibinfo
  {author} {\bibfnamefont {M.}~\bibnamefont {Yamada}}, \bibinfo {author}
  {\bibfnamefont {R.}~\bibnamefont {Kulsrud}}, \bibinfo {author} {\bibfnamefont
  {A.}~\bibnamefont {Kuritsyn}}, \ and\ \bibinfo {author} {\bibfnamefont
  {Y.}~\bibnamefont {Ren}},\ }\bibfield  {title} {\enquote {\bibinfo {title}
  {Electromagnetic fluctuation during fast reconnection in a laboratory
  plasma},}\ }\href@noop {} {\bibfield  {journal} {\bibinfo  {journal}
  {Phys.~Rev.~Lett.}\ }\textbf {\bibinfo {volume} {92}},\ \bibinfo {pages}
  {115001} (\bibinfo {year} {2004})}\BibitemShut {NoStop}%
\bibitem [{\citenamefont {Dorfman}\ \emph {et~al.}(2013)\citenamefont
  {Dorfman}, \citenamefont {Ji}, \citenamefont {Yamada}, \citenamefont {Yoo},
  \citenamefont {Lawrence}, \citenamefont {Myers},\ and\ \citenamefont
  {Tharp}}]{Dorfman:2013a}%
  \BibitemOpen
  \bibfield  {author} {\bibinfo {author} {\bibfnamefont {S.}~\bibnamefont
  {Dorfman}}, \bibinfo {author} {\bibfnamefont {H.}~\bibnamefont {Ji}},
  \bibinfo {author} {\bibfnamefont {M.}~\bibnamefont {Yamada}}, \bibinfo
  {author} {\bibfnamefont {J.}~\bibnamefont {Yoo}}, \bibinfo {author}
  {\bibfnamefont {E.}~\bibnamefont {Lawrence}}, \bibinfo {author}
  {\bibfnamefont {C.}~\bibnamefont {Myers}}, \ and\ \bibinfo {author}
  {\bibfnamefont {T.}~\bibnamefont {Tharp}},\ }\bibfield  {title} {\enquote
  {\bibinfo {title} {Three-dimensional, impulsive magnetic reconnection in a
  laboratory plasma},}\ }\href@noop {} {\bibfield  {journal} {\bibinfo
  {journal} {Geophys.~Res.~Lett.}\ }\textbf {\bibinfo {volume} {40}},\ \bibinfo
  {pages} {233--238} (\bibinfo {year} {2013})}\BibitemShut {NoStop}%
\bibitem [{\citenamefont {Lawrence}\ \emph {et~al.}(2013)\citenamefont
  {Lawrence}, \citenamefont {Ji}, \citenamefont {Yamada},\ and\ \citenamefont
  {Yoo}}]{Lawrence:2013}%
  \BibitemOpen
  \bibfield  {author} {\bibinfo {author} {\bibfnamefont {E.}~\bibnamefont
  {Lawrence}}, \bibinfo {author} {\bibfnamefont {H.}~\bibnamefont {Ji}},
  \bibinfo {author} {\bibfnamefont {M.}~\bibnamefont {Yamada}}, \ and\ \bibinfo
  {author} {\bibfnamefont {J.}~\bibnamefont {Yoo}},\ }\bibfield  {title}
  {\enquote {\bibinfo {title} {Laboratory study of hall reconnection in
  partially ionized plasmas},}\ }\href@noop {} {\bibfield  {journal} {\bibinfo
  {journal} {Phys.~Rev.~Lett.}\ }\textbf {\bibinfo {volume} {110}},\ \bibinfo
  {pages} {015001} (\bibinfo {year} {2013})}\BibitemShut {NoStop}%
\bibitem [{\citenamefont {Yamada}\ \emph {et~al.}(2014)\citenamefont {Yamada},
  \citenamefont {Yoo}, \citenamefont {Jara-Almonte}, \citenamefont {Ji},
  \citenamefont {Kulsrud},\ and\ \citenamefont {Myers}}]{Yamada:2014}%
  \BibitemOpen
  \bibfield  {author} {\bibinfo {author} {\bibfnamefont {M.}~\bibnamefont
  {Yamada}}, \bibinfo {author} {\bibfnamefont {J.}~\bibnamefont {Yoo}},
  \bibinfo {author} {\bibfnamefont {J.}~\bibnamefont {Jara-Almonte}}, \bibinfo
  {author} {\bibfnamefont {H.}~\bibnamefont {Ji}}, \bibinfo {author}
  {\bibfnamefont {R.~M.}\ \bibnamefont {Kulsrud}}, \ and\ \bibinfo {author}
  {\bibfnamefont {C.~E.}\ \bibnamefont {Myers}},\ }\bibfield  {title} {\enquote
  {\bibinfo {title} {Conversion of magnetic energy in the magnetic reconnection
  layer of a laboratory plasma},}\ }\href@noop {} {\bibfield  {journal}
  {\bibinfo  {journal} {Nat. Commun.}\ }\textbf {\bibinfo {volume} {5}},\
  \bibinfo {pages} {4774} (\bibinfo {year} {2014})}\BibitemShut {NoStop}%
\bibitem [{\citenamefont {{Yamada}}\ \emph {et~al.}(2015)\citenamefont
  {{Yamada}}, \citenamefont {{Yoo}}, \citenamefont {{Jara-Almonte}},
  \citenamefont {{Daughton}}, \citenamefont {{Ji}}, \citenamefont {{Kulsrud}},\
  and\ \citenamefont {{Myers}}}]{Yamada:2015}%
  \BibitemOpen
  \bibfield  {author} {\bibinfo {author} {\bibfnamefont {M.}~\bibnamefont
  {{Yamada}}}, \bibinfo {author} {\bibfnamefont {J.}~\bibnamefont {{Yoo}}},
  \bibinfo {author} {\bibfnamefont {J.}~\bibnamefont {{Jara-Almonte}}},
  \bibinfo {author} {\bibfnamefont {W.}~\bibnamefont {{Daughton}}}, \bibinfo
  {author} {\bibfnamefont {H.}~\bibnamefont {{Ji}}}, \bibinfo {author}
  {\bibfnamefont {R.~M.}\ \bibnamefont {{Kulsrud}}}, \ and\ \bibinfo {author}
  {\bibfnamefont {C.~E.}\ \bibnamefont {{Myers}}},\ }\bibfield  {title}
  {\enquote {\bibinfo {title} {{Study of energy conversion and partitioning in
  the magnetic reconnection layer of a laboratory plasmas)}},}\ }\href@noop {}
  {\bibfield  {journal} {\bibinfo  {journal} {Phys.~Plasmas}\ }\textbf
  {\bibinfo {volume} {22}},\ \bibinfo {eid} {056501} (\bibinfo {year}
  {2015})}\BibitemShut {NoStop}%
\bibitem [{\citenamefont {Yoo}\ \emph {et~al.}(2014)\citenamefont {Yoo},
  \citenamefont {Yamada}, \citenamefont {Ji}, \citenamefont {Jara-Almonte},
  \citenamefont {Myers},\ and\ \citenamefont {Chen}}]{Yoo:2014}%
  \BibitemOpen
  \bibfield  {author} {\bibinfo {author} {\bibfnamefont {J.}~\bibnamefont
  {Yoo}}, \bibinfo {author} {\bibfnamefont {M.}~\bibnamefont {Yamada}},
  \bibinfo {author} {\bibfnamefont {H.}~\bibnamefont {Ji}}, \bibinfo {author}
  {\bibfnamefont {J.}~\bibnamefont {Jara-Almonte}}, \bibinfo {author}
  {\bibfnamefont {C.~E.}\ \bibnamefont {Myers}}, \ and\ \bibinfo {author}
  {\bibfnamefont {L.-J.}\ \bibnamefont {Chen}},\ }\bibfield  {title} {\enquote
  {\bibinfo {title} {{Laboratory Study of Magnetic Reconnection with a Density
  Asymmetry across the Current Sheet}},}\ }\href@noop {} {\bibfield  {journal}
  {\bibinfo  {journal} {Phys.~Rev.~Lett.}\ }\textbf {\bibinfo {volume} {113}},\
  \bibinfo {pages} {095002} (\bibinfo {year} {2014})}\BibitemShut {NoStop}%
\bibitem [{\citenamefont {Yoo}\ \emph {et~al.}(2017)\citenamefont {Yoo},
  \citenamefont {Na}, \citenamefont {Jara-Almonte}, \citenamefont {Yamada},
  \citenamefont {Ji}, \citenamefont {Fox},\ and\ \citenamefont
  {Chen}}]{Yoo:2017}%
  \BibitemOpen
  \bibfield  {author} {\bibinfo {author} {\bibfnamefont {J.}~\bibnamefont
  {Yoo}}, \bibinfo {author} {\bibfnamefont {B.}~\bibnamefont {Na}}, \bibinfo
  {author} {\bibfnamefont {J.}~\bibnamefont {Jara-Almonte}}, \bibinfo {author}
  {\bibfnamefont {M.}~\bibnamefont {Yamada}}, \bibinfo {author} {\bibfnamefont
  {H.}~\bibnamefont {Ji}}, \bibinfo {author} {\bibfnamefont {W.}~\bibnamefont
  {Fox}}, \ and\ \bibinfo {author} {\bibfnamefont {L.-J.}\ \bibnamefont
  {Chen}},\ }\bibfield  {title} {\enquote {\bibinfo {title} {{Electron heating
  and energy inventory during asymmetric reconnection in a laboratory
  plasma}},}\ }\href@noop {} {\bibfield  {journal} {\bibinfo  {journal} {in
  press, J. Geophys. Res.}\ } (\bibinfo {year} {2017})}\BibitemShut {NoStop}%
\bibitem [{\citenamefont {Shibata}\ and\ \citenamefont
  {Tanuma}(2001)}]{Shibata:2001}%
  \BibitemOpen
  \bibfield  {author} {\bibinfo {author} {\bibfnamefont {K.}~\bibnamefont
  {Shibata}}\ and\ \bibinfo {author} {\bibfnamefont {S.}~\bibnamefont
  {Tanuma}},\ }\bibfield  {title} {\enquote {\bibinfo {title}
  {Plasmoid-induced-reconnection and fractal reconnection},}\ }\href@noop {}
  {\bibfield  {journal} {\bibinfo  {journal} {Earth Planets Space}\ }\textbf
  {\bibinfo {volume} {53}},\ \bibinfo {pages} {473} (\bibinfo {year}
  {2001})}\BibitemShut {NoStop}%
\bibitem [{\citenamefont {Loureiro}, \citenamefont {Schekochihin},\ and\
  \citenamefont {Cowley}(2007)}]{Loureiro:2007}%
  \BibitemOpen
  \bibfield  {author} {\bibinfo {author} {\bibfnamefont {N.}~\bibnamefont
  {Loureiro}}, \bibinfo {author} {\bibfnamefont {A.}~\bibnamefont
  {Schekochihin}}, \ and\ \bibinfo {author} {\bibfnamefont {S.}~\bibnamefont
  {Cowley}},\ }\bibfield  {title} {\enquote {\bibinfo {title} {Instability of
  current sheets and formation of plasmoid chains},}\ }\href@noop {} {\bibfield
   {journal} {\bibinfo  {journal} {Phys.~Plasmas}\ }\textbf {\bibinfo {volume}
  {14}},\ \bibinfo {pages} {100703} (\bibinfo {year} {2007})}\BibitemShut
  {NoStop}%
\bibitem [{\citenamefont {Bhattacharjee}\ \emph {et~al.}(2009)\citenamefont
  {Bhattacharjee}, \citenamefont {Huang}, \citenamefont {Yang},\ and\
  \citenamefont {Rogers}}]{Bhattacharjee:2009}%
  \BibitemOpen
  \bibfield  {author} {\bibinfo {author} {\bibfnamefont {A.}~\bibnamefont
  {Bhattacharjee}}, \bibinfo {author} {\bibfnamefont {Y.}~\bibnamefont
  {Huang}}, \bibinfo {author} {\bibfnamefont {H.}~\bibnamefont {Yang}}, \ and\
  \bibinfo {author} {\bibfnamefont {B.}~\bibnamefont {Rogers}},\ }\bibfield
  {title} {\enquote {\bibinfo {title} {Fast reconnection in
  high-lundquist-number plasmas due to the plasmoid instability},}\ }\href@noop
  {} {\bibfield  {journal} {\bibinfo  {journal} {Phys. Plasmas}\ }\textbf
  {\bibinfo {volume} {16}},\ \bibinfo {pages} {112102} (\bibinfo {year}
  {2009})}\BibitemShut {NoStop}%
\bibitem [{\citenamefont {{Samtaney}}\ \emph {et~al.}(2009)\citenamefont
  {{Samtaney}}, \citenamefont {{Loureiro}}, \citenamefont {{Uzdensky}},
  \citenamefont {{Schekochihin}},\ and\ \citenamefont
  {{Cowley}}}]{Samtaney:2009}%
  \BibitemOpen
  \bibfield  {author} {\bibinfo {author} {\bibfnamefont {R.}~\bibnamefont
  {{Samtaney}}}, \bibinfo {author} {\bibfnamefont {N.~F.}\ \bibnamefont
  {{Loureiro}}}, \bibinfo {author} {\bibfnamefont {D.~A.}\ \bibnamefont
  {{Uzdensky}}}, \bibinfo {author} {\bibfnamefont {A.~A.}\ \bibnamefont
  {{Schekochihin}}}, \ and\ \bibinfo {author} {\bibfnamefont {S.~C.}\
  \bibnamefont {{Cowley}}},\ }\bibfield  {title} {\enquote {\bibinfo {title}
  {{Formation of Plasmoid Chains in Magnetic Reconnection}},}\ }\href {\doibase
  10.1103/PhysRevLett.103.105004} {\bibfield  {journal} {\bibinfo  {journal}
  {Phys.~Rev.~Lett.}\ }\textbf {\bibinfo {volume} {103}},\ \bibinfo {eid}
  {105004} (\bibinfo {year} {2009})},\ \Eprint {http://arxiv.org/abs/0903.0542}
  {arXiv:0903.0542 [astro-ph.SR]} \BibitemShut {NoStop}%
\bibitem [{\citenamefont {{Uzdensky}}, \citenamefont {{Loureiro}},\ and\
  \citenamefont {{Schekochihin}}(2010)}]{Uzdensky:2010}%
  \BibitemOpen
  \bibfield  {author} {\bibinfo {author} {\bibfnamefont {D.~A.}\ \bibnamefont
  {{Uzdensky}}}, \bibinfo {author} {\bibfnamefont {N.~F.}\ \bibnamefont
  {{Loureiro}}}, \ and\ \bibinfo {author} {\bibfnamefont {A.~A.}\ \bibnamefont
  {{Schekochihin}}},\ }\bibfield  {title} {\enquote {\bibinfo {title} {{Fast
  Magnetic Reconnection in the Plasmoid-Dominated Regime}},}\ }\href {\doibase
  10.1103/PhysRevLett.105.235002} {\bibfield  {journal} {\bibinfo  {journal}
  {Phys.~Rev.~Lett.}\ }\textbf {\bibinfo {volume} {105}},\ \bibinfo {eid}
  {235002} (\bibinfo {year} {2010})},\ \Eprint {http://arxiv.org/abs/1008.3330}
  {arXiv:1008.3330 [astro-ph.SR]} \BibitemShut {NoStop}%
\bibitem [{\citenamefont {{Loureiro}}\ \emph {et~al.}(2012)\citenamefont
  {{Loureiro}}, \citenamefont {{Samtaney}}, \citenamefont {{Schekochihin}},\
  and\ \citenamefont {{Uzdensky}}}]{Loureiro:2012}%
  \BibitemOpen
  \bibfield  {author} {\bibinfo {author} {\bibfnamefont {N.~F.}\ \bibnamefont
  {{Loureiro}}}, \bibinfo {author} {\bibfnamefont {R.}~\bibnamefont
  {{Samtaney}}}, \bibinfo {author} {\bibfnamefont {A.~A.}\ \bibnamefont
  {{Schekochihin}}}, \ and\ \bibinfo {author} {\bibfnamefont {D.~A.}\
  \bibnamefont {{Uzdensky}}},\ }\bibfield  {title} {\enquote {\bibinfo {title}
  {{Magnetic reconnection and stochastic plasmoid chains in
  high-Lundquist-number plasmas}},}\ }\href {\doibase 10.1063/1.3703318}
  {\bibfield  {journal} {\bibinfo  {journal} {Phys.~Plasmas}\ }\textbf
  {\bibinfo {volume} {19}},\ \bibinfo {pages} {042303--042303} (\bibinfo {year}
  {2012})},\ \Eprint {http://arxiv.org/abs/1108.4040} {arXiv:1108.4040
  [astro-ph.SR]} \BibitemShut {NoStop}%
\bibitem [{\citenamefont {Daughton}\ \emph {et~al.}(2009)\citenamefont
  {Daughton}, \citenamefont {Roytershteyn}, \citenamefont {Albright},
  \citenamefont {Karimabadi}, \citenamefont {Yin},\ and\ \citenamefont
  {Bowers}}]{Daughton:2009b}%
  \BibitemOpen
  \bibfield  {author} {\bibinfo {author} {\bibfnamefont {W.}~\bibnamefont
  {Daughton}}, \bibinfo {author} {\bibfnamefont {V.}~\bibnamefont
  {Roytershteyn}}, \bibinfo {author} {\bibfnamefont {B.}~\bibnamefont
  {Albright}}, \bibinfo {author} {\bibfnamefont {H.}~\bibnamefont
  {Karimabadi}}, \bibinfo {author} {\bibfnamefont {L.}~\bibnamefont {Yin}}, \
  and\ \bibinfo {author} {\bibfnamefont {K.}~\bibnamefont {Bowers}},\
  }\bibfield  {title} {\enquote {\bibinfo {title} {Transition from collisional
  to kinetic reconnection in large-scale plasmas},}\ }\href@noop {} {\bibfield
  {journal} {\bibinfo  {journal} {Phys.~Rev.~Lett.}\ }\textbf {\bibinfo
  {volume} {103}},\ \bibinfo {pages} {065004} (\bibinfo {year}
  {2009})}\BibitemShut {NoStop}%
\bibitem [{\citenamefont {{Olson}}\ \emph {et~al.}(2016)\citenamefont
  {{Olson}}, \citenamefont {{Egedal}}, \citenamefont {{Greess}}, \citenamefont
  {{Myers}}, \citenamefont {{Clark}}, \citenamefont {{Endrizzi}}, \citenamefont
  {{Flanagan}}, \citenamefont {{Milhone}}, \citenamefont {{Peterson}},
  \citenamefont {{Wallace}}, \citenamefont {{Weisberg}},\ and\ \citenamefont
  {{Forest}}}]{Olson:2016}%
  \BibitemOpen
  \bibfield  {author} {\bibinfo {author} {\bibfnamefont {J.}~\bibnamefont
  {{Olson}}}, \bibinfo {author} {\bibfnamefont {J.}~\bibnamefont {{Egedal}}},
  \bibinfo {author} {\bibfnamefont {S.}~\bibnamefont {{Greess}}}, \bibinfo
  {author} {\bibfnamefont {R.}~\bibnamefont {{Myers}}}, \bibinfo {author}
  {\bibfnamefont {M.}~\bibnamefont {{Clark}}}, \bibinfo {author} {\bibfnamefont
  {D.}~\bibnamefont {{Endrizzi}}}, \bibinfo {author} {\bibfnamefont
  {K.}~\bibnamefont {{Flanagan}}}, \bibinfo {author} {\bibfnamefont
  {J.}~\bibnamefont {{Milhone}}}, \bibinfo {author} {\bibfnamefont
  {E.}~\bibnamefont {{Peterson}}}, \bibinfo {author} {\bibfnamefont
  {J.}~\bibnamefont {{Wallace}}}, \bibinfo {author} {\bibfnamefont
  {D.}~\bibnamefont {{Weisberg}}}, \ and\ \bibinfo {author} {\bibfnamefont
  {C.~B.}\ \bibnamefont {{Forest}}},\ }\bibfield  {title} {\enquote {\bibinfo
  {title} {{Experimental Demonstration of the Collisionless Plasmoid
  Instability below the Ion Kinetic Scale during Magnetic Reconnection}},}\
  }\href {\doibase 10.1103/PhysRevLett.116.255001} {\bibfield  {journal}
  {\bibinfo  {journal} {Phys.~Rev.~Lett.}\ }\textbf {\bibinfo {volume} {116}},\
  \bibinfo {eid} {255001} (\bibinfo {year} {2016})}\BibitemShut {NoStop}%
\bibitem [{\citenamefont {{Keiling}}(2009)}]{Keiling:2009}%
  \BibitemOpen
  \bibfield  {author} {\bibinfo {author} {\bibfnamefont {A.}~\bibnamefont
  {{Keiling}}},\ }\bibfield  {title} {\enquote {\bibinfo {title} {{Alfv{\'e}n
  Waves and Their Roles in the Dynamics of the Earth's Magnetotail: A
  Review}},}\ }\href {\doibase 10.1007/s11214-008-9463-8} {\bibfield  {journal}
  {\bibinfo  {journal} {Space Sci.~Rev.}\ }\textbf {\bibinfo {volume} {142}},\
  \bibinfo {pages} {73--156} (\bibinfo {year} {2009})}\BibitemShut {NoStop}%
\bibitem [{\citenamefont {{Hasegawa}}(1976)}]{Hasegawa:1976c}%
  \BibitemOpen
  \bibfield  {author} {\bibinfo {author} {\bibfnamefont {A.}~\bibnamefont
  {{Hasegawa}}},\ }\bibfield  {title} {\enquote {\bibinfo {title} {{Particle
  acceleration by MHD surface wave and formation of aurora}},}\ }\href
  {\doibase 10.1029/JA081i028p05083} {\bibfield  {journal} {\bibinfo  {journal}
  {J.~Geophys.~Res.}\ }\textbf {\bibinfo {volume} {81}},\ \bibinfo {pages}
  {5083--5090} (\bibinfo {year} {1976})}\BibitemShut {NoStop}%
\bibitem [{\citenamefont {{Goertz}}\ and\ \citenamefont
  {{Boswell}}(1979)}]{Goertz:1979}%
  \BibitemOpen
  \bibfield  {author} {\bibinfo {author} {\bibfnamefont {C.~K.}\ \bibnamefont
  {{Goertz}}}\ and\ \bibinfo {author} {\bibfnamefont {R.~W.}\ \bibnamefont
  {{Boswell}}},\ }\bibfield  {title} {\enquote {\bibinfo {title}
  {{Magnetosphere-ionosphere coupling}},}\ }\href {\doibase
  10.1029/JA084iA12p07239} {\bibfield  {journal} {\bibinfo  {journal}
  {J.~Geophys.~Res.}\ }\textbf {\bibinfo {volume} {84}},\ \bibinfo {pages}
  {7239--7246} (\bibinfo {year} {1979})}\BibitemShut {NoStop}%
\bibitem [{\citenamefont {{Kletzing}}(1994)}]{Kletzing:1994}%
  \BibitemOpen
  \bibfield  {author} {\bibinfo {author} {\bibfnamefont {C.~A.}\ \bibnamefont
  {{Kletzing}}},\ }\bibfield  {title} {\enquote {\bibinfo {title} {{Electron
  acceleration by kinetic Alfv{\'e}n waves}},}\ }\href {\doibase
  10.1029/94JA00345} {\bibfield  {journal} {\bibinfo  {journal}
  {J.~Geophys.~Res.}\ }\textbf {\bibinfo {volume} {99}},\ \bibinfo {pages}
  {11095--11104} (\bibinfo {year} {1994})}\BibitemShut {NoStop}%
\bibitem [{\citenamefont {{Kletzing}}\ and\ \citenamefont
  {{Hu}}(2001)}]{Kletzing:2001}%
  \BibitemOpen
  \bibfield  {author} {\bibinfo {author} {\bibfnamefont {C.~A.}\ \bibnamefont
  {{Kletzing}}}\ and\ \bibinfo {author} {\bibfnamefont {S.}~\bibnamefont
  {{Hu}}},\ }\bibfield  {title} {\enquote {\bibinfo {title} {{Alfv{\'e}n wave
  generated electron time dispersion}},}\ }\href {\doibase
  10.1029/2000GL012179} {\bibfield  {journal} {\bibinfo  {journal}
  {Geophys.~Res.~Lett.}\ }\textbf {\bibinfo {volume} {28}},\ \bibinfo {pages}
  {693--696} (\bibinfo {year} {2001})}\BibitemShut {NoStop}%
\bibitem [{\citenamefont {{Chaston}}\ \emph {et~al.}(2014)\citenamefont
  {{Chaston}}, \citenamefont {{Bonnell}}, \citenamefont {{Wygant}},
  \citenamefont {{Mozer}}, \citenamefont {{Bale}}, \citenamefont {{Kersten}},
  \citenamefont {{Breneman}}, \citenamefont {{Kletzing}}, \citenamefont
  {{Kurth}}, \citenamefont {{Hospodarsky}}, \citenamefont {{Smith}},\ and\
  \citenamefont {{MacDonald}}}]{Chaston:2014}%
  \BibitemOpen
  \bibfield  {author} {\bibinfo {author} {\bibfnamefont {C.~C.}\ \bibnamefont
  {{Chaston}}}, \bibinfo {author} {\bibfnamefont {J.~W.}\ \bibnamefont
  {{Bonnell}}}, \bibinfo {author} {\bibfnamefont {J.~R.}\ \bibnamefont
  {{Wygant}}}, \bibinfo {author} {\bibfnamefont {F.}~\bibnamefont {{Mozer}}},
  \bibinfo {author} {\bibfnamefont {S.~D.}\ \bibnamefont {{Bale}}}, \bibinfo
  {author} {\bibfnamefont {K.}~\bibnamefont {{Kersten}}}, \bibinfo {author}
  {\bibfnamefont {A.~W.}\ \bibnamefont {{Breneman}}}, \bibinfo {author}
  {\bibfnamefont {C.~A.}\ \bibnamefont {{Kletzing}}}, \bibinfo {author}
  {\bibfnamefont {W.~S.}\ \bibnamefont {{Kurth}}}, \bibinfo {author}
  {\bibfnamefont {G.~B.}\ \bibnamefont {{Hospodarsky}}}, \bibinfo {author}
  {\bibfnamefont {C.~W.}\ \bibnamefont {{Smith}}}, \ and\ \bibinfo {author}
  {\bibfnamefont {E.~A.}\ \bibnamefont {{MacDonald}}},\ }\bibfield  {title}
  {\enquote {\bibinfo {title} {{Observations of kinetic scale field line
  resonances}},}\ }\href {\doibase 10.1002/2013GL058507} {\bibfield  {journal}
  {\bibinfo  {journal} {Geophys.~Res.~Lett.}\ }\textbf {\bibinfo {volume}
  {41}},\ \bibinfo {pages} {209--215} (\bibinfo {year} {2014})}\BibitemShut
  {NoStop}%
\bibitem [{\citenamefont {{Chaston}}\ \emph {et~al.}(2015)\citenamefont
  {{Chaston}}, \citenamefont {{Bonnell}}, \citenamefont {{Wygant}},
  \citenamefont {{Kletzing}}, \citenamefont {{Reeves}}, \citenamefont
  {{Gerrard}}, \citenamefont {{Lanzerotti}},\ and\ \citenamefont
  {{Smith}}}]{Chaston:2015}%
  \BibitemOpen
  \bibfield  {author} {\bibinfo {author} {\bibfnamefont {C.~C.}\ \bibnamefont
  {{Chaston}}}, \bibinfo {author} {\bibfnamefont {J.~W.}\ \bibnamefont
  {{Bonnell}}}, \bibinfo {author} {\bibfnamefont {J.~R.}\ \bibnamefont
  {{Wygant}}}, \bibinfo {author} {\bibfnamefont {C.~A.}\ \bibnamefont
  {{Kletzing}}}, \bibinfo {author} {\bibfnamefont {G.~D.}\ \bibnamefont
  {{Reeves}}}, \bibinfo {author} {\bibfnamefont {A.}~\bibnamefont {{Gerrard}}},
  \bibinfo {author} {\bibfnamefont {L.}~\bibnamefont {{Lanzerotti}}}, \ and\
  \bibinfo {author} {\bibfnamefont {C.~W.}\ \bibnamefont {{Smith}}},\
  }\bibfield  {title} {\enquote {\bibinfo {title} {{Extreme ionospheric ion
  energization and electron heating in Alfv{\'e}n waves in the storm time inner
  magnetosphere}},}\ }\href {\doibase 10.1002/2015GL066674} {\bibfield
  {journal} {\bibinfo  {journal} {Geophys.~Res.~Lett.}\ }\textbf {\bibinfo
  {volume} {42}},\ \bibinfo {pages} {10} (\bibinfo {year} {2015})}\BibitemShut
  {NoStop}%
\bibitem [{\citenamefont {{Chaston}}\ \emph {et~al.}(2016)\citenamefont
  {{Chaston}}, \citenamefont {{Bonnell}}, \citenamefont {{Reeves}},\ and\
  \citenamefont {{Skoug}}}]{Chaston:2016}%
  \BibitemOpen
  \bibfield  {author} {\bibinfo {author} {\bibfnamefont {C.~C.}\ \bibnamefont
  {{Chaston}}}, \bibinfo {author} {\bibfnamefont {J.~W.}\ \bibnamefont
  {{Bonnell}}}, \bibinfo {author} {\bibfnamefont {G.~D.}\ \bibnamefont
  {{Reeves}}}, \ and\ \bibinfo {author} {\bibfnamefont {R.~M.}\ \bibnamefont
  {{Skoug}}},\ }\bibfield  {title} {\enquote {\bibinfo {title} {{Driving
  ionospheric outflows and magnetospheric O$^{+}$ energy density with
  Alfv{\'e}n waves}},}\ }\href {\doibase 10.1002/2016GL069008} {\bibfield
  {journal} {\bibinfo  {journal} {Geophys.~Res.~Lett.}\ }\textbf {\bibinfo
  {volume} {43}},\ \bibinfo {pages} {4825--4833} (\bibinfo {year}
  {2016})}\BibitemShut {NoStop}%
\bibitem [{\citenamefont {{Sagdeev}}(1966)}]{Sagdeev:1966}%
  \BibitemOpen
  \bibfield  {author} {\bibinfo {author} {\bibfnamefont {R.~Z.}\ \bibnamefont
  {{Sagdeev}}},\ }\bibfield  {title} {\enquote {\bibinfo {title} {{Cooperative
  Phenomena and Shock Waves in Collisionless Plasmas}},}\ }\href@noop {}
  {\bibfield  {journal} {\bibinfo  {journal} {Rev. Plasma Phys.}\ }\textbf
  {\bibinfo {volume} {4}},\ \bibinfo {pages} {23} (\bibinfo {year}
  {1966})}\BibitemShut {NoStop}%
\bibitem [{\citenamefont {{Lever}}, \citenamefont {{Quest}},\ and\
  \citenamefont {{Shapiro}}(2001)}]{Lever:2001}%
  \BibitemOpen
  \bibfield  {author} {\bibinfo {author} {\bibfnamefont {E.~L.}\ \bibnamefont
  {{Lever}}}, \bibinfo {author} {\bibfnamefont {K.~B.}\ \bibnamefont
  {{Quest}}}, \ and\ \bibinfo {author} {\bibfnamefont {V.~D.}\ \bibnamefont
  {{Shapiro}}},\ }\bibfield  {title} {\enquote {\bibinfo {title} {{Shock
  surfing vs. shock drift acceleration}},}\ }\href {\doibase
  10.1029/2000GL012516} {\bibfield  {journal} {\bibinfo  {journal} {Geophys.
  Res. Lett.}\ }\textbf {\bibinfo {volume} {28}},\ \bibinfo {pages}
  {1367--1370} (\bibinfo {year} {2001})}\BibitemShut {NoStop}%
\bibitem [{\citenamefont {{Shapiro}}\ and\ \citenamefont {{{\"U}{\c
  c}er}}(2003)}]{Shapiro:2003a}%
  \BibitemOpen
  \bibfield  {author} {\bibinfo {author} {\bibfnamefont {V.~D.}\ \bibnamefont
  {{Shapiro}}}\ and\ \bibinfo {author} {\bibfnamefont {D.}~\bibnamefont
  {{{\"U}{\c c}er}}},\ }\bibfield  {title} {\enquote {\bibinfo {title} {{Shock
  surfing acceleration}},}\ }\href@noop {} {\bibfield  {journal} {\bibinfo
  {journal} {Planet. Space Sci.}\ }\textbf {\bibinfo {volume} {51}},\ \bibinfo
  {pages} {665--680} (\bibinfo {year} {2003})}\BibitemShut {NoStop}%
\bibitem [{\citenamefont {{Anagnostopoulos}}\ \emph {et~al.}(1998)\citenamefont
  {{Anagnostopoulos}}, \citenamefont {{Rigas}}, \citenamefont {{Sarris}},\ and\
  \citenamefont {{Krimigis}}}]{Anagnostopoulos:1998a}%
  \BibitemOpen
  \bibfield  {author} {\bibinfo {author} {\bibfnamefont {G.~C.}\ \bibnamefont
  {{Anagnostopoulos}}}, \bibinfo {author} {\bibfnamefont {A.~G.}\ \bibnamefont
  {{Rigas}}}, \bibinfo {author} {\bibfnamefont {E.~T.}\ \bibnamefont
  {{Sarris}}}, \ and\ \bibinfo {author} {\bibfnamefont {S.~M.}\ \bibnamefont
  {{Krimigis}}},\ }\bibfield  {title} {\enquote {\bibinfo {title}
  {{Characteristics of upstream energetic (E$\geq$50keV) ion events during
  intense geomagnetic activity}},}\ }\href {\doibase 10.1029/97JA02925}
  {\bibfield  {journal} {\bibinfo  {journal} {J.~Geophys.~Res.}\ }\textbf
  {\bibinfo {volume} {103}},\ \bibinfo {pages} {9521--9534} (\bibinfo {year}
  {1998})}\BibitemShut {NoStop}%
\bibitem [{\citenamefont {{Park}}\ \emph {et~al.}(2013)\citenamefont {{Park}},
  \citenamefont {{Ren}}, \citenamefont {{Workman}},\ and\ \citenamefont
  {{Blackman}}}]{Park:2013a}%
  \BibitemOpen
  \bibfield  {author} {\bibinfo {author} {\bibfnamefont {J.}~\bibnamefont
  {{Park}}}, \bibinfo {author} {\bibfnamefont {C.}~\bibnamefont {{Ren}}},
  \bibinfo {author} {\bibfnamefont {J.~C.}\ \bibnamefont {{Workman}}}, \ and\
  \bibinfo {author} {\bibfnamefont {E.~G.}\ \bibnamefont {{Blackman}}},\
  }\bibfield  {title} {\enquote {\bibinfo {title} {{Particle-in-cell
  Simulations of Particle Energization via Shock Drift Acceleration from Low
  Mach Number Quasi-perpendicular Shocks in Solar Flares}},}\ }\href {\doibase
  10.1088/0004-637X/765/2/147} {\bibfield  {journal} {\bibinfo  {journal}
  {Astrophys.~J.}\ }\textbf {\bibinfo {volume} {765}},\ \bibinfo {eid} {147}
  (\bibinfo {year} {2013})},\ \Eprint {http://arxiv.org/abs/1210.5654}
  {arXiv:1210.5654 [astro-ph.SR]} \BibitemShut {NoStop}%
\bibitem [{\citenamefont {{Wu}}(1984)}]{Wu:1984b}%
  \BibitemOpen
  \bibfield  {author} {\bibinfo {author} {\bibfnamefont {C.~S.}\ \bibnamefont
  {{Wu}}},\ }\bibfield  {title} {\enquote {\bibinfo {title} {{A fast Fermi
  process - Energetic electrons accelerated by a nearly perpendicular bow
  shock}},}\ }\href {\doibase 10.1029/JA089iA10p08857} {\bibfield  {journal}
  {\bibinfo  {journal} {J.~Geophys.~Res.}\ }\textbf {\bibinfo {volume} {89}},\
  \bibinfo {pages} {8857--8862} (\bibinfo {year} {1984})}\BibitemShut {NoStop}%
\bibitem [{\citenamefont {{Leroy}}\ and\ \citenamefont
  {{Mangeney}}(1984)}]{Leroy:1984a}%
  \BibitemOpen
  \bibfield  {author} {\bibinfo {author} {\bibfnamefont {M.~M.}\ \bibnamefont
  {{Leroy}}}\ and\ \bibinfo {author} {\bibfnamefont {A.}~\bibnamefont
  {{Mangeney}}},\ }\bibfield  {title} {\enquote {\bibinfo {title} {{A theory of
  energization of solar wind electrons by the earth's bow shock}},}\
  }\href@noop {} {\bibfield  {journal} {\bibinfo  {journal} {Ann. Geophys.}\
  }\textbf {\bibinfo {volume} {2}},\ \bibinfo {pages} {449--456} (\bibinfo
  {year} {1984})}\BibitemShut {NoStop}%
\bibitem [{\citenamefont {{Drake}}\ \emph {et~al.}(2009)\citenamefont
  {{Drake}}, \citenamefont {{Cassak}}, \citenamefont {{Shay}}, \citenamefont
  {{Swisdak}},\ and\ \citenamefont {{Quataert}}}]{Drake:2009c}%
  \BibitemOpen
  \bibfield  {author} {\bibinfo {author} {\bibfnamefont {J.~F.}\ \bibnamefont
  {{Drake}}}, \bibinfo {author} {\bibfnamefont {P.~A.}\ \bibnamefont
  {{Cassak}}}, \bibinfo {author} {\bibfnamefont {M.~A.}\ \bibnamefont
  {{Shay}}}, \bibinfo {author} {\bibfnamefont {M.}~\bibnamefont {{Swisdak}}}, \
  and\ \bibinfo {author} {\bibfnamefont {E.}~\bibnamefont {{Quataert}}},\
  }\bibfield  {title} {\enquote {\bibinfo {title} {{A Magnetic Reconnection
  Mechanism for Ion Acceleration and Abundance Enhancements in Impulsive
  Flares}},}\ }\href {\doibase 10.1088/0004-637X/700/1/L16} {\bibfield
  {journal} {\bibinfo  {journal} {Astrophys. J. Lett.}\ }\textbf {\bibinfo
  {volume} {700}},\ \bibinfo {pages} {L16--L20} (\bibinfo {year}
  {2009})}\BibitemShut {NoStop}%
\bibitem [{\citenamefont {{Hoshino}}\ \emph {et~al.}(2001)\citenamefont
  {{Hoshino}}, \citenamefont {{Mukai}}, \citenamefont {{Terasawa}},\ and\
  \citenamefont {{Shinohara}}}]{Hoshino:2001}%
  \BibitemOpen
  \bibfield  {author} {\bibinfo {author} {\bibfnamefont {M.}~\bibnamefont
  {{Hoshino}}}, \bibinfo {author} {\bibfnamefont {T.}~\bibnamefont {{Mukai}}},
  \bibinfo {author} {\bibfnamefont {T.}~\bibnamefont {{Terasawa}}}, \ and\
  \bibinfo {author} {\bibfnamefont {I.}~\bibnamefont {{Shinohara}}},\
  }\bibfield  {title} {\enquote {\bibinfo {title} {{Suprathermal electron
  acceleration in magnetic reconnection}},}\ }\href {\doibase
  10.1029/2001JA900052} {\bibfield  {journal} {\bibinfo  {journal}
  {J.~Geophys.~Res.}\ }\textbf {\bibinfo {volume} {106}},\ \bibinfo {pages}
  {25979--25998} (\bibinfo {year} {2001})}\BibitemShut {NoStop}%
\bibitem [{\citenamefont {{Drake}}\ \emph {et~al.}(2006)\citenamefont
  {{Drake}}, \citenamefont {{Swisdak}}, \citenamefont {{Che}},\ and\
  \citenamefont {{Shay}}}]{Drake:2006a}%
  \BibitemOpen
  \bibfield  {author} {\bibinfo {author} {\bibfnamefont {J.~F.}\ \bibnamefont
  {{Drake}}}, \bibinfo {author} {\bibfnamefont {M.}~\bibnamefont {{Swisdak}}},
  \bibinfo {author} {\bibfnamefont {H.}~\bibnamefont {{Che}}}, \ and\ \bibinfo
  {author} {\bibfnamefont {M.~A.}\ \bibnamefont {{Shay}}},\ }\bibfield  {title}
  {\enquote {\bibinfo {title} {{Electron acceleration from contracting magnetic
  islands during reconnection}},}\ }\href {\doibase 10.1038/nature05116}
  {\bibfield  {journal} {\bibinfo  {journal} {Nature}\ }\textbf {\bibinfo
  {volume} {443}},\ \bibinfo {pages} {553--556} (\bibinfo {year}
  {2006})}\BibitemShut {NoStop}%
\bibitem [{\citenamefont {{Dahlin}}, \citenamefont {{Drake}},\ and\
  \citenamefont {{Swisdak}}(2014)}]{Dahlin:2014}%
  \BibitemOpen
  \bibfield  {author} {\bibinfo {author} {\bibfnamefont {J.~T.}\ \bibnamefont
  {{Dahlin}}}, \bibinfo {author} {\bibfnamefont {J.~F.}\ \bibnamefont
  {{Drake}}}, \ and\ \bibinfo {author} {\bibfnamefont {M.}~\bibnamefont
  {{Swisdak}}},\ }\bibfield  {title} {\enquote {\bibinfo {title} {{The
  mechanisms of electron heating and acceleration during magnetic
  reconnection}},}\ }\href {\doibase 10.1063/1.4894484} {\bibfield  {journal}
  {\bibinfo  {journal} {Phys.~Plasmas}\ }\textbf {\bibinfo {volume} {21}},\
  \bibinfo {eid} {092304} (\bibinfo {year} {2014})},\ \Eprint
  {http://arxiv.org/abs/1406.0831} {arXiv:1406.0831 [physics.plasm-ph]}
  \BibitemShut {NoStop}%
\bibitem [{\citenamefont {{Dahlin}}, \citenamefont {{Drake}},\ and\
  \citenamefont {{Swisdak}}(2015)}]{Dahlin:2015}%
  \BibitemOpen
  \bibfield  {author} {\bibinfo {author} {\bibfnamefont {J.~T.}\ \bibnamefont
  {{Dahlin}}}, \bibinfo {author} {\bibfnamefont {J.~F.}\ \bibnamefont
  {{Drake}}}, \ and\ \bibinfo {author} {\bibfnamefont {M.}~\bibnamefont
  {{Swisdak}}},\ }\bibfield  {title} {\enquote {\bibinfo {title} {{Electron
  acceleration in three-dimensional magnetic reconnection with a guide
  field}},}\ }\href {\doibase 10.1063/1.4933212} {\bibfield  {journal}
  {\bibinfo  {journal} {Phys.~Plasmas}\ }\textbf {\bibinfo {volume} {22}},\
  \bibinfo {eid} {100704} (\bibinfo {year} {2015})},\ \Eprint
  {http://arxiv.org/abs/1503.02218} {arXiv:1503.02218 [physics.plasm-ph]}
  \BibitemShut {NoStop}%
\bibitem [{\citenamefont {{Drake}}\ \emph {et~al.}(2010)\citenamefont
  {{Drake}}, \citenamefont {{Opher}}, \citenamefont {{Swisdak}},\ and\
  \citenamefont {{Chamoun}}}]{Drake:2010}%
  \BibitemOpen
  \bibfield  {author} {\bibinfo {author} {\bibfnamefont {J.~F.}\ \bibnamefont
  {{Drake}}}, \bibinfo {author} {\bibfnamefont {M.}~\bibnamefont {{Opher}}},
  \bibinfo {author} {\bibfnamefont {M.}~\bibnamefont {{Swisdak}}}, \ and\
  \bibinfo {author} {\bibfnamefont {J.~N.}\ \bibnamefont {{Chamoun}}},\
  }\bibfield  {title} {\enquote {\bibinfo {title} {{A Magnetic Reconnection
  Mechanism for the Generation of Anomalous Cosmic Rays}},}\ }\href {\doibase
  10.1088/0004-637X/709/2/963} {\bibfield  {journal} {\bibinfo  {journal}
  {Astrophys.~J.}\ }\textbf {\bibinfo {volume} {709}},\ \bibinfo {pages}
  {963--974} (\bibinfo {year} {2010})},\ \Eprint
  {http://arxiv.org/abs/0911.3098} {arXiv:0911.3098 [astro-ph.SR]} \BibitemShut
  {NoStop}%
\bibitem [{\citenamefont {{Aschwanden}}(2002)}]{Aschwanden:2002}%
  \BibitemOpen
  \bibfield  {author} {\bibinfo {author} {\bibfnamefont {M.~J.}\ \bibnamefont
  {{Aschwanden}}},\ }\bibfield  {title} {\enquote {\bibinfo {title} {{Particle
  acceleration and kinematics in solar flares - A Synthesis of Recent
  Observations and Theoretical Concepts (Invited Review)}},}\ }\href {\doibase
  10.1023/A:1019712124366} {\bibfield  {journal} {\bibinfo  {journal} {Space
  Sci.~Rev.}\ }\textbf {\bibinfo {volume} {101}},\ \bibinfo {pages} {1--227}
  (\bibinfo {year} {2002})}\BibitemShut {NoStop}%
\bibitem [{\citenamefont {{Lin}}\ \emph {et~al.}(2003)\citenamefont {{Lin}},
  \citenamefont {{Krucker}}, \citenamefont {{Hurford}}, \citenamefont
  {{Smith}}, \citenamefont {{Hudson}}, \citenamefont {{Holman}}, \citenamefont
  {{Schwartz}}, \citenamefont {{Dennis}}, \citenamefont {{Share}},
  \citenamefont {{Murphy}}, \citenamefont {{Emslie}}, \citenamefont
  {{Johns-Krull}},\ and\ \citenamefont {{Vilmer}}}]{Lin:2003}%
  \BibitemOpen
  \bibfield  {author} {\bibinfo {author} {\bibfnamefont {R.~P.}\ \bibnamefont
  {{Lin}}}, \bibinfo {author} {\bibfnamefont {S.}~\bibnamefont {{Krucker}}},
  \bibinfo {author} {\bibfnamefont {G.~J.}\ \bibnamefont {{Hurford}}}, \bibinfo
  {author} {\bibfnamefont {D.~M.}\ \bibnamefont {{Smith}}}, \bibinfo {author}
  {\bibfnamefont {H.~S.}\ \bibnamefont {{Hudson}}}, \bibinfo {author}
  {\bibfnamefont {G.~D.}\ \bibnamefont {{Holman}}}, \bibinfo {author}
  {\bibfnamefont {R.~A.}\ \bibnamefont {{Schwartz}}}, \bibinfo {author}
  {\bibfnamefont {B.~R.}\ \bibnamefont {{Dennis}}}, \bibinfo {author}
  {\bibfnamefont {G.~H.}\ \bibnamefont {{Share}}}, \bibinfo {author}
  {\bibfnamefont {R.~J.}\ \bibnamefont {{Murphy}}}, \bibinfo {author}
  {\bibfnamefont {A.~G.}\ \bibnamefont {{Emslie}}}, \bibinfo {author}
  {\bibfnamefont {C.}~\bibnamefont {{Johns-Krull}}}, \ and\ \bibinfo {author}
  {\bibfnamefont {N.}~\bibnamefont {{Vilmer}}},\ }\bibfield  {title} {\enquote
  {\bibinfo {title} {{RHESSI Observations of Particle Acceleration and Energy
  Release in an Intense Solar Gamma-Ray Line Flare}},}\ }\href {\doibase
  10.1086/378932} {\bibfield  {journal} {\bibinfo  {journal}
  {Astrophys.~J.~Lett.}\ }\textbf {\bibinfo {volume} {595}},\ \bibinfo {pages}
  {L69--L76} (\bibinfo {year} {2003})}\BibitemShut {NoStop}%
\bibitem [{\citenamefont {Hsu}\ \emph {et~al.}(2012)\citenamefont {Hsu},
  \citenamefont {Merritt}, \citenamefont {Moser}, \citenamefont {Awe},
  \citenamefont {Brockington}, \citenamefont {Davis}, \citenamefont {Adams},
  \citenamefont {Case}, \citenamefont {Cassibry}, \citenamefont {Dunn},
  \citenamefont {Gilmore}, \citenamefont {Lynn}, \citenamefont {Messer},\ and\
  \citenamefont {Witherspoon}}]{Hsu:2012}%
  \BibitemOpen
  \bibfield  {author} {\bibinfo {author} {\bibfnamefont {S.~C.}\ \bibnamefont
  {Hsu}}, \bibinfo {author} {\bibfnamefont {E.~C.}\ \bibnamefont {Merritt}},
  \bibinfo {author} {\bibfnamefont {A.~L.}\ \bibnamefont {Moser}}, \bibinfo
  {author} {\bibfnamefont {T.~J.}\ \bibnamefont {Awe}}, \bibinfo {author}
  {\bibfnamefont {S.~J.~E.}\ \bibnamefont {Brockington}}, \bibinfo {author}
  {\bibfnamefont {J.~S.}\ \bibnamefont {Davis}}, \bibinfo {author}
  {\bibfnamefont {C.~S.}\ \bibnamefont {Adams}}, \bibinfo {author}
  {\bibfnamefont {A.}~\bibnamefont {Case}}, \bibinfo {author} {\bibfnamefont
  {J.~T.}\ \bibnamefont {Cassibry}}, \bibinfo {author} {\bibfnamefont {J.~P.}\
  \bibnamefont {Dunn}}, \bibinfo {author} {\bibfnamefont {M.~A.}\ \bibnamefont
  {Gilmore}}, \bibinfo {author} {\bibfnamefont {A.~G.}\ \bibnamefont {Lynn}},
  \bibinfo {author} {\bibfnamefont {S.~J.}\ \bibnamefont {Messer}}, \ and\
  \bibinfo {author} {\bibfnamefont {F.~D.}\ \bibnamefont {Witherspoon}},\
  }\bibfield  {title} {\enquote {\bibinfo {title} {Experimental
  characterization of railgun-driven supersonic plasma jets motivated by high
  energy density physics applications},}\ }\href@noop {} {\bibfield  {journal}
  {\bibinfo  {journal} {Phys.~Plasmas}\ }\textbf {\bibinfo {volume} {19}},\
  \bibinfo {pages} {123514} (\bibinfo {year} {2012})}\BibitemShut {NoStop}%
\bibitem [{\citenamefont {{Merritt}}\ \emph {et~al.}(2013)\citenamefont
  {{Merritt}}, \citenamefont {{Moser}}, \citenamefont {{Hsu}}, \citenamefont
  {{Loverich}},\ and\ \citenamefont {{Gilmore}}}]{Merritt:2013}%
  \BibitemOpen
  \bibfield  {author} {\bibinfo {author} {\bibfnamefont {E.~C.}\ \bibnamefont
  {{Merritt}}}, \bibinfo {author} {\bibfnamefont {A.~L.}\ \bibnamefont
  {{Moser}}}, \bibinfo {author} {\bibfnamefont {S.~C.}\ \bibnamefont {{Hsu}}},
  \bibinfo {author} {\bibfnamefont {J.}~\bibnamefont {{Loverich}}}, \ and\
  \bibinfo {author} {\bibfnamefont {M.}~\bibnamefont {{Gilmore}}},\ }\bibfield
  {title} {\enquote {\bibinfo {title} {{Experimental Characterization of the
  Stagnation Layer between Two Obliquely Merging Supersonic Plasma Jets}},}\
  }\href {\doibase 10.1103/PhysRevLett.111.085003} {\bibfield  {journal}
  {\bibinfo  {journal} {Physical Review Letters}\ }\textbf {\bibinfo {volume}
  {111}},\ \bibinfo {eid} {085003} (\bibinfo {year} {2013})},\ \Eprint
  {http://arxiv.org/abs/1303.5821} {arXiv:1303.5821 [physics.plasm-ph]}
  \BibitemShut {NoStop}%
\bibitem [{\citenamefont {Merritt}\ \emph {et~al.}(2014)\citenamefont
  {Merritt}, \citenamefont {Moser}, \citenamefont {Hsu}, \citenamefont {Adams},
  \citenamefont {Dunn}, \citenamefont {Holgado},\ and\ \citenamefont
  {Gilmore}}]{Merritt:2014}%
  \BibitemOpen
  \bibfield  {author} {\bibinfo {author} {\bibfnamefont {E.~C.}\ \bibnamefont
  {Merritt}}, \bibinfo {author} {\bibfnamefont {A.~L.}\ \bibnamefont {Moser}},
  \bibinfo {author} {\bibfnamefont {S.~C.}\ \bibnamefont {Hsu}}, \bibinfo
  {author} {\bibfnamefont {C.~S.}\ \bibnamefont {Adams}}, \bibinfo {author}
  {\bibfnamefont {J.~P.}\ \bibnamefont {Dunn}}, \bibinfo {author}
  {\bibfnamefont {A.~M.}\ \bibnamefont {Holgado}}, \ and\ \bibinfo {author}
  {\bibfnamefont {M.~A.}\ \bibnamefont {Gilmore}},\ }\bibfield  {title}
  {\enquote {\bibinfo {title} {Experimental evidence for collisional shock
  formation via two obliquely merging supersonic plasma jets},}\ }\href@noop {}
  {\bibfield  {journal} {\bibinfo  {journal} {Phys.~Plasmas}\ }\textbf
  {\bibinfo {volume} {21}},\ \bibinfo {pages} {055703} (\bibinfo {year}
  {2014})}\BibitemShut {NoStop}%
\bibitem [{\citenamefont {Moser}\ and\ \citenamefont {Hsu}(2015)}]{Moser:2015}%
  \BibitemOpen
  \bibfield  {author} {\bibinfo {author} {\bibfnamefont {A.~L.}\ \bibnamefont
  {Moser}}\ and\ \bibinfo {author} {\bibfnamefont {S.~C.}\ \bibnamefont
  {Hsu}},\ }\bibfield  {title} {\enquote {\bibinfo {title} {{Experimental
  characterization of a transition from collisionless to collisional
  interaction between head-on-merging supersonic plasma jets}},}\ }\href@noop
  {} {\bibfield  {journal} {\bibinfo  {journal} {Phys.~Plasmas}\ }\textbf
  {\bibinfo {volume} {22}},\ \bibinfo {pages} {055707} (\bibinfo {year}
  {2015})}\BibitemShut {NoStop}%
\bibitem [{\citenamefont {{DuBois}}\ \emph {et~al.}(2017)\citenamefont
  {{DuBois}}, \citenamefont {{Almagri}}, \citenamefont {{Anderson}},
  \citenamefont {{Den Hartog}}, \citenamefont {{Lee}},\ and\ \citenamefont
  {{Sarff}}}]{DuBois:2017}%
  \BibitemOpen
  \bibfield  {author} {\bibinfo {author} {\bibfnamefont {A.~M.}\ \bibnamefont
  {{DuBois}}}, \bibinfo {author} {\bibfnamefont {A.~F.}\ \bibnamefont
  {{Almagri}}}, \bibinfo {author} {\bibfnamefont {J.~K.}\ \bibnamefont
  {{Anderson}}}, \bibinfo {author} {\bibfnamefont {D.~J.}\ \bibnamefont {{Den
  Hartog}}}, \bibinfo {author} {\bibfnamefont {J.~D.}\ \bibnamefont {{Lee}}}, \
  and\ \bibinfo {author} {\bibfnamefont {J.~S.}\ \bibnamefont {{Sarff}}},\
  }\bibfield  {title} {\enquote {\bibinfo {title} {{Anisotropic Electron Tail
  Generation during Tearing Mode Magnetic Reconnection}},}\ }\href {\doibase
  10.1103/PhysRevLett.118.075001} {\bibfield  {journal} {\bibinfo  {journal}
  {Phys.~Rev.~Lett.}\ }\textbf {\bibinfo {volume} {118}},\ \bibinfo {eid}
  {075001} (\bibinfo {year} {2017})}\BibitemShut {NoStop}%
\bibitem [{\citenamefont {Niemann}\ \emph {et~al.}(2014)\citenamefont
  {Niemann}, \citenamefont {Gekelman}, \citenamefont {Constantin},
  \citenamefont {Everson}, \citenamefont {Schaeffer}, \citenamefont
  {Bondarenko}, \citenamefont {Clark}, \citenamefont {Winske}, \citenamefont
  {Vincena}, \citenamefont {Van~Compernolle},\ and\ \citenamefont
  {Pribyl}}]{Niemann:2014}%
  \BibitemOpen
  \bibfield  {author} {\bibinfo {author} {\bibfnamefont {C.}~\bibnamefont
  {Niemann}}, \bibinfo {author} {\bibfnamefont {W.}~\bibnamefont {Gekelman}},
  \bibinfo {author} {\bibfnamefont {C.~G.}\ \bibnamefont {Constantin}},
  \bibinfo {author} {\bibfnamefont {E.~T.}\ \bibnamefont {Everson}}, \bibinfo
  {author} {\bibfnamefont {D.~B.}\ \bibnamefont {Schaeffer}}, \bibinfo {author}
  {\bibfnamefont {A.~S.}\ \bibnamefont {Bondarenko}}, \bibinfo {author}
  {\bibfnamefont {S.~E.}\ \bibnamefont {Clark}}, \bibinfo {author}
  {\bibfnamefont {D.}~\bibnamefont {Winske}}, \bibinfo {author} {\bibfnamefont
  {S.}~\bibnamefont {Vincena}}, \bibinfo {author} {\bibfnamefont
  {B.}~\bibnamefont {Van~Compernolle}}, \ and\ \bibinfo {author} {\bibfnamefont
  {P.}~\bibnamefont {Pribyl}},\ }\bibfield  {title} {\enquote {\bibinfo {title}
  {Observation of collisionless shocks in a large current-free laboratory
  plasma},}\ }\href {\doibase 10.1002/2014GL061820} {\bibfield  {journal}
  {\bibinfo  {journal} {Geophys.~Res.~Lett.}\ }\textbf {\bibinfo {volume}
  {41}},\ \bibinfo {pages} {7413--7418} (\bibinfo {year} {2014})}\BibitemShut
  {NoStop}%
\bibitem [{\citenamefont {Schaeffer}\ \emph {et~al.}(2017)\citenamefont
  {Schaeffer}, \citenamefont {Fox}, \citenamefont {Haberberger}, \citenamefont
  {Fiksel}, \citenamefont {Bhattacharjee}, \citenamefont {Barnak},
  \citenamefont {Hu},\ and\ \citenamefont {Germaschewski}}]{Schaeffer:2017}%
  \BibitemOpen
  \bibfield  {author} {\bibinfo {author} {\bibfnamefont {D.~B.}\ \bibnamefont
  {Schaeffer}}, \bibinfo {author} {\bibfnamefont {W.}~\bibnamefont {Fox}},
  \bibinfo {author} {\bibfnamefont {D.}~\bibnamefont {Haberberger}}, \bibinfo
  {author} {\bibfnamefont {G.}~\bibnamefont {Fiksel}}, \bibinfo {author}
  {\bibfnamefont {A.}~\bibnamefont {Bhattacharjee}}, \bibinfo {author}
  {\bibfnamefont {D.~H.}\ \bibnamefont {Barnak}}, \bibinfo {author}
  {\bibfnamefont {S.~X.}\ \bibnamefont {Hu}}, \ and\ \bibinfo {author}
  {\bibfnamefont {K.}~\bibnamefont {Germaschewski}},\ }\bibfield  {title}
  {\enquote {\bibinfo {title} {Generation and evolution of high-mach-number
  laser-driven magnetized collisionless shocks in the laboratory},}\ }\href
  {\doibase 10.1103/PhysRevLett.119.025001} {\bibfield  {journal} {\bibinfo
  {journal} {Phys.~Rev.~Lett.}\ }\textbf {\bibinfo {volume} {119}},\ \bibinfo
  {pages} {025001} (\bibinfo {year} {2017})}\BibitemShut {NoStop}%
\bibitem [{\citenamefont {Bondarenko}\ \emph {et~al.}(2016)\citenamefont
  {Bondarenko}, \citenamefont {Schaeffer}, \citenamefont {Everson},
  \citenamefont {Clark}, \citenamefont {Lee}, \citenamefont {Constantin},
  \citenamefont {Vincena}, \citenamefont {Van~Compernolle}, \citenamefont
  {Tripathi}, \citenamefont {Winske},\ and\ \citenamefont
  {Niemann}}]{Bondarenko:2016}%
  \BibitemOpen
  \bibfield  {author} {\bibinfo {author} {\bibfnamefont {A.}~\bibnamefont
  {Bondarenko}}, \bibinfo {author} {\bibfnamefont {D.}~\bibnamefont
  {Schaeffer}}, \bibinfo {author} {\bibfnamefont {E.}~\bibnamefont {Everson}},
  \bibinfo {author} {\bibfnamefont {S.}~\bibnamefont {Clark}}, \bibinfo
  {author} {\bibfnamefont {B.}~\bibnamefont {Lee}}, \bibinfo {author}
  {\bibfnamefont {C.}~\bibnamefont {Constantin}}, \bibinfo {author}
  {\bibfnamefont {S.}~\bibnamefont {Vincena}}, \bibinfo {author} {\bibfnamefont
  {B.}~\bibnamefont {Van~Compernolle}}, \bibinfo {author} {\bibfnamefont
  {S.}~\bibnamefont {Tripathi}}, \bibinfo {author} {\bibfnamefont
  {D.}~\bibnamefont {Winske}}, \ and\ \bibinfo {author} {\bibfnamefont
  {C.}~\bibnamefont {Niemann}},\ }\bibfield  {title} {\enquote {\bibinfo
  {title} {Collisionless momentum transfer in space and astrophysical
  explosions},}\ }\href@noop {} {\bibfield  {journal} {\bibinfo  {journal}
  {Nature Phys.}\ }\textbf {\bibinfo {volume} {13}},\ \bibinfo {pages}
  {573--577} (\bibinfo {year} {2016})}\BibitemShut {NoStop}%
\bibitem [{\citenamefont {Weidl}\ \emph {et~al.}(2016)\citenamefont {Weidl},
  \citenamefont {Winske}, \citenamefont {Jenko},\ and\ \citenamefont
  {Niemann}}]{Weidl:2016}%
  \BibitemOpen
  \bibfield  {author} {\bibinfo {author} {\bibfnamefont {M.~S.}\ \bibnamefont
  {Weidl}}, \bibinfo {author} {\bibfnamefont {D.}~\bibnamefont {Winske}},
  \bibinfo {author} {\bibfnamefont {F.}~\bibnamefont {Jenko}}, \ and\ \bibinfo
  {author} {\bibfnamefont {C.}~\bibnamefont {Niemann}},\ }\bibfield  {title}
  {\enquote {\bibinfo {title} {Hybrid simulations of a parallel collisionless
  shock in the large plasma device},}\ }\href {\doibase 10.1063/1.4971231}
  {\bibfield  {journal} {\bibinfo  {journal} {Phys.~Plasmas}\ }\textbf
  {\bibinfo {volume} {23}},\ \bibinfo {pages} {122102} (\bibinfo {year}
  {2016})}\BibitemShut {NoStop}%
\bibitem [{\citenamefont {Reville}, \citenamefont {Bell},\ and\ \citenamefont
  {Gregori}(2013)}]{Revile:2013}%
  \BibitemOpen
  \bibfield  {author} {\bibinfo {author} {\bibfnamefont {B.}~\bibnamefont
  {Reville}}, \bibinfo {author} {\bibfnamefont {A.}~\bibnamefont {Bell}}, \
  and\ \bibinfo {author} {\bibfnamefont {G.}~\bibnamefont {Gregori}},\
  }\bibfield  {title} {\enquote {\bibinfo {title} {Diffusive shock acceleration
  at laser-driven shocks: studying cosmic-ray accelerators in the
  laboratory},}\ }\href {http://stacks.iop.org/1367-2630/15/i=1/a=015015}
  {\bibfield  {journal} {\bibinfo  {journal} {New J.~Phys.}\ }\textbf {\bibinfo
  {volume} {15}},\ \bibinfo {pages} {015015} (\bibinfo {year}
  {2013})}\BibitemShut {NoStop}%
\bibitem [{\citenamefont {{Sandahl}}, \citenamefont {{Sergienko}},\ and\
  \citenamefont {{Br{\"a}ndstr{\"o}m}}(2008)}]{Sandahl:2008}%
  \BibitemOpen
  \bibfield  {author} {\bibinfo {author} {\bibfnamefont {I.}~\bibnamefont
  {{Sandahl}}}, \bibinfo {author} {\bibfnamefont {T.}~\bibnamefont
  {{Sergienko}}}, \ and\ \bibinfo {author} {\bibfnamefont {U.}~\bibnamefont
  {{Br{\"a}ndstr{\"o}m}}},\ }\bibfield  {title} {\enquote {\bibinfo {title}
  {{Fine structure of optical aurora}},}\ }\href {\doibase
  10.1016/j.jastp.2008.08.016} {\bibfield  {journal} {\bibinfo  {journal}
  {Journal of Atmospheric and Solar-Terrestrial Physics}\ }\textbf {\bibinfo
  {volume} {70}},\ \bibinfo {pages} {2275--2292} (\bibinfo {year}
  {2008})}\BibitemShut {NoStop}%
\bibitem [{\citenamefont {{Knudsen}}(1996)}]{Knudsen:1996}%
  \BibitemOpen
  \bibfield  {author} {\bibinfo {author} {\bibfnamefont {D.~J.}\ \bibnamefont
  {{Knudsen}}},\ }\bibfield  {title} {\enquote {\bibinfo {title} {{Spatial
  modulation of electron energy and density by nonlinear stationary inertial
  Alfv{\'e}n waves}},}\ }\href {\doibase 10.1029/96JA00429} {\bibfield
  {journal} {\bibinfo  {journal} {J.~Geophys.~Res.}\ }\textbf {\bibinfo
  {volume} {101}},\ \bibinfo {pages} {10761--10772} (\bibinfo {year}
  {1996})}\BibitemShut {NoStop}%
\bibitem [{\citenamefont {{Finnegan}}, \citenamefont {{Koepke}},\ and\
  \citenamefont {{Knudsen}}(2008)}]{Finnegan:2008}%
  \BibitemOpen
  \bibfield  {author} {\bibinfo {author} {\bibfnamefont {S.~M.}\ \bibnamefont
  {{Finnegan}}}, \bibinfo {author} {\bibfnamefont {M.~E.}\ \bibnamefont
  {{Koepke}}}, \ and\ \bibinfo {author} {\bibfnamefont {D.~J.}\ \bibnamefont
  {{Knudsen}}},\ }\bibfield  {title} {\enquote {\bibinfo {title} {{The
  dispersive Alfv{\'e}n wave in the time-stationary limit with a focus on
  collisional and warm-plasma effects}},}\ }\href {\doibase 10.1063/1.2890774}
  {\bibfield  {journal} {\bibinfo  {journal} {Phys.~Plasmas}\ }\textbf
  {\bibinfo {volume} {15}},\ \bibinfo {eid} {052108} (\bibinfo {year}
  {2008})}\BibitemShut {NoStop}%
\bibitem [{\citenamefont {{Koepke}}\ \emph {et~al.}(2016)\citenamefont
  {{Koepke}}, \citenamefont {{Finnegan}}, \citenamefont {{Vincena}},
  \citenamefont {{Knudsen}}, \citenamefont {{Nogami}},\ and\ \citenamefont
  {{Vassiliadis}}}]{Koepke:2016}%
  \BibitemOpen
  \bibfield  {author} {\bibinfo {author} {\bibfnamefont {M.~E.}\ \bibnamefont
  {{Koepke}}}, \bibinfo {author} {\bibfnamefont {S.~M.}\ \bibnamefont
  {{Finnegan}}}, \bibinfo {author} {\bibfnamefont {S.}~\bibnamefont
  {{Vincena}}}, \bibinfo {author} {\bibfnamefont {D.~J.}\ \bibnamefont
  {{Knudsen}}}, \bibinfo {author} {\bibfnamefont {S.~H.}\ \bibnamefont
  {{Nogami}}}, \ and\ \bibinfo {author} {\bibfnamefont {D.}~\bibnamefont
  {{Vassiliadis}}},\ }\bibfield  {title} {\enquote {\bibinfo {title}
  {{Laboratory evidence for stationary inertial Alfv{\'e}n waves}},}\ }\href
  {\doibase 10.1088/0741-3335/58/8/084006} {\bibfield  {journal} {\bibinfo
  {journal} {Plasma Phys.~Con.~Fus.}\ }\textbf {\bibinfo {volume} {58}},\
  \bibinfo {eid} {084006} (\bibinfo {year} {2016})}\BibitemShut {NoStop}%
\bibitem [{\citenamefont {{Thuecks}}, \citenamefont {{Skiff}},\ and\
  \citenamefont {{Kletzing}}(2012)}]{Thuecks:2012}%
  \BibitemOpen
  \bibfield  {author} {\bibinfo {author} {\bibfnamefont {D.~J.}\ \bibnamefont
  {{Thuecks}}}, \bibinfo {author} {\bibfnamefont {F.}~\bibnamefont {{Skiff}}},
  \ and\ \bibinfo {author} {\bibfnamefont {C.~A.}\ \bibnamefont {{Kletzing}}},\
  }\bibfield  {title} {\enquote {\bibinfo {title} {Measurements of parallel
  electron velocity distributions using whistler wave absorption},}\ }\href
  {\doibase 10.1063/1.4742766} {\bibfield  {journal} {\bibinfo  {journal} {Rev.
  Sci. Instr.}\ }\textbf {\bibinfo {volume} {83}},\ \bibinfo {eid} {083503}
  (\bibinfo {year} {2012})}\BibitemShut {NoStop}%
\bibitem [{\citenamefont {{Schroeder}}\ \emph {et~al.}(2015)\citenamefont
  {{Schroeder}}, \citenamefont {{Skiff}}, \citenamefont {{Howes}},
  \citenamefont {{Kletzing}}, \citenamefont {{Carter}},\ and\ \citenamefont
  {{Dorfman}}}]{Schroeder:2015}%
  \BibitemOpen
  \bibfield  {author} {\bibinfo {author} {\bibfnamefont {J.~W.~R.}\
  \bibnamefont {{Schroeder}}}, \bibinfo {author} {\bibfnamefont
  {F.}~\bibnamefont {{Skiff}}}, \bibinfo {author} {\bibfnamefont {G.~G.}\
  \bibnamefont {{Howes}}}, \bibinfo {author} {\bibfnamefont {C.~A.}\
  \bibnamefont {{Kletzing}}}, \bibinfo {author} {\bibfnamefont {T.~A.}\
  \bibnamefont {{Carter}}}, \ and\ \bibinfo {author} {\bibfnamefont
  {S.}~\bibnamefont {{Dorfman}}},\ }\bibfield  {title} {\enquote {\bibinfo
  {title} {Alfv{\'e}nic oscillations of the electron distribution function:
  Linear theory and experimental measurements},}\ }in\ \href {\doibase
  10.1063/1.4936466} {\emph {\bibinfo {booktitle} {American Institute of
  Physics Conference Series}}},\ \bibinfo {series} {American Institute of
  Physics Conference Series}, Vol.\ \bibinfo {volume} {1689}\ (\bibinfo {year}
  {2015})\ p.\ \bibinfo {pages} {030001}\BibitemShut {NoStop}%
\bibitem [{\citenamefont {{Schroeder}}\ \emph {et~al.}(2016)\citenamefont
  {{Schroeder}}, \citenamefont {{Skiff}}, \citenamefont {{Kletzing}},
  \citenamefont {{Howes}}, \citenamefont {{Carter}},\ and\ \citenamefont
  {{Dorfman}}}]{Schroeder:2016}%
  \BibitemOpen
  \bibfield  {author} {\bibinfo {author} {\bibfnamefont {J.~W.~R.}\
  \bibnamefont {{Schroeder}}}, \bibinfo {author} {\bibfnamefont
  {F.}~\bibnamefont {{Skiff}}}, \bibinfo {author} {\bibfnamefont {C.~A.}\
  \bibnamefont {{Kletzing}}}, \bibinfo {author} {\bibfnamefont {G.~G.}\
  \bibnamefont {{Howes}}}, \bibinfo {author} {\bibfnamefont {T.~A.}\
  \bibnamefont {{Carter}}}, \ and\ \bibinfo {author} {\bibfnamefont
  {S.}~\bibnamefont {{Dorfman}}},\ }\bibfield  {title} {\enquote {\bibinfo
  {title} {{Direct measurement of electron sloshing of an inertial Alfv{\'e}n
  wave}},}\ }\href {\doibase 10.1002/2016GL068865} {\bibfield  {journal}
  {\bibinfo  {journal} {Geophys.~Res.~Lett.}\ }\textbf {\bibinfo {volume}
  {43}},\ \bibinfo {pages} {4701--4707} (\bibinfo {year} {2016})}\BibitemShut
  {NoStop}%
\bibitem [{\citenamefont {{Schroeder}}\ \emph {et~al.}(2017)\citenamefont
  {{Schroeder}}, \citenamefont {{Skiff}}, \citenamefont {{Howes}},
  \citenamefont {{Kletzing}}, \citenamefont {{Carter}},\ and\ \citenamefont
  {{Dorfman}}}]{Schroeder:2017}%
  \BibitemOpen
  \bibfield  {author} {\bibinfo {author} {\bibfnamefont {J.~W.~R.}\
  \bibnamefont {{Schroeder}}}, \bibinfo {author} {\bibfnamefont
  {F.}~\bibnamefont {{Skiff}}}, \bibinfo {author} {\bibfnamefont {G.~G.}\
  \bibnamefont {{Howes}}}, \bibinfo {author} {\bibfnamefont {C.~A.}\
  \bibnamefont {{Kletzing}}}, \bibinfo {author} {\bibfnamefont {T.~A.}\
  \bibnamefont {{Carter}}}, \ and\ \bibinfo {author} {\bibfnamefont
  {S.}~\bibnamefont {{Dorfman}}},\ }\bibfield  {title} {\enquote {\bibinfo
  {title} {{Linear theory and measurements of electron oscillations in an
  inertial Alfv{\'e}n wave}},}\ }\href {\doibase 10.1063/1.4978293} {\bibfield
  {journal} {\bibinfo  {journal} {Phys.~Plasmas}\ }\textbf {\bibinfo {volume}
  {24}},\ \bibinfo {eid} {032902} (\bibinfo {year} {2017})}\BibitemShut
  {NoStop}%
\bibitem [{\citenamefont {Treumann}(2009)}]{Treumann:2009}%
  \BibitemOpen
  \bibfield  {author} {\bibinfo {author} {\bibfnamefont {R.~A.}\ \bibnamefont
  {Treumann}},\ }\bibfield  {title} {\enquote {\bibinfo {title} {Fundamentals
  of collisionless shocks for astrophysical application, 1. non-relativistic
  shocks},}\ }\href {\doibase 10.1007/s00159-009-0024-2} {\bibfield  {journal}
  {\bibinfo  {journal} {The Astronomy and Astrophysics Review}\ }\textbf
  {\bibinfo {volume} {17}},\ \bibinfo {pages} {409--535} (\bibinfo {year}
  {2009})}\BibitemShut {NoStop}%
\bibitem [{\citenamefont {{Russell}}\ and\ \citenamefont
  {{Greenstadt}}(1979)}]{Russell:1979}%
  \BibitemOpen
  \bibfield  {author} {\bibinfo {author} {\bibfnamefont {C.~T.}\ \bibnamefont
  {{Russell}}}\ and\ \bibinfo {author} {\bibfnamefont {E.~W.}\ \bibnamefont
  {{Greenstadt}}},\ }\bibfield  {title} {\enquote {\bibinfo {title} {{Initial
  ISEE magnetometer results - Shock observation}},}\ }\href {\doibase
  10.1007/BF00174109} {\bibfield  {journal} {\bibinfo  {journal} {Space
  Sci.~Rev.}\ }\textbf {\bibinfo {volume} {23}},\ \bibinfo {pages} {3--37}
  (\bibinfo {year} {1979})}\BibitemShut {NoStop}%
\bibitem [{\citenamefont {Burlaga}(2001)}]{Burlaga:2001}%
  \BibitemOpen
  \bibfield  {author} {\bibinfo {author} {\bibfnamefont {L.}~\bibnamefont
  {Burlaga}},\ }\bibfield  {title} {\enquote {\bibinfo {title} {Fast ejecta
  during the ascending phase of solar cycle 23: Ace observations},}\
  }\href@noop {} {\bibfield  {journal} {\bibinfo  {journal} {J. Geophys. Res.}\
  }\textbf {\bibinfo {volume} {106}},\ \bibinfo {pages} {20957} (\bibinfo
  {year} {2001})}\BibitemShut {NoStop}%
\bibitem [{\citenamefont {{Richardson}}\ \emph {et~al.}(2008)\citenamefont
  {{Richardson}}, \citenamefont {{Kasper}}, \citenamefont {{Wang}},
  \citenamefont {{Belcher}},\ and\ \citenamefont
  {{Lazarus}}}]{Richardson:2008}%
  \BibitemOpen
  \bibfield  {author} {\bibinfo {author} {\bibfnamefont {J.}~\bibnamefont
  {{Richardson}}}, \bibinfo {author} {\bibfnamefont {J.}~\bibnamefont
  {{Kasper}}}, \bibinfo {author} {\bibfnamefont {C.}~\bibnamefont {{Wang}}},
  \bibinfo {author} {\bibfnamefont {J.}~\bibnamefont {{Belcher}}}, \ and\
  \bibinfo {author} {\bibfnamefont {A.}~\bibnamefont {{Lazarus}}},\ }\bibfield
  {title} {\enquote {\bibinfo {title} {Cool heliosheath plasma and deceleration
  of the upstream solar wind at the termination shock},}\ }\href@noop {}
  {\bibfield  {journal} {\bibinfo  {journal} {Nature}\ }\textbf {\bibinfo
  {volume} {454}},\ \bibinfo {pages} {63--66} (\bibinfo {year}
  {2008})}\BibitemShut {NoStop}%
\bibitem [{\citenamefont {Schaeffer}\ \emph {et~al.}(2016)\citenamefont
  {Schaeffer}, \citenamefont {Winske}, \citenamefont {Larson}, \citenamefont
  {Cowee}, \citenamefont {Constantin}, \citenamefont {Bondarenko},
  \citenamefont {Clark},\ and\ \citenamefont {Niemann}}]{Schaeffer:2016}%
  \BibitemOpen
  \bibfield  {author} {\bibinfo {author} {\bibfnamefont {D.}~\bibnamefont
  {Schaeffer}}, \bibinfo {author} {\bibfnamefont {D.}~\bibnamefont {Winske}},
  \bibinfo {author} {\bibfnamefont {D.}~\bibnamefont {Larson}}, \bibinfo
  {author} {\bibfnamefont {M.}~\bibnamefont {Cowee}}, \bibinfo {author}
  {\bibfnamefont {C.}~\bibnamefont {Constantin}}, \bibinfo {author}
  {\bibfnamefont {A.}~\bibnamefont {Bondarenko}}, \bibinfo {author}
  {\bibfnamefont {S.}~\bibnamefont {Clark}}, \ and\ \bibinfo {author}
  {\bibfnamefont {C.}~\bibnamefont {Niemann}},\ }\bibfield  {title} {\enquote
  {\bibinfo {title} {On the generation of magnetized collisionless shocks in
  the large plasma device},}\ }\href@noop {} {\bibfield  {journal} {\bibinfo
  {journal} {Phys.~Plasmas}\ }\textbf {\bibinfo {volume} {24}},\ \bibinfo
  {pages} {041405} (\bibinfo {year} {2016})}\BibitemShut {NoStop}%
\bibitem [{\citenamefont {Jaffrin}\ and\ \citenamefont
  {Probstein}(1964)}]{Jaffrin:1964}%
  \BibitemOpen
  \bibfield  {author} {\bibinfo {author} {\bibfnamefont {M.~Y.}\ \bibnamefont
  {Jaffrin}}\ and\ \bibinfo {author} {\bibfnamefont {R.~F.}\ \bibnamefont
  {Probstein}},\ }\bibfield  {title} {\enquote {\bibinfo {title} {Structore of
  a plasma shock wave},}\ }\href@noop {} {\bibfield  {journal} {\bibinfo
  {journal} {Phys.~Fluids}\ }\textbf {\bibinfo {volume} {7}},\ \bibinfo {pages}
  {1658} (\bibinfo {year} {1964})}\BibitemShut {NoStop}%
\bibitem [{\citenamefont {Kagan}\ and\ \citenamefont
  {Tang}(2012)}]{Kagan:2012}%
  \BibitemOpen
  \bibfield  {author} {\bibinfo {author} {\bibfnamefont {G.}~\bibnamefont
  {Kagan}}\ and\ \bibinfo {author} {\bibfnamefont {X.-Z.}\ \bibnamefont
  {Tang}},\ }\bibfield  {title} {\enquote {\bibinfo {title} {Electro-diffusion
  in a plasma with two ion species},}\ }\href@noop {} {\bibfield  {journal}
  {\bibinfo  {journal} {Phys.~Plasmas}\ }\textbf {\bibinfo {volume} {19}},\
  \bibinfo {pages} {082709} (\bibinfo {year} {2012})}\BibitemShut {NoStop}%
\bibitem [{\citenamefont {Hsu}\ \emph {et~al.}(2016)\citenamefont {Hsu},
  \citenamefont {Joshi}, \citenamefont {Hakel}, \citenamefont {Vold},
  \citenamefont {Schmitt}, \citenamefont {Hoffman}, \citenamefont {Rauenzahn},
  \citenamefont {Kagan}, \citenamefont {Tang}, \citenamefont {Mancini},
  \citenamefont {Kim},\ and\ \citenamefont {Herrmann}}]{Hsu:2016}%
  \BibitemOpen
  \bibfield  {author} {\bibinfo {author} {\bibfnamefont {S.~C.}\ \bibnamefont
  {Hsu}}, \bibinfo {author} {\bibfnamefont {T.~R.}\ \bibnamefont {Joshi}},
  \bibinfo {author} {\bibfnamefont {P.}~\bibnamefont {Hakel}}, \bibinfo
  {author} {\bibfnamefont {E.~L.}\ \bibnamefont {Vold}}, \bibinfo {author}
  {\bibfnamefont {M.~J.}\ \bibnamefont {Schmitt}}, \bibinfo {author}
  {\bibfnamefont {N.~M.}\ \bibnamefont {Hoffman}}, \bibinfo {author}
  {\bibfnamefont {R.~M.}\ \bibnamefont {Rauenzahn}}, \bibinfo {author}
  {\bibfnamefont {G.}~\bibnamefont {Kagan}}, \bibinfo {author} {\bibfnamefont
  {X.-Z.}\ \bibnamefont {Tang}}, \bibinfo {author} {\bibfnamefont {R.~C.}\
  \bibnamefont {Mancini}}, \bibinfo {author} {\bibfnamefont {Y.}~\bibnamefont
  {Kim}}, \ and\ \bibinfo {author} {\bibfnamefont {H.~W.}\ \bibnamefont
  {Herrmann}},\ }\bibfield  {title} {\enquote {\bibinfo {title} {Observation of
  interspecies ion separation in inertial-confinement-fusion implosions},}\
  }\href@noop {} {\bibfield  {journal} {\bibinfo  {journal}
  {Euro.~Phys.~Lett.}\ }\textbf {\bibinfo {volume} {115}},\ \bibinfo {pages}
  {65001} (\bibinfo {year} {2016})}\BibitemShut {NoStop}%
\bibitem [{\citenamefont {{Kennel}}\ and\ \citenamefont
  {{Petscheck}}(1966)}]{Kennel:1966}%
  \BibitemOpen
  \bibfield  {author} {\bibinfo {author} {\bibfnamefont {C.~F.}\ \bibnamefont
  {{Kennel}}}\ and\ \bibinfo {author} {\bibfnamefont {H.~E.}\ \bibnamefont
  {{Petscheck}}},\ }\bibfield  {title} {\enquote {\bibinfo {title} {{Limit on
  stably trapped particle fluxes}},}\ }\href@noop {} {\bibfield  {journal}
  {\bibinfo  {journal} {J.~Geophys.~Res.}\ }\textbf {\bibinfo {volume} {71}},\
  \bibinfo {pages} {1--28} (\bibinfo {year} {1966})}\BibitemShut {NoStop}%
\bibitem [{\citenamefont {{Gary}}\ \emph {et~al.}(1976)\citenamefont {{Gary}},
  \citenamefont {{Montgomery}}, \citenamefont {{Feldman}},\ and\ \citenamefont
  {{Forslund}}}]{Gary:1976}%
  \BibitemOpen
  \bibfield  {author} {\bibinfo {author} {\bibfnamefont {S.~P.}\ \bibnamefont
  {{Gary}}}, \bibinfo {author} {\bibfnamefont {M.~D.}\ \bibnamefont
  {{Montgomery}}}, \bibinfo {author} {\bibfnamefont {W.~C.}\ \bibnamefont
  {{Feldman}}}, \ and\ \bibinfo {author} {\bibfnamefont {D.~W.}\ \bibnamefont
  {{Forslund}}},\ }\bibfield  {title} {\enquote {\bibinfo {title} {{Proton
  temperature anisotropy instabilities in the solar wind}},}\ }\href@noop {}
  {\bibfield  {journal} {\bibinfo  {journal} {J.~Geophys.~Res.}\ }\textbf
  {\bibinfo {volume} {81}},\ \bibinfo {pages} {1241--1246} (\bibinfo {year}
  {1976})}\BibitemShut {NoStop}%
\bibitem [{\citenamefont {{Hellinger}}\ and\ \citenamefont
  {{Matsumoto}}(2000)}]{Hellinger:2000}%
  \BibitemOpen
  \bibfield  {author} {\bibinfo {author} {\bibfnamefont {P.}~\bibnamefont
  {{Hellinger}}}\ and\ \bibinfo {author} {\bibfnamefont {H.}~\bibnamefont
  {{Matsumoto}}},\ }\bibfield  {title} {\enquote {\bibinfo {title} {{New
  kinetic instability: Oblique Alfv{\'e}n fire hose}},}\ }\href {\doibase
  10.1029/1999JA000297} {\bibfield  {journal} {\bibinfo  {journal}
  {J.~Geophys.~Res.}\ }\textbf {\bibinfo {volume} {105}},\ \bibinfo {pages}
  {10519--10526} (\bibinfo {year} {2000})}\BibitemShut {NoStop}%
\bibitem [{\citenamefont {{Chandrasekhar}}, \citenamefont {{Kaufman}},\ and\
  \citenamefont {{Watson}}(1958)}]{Chandrasekhar:1958}%
  \BibitemOpen
  \bibfield  {author} {\bibinfo {author} {\bibfnamefont {S.}~\bibnamefont
  {{Chandrasekhar}}}, \bibinfo {author} {\bibfnamefont {A.~N.}\ \bibnamefont
  {{Kaufman}}}, \ and\ \bibinfo {author} {\bibfnamefont {K.~M.}\ \bibnamefont
  {{Watson}}},\ }\bibfield  {title} {\enquote {\bibinfo {title} {{The stability
  of the pinch}},}\ }\href@noop {} {\bibfield  {journal} {\bibinfo  {journal}
  {Proc. Roy. Soc. London Ser. A}\ }\textbf {\bibinfo {volume} {245}},\
  \bibinfo {pages} {435} (\bibinfo {year} {1958})}\BibitemShut {NoStop}%
\bibitem [{\citenamefont {{Vedenov}}\ and\ \citenamefont
  {{Sagdeev}}(1958)}]{Vedenov:1958}%
  \BibitemOpen
  \bibfield  {author} {\bibinfo {author} {\bibfnamefont {A.~A.}\ \bibnamefont
  {{Vedenov}}}\ and\ \bibinfo {author} {\bibfnamefont {R.~Z.}\ \bibnamefont
  {{Sagdeev}}},\ }\bibfield  {title} {\enquote {\bibinfo {title} {Some
  properties of a plasma with an anisotropic ion velocity distribution in a
  magnetic field},}\ }in\ \href@noop {} {\emph {\bibinfo {booktitle} {Plasma
  Physics and the Problem of Controlled Thermonuclear Reactions}}},\
  Vol.~\bibinfo {volume} {3},\ \bibinfo {editor} {edited by\ \bibinfo {editor}
  {\bibfnamefont {M.~A.}\ \bibnamefont {Leontovich}}}\ (\bibinfo  {publisher}
  {Pergamon, New York},\ \bibinfo {year} {1958})\ p.\ \bibinfo {pages}
  {332}\BibitemShut {NoStop}%
\bibitem [{\citenamefont {{Tajiri}}(1967)}]{Tajiri:1967}%
  \BibitemOpen
  \bibfield  {author} {\bibinfo {author} {\bibfnamefont {M.}~\bibnamefont
  {{Tajiri}}},\ }\bibfield  {title} {\enquote {\bibinfo {title} {{Propagation
  of Hydromagnetic Waves in Collisionless Plasma. II. Kinetic Approach}},}\
  }\href@noop {} {\bibfield  {journal} {\bibinfo  {journal} {Journal of the
  Physical Society of Japan}\ }\textbf {\bibinfo {volume} {22}},\ \bibinfo
  {pages} {1482} (\bibinfo {year} {1967})}\BibitemShut {NoStop}%
\bibitem [{\citenamefont {{Southwood}}\ and\ \citenamefont
  {{Kivelson}}(1993)}]{Southwood:1993}%
  \BibitemOpen
  \bibfield  {author} {\bibinfo {author} {\bibfnamefont {D.~J.}\ \bibnamefont
  {{Southwood}}}\ and\ \bibinfo {author} {\bibfnamefont {M.~G.}\ \bibnamefont
  {{Kivelson}}},\ }\bibfield  {title} {\enquote {\bibinfo {title} {{Mirror
  instability. I - Physical mechanism of linear instability}},}\ }\href
  {\doibase 10.1029/92JA02837} {\bibfield  {journal} {\bibinfo  {journal}
  {J.~Geophys.~Res.}\ }\textbf {\bibinfo {volume} {98}},\ \bibinfo {pages}
  {9181--9187} (\bibinfo {year} {1993})}\BibitemShut {NoStop}%
\bibitem [{\citenamefont {{Davidson}}\ and\ \citenamefont
  {{Ogden}}(1975)}]{Davidson:1975}%
  \BibitemOpen
  \bibfield  {author} {\bibinfo {author} {\bibfnamefont {R.~C.}\ \bibnamefont
  {{Davidson}}}\ and\ \bibinfo {author} {\bibfnamefont {J.~M.}\ \bibnamefont
  {{Ogden}}},\ }\bibfield  {title} {\enquote {\bibinfo {title}
  {{Electromagnetic ion cyclotron instability driven by ion energy anisotropy
  in high-beta plasmas}},}\ }\href@noop {} {\bibfield  {journal} {\bibinfo
  {journal} {Physics of Fluids}\ }\textbf {\bibinfo {volume} {18}},\ \bibinfo
  {pages} {1045--1050} (\bibinfo {year} {1975})}\BibitemShut {NoStop}%
\bibitem [{\citenamefont {{Smith}}(1984)}]{Smith:1984}%
  \BibitemOpen
  \bibfield  {author} {\bibinfo {author} {\bibfnamefont {G.~R.}\ \bibnamefont
  {{Smith}}},\ }\bibfield  {title} {\enquote {\bibinfo {title} {{Alfv{\'e}n
  ion-cyclotron instability in tandem-mirror plasmas. I}},}\ }\href {\doibase
  10.1063/1.864773} {\bibfield  {journal} {\bibinfo  {journal} {Phys.~Fluids}\
  }\textbf {\bibinfo {volume} {27}},\ \bibinfo {pages} {1499--1513} (\bibinfo
  {year} {1984})}\BibitemShut {NoStop}%
\bibitem [{\citenamefont {{Kasper}}, \citenamefont {{Lazarus}},\ and\
  \citenamefont {{Gary}}(2002)}]{Kasper:2002}%
  \BibitemOpen
  \bibfield  {author} {\bibinfo {author} {\bibfnamefont {J.~C.}\ \bibnamefont
  {{Kasper}}}, \bibinfo {author} {\bibfnamefont {A.~J.}\ \bibnamefont
  {{Lazarus}}}, \ and\ \bibinfo {author} {\bibfnamefont {S.~P.}\ \bibnamefont
  {{Gary}}},\ }\bibfield  {title} {\enquote {\bibinfo {title} {{Wind/SWE
  observations of firehose constraint on solar wind proton temperature
  anisotropy}},}\ }\href@noop {} {\bibfield  {journal} {\bibinfo  {journal}
  {Geophys.~Res.~Lett.}\ }\textbf {\bibinfo {volume} {29}},\ \bibinfo {pages}
  {20--1} (\bibinfo {year} {2002})}\BibitemShut {NoStop}%
\bibitem [{\citenamefont {{Hellinger}}\ \emph {et~al.}(2006)\citenamefont
  {{Hellinger}}, \citenamefont {{Tr{\'a}vn{\'{\i}}{\v c}ek}}, \citenamefont
  {{Kasper}},\ and\ \citenamefont {{Lazarus}}}]{Hellinger:2006}%
  \BibitemOpen
  \bibfield  {author} {\bibinfo {author} {\bibfnamefont {P.}~\bibnamefont
  {{Hellinger}}}, \bibinfo {author} {\bibfnamefont {P.}~\bibnamefont
  {{Tr{\'a}vn{\'{\i}}{\v c}ek}}}, \bibinfo {author} {\bibfnamefont {J.~C.}\
  \bibnamefont {{Kasper}}}, \ and\ \bibinfo {author} {\bibfnamefont {A.~J.}\
  \bibnamefont {{Lazarus}}},\ }\bibfield  {title} {\enquote {\bibinfo {title}
  {{Solar wind proton temperature anisotropy: Linear theory and WIND/SWE
  observations}},}\ }\href {\doibase 10.1029/2006GL025925} {\bibfield
  {journal} {\bibinfo  {journal} {Geophys.~Res.~Lett.}\ }\textbf {\bibinfo
  {volume} {33}},\ \bibinfo {pages} {9101--+} (\bibinfo {year}
  {2006})}\BibitemShut {NoStop}%
\bibitem [{\citenamefont {{Bale}}\ \emph {et~al.}(2009)\citenamefont {{Bale}},
  \citenamefont {{Kasper}}, \citenamefont {{Howes}}, \citenamefont
  {{Quataert}}, \citenamefont {{Salem}},\ and\ \citenamefont
  {{Sundkvist}}}]{Bale:2009}%
  \BibitemOpen
  \bibfield  {author} {\bibinfo {author} {\bibfnamefont {S.~D.}\ \bibnamefont
  {{Bale}}}, \bibinfo {author} {\bibfnamefont {J.~C.}\ \bibnamefont
  {{Kasper}}}, \bibinfo {author} {\bibfnamefont {G.~G.}\ \bibnamefont
  {{Howes}}}, \bibinfo {author} {\bibfnamefont {E.}~\bibnamefont {{Quataert}}},
  \bibinfo {author} {\bibfnamefont {C.}~\bibnamefont {{Salem}}}, \ and\
  \bibinfo {author} {\bibfnamefont {D.}~\bibnamefont {{Sundkvist}}},\
  }\bibfield  {title} {\enquote {\bibinfo {title} {{Magnetic Fluctuation Power
  Near Proton Temperature Anisotropy Instability Thresholds in the Solar
  Wind}},}\ }\href {\doibase 10.1103/PhysRevLett.103.211101} {\bibfield
  {journal} {\bibinfo  {journal} {Phys.~Rev.~Lett.}\ }\textbf {\bibinfo
  {volume} {103}},\ \bibinfo {pages} {211101} (\bibinfo {year} {2009})},\
  \Eprint {http://arxiv.org/abs/0908.1274} {arXiv:0908.1274} \BibitemShut
  {NoStop}%
\bibitem [{\citenamefont {{Maruca}}, \citenamefont {{Kasper}},\ and\
  \citenamefont {{Bale}}(2011)}]{Maruca:2011}%
  \BibitemOpen
  \bibfield  {author} {\bibinfo {author} {\bibfnamefont {B.~A.}\ \bibnamefont
  {{Maruca}}}, \bibinfo {author} {\bibfnamefont {J.~C.}\ \bibnamefont
  {{Kasper}}}, \ and\ \bibinfo {author} {\bibfnamefont {S.~D.}\ \bibnamefont
  {{Bale}}},\ }\bibfield  {title} {\enquote {\bibinfo {title} {{What Are the
  Relative Roles of Heating and Cooling in Generating Solar Wind Temperature
  Anisotropies?}}}\ }\href {\doibase 10.1103/PhysRevLett.107.201101} {\bibfield
   {journal} {\bibinfo  {journal} {Phys.~Rev.~Lett.}\ }\textbf {\bibinfo
  {volume} {107}},\ \bibinfo {eid} {201101} (\bibinfo {year}
  {2011})}\BibitemShut {NoStop}%
\bibitem [{\citenamefont {{Chen}}\ \emph {et~al.}(2016)\citenamefont {{Chen}},
  \citenamefont {{Matteini}}, \citenamefont {{Schekochihin}}, \citenamefont
  {{Stevens}}, \citenamefont {{Salem}}, \citenamefont {{Maruca}}, \citenamefont
  {{Kunz}},\ and\ \citenamefont {{Bale}}}]{Chen:2016}%
  \BibitemOpen
  \bibfield  {author} {\bibinfo {author} {\bibfnamefont {C.~H.~K.}\
  \bibnamefont {{Chen}}}, \bibinfo {author} {\bibfnamefont {L.}~\bibnamefont
  {{Matteini}}}, \bibinfo {author} {\bibfnamefont {A.~A.}\ \bibnamefont
  {{Schekochihin}}}, \bibinfo {author} {\bibfnamefont {M.~L.}\ \bibnamefont
  {{Stevens}}}, \bibinfo {author} {\bibfnamefont {C.~S.}\ \bibnamefont
  {{Salem}}}, \bibinfo {author} {\bibfnamefont {B.~A.}\ \bibnamefont
  {{Maruca}}}, \bibinfo {author} {\bibfnamefont {M.~W.}\ \bibnamefont
  {{Kunz}}}, \ and\ \bibinfo {author} {\bibfnamefont {S.~D.}\ \bibnamefont
  {{Bale}}},\ }\bibfield  {title} {\enquote {\bibinfo {title} {{Multi-species
  Measurements of the Firehose and Mirror Instability Thresholds in the Solar
  Wind}},}\ }\href {\doibase 10.3847/2041-8205/825/2/L26} {\bibfield  {journal}
  {\bibinfo  {journal} {Astrophys.~J.~Lett.}\ }\textbf {\bibinfo {volume}
  {825}},\ \bibinfo {eid} {L26} (\bibinfo {year} {2016})},\ \Eprint
  {http://arxiv.org/abs/1606.02624} {arXiv:1606.02624 [physics.space-ph]}
  \BibitemShut {NoStop}%
\bibitem [{\citenamefont {{Klein}}\ and\ \citenamefont
  {{Howes}}(2015)}]{Klein:2015}%
  \BibitemOpen
  \bibfield  {author} {\bibinfo {author} {\bibfnamefont {K.~G.}\ \bibnamefont
  {{Klein}}}\ and\ \bibinfo {author} {\bibfnamefont {G.~G.}\ \bibnamefont
  {{Howes}}},\ }\bibfield  {title} {\enquote {\bibinfo {title} {{Predicted
  impacts of proton temperature anisotropy on solar wind turbulence}},}\ }\href
  {\doibase 10.1063/1.4914933} {\bibfield  {journal} {\bibinfo  {journal}
  {Phys.~Plasmas}\ }\textbf {\bibinfo {volume} {22}},\ \bibinfo {eid} {032903}
  (\bibinfo {year} {2015})},\ \Eprint {http://arxiv.org/abs/1503.00695}
  {arXiv:1503.00695 [physics.space-ph]} \BibitemShut {NoStop}%
\bibitem [{\citenamefont {{Kunz}}\ \emph {et~al.}(2015)\citenamefont {{Kunz}},
  \citenamefont {{Schekochihin}}, \citenamefont {{Chen}}, \citenamefont
  {{Abel}},\ and\ \citenamefont {{Cowley}}}]{Kunz:2015}%
  \BibitemOpen
  \bibfield  {author} {\bibinfo {author} {\bibfnamefont {M.~W.}\ \bibnamefont
  {{Kunz}}}, \bibinfo {author} {\bibfnamefont {A.~A.}\ \bibnamefont
  {{Schekochihin}}}, \bibinfo {author} {\bibfnamefont {C.~H.~K.}\ \bibnamefont
  {{Chen}}}, \bibinfo {author} {\bibfnamefont {I.~G.}\ \bibnamefont {{Abel}}},
  \ and\ \bibinfo {author} {\bibfnamefont {S.~C.}\ \bibnamefont {{Cowley}}},\
  }\bibfield  {title} {\enquote {\bibinfo {title} {{Inertial-range kinetic
  turbulence in pressure-anisotropic astrophysical plasmas}},}\ }\href
  {\doibase 10.1017/S0022377815000811} {\bibfield  {journal} {\bibinfo
  {journal} {J.~Plasma Phys.}\ }\textbf {\bibinfo {volume} {81}},\ \bibinfo
  {eid} {325810501} (\bibinfo {year} {2015})},\ \Eprint
  {http://arxiv.org/abs/1501.06771} {arXiv:1501.06771 [astro-ph.HE]}
  \BibitemShut {NoStop}%
\bibitem [{\citenamefont {{Kunz}}\ \emph {et~al.}(2017)\citenamefont {{Kunz}},
  \citenamefont {{Abel}}, \citenamefont {{Klein}},\ and\ \citenamefont
  {{Schekochihin}}}]{Kunz:2018}%
  \BibitemOpen
  \bibfield  {author} {\bibinfo {author} {\bibfnamefont {M.~W.}\ \bibnamefont
  {{Kunz}}}, \bibinfo {author} {\bibfnamefont {I.~G.}\ \bibnamefont {{Abel}}},
  \bibinfo {author} {\bibfnamefont {K.~G.}\ \bibnamefont {{Klein}}}, \ and\
  \bibinfo {author} {\bibfnamefont {A.~A.}\ \bibnamefont {{Schekochihin}}},\
  }\bibfield  {title} {\enquote {\bibinfo {title} {{Astrophysical gyrokinetics:
  Turbulence in pressure-anisotropic plasmas at ion scales and beyond}},}\
  }\href@noop {} {\bibfield  {journal} {\bibinfo  {journal} {ArXiv e-prints}\ }
  (\bibinfo {year} {2017})},\ \Eprint {http://arxiv.org/abs/1712.02269}
  {arXiv:1712.02269 [astro-ph.HE]} \BibitemShut {NoStop}%
\bibitem [{\citenamefont {{Squire}}, \citenamefont {{Quataert}},\ and\
  \citenamefont {{Schekochihin}}(2016)}]{Squire:2016}%
  \BibitemOpen
  \bibfield  {author} {\bibinfo {author} {\bibfnamefont {J.}~\bibnamefont
  {{Squire}}}, \bibinfo {author} {\bibfnamefont {E.}~\bibnamefont
  {{Quataert}}}, \ and\ \bibinfo {author} {\bibfnamefont {A.~A.}\ \bibnamefont
  {{Schekochihin}}},\ }\bibfield  {title} {\enquote {\bibinfo {title} {{A
  Stringent Limit on the Amplitude of Alfv{\'e}nic Perturbations in High-beta
  Low-collisionality Plasmas}},}\ }\href {\doibase 10.3847/2041-8205/830/2/L25}
  {\bibfield  {journal} {\bibinfo  {journal} {Astrophys.~J.~Lett.}\ }\textbf
  {\bibinfo {volume} {830}},\ \bibinfo {eid} {L25} (\bibinfo {year} {2016})},\
  \Eprint {http://arxiv.org/abs/1605.02759} {arXiv:1605.02759
  [physics.plasm-ph]} \BibitemShut {NoStop}%
\bibitem [{\citenamefont {{Squire}}, \citenamefont {{Schekochihin}},\ and\
  \citenamefont {{Quataert}}(2017)}]{Squire:2017a}%
  \BibitemOpen
  \bibfield  {author} {\bibinfo {author} {\bibfnamefont {J.}~\bibnamefont
  {{Squire}}}, \bibinfo {author} {\bibfnamefont {A.~A.}\ \bibnamefont
  {{Schekochihin}}}, \ and\ \bibinfo {author} {\bibfnamefont {E.}~\bibnamefont
  {{Quataert}}},\ }\bibfield  {title} {\enquote {\bibinfo {title} {{Amplitude
  limits and nonlinear damping of shear-Alfv{\'e}n waves in high-beta
  low-collisionality plasmas}},}\ }\href {\doibase 10.1088/1367-2630/aa6bb1}
  {\bibfield  {journal} {\bibinfo  {journal} {New J. Phys.}\ }\textbf {\bibinfo
  {volume} {19}},\ \bibinfo {eid} {055005} (\bibinfo {year} {2017})},\ \Eprint
  {http://arxiv.org/abs/1701.03175} {arXiv:1701.03175 [physics.plasm-ph]}
  \BibitemShut {NoStop}%
\bibitem [{\citenamefont {{Squire}}\ \emph {et~al.}(2017)\citenamefont
  {{Squire}}, \citenamefont {{Kunz}}, \citenamefont {{Quataert}},\ and\
  \citenamefont {{Schekochihin}}}]{Squire:2017b}%
  \BibitemOpen
  \bibfield  {author} {\bibinfo {author} {\bibfnamefont {J.}~\bibnamefont
  {{Squire}}}, \bibinfo {author} {\bibfnamefont {M.~W.}\ \bibnamefont
  {{Kunz}}}, \bibinfo {author} {\bibfnamefont {E.}~\bibnamefont {{Quataert}}},
  \ and\ \bibinfo {author} {\bibfnamefont {A.~A.}\ \bibnamefont
  {{Schekochihin}}},\ }\bibfield  {title} {\enquote {\bibinfo {title} {{Kinetic
  Simulations of the Interruption of Large-Amplitude Shear-Alfv{\'e}n Waves in
  a High-{$\beta$} Plasma}},}\ }\href {\doibase 10.1103/PhysRevLett.119.155101}
  {\bibfield  {journal} {\bibinfo  {journal} {Phys.~Rev.~Lett.}\ }\textbf
  {\bibinfo {volume} {119}},\ \bibinfo {eid} {155101} (\bibinfo {year}
  {2017})},\ \Eprint {http://arxiv.org/abs/1705.01956} {arXiv:1705.01956
  [physics.plasm-ph]} \BibitemShut {NoStop}%
\bibitem [{\citenamefont {{Keiter}}\ \emph {et~al.}(2000)\citenamefont
  {{Keiter}}, \citenamefont {{Scime}}, \citenamefont {{Balkey}}, \citenamefont
  {{Boivin}}, \citenamefont {{Kline}},\ and\ \citenamefont
  {{Gary}}}]{Keiter:2000}%
  \BibitemOpen
  \bibfield  {author} {\bibinfo {author} {\bibfnamefont {P.~A.}\ \bibnamefont
  {{Keiter}}}, \bibinfo {author} {\bibfnamefont {E.~E.}\ \bibnamefont
  {{Scime}}}, \bibinfo {author} {\bibfnamefont {M.~M.}\ \bibnamefont
  {{Balkey}}}, \bibinfo {author} {\bibfnamefont {R.}~\bibnamefont {{Boivin}}},
  \bibinfo {author} {\bibfnamefont {J.~L.}\ \bibnamefont {{Kline}}}, \ and\
  \bibinfo {author} {\bibfnamefont {S.~P.}\ \bibnamefont {{Gary}}},\ }\bibfield
   {title} {\enquote {\bibinfo {title} {{Beta-dependent upper bound on ion
  temperature anisotropy in a laboratory plasma}},}\ }\href {\doibase
  10.1063/1.873872} {\bibfield  {journal} {\bibinfo  {journal} {Phys.~Plasmas}\
  }\textbf {\bibinfo {volume} {7}},\ \bibinfo {pages} {779--783} (\bibinfo
  {year} {2000})}\BibitemShut {NoStop}%
\bibitem [{\citenamefont {{Scime}}\ \emph {et~al.}(2000)\citenamefont
  {{Scime}}, \citenamefont {{Keiter}}, \citenamefont {{Balkey}}, \citenamefont
  {{Boivin}}, \citenamefont {{Kline}}, \citenamefont {{Blackburn}},\ and\
  \citenamefont {{Gary}}}]{Scime:2000}%
  \BibitemOpen
  \bibfield  {author} {\bibinfo {author} {\bibfnamefont {E.~E.}\ \bibnamefont
  {{Scime}}}, \bibinfo {author} {\bibfnamefont {P.~A.}\ \bibnamefont
  {{Keiter}}}, \bibinfo {author} {\bibfnamefont {M.~M.}\ \bibnamefont
  {{Balkey}}}, \bibinfo {author} {\bibfnamefont {R.~F.}\ \bibnamefont
  {{Boivin}}}, \bibinfo {author} {\bibfnamefont {J.~L.}\ \bibnamefont
  {{Kline}}}, \bibinfo {author} {\bibfnamefont {M.}~\bibnamefont
  {{Blackburn}}}, \ and\ \bibinfo {author} {\bibfnamefont {S.~P.}\ \bibnamefont
  {{Gary}}},\ }\bibfield  {title} {\enquote {\bibinfo {title} {{Ion temperature
  anisotropy limitation in high beta plasmas}},}\ }\href {\doibase
  10.1063/1.874036} {\bibfield  {journal} {\bibinfo  {journal} {Phys.~Plasmas}\
  }\textbf {\bibinfo {volume} {7}},\ \bibinfo {pages} {2157--2165} (\bibinfo
  {year} {2000})}\BibitemShut {NoStop}%
\bibitem [{\citenamefont {{Summers}}, \citenamefont {{Ni}},\ and\ \citenamefont
  {{Meredith}}(2007{\natexlab{a}})}]{Summers:2007a}%
  \BibitemOpen
  \bibfield  {author} {\bibinfo {author} {\bibfnamefont {D.}~\bibnamefont
  {{Summers}}}, \bibinfo {author} {\bibfnamefont {B.}~\bibnamefont {{Ni}}}, \
  and\ \bibinfo {author} {\bibfnamefont {N.~P.}\ \bibnamefont {{Meredith}}},\
  }\bibfield  {title} {\enquote {\bibinfo {title} {{Timescales for radiation
  belt electron acceleration and loss due to resonant wave-particle
  interactions: 1. Theory}},}\ }\href {\doibase 10.1029/2006JA011801}
  {\bibfield  {journal} {\bibinfo  {journal} {Journal of Geophysical Research
  (Space Physics)}\ }\textbf {\bibinfo {volume} {112}},\ \bibinfo {eid}
  {A04206} (\bibinfo {year} {2007}{\natexlab{a}})}\BibitemShut {NoStop}%
\bibitem [{\citenamefont {{Summers}}, \citenamefont {{Ni}},\ and\ \citenamefont
  {{Meredith}}(2007{\natexlab{b}})}]{Summers:2007b}%
  \BibitemOpen
  \bibfield  {author} {\bibinfo {author} {\bibfnamefont {D.}~\bibnamefont
  {{Summers}}}, \bibinfo {author} {\bibfnamefont {B.}~\bibnamefont {{Ni}}}, \
  and\ \bibinfo {author} {\bibfnamefont {N.~P.}\ \bibnamefont {{Meredith}}},\
  }\bibfield  {title} {\enquote {\bibinfo {title} {{Timescales for radiation
  belt electron acceleration and loss due to resonant wave-particle
  interactions: 2. Evaluation for VLF chorus, ELF hiss, and electromagnetic ion
  cyclotron waves}},}\ }\href {\doibase 10.1029/2006JA011993} {\bibfield
  {journal} {\bibinfo  {journal} {J.~Geophys.~Res.}\ }\textbf {\bibinfo
  {volume} {112}},\ \bibinfo {eid} {A04207} (\bibinfo {year}
  {2007}{\natexlab{b}})}\BibitemShut {NoStop}%
\bibitem [{\citenamefont {{Summers}}, \citenamefont {{Thorne}},\ and\
  \citenamefont {{Xiao}}(1998)}]{Summers:1998}%
  \BibitemOpen
  \bibfield  {author} {\bibinfo {author} {\bibfnamefont {D.}~\bibnamefont
  {{Summers}}}, \bibinfo {author} {\bibfnamefont {R.~M.}\ \bibnamefont
  {{Thorne}}}, \ and\ \bibinfo {author} {\bibfnamefont {F.}~\bibnamefont
  {{Xiao}}},\ }\bibfield  {title} {\enquote {\bibinfo {title} {{Relativistic
  theory of wave-particle resonant diffusion with application to electron
  acceleration in the magnetosphere}},}\ }\href {\doibase 10.1029/98JA01740}
  {\bibfield  {journal} {\bibinfo  {journal} {J.~Geophys.~Res.}\ }\textbf
  {\bibinfo {volume} {103}},\ \bibinfo {pages} {20487--20500} (\bibinfo {year}
  {1998})}\BibitemShut {NoStop}%
\bibitem [{\citenamefont {{Horne}}\ \emph {et~al.}(2005)\citenamefont
  {{Horne}}, \citenamefont {{Thorne}}, \citenamefont {{Glauert}}, \citenamefont
  {{Albert}}, \citenamefont {{Meredith}},\ and\ \citenamefont
  {{Anderson}}}]{Horne:2005}%
  \BibitemOpen
  \bibfield  {author} {\bibinfo {author} {\bibfnamefont {R.~B.}\ \bibnamefont
  {{Horne}}}, \bibinfo {author} {\bibfnamefont {R.~M.}\ \bibnamefont
  {{Thorne}}}, \bibinfo {author} {\bibfnamefont {S.~A.}\ \bibnamefont
  {{Glauert}}}, \bibinfo {author} {\bibfnamefont {J.~M.}\ \bibnamefont
  {{Albert}}}, \bibinfo {author} {\bibfnamefont {N.~P.}\ \bibnamefont
  {{Meredith}}}, \ and\ \bibinfo {author} {\bibfnamefont {R.~R.}\ \bibnamefont
  {{Anderson}}},\ }\bibfield  {title} {\enquote {\bibinfo {title} {{Timescale
  for radiation belt electron acceleration by whistler mode chorus waves}},}\
  }\href {\doibase 10.1029/2004JA010811} {\bibfield  {journal} {\bibinfo
  {journal} {J.~Geophys.~Res.}\ }\textbf {\bibinfo {volume} {110}},\ \bibinfo
  {eid} {A03225} (\bibinfo {year} {2005})}\BibitemShut {NoStop}%
\bibitem [{\citenamefont {{Omura}}, \citenamefont {{Katoh}},\ and\
  \citenamefont {{Summers}}(2008)}]{Omura:2008}%
  \BibitemOpen
  \bibfield  {author} {\bibinfo {author} {\bibfnamefont {Y.}~\bibnamefont
  {{Omura}}}, \bibinfo {author} {\bibfnamefont {Y.}~\bibnamefont {{Katoh}}}, \
  and\ \bibinfo {author} {\bibfnamefont {D.}~\bibnamefont {{Summers}}},\
  }\bibfield  {title} {\enquote {\bibinfo {title} {{Theory and simulation of
  the generation of whistler-mode chorus}},}\ }\href {\doibase
  10.1029/2007JA012622} {\bibfield  {journal} {\bibinfo  {journal}
  {J.~Geophys.~Res.}\ }\textbf {\bibinfo {volume} {113}},\ \bibinfo {eid}
  {A04223} (\bibinfo {year} {2008})}\BibitemShut {NoStop}%
\bibitem [{\citenamefont {{Cornwall}}(1965)}]{Cornwall:1965}%
  \BibitemOpen
  \bibfield  {author} {\bibinfo {author} {\bibfnamefont {J.~M.}\ \bibnamefont
  {{Cornwall}}},\ }\bibfield  {title} {\enquote {\bibinfo {title} {{Cyclotron
  Instabilities and Electromagnetic Emission in the Ultra Low Frequency and
  Very Low Frequency Ranges}},}\ }\href {\doibase 10.1029/JZ070i001p00061}
  {\bibfield  {journal} {\bibinfo  {journal} {J.~Geophys.~Res.}\ }\textbf
  {\bibinfo {volume} {70}},\ \bibinfo {pages} {61--69} (\bibinfo {year}
  {1965})}\BibitemShut {NoStop}%
\bibitem [{\citenamefont {{Horne}}\ and\ \citenamefont
  {{Thorne}}(1994)}]{Horne:1994}%
  \BibitemOpen
  \bibfield  {author} {\bibinfo {author} {\bibfnamefont {R.~B.}\ \bibnamefont
  {{Horne}}}\ and\ \bibinfo {author} {\bibfnamefont {R.~M.}\ \bibnamefont
  {{Thorne}}},\ }\bibfield  {title} {\enquote {\bibinfo {title} {{Convective
  instabilities of electromagnetic ion cyclotron waves in the outer
  magnetosphere}},}\ }\href {\doibase 10.1029/94JA01259} {\bibfield  {journal}
  {\bibinfo  {journal} {J.~Geophys.~Res.}\ }\textbf {\bibinfo {volume} {99}},\
  \bibinfo {pages} {17} (\bibinfo {year} {1994})}\BibitemShut {NoStop}%
\bibitem [{\citenamefont {{Anderson}}\ \emph {et~al.}(1996)\citenamefont
  {{Anderson}}, \citenamefont {{Denton}}, \citenamefont {{Ho}}, \citenamefont
  {{Hamilton}}, \citenamefont {{Fuselier}},\ and\ \citenamefont
  {{Strangeway}}}]{Anderson:1996}%
  \BibitemOpen
  \bibfield  {author} {\bibinfo {author} {\bibfnamefont {B.~J.}\ \bibnamefont
  {{Anderson}}}, \bibinfo {author} {\bibfnamefont {R.~E.}\ \bibnamefont
  {{Denton}}}, \bibinfo {author} {\bibfnamefont {G.}~\bibnamefont {{Ho}}},
  \bibinfo {author} {\bibfnamefont {D.~C.}\ \bibnamefont {{Hamilton}}},
  \bibinfo {author} {\bibfnamefont {S.~A.}\ \bibnamefont {{Fuselier}}}, \ and\
  \bibinfo {author} {\bibfnamefont {R.~J.}\ \bibnamefont {{Strangeway}}},\
  }\bibfield  {title} {\enquote {\bibinfo {title} {{Observational test of local
  proton cyclotron instability in the Earth's magnetosphere}},}\ }\href
  {\doibase 10.1029/96JA01251} {\bibfield  {journal} {\bibinfo  {journal}
  {J.~Geophys.~Res.}\ }\textbf {\bibinfo {volume} {101}},\ \bibinfo {pages}
  {21527--21544} (\bibinfo {year} {1996})}\BibitemShut {NoStop}%
\bibitem [{\citenamefont {{Fraser}}\ and\ \citenamefont
  {{Nguyen}}(2001)}]{Fraser:2001}%
  \BibitemOpen
  \bibfield  {author} {\bibinfo {author} {\bibfnamefont {B.~J.}\ \bibnamefont
  {{Fraser}}}\ and\ \bibinfo {author} {\bibfnamefont {T.~S.}\ \bibnamefont
  {{Nguyen}}},\ }\bibfield  {title} {\enquote {\bibinfo {title} {{Is the
  plasmapause a preferred source region of electromagnetic ion cyclotron waves
  in the magnetosphere?}}}\ }\href {\doibase 10.1016/S1364-6826(00)00225-X}
  {\bibfield  {journal} {\bibinfo  {journal} {J Atmos. Solar-Terrest. Phys.}\
  }\textbf {\bibinfo {volume} {63}},\ \bibinfo {pages} {1225--1247} (\bibinfo
  {year} {2001})}\BibitemShut {NoStop}%
\bibitem [{\citenamefont {{Min}}\ \emph {et~al.}(2012)\citenamefont {{Min}},
  \citenamefont {{Lee}}, \citenamefont {{Keika}},\ and\ \citenamefont
  {{Li}}}]{Min:2012}%
  \BibitemOpen
  \bibfield  {author} {\bibinfo {author} {\bibfnamefont {K.}~\bibnamefont
  {{Min}}}, \bibinfo {author} {\bibfnamefont {J.}~\bibnamefont {{Lee}}},
  \bibinfo {author} {\bibfnamefont {K.}~\bibnamefont {{Keika}}}, \ and\
  \bibinfo {author} {\bibfnamefont {W.}~\bibnamefont {{Li}}},\ }\bibfield
  {title} {\enquote {\bibinfo {title} {{Global distribution of EMIC waves
  derived from THEMIS observations}},}\ }\href {\doibase 10.1029/2012JA017515}
  {\bibfield  {journal} {\bibinfo  {journal} {J.~Geophys.~Res.}\ }\textbf
  {\bibinfo {volume} {117}},\ \bibinfo {eid} {A05219} (\bibinfo {year}
  {2012})}\BibitemShut {NoStop}%
\bibitem [{\citenamefont {{Russell}}, \citenamefont {{Holzer}},\ and\
  \citenamefont {{Smith}}(1970)}]{Russell:1970}%
  \BibitemOpen
  \bibfield  {author} {\bibinfo {author} {\bibfnamefont {C.~T.}\ \bibnamefont
  {{Russell}}}, \bibinfo {author} {\bibfnamefont {R.~E.}\ \bibnamefont
  {{Holzer}}}, \ and\ \bibinfo {author} {\bibfnamefont {E.~J.}\ \bibnamefont
  {{Smith}}},\ }\bibfield  {title} {\enquote {\bibinfo {title} {{OGO 3
  observations of ELF noise in the magnetosphere: 2. The nature of the
  equatorial noise}},}\ }\href {\doibase 10.1029/JA075i004p00755} {\bibfield
  {journal} {\bibinfo  {journal} {J.~Geophys.~Res.}\ }\textbf {\bibinfo
  {volume} {75}},\ \bibinfo {pages} {755} (\bibinfo {year} {1970})}\BibitemShut
  {NoStop}%
\bibitem [{\citenamefont {{Gurnett}}(1976)}]{Gurnett:1976}%
  \BibitemOpen
  \bibfield  {author} {\bibinfo {author} {\bibfnamefont {D.~A.}\ \bibnamefont
  {{Gurnett}}},\ }\bibfield  {title} {\enquote {\bibinfo {title} {{Plasma wave
  interactions with energetic ions near the magnetic equator}},}\ }\href
  {\doibase 10.1029/JA081i016p02765} {\bibfield  {journal} {\bibinfo  {journal}
  {J.~Geophys.~Res.}\ }\textbf {\bibinfo {volume} {81}},\ \bibinfo {pages}
  {2765--2770} (\bibinfo {year} {1976})}\BibitemShut {NoStop}%
\bibitem [{\citenamefont {{Olsen}}\ \emph {et~al.}(1987)\citenamefont
  {{Olsen}}, \citenamefont {{Shawhan}}, \citenamefont {{Gallagher}},
  \citenamefont {{Chappell}},\ and\ \citenamefont {{Green}}}]{Olsen:1987}%
  \BibitemOpen
  \bibfield  {author} {\bibinfo {author} {\bibfnamefont {R.~C.}\ \bibnamefont
  {{Olsen}}}, \bibinfo {author} {\bibfnamefont {S.~D.}\ \bibnamefont
  {{Shawhan}}}, \bibinfo {author} {\bibfnamefont {D.~L.}\ \bibnamefont
  {{Gallagher}}}, \bibinfo {author} {\bibfnamefont {C.~R.}\ \bibnamefont
  {{Chappell}}}, \ and\ \bibinfo {author} {\bibfnamefont {J.~L.}\ \bibnamefont
  {{Green}}},\ }\bibfield  {title} {\enquote {\bibinfo {title} {{Plasma
  observations at the earth's magnetic equator}},}\ }\href {\doibase
  10.1029/JA092iA03p02385} {\bibfield  {journal} {\bibinfo  {journal}
  {J.~Geophys.~Res.}\ }\textbf {\bibinfo {volume} {92}},\ \bibinfo {pages}
  {2385--2407} (\bibinfo {year} {1987})}\BibitemShut {NoStop}%
\bibitem [{\citenamefont {{Liu}}, \citenamefont {{Gary}},\ and\ \citenamefont
  {{Winske}}(2011)}]{Liu:2011}%
  \BibitemOpen
  \bibfield  {author} {\bibinfo {author} {\bibfnamefont {K.}~\bibnamefont
  {{Liu}}}, \bibinfo {author} {\bibfnamefont {S.~P.}\ \bibnamefont {{Gary}}}, \
  and\ \bibinfo {author} {\bibfnamefont {D.}~\bibnamefont {{Winske}}},\
  }\bibfield  {title} {\enquote {\bibinfo {title} {{Excitation of magnetosonic
  waves in the terrestrial magnetosphere: Particle-in-cell simulations}},}\
  }\href {\doibase 10.1029/2010JA016372} {\bibfield  {journal} {\bibinfo
  {journal} {J.~Geophys.~Res.}\ }\textbf {\bibinfo {volume} {116}},\ \bibinfo
  {eid} {A07212} (\bibinfo {year} {2011})}\BibitemShut {NoStop}%
\bibitem [{\citenamefont {{Thorne}}\ \emph {et~al.}(1973)\citenamefont
  {{Thorne}}, \citenamefont {{Smith}}, \citenamefont {{Burton}},\ and\
  \citenamefont {{Holzer}}}]{Thorne:1973}%
  \BibitemOpen
  \bibfield  {author} {\bibinfo {author} {\bibfnamefont {R.~M.}\ \bibnamefont
  {{Thorne}}}, \bibinfo {author} {\bibfnamefont {E.~J.}\ \bibnamefont
  {{Smith}}}, \bibinfo {author} {\bibfnamefont {R.~K.}\ \bibnamefont
  {{Burton}}}, \ and\ \bibinfo {author} {\bibfnamefont {R.~E.}\ \bibnamefont
  {{Holzer}}},\ }\bibfield  {title} {\enquote {\bibinfo {title} {{Plasmaspheric
  hiss}},}\ }\href {\doibase 10.1029/JA078i010p01581} {\bibfield  {journal}
  {\bibinfo  {journal} {J.~Geophys.~Res.}\ }\textbf {\bibinfo {volume} {78}},\
  \bibinfo {pages} {1581--1596} (\bibinfo {year} {1973})}\BibitemShut {NoStop}%
\bibitem [{\citenamefont {{Church}}\ and\ \citenamefont
  {{Thorne}}(1983)}]{Church:1983}%
  \BibitemOpen
  \bibfield  {author} {\bibinfo {author} {\bibfnamefont {S.~R.}\ \bibnamefont
  {{Church}}}\ and\ \bibinfo {author} {\bibfnamefont {R.~M.}\ \bibnamefont
  {{Thorne}}},\ }\bibfield  {title} {\enquote {\bibinfo {title} {{On the origin
  of plasmaspheric hiss - Ray path integrated amplification}},}\ }\href
  {\doibase 10.1029/JA088iA10p07941} {\bibfield  {journal} {\bibinfo  {journal}
  {J.~Geophys.~Res.}\ }\textbf {\bibinfo {volume} {88}},\ \bibinfo {pages}
  {7941--7957} (\bibinfo {year} {1983})}\BibitemShut {NoStop}%
\bibitem [{\citenamefont {{Sonwalkar}}\ and\ \citenamefont
  {{Inan}}(1989)}]{Sonwalkar:1989}%
  \BibitemOpen
  \bibfield  {author} {\bibinfo {author} {\bibfnamefont {V.~S.}\ \bibnamefont
  {{Sonwalkar}}}\ and\ \bibinfo {author} {\bibfnamefont {U.~S.}\ \bibnamefont
  {{Inan}}},\ }\bibfield  {title} {\enquote {\bibinfo {title} {{Lightning as an
  embryonic source of VLF hiss}},}\ }\href {\doibase 10.1029/JA094iA06p06986}
  {\bibfield  {journal} {\bibinfo  {journal} {J.~Geophys.~Res.}\ }\textbf
  {\bibinfo {volume} {94}},\ \bibinfo {pages} {6986--6994} (\bibinfo {year}
  {1989})}\BibitemShut {NoStop}%
\bibitem [{\citenamefont {{Hayakawa}}\ and\ \citenamefont
  {{Sazhin}}(1992)}]{Hayakawa:1992}%
  \BibitemOpen
  \bibfield  {author} {\bibinfo {author} {\bibfnamefont {M.}~\bibnamefont
  {{Hayakawa}}}\ and\ \bibinfo {author} {\bibfnamefont {S.~S.}\ \bibnamefont
  {{Sazhin}}},\ }\bibfield  {title} {\enquote {\bibinfo {title} {{Mid-latitude
  and plasmaspheric HISS - A review}},}\ }\href {\doibase
  10.1016/0032-0633(92)90089-7} {\bibfield  {journal} {\bibinfo  {journal}
  {Planet. Space Sci.}\ }\textbf {\bibinfo {volume} {40}},\ \bibinfo {pages}
  {1325--1338} (\bibinfo {year} {1992})}\BibitemShut {NoStop}%
\bibitem [{\citenamefont {{Abel}}\ and\ \citenamefont
  {{Thorne}}(1998)}]{Abel:1998}%
  \BibitemOpen
  \bibfield  {author} {\bibinfo {author} {\bibfnamefont {B.}~\bibnamefont
  {{Abel}}}\ and\ \bibinfo {author} {\bibfnamefont {R.~M.}\ \bibnamefont
  {{Thorne}}},\ }\bibfield  {title} {\enquote {\bibinfo {title} {{Electron
  scattering loss in Earth's inner magnetosphere 1. Dominant physical
  processes}},}\ }\href {\doibase 10.1029/97JA02919} {\bibfield  {journal}
  {\bibinfo  {journal} {J.~Geophys.~Res.}\ }\textbf {\bibinfo {volume} {103}},\
  \bibinfo {pages} {2385--2396} (\bibinfo {year} {1998})}\BibitemShut {NoStop}%
\bibitem [{\citenamefont {{Green}}\ \emph {et~al.}(2005)\citenamefont
  {{Green}}, \citenamefont {{Boardsen}}, \citenamefont {{Garcia}},
  \citenamefont {{Taylor}}, \citenamefont {{Fung}},\ and\ \citenamefont
  {{Reinisch}}}]{Green:2005}%
  \BibitemOpen
  \bibfield  {author} {\bibinfo {author} {\bibfnamefont {J.~L.}\ \bibnamefont
  {{Green}}}, \bibinfo {author} {\bibfnamefont {S.}~\bibnamefont {{Boardsen}}},
  \bibinfo {author} {\bibfnamefont {L.}~\bibnamefont {{Garcia}}}, \bibinfo
  {author} {\bibfnamefont {W.~W.~L.}\ \bibnamefont {{Taylor}}}, \bibinfo
  {author} {\bibfnamefont {S.~F.}\ \bibnamefont {{Fung}}}, \ and\ \bibinfo
  {author} {\bibfnamefont {B.~W.}\ \bibnamefont {{Reinisch}}},\ }\bibfield
  {title} {\enquote {\bibinfo {title} {{On the origin of whistler mode
  radiation in the plasmasphere}},}\ }\href {\doibase 10.1029/2004JA010495}
  {\bibfield  {journal} {\bibinfo  {journal} {J.~Geophys.~Res.}\ }\textbf
  {\bibinfo {volume} {110}},\ \bibinfo {eid} {A03201} (\bibinfo {year}
  {2005})}\BibitemShut {NoStop}%
\bibitem [{\citenamefont {{Meredith}}\ \emph {et~al.}(2006)\citenamefont
  {{Meredith}}, \citenamefont {{Horne}}, \citenamefont {{Clilverd}},
  \citenamefont {{Horsfall}}, \citenamefont {{Thorne}},\ and\ \citenamefont
  {{Anderson}}}]{Meredith:2006}%
  \BibitemOpen
  \bibfield  {author} {\bibinfo {author} {\bibfnamefont {N.~P.}\ \bibnamefont
  {{Meredith}}}, \bibinfo {author} {\bibfnamefont {R.~B.}\ \bibnamefont
  {{Horne}}}, \bibinfo {author} {\bibfnamefont {M.~A.}\ \bibnamefont
  {{Clilverd}}}, \bibinfo {author} {\bibfnamefont {D.}~\bibnamefont
  {{Horsfall}}}, \bibinfo {author} {\bibfnamefont {R.~M.}\ \bibnamefont
  {{Thorne}}}, \ and\ \bibinfo {author} {\bibfnamefont {R.~R.}\ \bibnamefont
  {{Anderson}}},\ }\bibfield  {title} {\enquote {\bibinfo {title} {{Origins of
  plasmaspheric hiss}},}\ }\href {\doibase 10.1029/2006JA011707} {\bibfield
  {journal} {\bibinfo  {journal} {J.~Geophys.~Res.}\ }\textbf {\bibinfo
  {volume} {111}},\ \bibinfo {eid} {A09217} (\bibinfo {year}
  {2006})}\BibitemShut {NoStop}%
\bibitem [{\citenamefont {{Bortnik}}, \citenamefont {{Thorne}},\ and\
  \citenamefont {{Meredith}}(2008)}]{Bortnik:2008}%
  \BibitemOpen
  \bibfield  {author} {\bibinfo {author} {\bibfnamefont {J.}~\bibnamefont
  {{Bortnik}}}, \bibinfo {author} {\bibfnamefont {R.~M.}\ \bibnamefont
  {{Thorne}}}, \ and\ \bibinfo {author} {\bibfnamefont {N.~P.}\ \bibnamefont
  {{Meredith}}},\ }\bibfield  {title} {\enquote {\bibinfo {title} {{The
  unexpected origin of plasmaspheric hiss from discrete chorus emissions}},}\
  }\href {\doibase 10.1038/nature06741} {\bibfield  {journal} {\bibinfo
  {journal} {Nature}\ }\textbf {\bibinfo {volume} {452}},\ \bibinfo {pages}
  {62--66} (\bibinfo {year} {2008})}\BibitemShut {NoStop}%
\bibitem [{\citenamefont {Kintner}(1992)}]{Kintner:1992}%
  \BibitemOpen
  \bibfield  {author} {\bibinfo {author} {\bibfnamefont {P.~M.}\ \bibnamefont
  {Kintner}},\ }\bibfield  {title} {\enquote {\bibinfo {title} {Plasma waves
  and transversely accelerated ions in the terrestrial ionosphere},}\ }\href
  {\doibase 10.1063/1.4729441} {\bibfield  {journal} {\bibinfo  {journal}
  {Phys.~Fluids B}\ }\textbf {\bibinfo {volume} {4}},\ \bibinfo {pages} {2264}
  (\bibinfo {year} {1992})}\BibitemShut {NoStop}%
\bibitem [{\citenamefont {Ganguli}, \citenamefont {Lee},\ and\ \citenamefont
  {Palmadesso}(1985)}]{Ganguli:1985a}%
  \BibitemOpen
  \bibfield  {author} {\bibinfo {author} {\bibfnamefont {G.}~\bibnamefont
  {Ganguli}}, \bibinfo {author} {\bibfnamefont {Y.~C.}\ \bibnamefont {Lee}}, \
  and\ \bibinfo {author} {\bibfnamefont {P.}~\bibnamefont {Palmadesso}},\
  }\bibfield  {title} {\enquote {\bibinfo {title} {Electrostatic ion-cyclotron
  instability caused by a nonuniform electric field perpendicular to the
  external magnetic field},}\ }\href@noop {} {\bibfield  {journal} {\bibinfo
  {journal} {Phys.~Fluids}\ }\textbf {\bibinfo {volume} {28}},\ \bibinfo
  {pages} {761} (\bibinfo {year} {1985})}\BibitemShut {NoStop}%
\bibitem [{\citenamefont {Ganguli}, \citenamefont {Lee},\ and\ \citenamefont
  {Palmadesso}(1988)}]{Ganguli:1988a}%
  \BibitemOpen
  \bibfield  {author} {\bibinfo {author} {\bibfnamefont {G.}~\bibnamefont
  {Ganguli}}, \bibinfo {author} {\bibfnamefont {Y.~C.}\ \bibnamefont {Lee}}, \
  and\ \bibinfo {author} {\bibfnamefont {P.~J.}\ \bibnamefont {Palmadesso}},\
  }\bibfield  {title} {\enquote {\bibinfo {title} {Kinetic theory for
  electrostatic waves due to transverse velocity shears},}\ }\href@noop {}
  {\bibfield  {journal} {\bibinfo  {journal} {Phys.~Fluids}\ }\textbf {\bibinfo
  {volume} {31}},\ \bibinfo {pages} {823} (\bibinfo {year} {1988})}\BibitemShut
  {NoStop}%
\bibitem [{\citenamefont {Pe\~{n}ano}\ and\ \citenamefont
  {Ganguli}(1999)}]{Penano:1999}%
  \BibitemOpen
  \bibfield  {author} {\bibinfo {author} {\bibfnamefont {J.~R.}\ \bibnamefont
  {Pe\~{n}ano}}\ and\ \bibinfo {author} {\bibfnamefont {G.}~\bibnamefont
  {Ganguli}},\ }\bibfield  {title} {\enquote {\bibinfo {title} {Ionospheric
  source for low-frequency broadband electromagnetic signatures},}\ }\href@noop
  {} {\bibfield  {journal} {\bibinfo  {journal} {Phys.~Rev.~Lett.}\ }\textbf
  {\bibinfo {volume} {83}},\ \bibinfo {pages} {1343} (\bibinfo {year}
  {1999})}\BibitemShut {NoStop}%
\bibitem [{\citenamefont {Pollock}\ \emph {et~al.}(1990)\citenamefont
  {Pollock}, \citenamefont {Chandler}, \citenamefont {Moore}, \citenamefont
  {Waite}, \citenamefont {Chappell},\ and\ \citenamefont
  {Gurnett}}]{Pollock:1990}%
  \BibitemOpen
  \bibfield  {author} {\bibinfo {author} {\bibfnamefont {C.~J.}\ \bibnamefont
  {Pollock}}, \bibinfo {author} {\bibfnamefont {M.~O.}\ \bibnamefont
  {Chandler}}, \bibinfo {author} {\bibfnamefont {T.~E.}\ \bibnamefont {Moore}},
  \bibinfo {author} {\bibfnamefont {J.~H.}\ \bibnamefont {Waite}}, \bibinfo
  {author} {\bibfnamefont {C.~R.}\ \bibnamefont {Chappell}}, \ and\ \bibinfo
  {author} {\bibfnamefont {D.~A.}\ \bibnamefont {Gurnett}},\ }\bibfield
  {title} {\enquote {\bibinfo {title} {A survey of upwelling ion event
  characteristics},}\ }\href {\doibase 10.1029/JA095iA11p18969} {\bibfield
  {journal} {\bibinfo  {journal} {J.~Geophys.~Res.}\ }\textbf {\bibinfo
  {volume} {95}},\ \bibinfo {pages} {18969--18980} (\bibinfo {year}
  {1990})}\BibitemShut {NoStop}%
\bibitem [{\citenamefont {{Koepke}}\ \emph {et~al.}(1994)\citenamefont
  {{Koepke}}, \citenamefont {{Amatucci}}, \citenamefont {{Carroll}},\ and\
  \citenamefont {{Sheridan}}}]{Koepke:1994}%
  \BibitemOpen
  \bibfield  {author} {\bibinfo {author} {\bibfnamefont {M.~E.}\ \bibnamefont
  {{Koepke}}}, \bibinfo {author} {\bibfnamefont {W.~E.}\ \bibnamefont
  {{Amatucci}}}, \bibinfo {author} {\bibfnamefont {J.~J.}\ \bibnamefont
  {{Carroll}}, \bibfnamefont {III}}, \ and\ \bibinfo {author} {\bibfnamefont
  {T.~E.}\ \bibnamefont {{Sheridan}}},\ }\bibfield  {title} {\enquote {\bibinfo
  {title} {{Experimental verification of the inhomogeneous energy-density
  driven instability}},}\ }\href {\doibase 10.1103/PhysRevLett.72.3355}
  {\bibfield  {journal} {\bibinfo  {journal} {Phys.~Rev.~Lett.}\ }\textbf
  {\bibinfo {volume} {72}},\ \bibinfo {pages} {3355--3358} (\bibinfo {year}
  {1994})}\BibitemShut {NoStop}%
\bibitem [{\citenamefont {Amatucci}\ \emph {et~al.}(1994)\citenamefont
  {Amatucci}, \citenamefont {Koepke}, \citenamefont {Carroll~III},\ and\
  \citenamefont {Sheridan}}]{Amatucci:1994}%
  \BibitemOpen
  \bibfield  {author} {\bibinfo {author} {\bibfnamefont {W.~E.}\ \bibnamefont
  {Amatucci}}, \bibinfo {author} {\bibfnamefont {M.~E.}\ \bibnamefont
  {Koepke}}, \bibinfo {author} {\bibfnamefont {J.~J.}\ \bibnamefont
  {Carroll~III}}, \ and\ \bibinfo {author} {\bibfnamefont {T.~E.}\ \bibnamefont
  {Sheridan}},\ }\bibfield  {title} {\enquote {\bibinfo {title} {Observation of
  ion-cyclotron turbulence at small values of magentic-field-aligned
  current},}\ }\href@noop {} {\bibfield  {journal} {\bibinfo  {journal}
  {Geophys.~Res.~Lett.}\ }\textbf {\bibinfo {volume} {21}},\ \bibinfo {pages}
  {1595} (\bibinfo {year} {1994})}\BibitemShut {NoStop}%
\bibitem [{\citenamefont {Amatucci}\ \emph {et~al.}(1996)\citenamefont
  {Amatucci}, \citenamefont {Walker}, \citenamefont {Ganguli}, \citenamefont
  {Antoniades}, \citenamefont {Duncan}, \citenamefont {Bowles}, \citenamefont
  {Gavrishchaka},\ and\ \citenamefont {Koepke}}]{Amatucci:1996}%
  \BibitemOpen
  \bibfield  {author} {\bibinfo {author} {\bibfnamefont {W.~E.}\ \bibnamefont
  {Amatucci}}, \bibinfo {author} {\bibfnamefont {D.~N.}\ \bibnamefont
  {Walker}}, \bibinfo {author} {\bibfnamefont {G.}~\bibnamefont {Ganguli}},
  \bibinfo {author} {\bibfnamefont {J.~A.}\ \bibnamefont {Antoniades}},
  \bibinfo {author} {\bibfnamefont {D.}~\bibnamefont {Duncan}}, \bibinfo
  {author} {\bibfnamefont {J.~H.}\ \bibnamefont {Bowles}}, \bibinfo {author}
  {\bibfnamefont {V.}~\bibnamefont {Gavrishchaka}}, \ and\ \bibinfo {author}
  {\bibfnamefont {M.~E.}\ \bibnamefont {Koepke}},\ }\bibfield  {title}
  {\enquote {\bibinfo {title} {Plasma response to strongly sheared flow},}\
  }\href@noop {} {\bibfield  {journal} {\bibinfo  {journal} {Phys.~Rev.~Lett.}\
  }\textbf {\bibinfo {volume} {77}},\ \bibinfo {pages} {1978} (\bibinfo {year}
  {1996})}\BibitemShut {NoStop}%
\bibitem [{\citenamefont {{Walker}}\ \emph {et~al.}(1997)\citenamefont
  {{Walker}}, \citenamefont {{Amatucci}}, \citenamefont {{Ganguli}},
  \citenamefont {{Antoniades}}, \citenamefont {{Bowles}}, \citenamefont
  {{Duncan}}, \citenamefont {{Gavrishchaka}},\ and\ \citenamefont
  {{Koepke}}}]{Walker:1997}%
  \BibitemOpen
  \bibfield  {author} {\bibinfo {author} {\bibfnamefont {D.~N.}\ \bibnamefont
  {{Walker}}}, \bibinfo {author} {\bibfnamefont {W.~E.}\ \bibnamefont
  {{Amatucci}}}, \bibinfo {author} {\bibfnamefont {G.}~\bibnamefont
  {{Ganguli}}}, \bibinfo {author} {\bibfnamefont {J.~A.}\ \bibnamefont
  {{Antoniades}}}, \bibinfo {author} {\bibfnamefont {J.~H.}\ \bibnamefont
  {{Bowles}}}, \bibinfo {author} {\bibfnamefont {D.}~\bibnamefont {{Duncan}}},
  \bibinfo {author} {\bibfnamefont {V.}~\bibnamefont {{Gavrishchaka}}}, \ and\
  \bibinfo {author} {\bibfnamefont {M.~E.}\ \bibnamefont {{Koepke}}},\
  }\bibfield  {title} {\enquote {\bibinfo {title} {{Perpendicular ion heating
  by velocity-shear-driven waves}},}\ }\href {\doibase 10.1029/97GL01102}
  {\bibfield  {journal} {\bibinfo  {journal} {Geophys.~Res.~Lett.}\ }\textbf
  {\bibinfo {volume} {24}},\ \bibinfo {pages} {1187--1190} (\bibinfo {year}
  {1997})}\BibitemShut {NoStop}%
\bibitem [{\citenamefont {Tejero}\ \emph {et~al.}(2011)\citenamefont {Tejero},
  \citenamefont {Amatucci}, \citenamefont {Ganguli}, \citenamefont {Cothran},
  \citenamefont {Crabtree},\ and\ \citenamefont {Thomas}}]{Tejero:2011}%
  \BibitemOpen
  \bibfield  {author} {\bibinfo {author} {\bibfnamefont {E.~M.}\ \bibnamefont
  {Tejero}}, \bibinfo {author} {\bibfnamefont {W.~E.}\ \bibnamefont
  {Amatucci}}, \bibinfo {author} {\bibfnamefont {G.}~\bibnamefont {Ganguli}},
  \bibinfo {author} {\bibfnamefont {C.~D.}\ \bibnamefont {Cothran}}, \bibinfo
  {author} {\bibfnamefont {C.}~\bibnamefont {Crabtree}}, \ and\ \bibinfo
  {author} {\bibfnamefont {E.}~\bibnamefont {Thomas}},\ }\bibfield  {title}
  {\enquote {\bibinfo {title} {Spontaneous electromagnetic emission from a
  strongly localized plasma flow},}\ }\href {\doibase
  10.1103/PhysRevLett.106.185001} {\bibfield  {journal} {\bibinfo  {journal}
  {Phys.~Rev.~Lett.}\ }\textbf {\bibinfo {volume} {106}},\ \bibinfo {pages}
  {185001} (\bibinfo {year} {2011})}\BibitemShut {NoStop}%
\bibitem [{\citenamefont {Ganguli}\ \emph {et~al.}(1994)\citenamefont
  {Ganguli}, \citenamefont {Keskinen}, \citenamefont {Romero}, \citenamefont
  {Heelis}, \citenamefont {Moore},\ and\ \citenamefont
  {Pollock}}]{Ganguli:1994a}%
  \BibitemOpen
  \bibfield  {author} {\bibinfo {author} {\bibfnamefont {G.}~\bibnamefont
  {Ganguli}}, \bibinfo {author} {\bibfnamefont {M.~J.}\ \bibnamefont
  {Keskinen}}, \bibinfo {author} {\bibfnamefont {H.}~\bibnamefont {Romero}},
  \bibinfo {author} {\bibfnamefont {R.}~\bibnamefont {Heelis}}, \bibinfo
  {author} {\bibfnamefont {T.}~\bibnamefont {Moore}}, \ and\ \bibinfo {author}
  {\bibfnamefont {C.}~\bibnamefont {Pollock}},\ }\bibfield  {title} {\enquote
  {\bibinfo {title} {Coupling of microprocesses and macroprocesses due to
  velocity shear: An application to the low-altitude ionosphere},}\ }\href@noop
  {} {\bibfield  {journal} {\bibinfo  {journal} {J.~Geophys.~Res.}\ }\textbf
  {\bibinfo {volume} {99}},\ \bibinfo {pages} {8873} (\bibinfo {year}
  {1994})}\BibitemShut {NoStop}%
\bibitem [{\citenamefont {Earle}, \citenamefont {Kelley},\ and\ \citenamefont
  {Ganguli}(1989)}]{Earle:1989}%
  \BibitemOpen
  \bibfield  {author} {\bibinfo {author} {\bibfnamefont {G.~D.}\ \bibnamefont
  {Earle}}, \bibinfo {author} {\bibfnamefont {M.~C.}\ \bibnamefont {Kelley}}, \
  and\ \bibinfo {author} {\bibfnamefont {G.}~\bibnamefont {Ganguli}},\
  }\bibfield  {title} {\enquote {\bibinfo {title} {Large velocity shears and
  associated electrostatic waves and turbulence in the auroral f region},}\
  }\href {\doibase 10.1029/JA094iA11p15321} {\bibfield  {journal} {\bibinfo
  {journal} {J.~Geophys.~Res.}\ }\textbf {\bibinfo {volume} {94}},\ \bibinfo
  {pages} {15321--15333} (\bibinfo {year} {1989})}\BibitemShut {NoStop}%
\bibitem [{\citenamefont {Bonnell}\ \emph {et~al.}(1996)\citenamefont
  {Bonnell}, \citenamefont {Kintner}, \citenamefont {Wahlund}, \citenamefont
  {Lynch},\ and\ \citenamefont {Arnoldy}}]{Bonnell:1996}%
  \BibitemOpen
  \bibfield  {author} {\bibinfo {author} {\bibfnamefont {J.}~\bibnamefont
  {Bonnell}}, \bibinfo {author} {\bibfnamefont {P.}~\bibnamefont {Kintner}},
  \bibinfo {author} {\bibfnamefont {J.-E.}\ \bibnamefont {Wahlund}}, \bibinfo
  {author} {\bibfnamefont {K.}~\bibnamefont {Lynch}}, \ and\ \bibinfo {author}
  {\bibfnamefont {R.}~\bibnamefont {Arnoldy}},\ }\bibfield  {title} {\enquote
  {\bibinfo {title} {Interferometric determination of broadband elf wave phase
  velocity within a region of transverse auroral ion acceleration},}\ }\href
  {\doibase 10.1029/96GL03238} {\bibfield  {journal} {\bibinfo  {journal}
  {Geophys.~Res.~Lett.}\ }\textbf {\bibinfo {volume} {23}},\ \bibinfo {pages}
  {3297--3300} (\bibinfo {year} {1996})}\BibitemShut {NoStop}%
\bibitem [{\citenamefont {Bonnell}(1997)}]{Bonnell:1997}%
  \BibitemOpen
  \bibfield  {author} {\bibinfo {author} {\bibfnamefont {J.}~\bibnamefont
  {Bonnell}},\ }\emph {\bibinfo {title} {Identification of broadband ELF waves
  observed during transverse ion acceleration in the auroral ionosphere}},\
  \href@noop {} {Ph.D. thesis},\ \bibinfo  {school} {Cornell University}
  (\bibinfo {year} {1997})\BibitemShut {NoStop}%
\bibitem [{\citenamefont {Lundberg}\ \emph {et~al.}(2012)\citenamefont
  {Lundberg}, \citenamefont {Kintner}, \citenamefont {Lynch},\ and\
  \citenamefont {Mella}}]{Lundberg:2012}%
  \BibitemOpen
  \bibfield  {author} {\bibinfo {author} {\bibfnamefont {E.~T.}\ \bibnamefont
  {Lundberg}}, \bibinfo {author} {\bibfnamefont {P.~M.}\ \bibnamefont
  {Kintner}}, \bibinfo {author} {\bibfnamefont {K.~A.}\ \bibnamefont {Lynch}},
  \ and\ \bibinfo {author} {\bibfnamefont {M.~R.}\ \bibnamefont {Mella}},\
  }\bibfield  {title} {\enquote {\bibinfo {title} {Multi-payload measurement of
  transverse velocity shears in the topside ionosphere},}\ }\href {\doibase
  10.1029/2011GL050018} {\bibfield  {journal} {\bibinfo  {journal}
  {Geophys.~Res.~Lett.}\ }\textbf {\bibinfo {volume} {39}},\ \bibinfo {pages}
  {n/a--n/a} (\bibinfo {year} {2012})},\ \bibinfo {note} {l01107}\BibitemShut
  {NoStop}%
\bibitem [{\citenamefont {Romero}, \citenamefont {Ganguli},\ and\ \citenamefont
  {Lee}(1992)}]{Romero:1992}%
  \BibitemOpen
  \bibfield  {author} {\bibinfo {author} {\bibfnamefont {H.}~\bibnamefont
  {Romero}}, \bibinfo {author} {\bibfnamefont {G.}~\bibnamefont {Ganguli}}, \
  and\ \bibinfo {author} {\bibfnamefont {Y.~C.}\ \bibnamefont {Lee}},\
  }\bibfield  {title} {\enquote {\bibinfo {title} {Ion acceleration and
  coherent structures generated by lower hybrid shear-driven instabilities},}\
  }\href {\doibase 10.1103/PhysRevLett.69.3503} {\bibfield  {journal} {\bibinfo
   {journal} {Phys.~Rev.~Lett.}\ }\textbf {\bibinfo {volume} {69}},\ \bibinfo
  {pages} {3503--3506} (\bibinfo {year} {1992})}\BibitemShut {NoStop}%
\bibitem [{\citenamefont {Romero}\ and\ \citenamefont
  {Ganguli}(1993)}]{Romero:1993}%
  \BibitemOpen
  \bibfield  {author} {\bibinfo {author} {\bibfnamefont {H.}~\bibnamefont
  {Romero}}\ and\ \bibinfo {author} {\bibfnamefont {G.}~\bibnamefont
  {Ganguli}},\ }\bibfield  {title} {\enquote {\bibinfo {title} {Nonlinear
  evolution of a strongly sheared cross‐field plasma flow},}\ }\href
  {\doibase 10.1063/1.860653} {\bibfield  {journal} {\bibinfo  {journal}
  {Phys.~Fluids B}\ }\textbf {\bibinfo {volume} {5}},\ \bibinfo {pages}
  {3163--3181} (\bibinfo {year} {1993})}\BibitemShut {NoStop}%
\bibitem [{\citenamefont {Romero}\ and\ \citenamefont
  {Ganguli}(1994)}]{Romero:1994}%
  \BibitemOpen
  \bibfield  {author} {\bibinfo {author} {\bibfnamefont {H.}~\bibnamefont
  {Romero}}\ and\ \bibinfo {author} {\bibfnamefont {G.}~\bibnamefont
  {Ganguli}},\ }\bibfield  {title} {\enquote {\bibinfo {title} {Relaxation of
  the stressed plasma sheet boundary layer},}\ }\href {\doibase
  10.1029/93GL03385} {\bibfield  {journal} {\bibinfo  {journal}
  {Geophys.~Res.~Lett.}\ }\textbf {\bibinfo {volume} {21}},\ \bibinfo {pages}
  {645--648} (\bibinfo {year} {1994})}\BibitemShut {NoStop}%
\bibitem [{\citenamefont {Forbes}(1981)}]{Forbes:1981}%
  \BibitemOpen
  \bibfield  {author} {\bibinfo {author} {\bibfnamefont {J.~M.}\ \bibnamefont
  {Forbes}},\ }\bibfield  {title} {\enquote {\bibinfo {title} {The equatorial
  electrojet},}\ }\href {\doibase 10.1029/RG019i003p00469} {\bibfield
  {journal} {\bibinfo  {journal} {Rev. of Geophys.}\ }\textbf {\bibinfo
  {volume} {19}},\ \bibinfo {pages} {469--504} (\bibinfo {year}
  {1981})}\BibitemShut {NoStop}%
\bibitem [{\citenamefont {Parks}\ \emph {et~al.}(1984)\citenamefont {Parks},
  \citenamefont {McCarthy}, \citenamefont {Fitzenreiter}, \citenamefont
  {Etcheto}, \citenamefont {Anderson}, \citenamefont {Anderson}, \citenamefont
  {Eastman}, \citenamefont {Frank}, \citenamefont {Gurnett}, \citenamefont
  {Huang}, \citenamefont {Lin}, \citenamefont {Lui}, \citenamefont {Ogilvie},
  \citenamefont {Pedersen}, \citenamefont {Reme},\ and\ \citenamefont
  {Williams}}]{Parks:1984}%
  \BibitemOpen
  \bibfield  {author} {\bibinfo {author} {\bibfnamefont {G.~K.}\ \bibnamefont
  {Parks}}, \bibinfo {author} {\bibfnamefont {M.}~\bibnamefont {McCarthy}},
  \bibinfo {author} {\bibfnamefont {R.~J.}\ \bibnamefont {Fitzenreiter}},
  \bibinfo {author} {\bibfnamefont {J.}~\bibnamefont {Etcheto}}, \bibinfo
  {author} {\bibfnamefont {K.~A.}\ \bibnamefont {Anderson}}, \bibinfo {author}
  {\bibfnamefont {R.~R.}\ \bibnamefont {Anderson}}, \bibinfo {author}
  {\bibfnamefont {T.~E.}\ \bibnamefont {Eastman}}, \bibinfo {author}
  {\bibfnamefont {L.~A.}\ \bibnamefont {Frank}}, \bibinfo {author}
  {\bibfnamefont {D.~A.}\ \bibnamefont {Gurnett}}, \bibinfo {author}
  {\bibfnamefont {C.}~\bibnamefont {Huang}}, \bibinfo {author} {\bibfnamefont
  {R.~P.}\ \bibnamefont {Lin}}, \bibinfo {author} {\bibfnamefont {A.~T.~Y.}\
  \bibnamefont {Lui}}, \bibinfo {author} {\bibfnamefont {K.~W.}\ \bibnamefont
  {Ogilvie}}, \bibinfo {author} {\bibfnamefont {A.}~\bibnamefont {Pedersen}},
  \bibinfo {author} {\bibfnamefont {H.}~\bibnamefont {Reme}}, \ and\ \bibinfo
  {author} {\bibfnamefont {D.~J.}\ \bibnamefont {Williams}},\ }\bibfield
  {title} {\enquote {\bibinfo {title} {Particle and field characteristics of
  the high-latitude plasma sheet boundary layer},}\ }\href {\doibase
  10.1029/JA089iA10p08885} {\bibfield  {journal} {\bibinfo  {journal}
  {J.~Geophys.~Res.}\ }\textbf {\bibinfo {volume} {89}},\ \bibinfo {pages}
  {8885--8906} (\bibinfo {year} {1984})}\BibitemShut {NoStop}%
\bibitem [{\citenamefont {Amatucci}\ \emph {et~al.}(2003)\citenamefont
  {Amatucci}, \citenamefont {Ganguli}, \citenamefont {Walker}, \citenamefont
  {Gatling}, \citenamefont {Balkey},\ and\ \citenamefont
  {McCulloch}}]{Amatucci:2003}%
  \BibitemOpen
  \bibfield  {author} {\bibinfo {author} {\bibfnamefont {W.~E.}\ \bibnamefont
  {Amatucci}}, \bibinfo {author} {\bibfnamefont {G.}~\bibnamefont {Ganguli}},
  \bibinfo {author} {\bibfnamefont {D.~N.}\ \bibnamefont {Walker}}, \bibinfo
  {author} {\bibfnamefont {G.}~\bibnamefont {Gatling}}, \bibinfo {author}
  {\bibfnamefont {M.}~\bibnamefont {Balkey}}, \ and\ \bibinfo {author}
  {\bibfnamefont {T.}~\bibnamefont {McCulloch}},\ }\bibfield  {title} {\enquote
  {\bibinfo {title} {Laboratory investigation of boundary layer processes due
  to strong spatial inhomogeneity},}\ }\href {\doibase 10.1063/1.1562631}
  {\bibfield  {journal} {\bibinfo  {journal} {Phys.~Plasmas}\ }\textbf
  {\bibinfo {volume} {10}},\ \bibinfo {pages} {1963--1970} (\bibinfo {year}
  {2003})}\BibitemShut {NoStop}%
\bibitem [{\citenamefont {Enloe}\ \emph {et~al.}(2017)\citenamefont {Enloe},
  \citenamefont {Tejero}, \citenamefont {Crabtree}, \citenamefont {Ganguli},\
  and\ \citenamefont {Amatucci}}]{Enloe:2017}%
  \BibitemOpen
  \bibfield  {author} {\bibinfo {author} {\bibfnamefont {C.~L.}\ \bibnamefont
  {Enloe}}, \bibinfo {author} {\bibfnamefont {E.~M.}\ \bibnamefont {Tejero}},
  \bibinfo {author} {\bibfnamefont {C.}~\bibnamefont {Crabtree}}, \bibinfo
  {author} {\bibfnamefont {G.}~\bibnamefont {Ganguli}}, \ and\ \bibinfo
  {author} {\bibfnamefont {W.~E.}\ \bibnamefont {Amatucci}},\ }\bibfield
  {title} {\enquote {\bibinfo {title} {Electromagnetic fluctuations in the
  intermediate frequency range originating from a plasma boundary layer},}\
  }\href {\doibase 10.1063/1.4981923} {\bibfield  {journal} {\bibinfo
  {journal} {Phys.~Plasmas}\ }\textbf {\bibinfo {volume} {24}},\ \bibinfo
  {pages} {052107} (\bibinfo {year} {2017})}\BibitemShut {NoStop}%
\bibitem [{\citenamefont {{Schmidt}}\ and\ \citenamefont
  {{Marsch}}(1995)}]{Schmidt:1995}%
  \BibitemOpen
  \bibfield  {author} {\bibinfo {author} {\bibfnamefont {J.~M.}\ \bibnamefont
  {{Schmidt}}}\ and\ \bibinfo {author} {\bibfnamefont {E.}~\bibnamefont
  {{Marsch}}},\ }\bibfield  {title} {\enquote {\bibinfo {title} {{Spatial
  transport and spectral transfer of solar wind turbulence composed of Alfven
  waves and convective structures 1: The theoretical model}},}\ }\href
  {\doibase 10.1007/s00585-995-0459-3} {\bibfield  {journal} {\bibinfo
  {journal} {Ann.~Geophys.}\ }\textbf {\bibinfo {volume} {13}},\ \bibinfo
  {pages} {459--474} (\bibinfo {year} {1995})}\BibitemShut {NoStop}%
\bibitem [{\citenamefont {{Del Zanna}}, \citenamefont {{Velli}},\ and\
  \citenamefont {{Londrillo}}(2001)}]{DelZanna:2001}%
  \BibitemOpen
  \bibfield  {author} {\bibinfo {author} {\bibfnamefont {L.}~\bibnamefont {{Del
  Zanna}}}, \bibinfo {author} {\bibfnamefont {M.}~\bibnamefont {{Velli}}}, \
  and\ \bibinfo {author} {\bibfnamefont {P.}~\bibnamefont {{Londrillo}}},\
  }\bibfield  {title} {\enquote {\bibinfo {title} {{Parametric decay of
  circularly polarized Alfv{\'e}n waves: Multidimensional simulations in
  periodic and open domains}},}\ }\href {\doibase 10.1051/0004-6361:20000455}
  {\bibfield  {journal} {\bibinfo  {journal} {Astron.~Astrophys.}\ }\textbf
  {\bibinfo {volume} {367}},\ \bibinfo {pages} {705--718} (\bibinfo {year}
  {2001})}\BibitemShut {NoStop}%
\bibitem [{\citenamefont {{Dorfman}}\ and\ \citenamefont
  {{Carter}}(2013)}]{Dorfman:2013b}%
  \BibitemOpen
  \bibfield  {author} {\bibinfo {author} {\bibfnamefont {S.}~\bibnamefont
  {{Dorfman}}}\ and\ \bibinfo {author} {\bibfnamefont {T.~A.}\ \bibnamefont
  {{Carter}}},\ }\bibfield  {title} {\enquote {\bibinfo {title} {{Nonlinear
  Excitation of Acoustic Modes by Large-Amplitude Alfv{\'e}n Waves in a
  Laboratory Plasma}},}\ }\href {\doibase 10.1103/PhysRevLett.110.195001}
  {\bibfield  {journal} {\bibinfo  {journal} {Phys.~Rev.~Lett.}\ }\textbf
  {\bibinfo {volume} {110}},\ \bibinfo {eid} {195001} (\bibinfo {year}
  {2013})},\ \Eprint {http://arxiv.org/abs/1304.3379} {arXiv:1304.3379
  [physics.plasm-ph]} \BibitemShut {NoStop}%
\bibitem [{\citenamefont {{Dorfman}}\ and\ \citenamefont
  {{Carter}}(2015)}]{Dorfman:2015}%
  \BibitemOpen
  \bibfield  {author} {\bibinfo {author} {\bibfnamefont {S.}~\bibnamefont
  {{Dorfman}}}\ and\ \bibinfo {author} {\bibfnamefont {T.~A.}\ \bibnamefont
  {{Carter}}},\ }\bibfield  {title} {\enquote {\bibinfo {title} {{Non-linear
  Alfv{\'e}n wave interaction leading to resonant excitation of an acoustic
  mode in the laboratory}},}\ }\href {\doibase 10.1063/1.4919275} {\bibfield
  {journal} {\bibinfo  {journal} {Phys.~Plasmas}\ }\textbf {\bibinfo {volume}
  {22}},\ \bibinfo {eid} {055706} (\bibinfo {year} {2015})}\BibitemShut
  {NoStop}%
\bibitem [{\citenamefont {{Dorfman}}\ and\ \citenamefont
  {{Carter}}(2016)}]{Dorfman:2016}%
  \BibitemOpen
  \bibfield  {author} {\bibinfo {author} {\bibfnamefont {S.}~\bibnamefont
  {{Dorfman}}}\ and\ \bibinfo {author} {\bibfnamefont {T.~A.}\ \bibnamefont
  {{Carter}}},\ }\bibfield  {title} {\enquote {\bibinfo {title} {{Observation
  of an Alfv{\'e}n Wave Parametric Instability in a Laboratory Plasma}},}\
  }\href {\doibase 10.1103/PhysRevLett.116.195002} {\bibfield  {journal}
  {\bibinfo  {journal} {Phys.~Rev.~Lett.}\ }\textbf {\bibinfo {volume} {116}},\
  \bibinfo {eid} {195002} (\bibinfo {year} {2016})},\ \Eprint
  {http://arxiv.org/abs/1606.05055} {arXiv:1606.05055 [physics.plasm-ph]}
  \BibitemShut {NoStop}%
\bibitem [{\citenamefont {{Parker}}(1966)}]{Parker:1966}%
  \BibitemOpen
  \bibfield  {author} {\bibinfo {author} {\bibfnamefont {E.~N.}\ \bibnamefont
  {{Parker}}},\ }\bibfield  {title} {\enquote {\bibinfo {title} {{The Dynamical
  State of the Interstellar Gas and Field}},}\ }\href {\doibase 10.1086/148828}
  {\bibfield  {journal} {\bibinfo  {journal} {Astrophys.~J.}\ }\textbf
  {\bibinfo {volume} {145}},\ \bibinfo {pages} {811} (\bibinfo {year}
  {1966})}\BibitemShut {NoStop}%
\bibitem [{\citenamefont {{Pueschel}}\ \emph {et~al.}(2015)\citenamefont
  {{Pueschel}}, \citenamefont {{Terry}}, \citenamefont {{Told}},\ and\
  \citenamefont {{Jenko}}}]{Pueschel:2015}%
  \BibitemOpen
  \bibfield  {author} {\bibinfo {author} {\bibfnamefont {M.~J.}\ \bibnamefont
  {{Pueschel}}}, \bibinfo {author} {\bibfnamefont {P.~W.}\ \bibnamefont
  {{Terry}}}, \bibinfo {author} {\bibfnamefont {D.}~\bibnamefont {{Told}}}, \
  and\ \bibinfo {author} {\bibfnamefont {F.}~\bibnamefont {{Jenko}}},\
  }\bibfield  {title} {\enquote {\bibinfo {title} {{Enhanced magnetic
  reconnection in the presence of pressure gradients}},}\ }\href {\doibase
  10.1063/1.4922064} {\bibfield  {journal} {\bibinfo  {journal} {Physics of
  Plasmas}\ }\textbf {\bibinfo {volume} {22}},\ \bibinfo {eid} {062105}
  (\bibinfo {year} {2015})}\BibitemShut {NoStop}%
\bibitem [{\citenamefont {{Pueschel}}\ \emph {et~al.}(2017)\citenamefont
  {{Pueschel}}, \citenamefont {{Rossi}}, \citenamefont {{Told}}, \citenamefont
  {{Terry}}, \citenamefont {{Jenko}},\ and\ \citenamefont
  {{Carter}}}]{Pueschel:2017}%
  \BibitemOpen
  \bibfield  {author} {\bibinfo {author} {\bibfnamefont {M.~J.}\ \bibnamefont
  {{Pueschel}}}, \bibinfo {author} {\bibfnamefont {G.}~\bibnamefont {{Rossi}}},
  \bibinfo {author} {\bibfnamefont {D.}~\bibnamefont {{Told}}}, \bibinfo
  {author} {\bibfnamefont {P.~W.}\ \bibnamefont {{Terry}}}, \bibinfo {author}
  {\bibfnamefont {F.}~\bibnamefont {{Jenko}}}, \ and\ \bibinfo {author}
  {\bibfnamefont {T.~A.}\ \bibnamefont {{Carter}}},\ }\bibfield  {title}
  {\enquote {\bibinfo {title} {{A basic plasma test for gyrokinetics: GDC
  turbulence in LAPD}},}\ }\href {\doibase 10.1088/1361-6587/aa52e6} {\bibfield
   {journal} {\bibinfo  {journal} {Plasma Phys.~Con.~Fus.}\ }\textbf {\bibinfo
  {volume} {59}},\ \bibinfo {eid} {024006} (\bibinfo {year}
  {2017})}\BibitemShut {NoStop}%
\bibitem [{\citenamefont {{Moser}}\ and\ \citenamefont
  {{Bellan}}(2012)}]{Moser:2012}%
  \BibitemOpen
  \bibfield  {author} {\bibinfo {author} {\bibfnamefont {A.~L.}\ \bibnamefont
  {{Moser}}}\ and\ \bibinfo {author} {\bibfnamefont {P.~M.}\ \bibnamefont
  {{Bellan}}},\ }\bibfield  {title} {\enquote {\bibinfo {title} {{Magnetic
  reconnection from a multiscale instability cascade}},}\ }\href {\doibase
  10.1038/nature10827} {\bibfield  {journal} {\bibinfo  {journal} {Nature}\
  }\textbf {\bibinfo {volume} {482}},\ \bibinfo {pages} {379--381} (\bibinfo
  {year} {2012})}\BibitemShut {NoStop}%
\bibitem [{\citenamefont {{Bullard}}\ and\ \citenamefont
  {{Gellman}}(1954)}]{Bullard:1954}%
  \BibitemOpen
  \bibfield  {author} {\bibinfo {author} {\bibfnamefont {E.}~\bibnamefont
  {{Bullard}}}\ and\ \bibinfo {author} {\bibfnamefont {H.}~\bibnamefont
  {{Gellman}}},\ }\bibfield  {title} {\enquote {\bibinfo {title} {{Homogeneous
  Dynamos and Terrestrial Magnetism}},}\ }\href {\doibase
  10.1098/rsta.1954.0018} {\bibfield  {journal} {\bibinfo  {journal}
  {Philosophical Transactions of the Royal Society of London Series A}\
  }\textbf {\bibinfo {volume} {247}},\ \bibinfo {pages} {213--278} (\bibinfo
  {year} {1954})}\BibitemShut {NoStop}%
\bibitem [{\citenamefont {{Kuang}}\ and\ \citenamefont
  {{Bloxham}}(1997)}]{Kuang:1997}%
  \BibitemOpen
  \bibfield  {author} {\bibinfo {author} {\bibfnamefont {W.}~\bibnamefont
  {{Kuang}}}\ and\ \bibinfo {author} {\bibfnamefont {J.}~\bibnamefont
  {{Bloxham}}},\ }\bibfield  {title} {\enquote {\bibinfo {title} {{An
  Earth-like numerical dynamo model}},}\ }\href {\doibase 10.1038/38712}
  {\bibfield  {journal} {\bibinfo  {journal} {Nature}\ }\textbf {\bibinfo
  {volume} {389}},\ \bibinfo {pages} {371--374} (\bibinfo {year}
  {1997})}\BibitemShut {NoStop}%
\bibitem [{\citenamefont {{Gailitis}}\ \emph {et~al.}(2001)\citenamefont
  {{Gailitis}}, \citenamefont {{Lielausis}}, \citenamefont {{Platacis}},
  \citenamefont {{Dement'ev}}, \citenamefont {{Cifersons}}, \citenamefont
  {{Gerbeth}}, \citenamefont {{Gundrum}}, \citenamefont {{Stefani}},
  \citenamefont {{Christen}},\ and\ \citenamefont {{Will}}}]{Gailitis:2001}%
  \BibitemOpen
  \bibfield  {author} {\bibinfo {author} {\bibfnamefont {A.}~\bibnamefont
  {{Gailitis}}}, \bibinfo {author} {\bibfnamefont {O.}~\bibnamefont
  {{Lielausis}}}, \bibinfo {author} {\bibfnamefont {E.}~\bibnamefont
  {{Platacis}}}, \bibinfo {author} {\bibfnamefont {S.}~\bibnamefont
  {{Dement'ev}}}, \bibinfo {author} {\bibfnamefont {A.}~\bibnamefont
  {{Cifersons}}}, \bibinfo {author} {\bibfnamefont {G.}~\bibnamefont
  {{Gerbeth}}}, \bibinfo {author} {\bibfnamefont {T.}~\bibnamefont
  {{Gundrum}}}, \bibinfo {author} {\bibfnamefont {F.}~\bibnamefont
  {{Stefani}}}, \bibinfo {author} {\bibfnamefont {M.}~\bibnamefont
  {{Christen}}}, \ and\ \bibinfo {author} {\bibfnamefont {G.}~\bibnamefont
  {{Will}}},\ }\bibfield  {title} {\enquote {\bibinfo {title} {{Magnetic Field
  Saturation in the Riga Dynamo Experiment}},}\ }\href {\doibase
  10.1103/PhysRevLett.86.3024} {\bibfield  {journal} {\bibinfo  {journal}
  {Phys.~Rev.~Lett.}\ }\textbf {\bibinfo {volume} {86}},\ \bibinfo {pages}
  {3024--3027} (\bibinfo {year} {2001})},\ \Eprint
  {http://arxiv.org/abs/physics/0010047} {physics/0010047} \BibitemShut
  {NoStop}%
\bibitem [{\citenamefont {{Stieglitz}}\ and\ \citenamefont
  {{M{\"u}ller}}(2001)}]{Stieglitz:2001}%
  \BibitemOpen
  \bibfield  {author} {\bibinfo {author} {\bibfnamefont {R.}~\bibnamefont
  {{Stieglitz}}}\ and\ \bibinfo {author} {\bibfnamefont {U.}~\bibnamefont
  {{M{\"u}ller}}},\ }\bibfield  {title} {\enquote {\bibinfo {title}
  {{Experimental demonstration of a homogeneous two-scale dynamo}},}\ }\href
  {\doibase 10.1063/1.1331315} {\bibfield  {journal} {\bibinfo  {journal}
  {Physics of Fluids}\ }\textbf {\bibinfo {volume} {13}},\ \bibinfo {pages}
  {561--564} (\bibinfo {year} {2001})}\BibitemShut {NoStop}%
\bibitem [{\citenamefont {{Reighard}}\ and\ \citenamefont
  {{Brown}}(2001)}]{Reighard:2001}%
  \BibitemOpen
  \bibfield  {author} {\bibinfo {author} {\bibfnamefont {A.~B.}\ \bibnamefont
  {{Reighard}}}\ and\ \bibinfo {author} {\bibfnamefont {M.~R.}\ \bibnamefont
  {{Brown}}},\ }\bibfield  {title} {\enquote {\bibinfo {title} {{Turbulent
  Conductivity Measurements in a Spherical Liquid Sodium Flow}},}\ }\href
  {\doibase 10.1103/PhysRevLett.86.2794} {\bibfield  {journal} {\bibinfo
  {journal} {Phys.~Rev.~Lett.}\ }\textbf {\bibinfo {volume} {86}},\ \bibinfo
  {pages} {2794--2797} (\bibinfo {year} {2001})}\BibitemShut {NoStop}%
\bibitem [{\citenamefont {{Noguchi}}\ \emph {et~al.}(2002)\citenamefont
  {{Noguchi}}, \citenamefont {{Pariev}}, \citenamefont {{Colgate}},
  \citenamefont {{Beckley}},\ and\ \citenamefont {{Nordhaus}}}]{Noguchi:2002}%
  \BibitemOpen
  \bibfield  {author} {\bibinfo {author} {\bibfnamefont {K.}~\bibnamefont
  {{Noguchi}}}, \bibinfo {author} {\bibfnamefont {V.~I.}\ \bibnamefont
  {{Pariev}}}, \bibinfo {author} {\bibfnamefont {S.~A.}\ \bibnamefont
  {{Colgate}}}, \bibinfo {author} {\bibfnamefont {H.~F.}\ \bibnamefont
  {{Beckley}}}, \ and\ \bibinfo {author} {\bibfnamefont {J.}~\bibnamefont
  {{Nordhaus}}},\ }\bibfield  {title} {\enquote {\bibinfo {title}
  {{Magnetorotational Instability in Liquid Metal Couette Flow}},}\ }\href
  {\doibase 10.1086/341502} {\bibfield  {journal} {\bibinfo  {journal}
  {Astrophys.~J.}\ }\textbf {\bibinfo {volume} {575}},\ \bibinfo {pages}
  {1151--1162} (\bibinfo {year} {2002})},\ \Eprint
  {http://arxiv.org/abs/astro-ph/0204299} {astro-ph/0204299} \BibitemShut
  {NoStop}%
\bibitem [{\citenamefont {{Nornberg}}\ \emph {et~al.}(2006)\citenamefont
  {{Nornberg}}, \citenamefont {{Spence}}, \citenamefont {{Kendrick}},
  \citenamefont {{Jacobson}},\ and\ \citenamefont {{Forest}}}]{Nornberg:2006}%
  \BibitemOpen
  \bibfield  {author} {\bibinfo {author} {\bibfnamefont {M.~D.}\ \bibnamefont
  {{Nornberg}}}, \bibinfo {author} {\bibfnamefont {E.~J.}\ \bibnamefont
  {{Spence}}}, \bibinfo {author} {\bibfnamefont {R.~D.}\ \bibnamefont
  {{Kendrick}}}, \bibinfo {author} {\bibfnamefont {C.~M.}\ \bibnamefont
  {{Jacobson}}}, \ and\ \bibinfo {author} {\bibfnamefont {C.~B.}\ \bibnamefont
  {{Forest}}},\ }\bibfield  {title} {\enquote {\bibinfo {title} {{Intermittent
  Magnetic Field Excitation by a Turbulent Flow of Liquid Sodium}},}\ }\href
  {\doibase 10.1103/PhysRevLett.97.044503} {\bibfield  {journal} {\bibinfo
  {journal} {Physical Review Letters}\ }\textbf {\bibinfo {volume} {97}},\
  \bibinfo {eid} {044503} (\bibinfo {year} {2006})},\ \Eprint
  {http://arxiv.org/abs/physics/0606239} {physics/0606239} \BibitemShut
  {NoStop}%
\bibitem [{\citenamefont {{Monchaux}}\ \emph {et~al.}(2007)\citenamefont
  {{Monchaux}}, \citenamefont {{Berhanu}}, \citenamefont {{Bourgoin}},
  \citenamefont {{Moulin}}, \citenamefont {{Odier}}, \citenamefont {{Pinton}},
  \citenamefont {{Volk}}, \citenamefont {{Fauve}}, \citenamefont {{Mordant}},
  \citenamefont {{P{\'e}tr{\'e}lis}}, \citenamefont {{Chiffaudel}},
  \citenamefont {{Daviaud}}, \citenamefont {{Dubrulle}}, \citenamefont
  {{Gasquet}}, \citenamefont {{Mari{\'e}}},\ and\ \citenamefont
  {{Ravelet}}}]{Monchaux:2007}%
  \BibitemOpen
  \bibfield  {author} {\bibinfo {author} {\bibfnamefont {R.}~\bibnamefont
  {{Monchaux}}}, \bibinfo {author} {\bibfnamefont {M.}~\bibnamefont
  {{Berhanu}}}, \bibinfo {author} {\bibfnamefont {M.}~\bibnamefont
  {{Bourgoin}}}, \bibinfo {author} {\bibfnamefont {M.}~\bibnamefont
  {{Moulin}}}, \bibinfo {author} {\bibfnamefont {P.}~\bibnamefont {{Odier}}},
  \bibinfo {author} {\bibfnamefont {J.-F.}\ \bibnamefont {{Pinton}}}, \bibinfo
  {author} {\bibfnamefont {R.}~\bibnamefont {{Volk}}}, \bibinfo {author}
  {\bibfnamefont {S.}~\bibnamefont {{Fauve}}}, \bibinfo {author} {\bibfnamefont
  {N.}~\bibnamefont {{Mordant}}}, \bibinfo {author} {\bibfnamefont
  {F.}~\bibnamefont {{P{\'e}tr{\'e}lis}}}, \bibinfo {author} {\bibfnamefont
  {A.}~\bibnamefont {{Chiffaudel}}}, \bibinfo {author} {\bibfnamefont
  {F.}~\bibnamefont {{Daviaud}}}, \bibinfo {author} {\bibfnamefont
  {B.}~\bibnamefont {{Dubrulle}}}, \bibinfo {author} {\bibfnamefont
  {C.}~\bibnamefont {{Gasquet}}}, \bibinfo {author} {\bibfnamefont
  {L.}~\bibnamefont {{Mari{\'e}}}}, \ and\ \bibinfo {author} {\bibfnamefont
  {F.}~\bibnamefont {{Ravelet}}},\ }\bibfield  {title} {\enquote {\bibinfo
  {title} {{Generation of a Magnetic Field by Dynamo Action in a Turbulent Flow
  of Liquid Sodium}},}\ }\href {\doibase 10.1103/PhysRevLett.98.044502}
  {\bibfield  {journal} {\bibinfo  {journal} {Physical Review Letters}\
  }\textbf {\bibinfo {volume} {98}},\ \bibinfo {eid} {044502} (\bibinfo {year}
  {2007})},\ \Eprint {http://arxiv.org/abs/physics/0701075} {physics/0701075}
  \BibitemShut {NoStop}%
\bibitem [{\citenamefont {{Zimmerman}}\ \emph {et~al.}(2014)\citenamefont
  {{Zimmerman}}, \citenamefont {{Triana}}, \citenamefont {{Nataf}},\ and\
  \citenamefont {{Lathrop}}}]{Zimmerman:2014}%
  \BibitemOpen
  \bibfield  {author} {\bibinfo {author} {\bibfnamefont {D.~S.}\ \bibnamefont
  {{Zimmerman}}}, \bibinfo {author} {\bibfnamefont {S.~A.}\ \bibnamefont
  {{Triana}}}, \bibinfo {author} {\bibfnamefont {H.-C.}\ \bibnamefont
  {{Nataf}}}, \ and\ \bibinfo {author} {\bibfnamefont {D.~P.}\ \bibnamefont
  {{Lathrop}}},\ }\bibfield  {title} {\enquote {\bibinfo {title} {{A turbulent,
  high magnetic Reynolds number experimental model of Earth's core}},}\ }\href
  {\doibase 10.1002/2013JB010733} {\bibfield  {journal} {\bibinfo  {journal}
  {Journal of Geophysical Research (Solid Earth)}\ }\textbf {\bibinfo {volume}
  {119}},\ \bibinfo {pages} {4538--4557} (\bibinfo {year} {2014})}\BibitemShut
  {NoStop}%
\bibitem [{\citenamefont {{Weisberg}}\ \emph {et~al.}(2017)\citenamefont
  {{Weisberg}}, \citenamefont {{Peterson}}, \citenamefont {{Milhone}},
  \citenamefont {{Endrizzi}}, \citenamefont {{Cooper}}, \citenamefont
  {{D{\'e}sangles}}, \citenamefont {{Khalzov}}, \citenamefont {{Siller}},\ and\
  \citenamefont {{Forest}}}]{Weisberg:2017}%
  \BibitemOpen
  \bibfield  {author} {\bibinfo {author} {\bibfnamefont {D.~B.}\ \bibnamefont
  {{Weisberg}}}, \bibinfo {author} {\bibfnamefont {E.}~\bibnamefont
  {{Peterson}}}, \bibinfo {author} {\bibfnamefont {J.}~\bibnamefont
  {{Milhone}}}, \bibinfo {author} {\bibfnamefont {D.}~\bibnamefont
  {{Endrizzi}}}, \bibinfo {author} {\bibfnamefont {C.}~\bibnamefont
  {{Cooper}}}, \bibinfo {author} {\bibfnamefont {V.}~\bibnamefont
  {{D{\'e}sangles}}}, \bibinfo {author} {\bibfnamefont {I.}~\bibnamefont
  {{Khalzov}}}, \bibinfo {author} {\bibfnamefont {R.}~\bibnamefont {{Siller}}},
  \ and\ \bibinfo {author} {\bibfnamefont {C.~B.}\ \bibnamefont {{Forest}}},\
  }\bibfield  {title} {\enquote {\bibinfo {title} {{Driving large magnetic
  Reynolds number flow in highly ionized, unmagnetized plasmas}},}\ }\href
  {\doibase 10.1063/1.4978889} {\bibfield  {journal} {\bibinfo  {journal}
  {Phys.~Plasmas}\ }\textbf {\bibinfo {volume} {24}},\ \bibinfo {eid} {056502}
  (\bibinfo {year} {2017})}\BibitemShut {NoStop}%
\bibitem [{\citenamefont {{Weibel}}(1959)}]{Weibel:1959}%
  \BibitemOpen
  \bibfield  {author} {\bibinfo {author} {\bibfnamefont {E.~S.}\ \bibnamefont
  {{Weibel}}},\ }\bibfield  {title} {\enquote {\bibinfo {title} {{Spontaneously
  Growing Transverse Waves in a Plasma Due to an Anisotropic Velocity
  Distribution}},}\ }\href {\doibase 10.1103/PhysRevLett.2.83} {\bibfield
  {journal} {\bibinfo  {journal} {Phys.~Rev.~Lett.}\ }\textbf {\bibinfo
  {volume} {2}},\ \bibinfo {pages} {83--84} (\bibinfo {year}
  {1959})}\BibitemShut {NoStop}%
\bibitem [{\citenamefont {{Huntington}}\ \emph {et~al.}(2015)\citenamefont
  {{Huntington}}, \citenamefont {{Fiuza}}, \citenamefont {{Ross}},
  \citenamefont {{Zylstra}}, \citenamefont {{Drake}}, \citenamefont {{Froula}},
  \citenamefont {{Gregori}}, \citenamefont {{Kugland}}, \citenamefont
  {{Kuranz}}, \citenamefont {{Levy}}, \citenamefont {{Li}}, \citenamefont
  {{Meinecke}}, \citenamefont {{Morita}}, \citenamefont {{Petrasso}},
  \citenamefont {{Plechaty}}, \citenamefont {{Remington}}, \citenamefont
  {{Ryutov}}, \citenamefont {{Sakawa}}, \citenamefont {{Spitkovsky}},
  \citenamefont {{Takabe}},\ and\ \citenamefont {{Park}}}]{Huntington:2015}%
  \BibitemOpen
  \bibfield  {author} {\bibinfo {author} {\bibfnamefont {C.~M.}\ \bibnamefont
  {{Huntington}}}, \bibinfo {author} {\bibfnamefont {F.}~\bibnamefont
  {{Fiuza}}}, \bibinfo {author} {\bibfnamefont {J.~S.}\ \bibnamefont {{Ross}}},
  \bibinfo {author} {\bibfnamefont {A.~B.}\ \bibnamefont {{Zylstra}}}, \bibinfo
  {author} {\bibfnamefont {R.~P.}\ \bibnamefont {{Drake}}}, \bibinfo {author}
  {\bibfnamefont {D.~H.}\ \bibnamefont {{Froula}}}, \bibinfo {author}
  {\bibfnamefont {G.}~\bibnamefont {{Gregori}}}, \bibinfo {author}
  {\bibfnamefont {N.~L.}\ \bibnamefont {{Kugland}}}, \bibinfo {author}
  {\bibfnamefont {C.~C.}\ \bibnamefont {{Kuranz}}}, \bibinfo {author}
  {\bibfnamefont {M.~C.}\ \bibnamefont {{Levy}}}, \bibinfo {author}
  {\bibfnamefont {C.~K.}\ \bibnamefont {{Li}}}, \bibinfo {author}
  {\bibfnamefont {J.}~\bibnamefont {{Meinecke}}}, \bibinfo {author}
  {\bibfnamefont {T.}~\bibnamefont {{Morita}}}, \bibinfo {author}
  {\bibfnamefont {R.}~\bibnamefont {{Petrasso}}}, \bibinfo {author}
  {\bibfnamefont {C.}~\bibnamefont {{Plechaty}}}, \bibinfo {author}
  {\bibfnamefont {B.~A.}\ \bibnamefont {{Remington}}}, \bibinfo {author}
  {\bibfnamefont {D.~D.}\ \bibnamefont {{Ryutov}}}, \bibinfo {author}
  {\bibfnamefont {Y.}~\bibnamefont {{Sakawa}}}, \bibinfo {author}
  {\bibfnamefont {A.}~\bibnamefont {{Spitkovsky}}}, \bibinfo {author}
  {\bibfnamefont {H.}~\bibnamefont {{Takabe}}}, \ and\ \bibinfo {author}
  {\bibfnamefont {H.-S.}\ \bibnamefont {{Park}}},\ }\bibfield  {title}
  {\enquote {\bibinfo {title} {{Observation of magnetic field generation via
  the Weibel instability in interpenetrating plasma flows}},}\ }\href {\doibase
  10.1038/nphys3178} {\bibfield  {journal} {\bibinfo  {journal} {Nature Phys.}\
  }\textbf {\bibinfo {volume} {11}},\ \bibinfo {pages} {173--176} (\bibinfo
  {year} {2015})},\ \Eprint {http://arxiv.org/abs/1310.3337} {arXiv:1310.3337
  [astro-ph.HE]} \BibitemShut {NoStop}%
\bibitem [{\citenamefont {Bellan}(2000)}]{Bellan:2000}%
  \BibitemOpen
  \bibfield  {author} {\bibinfo {author} {\bibfnamefont {P.~M.}\ \bibnamefont
  {Bellan}},\ }\href@noop {} {\emph {\bibinfo {title} {Spheromaks}}}\ (\bibinfo
   {publisher} {Imperial College Press},\ \bibinfo {address} {London},\
  \bibinfo {year} {2000})\BibitemShut {NoStop}%
\bibitem [{\citenamefont {Hsu}\ and\ \citenamefont {Bellan}(2003)}]{Hsu:2003}%
  \BibitemOpen
  \bibfield  {author} {\bibinfo {author} {\bibfnamefont {S.~C.}\ \bibnamefont
  {Hsu}}\ and\ \bibinfo {author} {\bibfnamefont {P.~M.}\ \bibnamefont
  {Bellan}},\ }\bibfield  {title} {\enquote {\bibinfo {title} {Experimental
  identification of the kink instability as a poloidal flux amplification
  mechanism for coaxial gun spheromak formation},}\ }\href@noop {} {\bibfield
  {journal} {\bibinfo  {journal} {Phys.~Rev.~Lett.}\ }\textbf {\bibinfo
  {volume} {90}},\ \bibinfo {pages} {215002} (\bibinfo {year}
  {2003})}\BibitemShut {NoStop}%
\bibitem [{\citenamefont {Hsu}\ and\ \citenamefont {Bellan}(2005)}]{Hsu:2005}%
  \BibitemOpen
  \bibfield  {author} {\bibinfo {author} {\bibfnamefont {S.~C.}\ \bibnamefont
  {Hsu}}\ and\ \bibinfo {author} {\bibfnamefont {P.~M.}\ \bibnamefont
  {Bellan}},\ }\bibfield  {title} {\enquote {\bibinfo {title} {On the jets,
  kinks, and spheromaks formed by a planar magnetized coaxial gun},}\
  }\href@noop {} {\bibfield  {journal} {\bibinfo  {journal} {Phys.~Plasmas}\
  }\textbf {\bibinfo {volume} {12}},\ \bibinfo {pages} {032103} (\bibinfo
  {year} {2005})}\BibitemShut {NoStop}%
\bibitem [{\citenamefont {{Stenson}}\ and\ \citenamefont
  {{Bellan}}(2012)}]{Stenson:2012}%
  \BibitemOpen
  \bibfield  {author} {\bibinfo {author} {\bibfnamefont {E.~V.}\ \bibnamefont
  {{Stenson}}}\ and\ \bibinfo {author} {\bibfnamefont {P.~M.}\ \bibnamefont
  {{Bellan}}},\ }\bibfield  {title} {\enquote {\bibinfo {title} {{Magnetically
  Driven Flows in Arched Plasma Structures}},}\ }\href {\doibase
  10.1103/PhysRevLett.109.075001} {\bibfield  {journal} {\bibinfo  {journal}
  {Phys.~Rev.~Lett.}\ }\textbf {\bibinfo {volume} {109}},\ \bibinfo {eid}
  {075001} (\bibinfo {year} {2012})}\BibitemShut {NoStop}%
\bibitem [{\citenamefont {Hsu}\ and\ \citenamefont {Bellan}(2002)}]{Hsu:2002}%
  \BibitemOpen
  \bibfield  {author} {\bibinfo {author} {\bibfnamefont {S.~C.}\ \bibnamefont
  {Hsu}}\ and\ \bibinfo {author} {\bibfnamefont {P.~M.}\ \bibnamefont
  {Bellan}},\ }\bibfield  {title} {\enquote {\bibinfo {title} {A laboratory
  plasma experiment for studying magnetic dynamics of accretion discs and
  jets},}\ }\href@noop {} {\bibfield  {journal} {\bibinfo  {journal}
  {Mon.~Not.~Roy.~Astron.~Soc.}\ }\textbf {\bibinfo {volume} {334}},\ \bibinfo
  {pages} {257} (\bibinfo {year} {2002})}\BibitemShut {NoStop}%
\bibitem [{\citenamefont {{You}}\ \emph {et~al.}(2017)\citenamefont {{You}},
  \citenamefont {{von der Linden}}, \citenamefont {{Sander Lavine}},
  \citenamefont {{Carroll}}, \citenamefont {{Card}}, \citenamefont
  {{Quinley}},\ and\ \citenamefont {{Azuara-Rosales}}}]{You:2018}%
  \BibitemOpen
  \bibfield  {author} {\bibinfo {author} {\bibfnamefont {S.}~\bibnamefont
  {{You}}}, \bibinfo {author} {\bibfnamefont {J.}~\bibnamefont {{von der
  Linden}}}, \bibinfo {author} {\bibfnamefont {E.}~\bibnamefont {{Sander
  Lavine}}}, \bibinfo {author} {\bibfnamefont {E.~G.}\ \bibnamefont
  {{Carroll}}}, \bibinfo {author} {\bibfnamefont {A.}~\bibnamefont {{Card}}},
  \bibinfo {author} {\bibfnamefont {M.}~\bibnamefont {{Quinley}}}, \ and\
  \bibinfo {author} {\bibfnamefont {M.}~\bibnamefont {{Azuara-Rosales}}},\
  }\bibfield  {title} {\enquote {\bibinfo {title} {{The Mochi.LabJet experiment
  for measurements of canonical helicity injection in a laboratory
  astrophysical jet}},}\ }\href@noop {} {\bibfield  {journal} {\bibinfo
  {journal} {ArXiv e-prints}\ } (\bibinfo {year} {2017})},\ \Eprint
  {http://arxiv.org/abs/1711.07213} {arXiv:1711.07213 [physics.plasm-ph]}
  \BibitemShut {NoStop}%
\bibitem [{\citenamefont {{Albertazzi}}\ \emph {et~al.}(2013)\citenamefont
  {{Albertazzi}}, \citenamefont {{B{\'e}ard}}, \citenamefont {{Ciardi}},
  \citenamefont {{Vinci}}, \citenamefont {{Albrecht}}, \citenamefont
  {{Billette}}, \citenamefont {{Burris-Mog}}, \citenamefont {{Chen}},
  \citenamefont {{Da Silva}}, \citenamefont {{Dittrich}}, \citenamefont
  {{Herrmannsd{\"o}rfer}}, \citenamefont {{Hirardin}}, \citenamefont {{Kroll}},
  \citenamefont {{Nakatsutsumi}}, \citenamefont {{Nitsche}}, \citenamefont
  {{Riconda}}, \citenamefont {{Romagnagni}}, \citenamefont {{Schlenvoigt}},
  \citenamefont {{Simond}}, \citenamefont {{Veuillot}}, \citenamefont
  {{Cowan}}, \citenamefont {{Portugall}}, \citenamefont {{P{\'e}pin}},\ and\
  \citenamefont {{Fuchs}}}]{Albertazzi:2013}%
  \BibitemOpen
  \bibfield  {author} {\bibinfo {author} {\bibfnamefont {B.}~\bibnamefont
  {{Albertazzi}}}, \bibinfo {author} {\bibfnamefont {J.}~\bibnamefont
  {{B{\'e}ard}}}, \bibinfo {author} {\bibfnamefont {A.}~\bibnamefont
  {{Ciardi}}}, \bibinfo {author} {\bibfnamefont {T.}~\bibnamefont {{Vinci}}},
  \bibinfo {author} {\bibfnamefont {J.}~\bibnamefont {{Albrecht}}}, \bibinfo
  {author} {\bibfnamefont {J.}~\bibnamefont {{Billette}}}, \bibinfo {author}
  {\bibfnamefont {T.}~\bibnamefont {{Burris-Mog}}}, \bibinfo {author}
  {\bibfnamefont {S.~N.}\ \bibnamefont {{Chen}}}, \bibinfo {author}
  {\bibfnamefont {D.}~\bibnamefont {{Da Silva}}}, \bibinfo {author}
  {\bibfnamefont {S.}~\bibnamefont {{Dittrich}}}, \bibinfo {author}
  {\bibfnamefont {T.}~\bibnamefont {{Herrmannsd{\"o}rfer}}}, \bibinfo {author}
  {\bibfnamefont {B.}~\bibnamefont {{Hirardin}}}, \bibinfo {author}
  {\bibfnamefont {F.}~\bibnamefont {{Kroll}}}, \bibinfo {author} {\bibfnamefont
  {M.}~\bibnamefont {{Nakatsutsumi}}}, \bibinfo {author} {\bibfnamefont
  {S.}~\bibnamefont {{Nitsche}}}, \bibinfo {author} {\bibfnamefont
  {C.}~\bibnamefont {{Riconda}}}, \bibinfo {author} {\bibfnamefont
  {L.}~\bibnamefont {{Romagnagni}}}, \bibinfo {author} {\bibfnamefont {H.-P.}\
  \bibnamefont {{Schlenvoigt}}}, \bibinfo {author} {\bibfnamefont
  {S.}~\bibnamefont {{Simond}}}, \bibinfo {author} {\bibfnamefont
  {E.}~\bibnamefont {{Veuillot}}}, \bibinfo {author} {\bibfnamefont {T.~E.}\
  \bibnamefont {{Cowan}}}, \bibinfo {author} {\bibfnamefont {O.}~\bibnamefont
  {{Portugall}}}, \bibinfo {author} {\bibfnamefont {H.}~\bibnamefont
  {{P{\'e}pin}}}, \ and\ \bibinfo {author} {\bibfnamefont {J.}~\bibnamefont
  {{Fuchs}}},\ }\bibfield  {title} {\enquote {\bibinfo {title} {{Production of
  large volume, strongly magnetized laser-produced plasmas by use of pulsed
  external magnetic fields}},}\ }\href {\doibase 10.1063/1.4795551} {\bibfield
  {journal} {\bibinfo  {journal} {Rev. Sci. Instrum.}\ }\textbf {\bibinfo
  {volume} {84}},\ \bibinfo {eid} {043505-043505-6} (\bibinfo {year}
  {2013})}\BibitemShut {NoStop}%
\bibitem [{\citenamefont {{Ciardi}}\ \emph {et~al.}(2013)\citenamefont
  {{Ciardi}}, \citenamefont {{Vinci}}, \citenamefont {{Fuchs}}, \citenamefont
  {{Albertazzi}}, \citenamefont {{Riconda}}, \citenamefont {{P{\'e}pin}},\ and\
  \citenamefont {{Portugall}}}]{Ciardi:2013}%
  \BibitemOpen
  \bibfield  {author} {\bibinfo {author} {\bibfnamefont {A.}~\bibnamefont
  {{Ciardi}}}, \bibinfo {author} {\bibfnamefont {T.}~\bibnamefont {{Vinci}}},
  \bibinfo {author} {\bibfnamefont {J.}~\bibnamefont {{Fuchs}}}, \bibinfo
  {author} {\bibfnamefont {B.}~\bibnamefont {{Albertazzi}}}, \bibinfo {author}
  {\bibfnamefont {C.}~\bibnamefont {{Riconda}}}, \bibinfo {author}
  {\bibfnamefont {H.}~\bibnamefont {{P{\'e}pin}}}, \ and\ \bibinfo {author}
  {\bibfnamefont {O.}~\bibnamefont {{Portugall}}},\ }\bibfield  {title}
  {\enquote {\bibinfo {title} {{Astrophysics of Magnetically Collimated Jets
  Generated from Laser-Produced Plasmas}},}\ }\href {\doibase
  10.1103/PhysRevLett.110.025002} {\bibfield  {journal} {\bibinfo  {journal}
  {Phys.~Rev.~Lett.}\ }\textbf {\bibinfo {volume} {110}},\ \bibinfo {eid}
  {025002} (\bibinfo {year} {2013})},\ \Eprint {http://arxiv.org/abs/1212.2805}
  {arXiv:1212.2805 [physics.plasm-ph]} \BibitemShut {NoStop}%
\bibitem [{\citenamefont {{Albertazzi}}\ \emph {et~al.}(2014)\citenamefont
  {{Albertazzi}}, \citenamefont {{Ciardi}}, \citenamefont {{Nakatsutsumi}},
  \citenamefont {{Vinci}}, \citenamefont {{B{\'e}ard}}, \citenamefont
  {{Bonito}}, \citenamefont {{Billette}}, \citenamefont {{Borghesi}},
  \citenamefont {{Burkley}}, \citenamefont {{Chen}}, \citenamefont {{Cowan}},
  \citenamefont {{Herrmannsd{\"o}rfer}}, \citenamefont {{Higginson}},
  \citenamefont {{Kroll}}, \citenamefont {{Pikuz}}, \citenamefont {{Naughton}},
  \citenamefont {{Romagnani}}, \citenamefont {{Riconda}}, \citenamefont
  {{Revet}}, \citenamefont {{Riquier}}, \citenamefont {{Schlenvoigt}},
  \citenamefont {{Skobelev}}, \citenamefont {{Faenov}}, \citenamefont
  {{Soloviev}}, \citenamefont {{Huarte-Espinosa}}, \citenamefont {{Frank}},
  \citenamefont {{Portugall}}, \citenamefont {{P{\'e}pin}},\ and\ \citenamefont
  {{Fuchs}}}]{Albertazzi:2014}%
  \BibitemOpen
  \bibfield  {author} {\bibinfo {author} {\bibfnamefont {B.}~\bibnamefont
  {{Albertazzi}}}, \bibinfo {author} {\bibfnamefont {A.}~\bibnamefont
  {{Ciardi}}}, \bibinfo {author} {\bibfnamefont {M.}~\bibnamefont
  {{Nakatsutsumi}}}, \bibinfo {author} {\bibfnamefont {T.}~\bibnamefont
  {{Vinci}}}, \bibinfo {author} {\bibfnamefont {J.}~\bibnamefont
  {{B{\'e}ard}}}, \bibinfo {author} {\bibfnamefont {R.}~\bibnamefont
  {{Bonito}}}, \bibinfo {author} {\bibfnamefont {J.}~\bibnamefont
  {{Billette}}}, \bibinfo {author} {\bibfnamefont {M.}~\bibnamefont
  {{Borghesi}}}, \bibinfo {author} {\bibfnamefont {Z.}~\bibnamefont
  {{Burkley}}}, \bibinfo {author} {\bibfnamefont {S.~N.}\ \bibnamefont
  {{Chen}}}, \bibinfo {author} {\bibfnamefont {T.~E.}\ \bibnamefont {{Cowan}}},
  \bibinfo {author} {\bibfnamefont {T.}~\bibnamefont {{Herrmannsd{\"o}rfer}}},
  \bibinfo {author} {\bibfnamefont {D.~P.}\ \bibnamefont {{Higginson}}},
  \bibinfo {author} {\bibfnamefont {F.}~\bibnamefont {{Kroll}}}, \bibinfo
  {author} {\bibfnamefont {S.~A.}\ \bibnamefont {{Pikuz}}}, \bibinfo {author}
  {\bibfnamefont {K.}~\bibnamefont {{Naughton}}}, \bibinfo {author}
  {\bibfnamefont {L.}~\bibnamefont {{Romagnani}}}, \bibinfo {author}
  {\bibfnamefont {C.}~\bibnamefont {{Riconda}}}, \bibinfo {author}
  {\bibfnamefont {G.}~\bibnamefont {{Revet}}}, \bibinfo {author} {\bibfnamefont
  {R.}~\bibnamefont {{Riquier}}}, \bibinfo {author} {\bibfnamefont {H.-P.}\
  \bibnamefont {{Schlenvoigt}}}, \bibinfo {author} {\bibfnamefont {I.~Y.}\
  \bibnamefont {{Skobelev}}}, \bibinfo {author} {\bibfnamefont {A.~Y.}\
  \bibnamefont {{Faenov}}}, \bibinfo {author} {\bibfnamefont {A.}~\bibnamefont
  {{Soloviev}}}, \bibinfo {author} {\bibfnamefont {M.}~\bibnamefont
  {{Huarte-Espinosa}}}, \bibinfo {author} {\bibfnamefont {A.}~\bibnamefont
  {{Frank}}}, \bibinfo {author} {\bibfnamefont {O.}~\bibnamefont
  {{Portugall}}}, \bibinfo {author} {\bibfnamefont {H.}~\bibnamefont
  {{P{\'e}pin}}}, \ and\ \bibinfo {author} {\bibfnamefont {J.}~\bibnamefont
  {{Fuchs}}},\ }\bibfield  {title} {\enquote {\bibinfo {title} {{Laboratory
  formation of a scaled protostellar jet by coaligned poloidal magnetic
  field}},}\ }\href {\doibase 10.1126/science.1259694} {\bibfield  {journal}
  {\bibinfo  {journal} {Science}\ }\textbf {\bibinfo {volume} {346}},\ \bibinfo
  {pages} {325--328} (\bibinfo {year} {2014})}\BibitemShut {NoStop}%
\bibitem [{\citenamefont {{Mestel}}\ and\ \citenamefont
  {{Spitzer}}(1956)}]{Mestel:1956}%
  \BibitemOpen
  \bibfield  {author} {\bibinfo {author} {\bibfnamefont {L.}~\bibnamefont
  {{Mestel}}}\ and\ \bibinfo {author} {\bibfnamefont {L.}~\bibnamefont
  {{Spitzer}}, \bibfnamefont {Jr.}},\ }\bibfield  {title} {\enquote {\bibinfo
  {title} {{Star formation in magnetic dust clouds}},}\ }\href {\doibase
  10.1093/mnras/116.5.503} {\bibfield  {journal} {\bibinfo  {journal}
  {Mon.~Not.~Roy.~Astron.~Soc.}\ }\textbf {\bibinfo {volume} {116}},\ \bibinfo
  {pages} {503} (\bibinfo {year} {1956})}\BibitemShut {NoStop}%
\bibitem [{\citenamefont {{Goertz}}\ and\ \citenamefont
  {{Morfill}}(1983)}]{Goertz:1983}%
  \BibitemOpen
  \bibfield  {author} {\bibinfo {author} {\bibfnamefont {C.~K.}\ \bibnamefont
  {{Goertz}}}\ and\ \bibinfo {author} {\bibfnamefont {G.}~\bibnamefont
  {{Morfill}}},\ }\bibfield  {title} {\enquote {\bibinfo {title} {{A model for
  the formation of spokes in Saturn's rings}},}\ }\href {\doibase
  10.1016/0019-1035(83)90143-4} {\bibfield  {journal} {\bibinfo  {journal}
  {Icarus}\ }\textbf {\bibinfo {volume} {53}},\ \bibinfo {pages} {219--229}
  (\bibinfo {year} {1983})}\BibitemShut {NoStop}%
\bibitem [{\citenamefont {{Gurnett}}\ \emph {et~al.}(1983)\citenamefont
  {{Gurnett}}, \citenamefont {{Grun}}, \citenamefont {{Gallagher}},
  \citenamefont {{Kurth}},\ and\ \citenamefont {{Scarf}}}]{Gurnett:1983}%
  \BibitemOpen
  \bibfield  {author} {\bibinfo {author} {\bibfnamefont {D.~A.}\ \bibnamefont
  {{Gurnett}}}, \bibinfo {author} {\bibfnamefont {E.}~\bibnamefont {{Grun}}},
  \bibinfo {author} {\bibfnamefont {D.}~\bibnamefont {{Gallagher}}}, \bibinfo
  {author} {\bibfnamefont {W.~S.}\ \bibnamefont {{Kurth}}}, \ and\ \bibinfo
  {author} {\bibfnamefont {F.~L.}\ \bibnamefont {{Scarf}}},\ }\bibfield
  {title} {\enquote {\bibinfo {title} {{Micron-sized particles detected near
  Saturn by the Voyager plasma wave instrument}},}\ }\href {\doibase
  10.1016/0019-1035(83)90145-8} {\bibfield  {journal} {\bibinfo  {journal}
  {Icarus}\ }\textbf {\bibinfo {volume} {53}},\ \bibinfo {pages} {236--254}
  (\bibinfo {year} {1983})}\BibitemShut {NoStop}%
\bibitem [{\citenamefont {{Goertz}}(1989)}]{Goertz:1989}%
  \BibitemOpen
  \bibfield  {author} {\bibinfo {author} {\bibfnamefont {C.~K.}\ \bibnamefont
  {{Goertz}}},\ }\bibfield  {title} {\enquote {\bibinfo {title} {{Dusty plasmas
  in the solar system}},}\ }\href {\doibase 10.1029/RG027i002p00271} {\bibfield
   {journal} {\bibinfo  {journal} {Rev. Geophys.}\ }\textbf {\bibinfo {volume}
  {27}},\ \bibinfo {pages} {271--292} (\bibinfo {year} {1989})}\BibitemShut
  {NoStop}%
\bibitem [{\citenamefont {{Ahn}}\ \emph {et~al.}(1983)\citenamefont {{Ahn}},
  \citenamefont {{Akasofu}}, \citenamefont {{Robinson}},\ and\ \citenamefont
  {{Kamide}}}]{Ahn:1983}%
  \BibitemOpen
  \bibfield  {author} {\bibinfo {author} {\bibfnamefont {B.-H.}\ \bibnamefont
  {{Ahn}}}, \bibinfo {author} {\bibfnamefont {S.-I.}\ \bibnamefont
  {{Akasofu}}}, \bibinfo {author} {\bibfnamefont {R.~M.}\ \bibnamefont
  {{Robinson}}}, \ and\ \bibinfo {author} {\bibfnamefont {Y.}~\bibnamefont
  {{Kamide}}},\ }\bibfield  {title} {\enquote {\bibinfo {title} {{Electric
  conductivities, electric fields and auroral particle energy injection rate in
  the auroral ionosphere and their empirical relations to the horizontal
  magnetic disturbances}},}\ }\href {\doibase 10.1016/0032-0633(83)90005-3}
  {\bibfield  {journal} {\bibinfo  {journal} {Planet. Space Sci.}\ }\textbf
  {\bibinfo {volume} {31}},\ \bibinfo {pages} {641--653} (\bibinfo {year}
  {1983})}\BibitemShut {NoStop}%
\bibitem [{\citenamefont {{Luhmann}}(1983)}]{Luhmann:1983}%
  \BibitemOpen
  \bibfield  {author} {\bibinfo {author} {\bibfnamefont {J.~G.}\ \bibnamefont
  {{Luhmann}}},\ }\bibfield  {title} {\enquote {\bibinfo {title} {{Ionospheric
  disturbances resulting from ion-neutral coupling}},}\ }\href {\doibase
  10.1007/BF00175288} {\bibfield  {journal} {\bibinfo  {journal} {Space
  Sci.~Rev.}\ }\textbf {\bibinfo {volume} {34}},\ \bibinfo {pages} {337--346}
  (\bibinfo {year} {1983})}\BibitemShut {NoStop}%
\bibitem [{\citenamefont {{Lysak}}\ and\ \citenamefont
  {{Dum}}(1983)}]{Lysak:1983}%
  \BibitemOpen
  \bibfield  {author} {\bibinfo {author} {\bibfnamefont {R.~L.}\ \bibnamefont
  {{Lysak}}}\ and\ \bibinfo {author} {\bibfnamefont {C.~T.}\ \bibnamefont
  {{Dum}}},\ }\bibfield  {title} {\enquote {\bibinfo {title} {{Dynamics of
  magnetosphere-ionosphere coupling including turbulent transport}},}\ }\href
  {\doibase 10.1029/JA088iA01p00365} {\bibfield  {journal} {\bibinfo  {journal}
  {J.~Geophys.~Res.}\ }\textbf {\bibinfo {volume} {88}},\ \bibinfo {pages}
  {365--380} (\bibinfo {year} {1983})}\BibitemShut {NoStop}%
\bibitem [{\citenamefont {{Horanyi}}(1996)}]{Horanyi:1996}%
  \BibitemOpen
  \bibfield  {author} {\bibinfo {author} {\bibfnamefont {M.}~\bibnamefont
  {{Horanyi}}},\ }\bibfield  {title} {\enquote {\bibinfo {title} {{Charged Dust
  Dynamics in the Solar System}},}\ }\href {\doibase
  10.1146/annurev.astro.34.1.383} {\bibfield  {journal} {\bibinfo  {journal}
  {Ann.~Rev.~Astron.~Astrophys.}\ }\textbf {\bibinfo {volume} {34}},\ \bibinfo
  {pages} {383--418} (\bibinfo {year} {1996})}\BibitemShut {NoStop}%
\bibitem [{\citenamefont {{Mendis}}\ and\ \citenamefont
  {{Rosenberg}}(1994)}]{Mendis:1994}%
  \BibitemOpen
  \bibfield  {author} {\bibinfo {author} {\bibfnamefont {D.~A.}\ \bibnamefont
  {{Mendis}}}\ and\ \bibinfo {author} {\bibfnamefont {M.}~\bibnamefont
  {{Rosenberg}}},\ }\bibfield  {title} {\enquote {\bibinfo {title} {{Cosmic
  Dusty Plasmas}},}\ }\href {\doibase 10.1146/annurev.aa.32.090194.002223}
  {\bibfield  {journal} {\bibinfo  {journal} {Ann.~Rev.~Astron.~Astrophys.}\
  }\textbf {\bibinfo {volume} {32}},\ \bibinfo {pages} {419--463} (\bibinfo
  {year} {1994})}\BibitemShut {NoStop}%
\bibitem [{\citenamefont {{Verheest}}(1996)}]{Verheest:1996}%
  \BibitemOpen
  \bibfield  {author} {\bibinfo {author} {\bibfnamefont {F.}~\bibnamefont
  {{Verheest}}},\ }\bibfield  {title} {\enquote {\bibinfo {title} {{Waves and
  Instabilities in Dusty Space Plasmas}},}\ }\href {\doibase
  10.1007/BF00226225} {\bibfield  {journal} {\bibinfo  {journal} {Space
  Sci.~Rev.}\ }\textbf {\bibinfo {volume} {77}},\ \bibinfo {pages} {267--302}
  (\bibinfo {year} {1996})}\BibitemShut {NoStop}%
\bibitem [{\citenamefont {{Fortov}}\ \emph {et~al.}(1998)\citenamefont
  {{Fortov}}, \citenamefont {{Nefedov}}, \citenamefont {{Vaulina}},
  \citenamefont {{Lipaev}}, \citenamefont {{Molotkov}}, \citenamefont
  {{Samaryan}}, \citenamefont {{Nikitskii}}, \citenamefont {{Ivanov}},
  \citenamefont {{Savin}}, \citenamefont {{Kalmykov}}, \citenamefont
  {{Solov'Ev}},\ and\ \citenamefont {{Vinogradov}}}]{Fortov:1998}%
  \BibitemOpen
  \bibfield  {author} {\bibinfo {author} {\bibfnamefont {V.~E.}\ \bibnamefont
  {{Fortov}}}, \bibinfo {author} {\bibfnamefont {A.~P.}\ \bibnamefont
  {{Nefedov}}}, \bibinfo {author} {\bibfnamefont {O.~S.}\ \bibnamefont
  {{Vaulina}}}, \bibinfo {author} {\bibfnamefont {A.~M.}\ \bibnamefont
  {{Lipaev}}}, \bibinfo {author} {\bibfnamefont {V.~I.}\ \bibnamefont
  {{Molotkov}}}, \bibinfo {author} {\bibfnamefont {A.~A.}\ \bibnamefont
  {{Samaryan}}}, \bibinfo {author} {\bibfnamefont {V.~P.}\ \bibnamefont
  {{Nikitskii}}}, \bibinfo {author} {\bibfnamefont {A.~I.}\ \bibnamefont
  {{Ivanov}}}, \bibinfo {author} {\bibfnamefont {S.~F.}\ \bibnamefont
  {{Savin}}}, \bibinfo {author} {\bibfnamefont {A.~V.}\ \bibnamefont
  {{Kalmykov}}}, \bibinfo {author} {\bibfnamefont {A.~Y.}\ \bibnamefont
  {{Solov'Ev}}}, \ and\ \bibinfo {author} {\bibfnamefont {P.~V.}\ \bibnamefont
  {{Vinogradov}}},\ }\bibfield  {title} {\enquote {\bibinfo {title} {{Dusty
  plasma induced by solar radiation under microgravitational conditions: An
  experiment on board the Mir orbiting space station}},}\ }\href {\doibase
  10.1134/1.558598} {\bibfield  {journal} {\bibinfo  {journal} {Soviet Journal
  of Experimental and Theoretical Physics}\ }\textbf {\bibinfo {volume} {87}},\
  \bibinfo {pages} {1087--1097} (\bibinfo {year} {1998})}\BibitemShut {NoStop}%
\bibitem [{\citenamefont {{Nefedov}}\ \emph {et~al.}(2003)\citenamefont
  {{Nefedov}}, \citenamefont {{Morfill}}, \citenamefont {{Fortov}},
  \citenamefont {{Thomas}}, \citenamefont {{Rothermel}}, \citenamefont
  {{Hagl}}, \citenamefont {{Ivlev}}, \citenamefont {{Zuzic}}, \citenamefont
  {{Klumov}}, \citenamefont {{Lipaev}}, \citenamefont {{Molotkov}},
  \citenamefont {{Petrov}}, \citenamefont {{Gidzenko}}, \citenamefont
  {{Krikalev}}, \citenamefont {{Shepherd}}, \citenamefont {{Ivanov}},
  \citenamefont {{Roth}}, \citenamefont {{Binnenbruck}}, \citenamefont
  {{Goree}},\ and\ \citenamefont {{Semenov}}}]{Nefedov:2003}%
  \BibitemOpen
  \bibfield  {author} {\bibinfo {author} {\bibfnamefont {A.~P.}\ \bibnamefont
  {{Nefedov}}}, \bibinfo {author} {\bibfnamefont {G.~E.}\ \bibnamefont
  {{Morfill}}}, \bibinfo {author} {\bibfnamefont {V.~E.}\ \bibnamefont
  {{Fortov}}}, \bibinfo {author} {\bibfnamefont {H.~M.}\ \bibnamefont
  {{Thomas}}}, \bibinfo {author} {\bibfnamefont {H.}~\bibnamefont
  {{Rothermel}}}, \bibinfo {author} {\bibfnamefont {T.}~\bibnamefont {{Hagl}}},
  \bibinfo {author} {\bibfnamefont {A.~V.}\ \bibnamefont {{Ivlev}}}, \bibinfo
  {author} {\bibfnamefont {M.}~\bibnamefont {{Zuzic}}}, \bibinfo {author}
  {\bibfnamefont {B.~A.}\ \bibnamefont {{Klumov}}}, \bibinfo {author}
  {\bibfnamefont {A.~M.}\ \bibnamefont {{Lipaev}}}, \bibinfo {author}
  {\bibfnamefont {V.~I.}\ \bibnamefont {{Molotkov}}}, \bibinfo {author}
  {\bibfnamefont {O.~F.}\ \bibnamefont {{Petrov}}}, \bibinfo {author}
  {\bibfnamefont {Y.~P.}\ \bibnamefont {{Gidzenko}}}, \bibinfo {author}
  {\bibfnamefont {S.~K.}\ \bibnamefont {{Krikalev}}}, \bibinfo {author}
  {\bibfnamefont {W.}~\bibnamefont {{Shepherd}}}, \bibinfo {author}
  {\bibfnamefont {A.~I.}\ \bibnamefont {{Ivanov}}}, \bibinfo {author}
  {\bibfnamefont {M.}~\bibnamefont {{Roth}}}, \bibinfo {author} {\bibfnamefont
  {H.}~\bibnamefont {{Binnenbruck}}}, \bibinfo {author} {\bibfnamefont {J.~A.}\
  \bibnamefont {{Goree}}}, \ and\ \bibinfo {author} {\bibfnamefont {Y.~P.}\
  \bibnamefont {{Semenov}}},\ }\bibfield  {title} {\enquote {\bibinfo {title}
  {{PKE-Nefedov: plasma crystal experiments on the International Space
  Station}},}\ }\href {\doibase 10.1088/1367-2630/5/1/333} {\bibfield
  {journal} {\bibinfo  {journal} {New J.~Phys.}\ }\textbf {\bibinfo {volume}
  {5}},\ \bibinfo {pages} {33} (\bibinfo {year} {2003})}\BibitemShut {NoStop}%
\bibitem [{\citenamefont {{Thomas}}\ \emph {et~al.}(2008)\citenamefont
  {{Thomas}}, \citenamefont {{Morfill}}, \citenamefont {{Fortov}},
  \citenamefont {{Ivlev}}, \citenamefont {{Molotkov}}, \citenamefont
  {{Lipaev}}, \citenamefont {{Hagl}}, \citenamefont {{Rothermel}},
  \citenamefont {{Khrapak}}, \citenamefont {{Suetterlin}}, \citenamefont
  {{Rubin-Zuzic}}, \citenamefont {{Petrov}}, \citenamefont {{Tokarev}},\ and\
  \citenamefont {{Krikalev}}}]{Thomas:2008}%
  \BibitemOpen
  \bibfield  {author} {\bibinfo {author} {\bibfnamefont {H.~M.}\ \bibnamefont
  {{Thomas}}}, \bibinfo {author} {\bibfnamefont {G.~E.}\ \bibnamefont
  {{Morfill}}}, \bibinfo {author} {\bibfnamefont {V.~E.}\ \bibnamefont
  {{Fortov}}}, \bibinfo {author} {\bibfnamefont {A.~V.}\ \bibnamefont
  {{Ivlev}}}, \bibinfo {author} {\bibfnamefont {V.~I.}\ \bibnamefont
  {{Molotkov}}}, \bibinfo {author} {\bibfnamefont {A.~M.}\ \bibnamefont
  {{Lipaev}}}, \bibinfo {author} {\bibfnamefont {T.}~\bibnamefont {{Hagl}}},
  \bibinfo {author} {\bibfnamefont {H.}~\bibnamefont {{Rothermel}}}, \bibinfo
  {author} {\bibfnamefont {S.~A.}\ \bibnamefont {{Khrapak}}}, \bibinfo {author}
  {\bibfnamefont {R.~K.}\ \bibnamefont {{Suetterlin}}}, \bibinfo {author}
  {\bibfnamefont {M.}~\bibnamefont {{Rubin-Zuzic}}}, \bibinfo {author}
  {\bibfnamefont {O.~F.}\ \bibnamefont {{Petrov}}}, \bibinfo {author}
  {\bibfnamefont {V.~I.}\ \bibnamefont {{Tokarev}}}, \ and\ \bibinfo {author}
  {\bibfnamefont {S.~K.}\ \bibnamefont {{Krikalev}}},\ }\bibfield  {title}
  {\enquote {\bibinfo {title} {{Complex plasma laboratory PK-3 Plus on the
  International Space Station}},}\ }\href {\doibase
  10.1088/1367-2630/10/3/033036} {\bibfield  {journal} {\bibinfo  {journal}
  {New J.~Phys.}\ }\textbf {\bibinfo {volume} {10}},\ \bibinfo {eid} {033036}
  (\bibinfo {year} {2008})}\BibitemShut {NoStop}%
\bibitem [{\citenamefont {{Fortov}}\ \emph {et~al.}(2005)\citenamefont
  {{Fortov}}, \citenamefont {{Morfill}}, \citenamefont {{Petrov}},
  \citenamefont {{Thoma}}, \citenamefont {{Usachev}}, \citenamefont
  {{Hoefner}}, \citenamefont {{Zobnin}}, \citenamefont {{Kretschmer}},
  \citenamefont {{Ratynskaia}}, \citenamefont {{Fink}}, \citenamefont
  {{Tarantik}}, \citenamefont {{Gerasimov}},\ and\ \citenamefont
  {{Esenkov}}}]{Fortov:2005}%
  \BibitemOpen
  \bibfield  {author} {\bibinfo {author} {\bibfnamefont {V.}~\bibnamefont
  {{Fortov}}}, \bibinfo {author} {\bibfnamefont {G.}~\bibnamefont {{Morfill}}},
  \bibinfo {author} {\bibfnamefont {O.}~\bibnamefont {{Petrov}}}, \bibinfo
  {author} {\bibfnamefont {M.}~\bibnamefont {{Thoma}}}, \bibinfo {author}
  {\bibfnamefont {A.}~\bibnamefont {{Usachev}}}, \bibinfo {author}
  {\bibfnamefont {H.}~\bibnamefont {{Hoefner}}}, \bibinfo {author}
  {\bibfnamefont {A.}~\bibnamefont {{Zobnin}}}, \bibinfo {author}
  {\bibfnamefont {M.}~\bibnamefont {{Kretschmer}}}, \bibinfo {author}
  {\bibfnamefont {S.}~\bibnamefont {{Ratynskaia}}}, \bibinfo {author}
  {\bibfnamefont {M.}~\bibnamefont {{Fink}}}, \bibinfo {author} {\bibfnamefont
  {K.}~\bibnamefont {{Tarantik}}}, \bibinfo {author} {\bibfnamefont
  {Y.}~\bibnamefont {{Gerasimov}}}, \ and\ \bibinfo {author} {\bibfnamefont
  {V.}~\bibnamefont {{Esenkov}}},\ }\bibfield  {title} {\enquote {\bibinfo
  {title} {{The project 'Plasmakristall-4' (PK-4)---a new stage in
  investigations of dusty plasmas under microgravity conditions: first results
  and future plans}},}\ }\href {\doibase 10.1088/0741-3335/47/12B/S39}
  {\bibfield  {journal} {\bibinfo  {journal} {Plasma Phys.~Con.~Fus.}\ }\textbf
  {\bibinfo {volume} {47}},\ \bibinfo {pages} {B537--B549} (\bibinfo {year}
  {2005})}\BibitemShut {NoStop}%
\bibitem [{\citenamefont {{Thomas}}\ \emph
  {et~al.}(2015{\natexlab{a}})\citenamefont {{Thomas}}, \citenamefont
  {{Konopka}}, \citenamefont {{Artis}}, \citenamefont {{Lynch}}, \citenamefont
  {{Leblanc}}, \citenamefont {{Adams}}, \citenamefont {{Merlino}},\ and\
  \citenamefont {{Rosenberg}}}]{Thomas:2015a}%
  \BibitemOpen
  \bibfield  {author} {\bibinfo {author} {\bibfnamefont {E.}~\bibnamefont
  {{Thomas}}}, \bibinfo {author} {\bibfnamefont {U.}~\bibnamefont {{Konopka}}},
  \bibinfo {author} {\bibfnamefont {D.}~\bibnamefont {{Artis}}}, \bibinfo
  {author} {\bibfnamefont {B.}~\bibnamefont {{Lynch}}}, \bibinfo {author}
  {\bibfnamefont {S.}~\bibnamefont {{Leblanc}}}, \bibinfo {author}
  {\bibfnamefont {S.}~\bibnamefont {{Adams}}}, \bibinfo {author} {\bibfnamefont
  {R.~L.}\ \bibnamefont {{Merlino}}}, \ and\ \bibinfo {author} {\bibfnamefont
  {M.}~\bibnamefont {{Rosenberg}}},\ }\bibfield  {title} {\enquote {\bibinfo
  {title} {{The magnetized dusty plasma experiment (MDPX)}},}\ }\href {\doibase
  10.1017/S0022377815000148} {\bibfield  {journal} {\bibinfo  {journal}
  {J.~Plasma Phys.}\ }\textbf {\bibinfo {volume} {81}},\ \bibinfo {eid}
  {345810206} (\bibinfo {year} {2015}{\natexlab{a}})}\BibitemShut {NoStop}%
\bibitem [{\citenamefont {{Bellan}}\ \emph {et~al.}(2015)\citenamefont
  {{Bellan}}, \citenamefont {{Zhai}}, \citenamefont {{Chai}},\ and\
  \citenamefont {{Ha}}}]{Bellan:2015}%
  \BibitemOpen
  \bibfield  {author} {\bibinfo {author} {\bibfnamefont {P.~M.}\ \bibnamefont
  {{Bellan}}}, \bibinfo {author} {\bibfnamefont {X.}~\bibnamefont {{Zhai}}},
  \bibinfo {author} {\bibfnamefont {K.~B.}\ \bibnamefont {{Chai}}}, \ and\
  \bibinfo {author} {\bibfnamefont {B.~N.}\ \bibnamefont {{Ha}}},\ }\bibfield
  {title} {\enquote {\bibinfo {title} {{Complex astrophysical experiments
  relating to jets, solar loops, and water ice dusty plasma}},}\ }\href
  {\doibase 10.1017/S0022377815000604} {\bibfield  {journal} {\bibinfo
  {journal} {J.~Plasma Phys.}\ }\textbf {\bibinfo {volume} {81}},\ \bibinfo
  {eid} {495810502} (\bibinfo {year} {2015})}\BibitemShut {NoStop}%
\bibitem [{\citenamefont {{Marshall}}, \citenamefont {{Chai}},\ and\
  \citenamefont {{Bellan}}(2017)}]{Marshall:2017}%
  \BibitemOpen
  \bibfield  {author} {\bibinfo {author} {\bibfnamefont {R.~S.}\ \bibnamefont
  {{Marshall}}}, \bibinfo {author} {\bibfnamefont {K.-B.}\ \bibnamefont
  {{Chai}}}, \ and\ \bibinfo {author} {\bibfnamefont {P.~M.}\ \bibnamefont
  {{Bellan}}},\ }\bibfield  {title} {\enquote {\bibinfo {title}
  {{Identification of Accretion as Grain Growth Mechanism in Astrophysically
  Relevant Water-ice Dusty Plasma Experiment}},}\ }\href {\doibase
  10.3847/1538-4357/aa5d11} {\bibfield  {journal} {\bibinfo  {journal}
  {Astrophys.~J.}\ }\textbf {\bibinfo {volume} {837}},\ \bibinfo {eid} {56}
  (\bibinfo {year} {2017})}\BibitemShut {NoStop}%
\bibitem [{\citenamefont {{Kindel}}\ and\ \citenamefont
  {{Kennel}}(1971)}]{Kindel:1971}%
  \BibitemOpen
  \bibfield  {author} {\bibinfo {author} {\bibfnamefont {J.~M.}\ \bibnamefont
  {{Kindel}}}\ and\ \bibinfo {author} {\bibfnamefont {C.~F.}\ \bibnamefont
  {{Kennel}}},\ }\bibfield  {title} {\enquote {\bibinfo {title} {{Topside
  current instabilities}},}\ }\href {\doibase 10.1029/JA076i013p03055}
  {\bibfield  {journal} {\bibinfo  {journal} {J.~Geophys.~Res.}\ }\textbf
  {\bibinfo {volume} {76}},\ \bibinfo {pages} {3055--3078} (\bibinfo {year}
  {1971})}\BibitemShut {NoStop}%
\bibitem [{\citenamefont {{Papadopoulos}}, \citenamefont {{Gaffey}},\ and\
  \citenamefont {{Palmadesso}}(1980)}]{Papadopoulos:1980}%
  \BibitemOpen
  \bibfield  {author} {\bibinfo {author} {\bibfnamefont {K.}~\bibnamefont
  {{Papadopoulos}}}, \bibinfo {author} {\bibfnamefont {J.~D.}\ \bibnamefont
  {{Gaffey}}, \bibfnamefont {Jr.}}, \ and\ \bibinfo {author} {\bibfnamefont
  {P.~J.}\ \bibnamefont {{Palmadesso}}},\ }\bibfield  {title} {\enquote
  {\bibinfo {title} {{Stochastic acceleration of large M/Q ions by hydrogen
  cyclotron waves in the magnetosphere}},}\ }\href {\doibase
  10.1029/GL007i011p01014} {\bibfield  {journal} {\bibinfo  {journal}
  {Geophys.~Res.~Lett.}\ }\textbf {\bibinfo {volume} {7}},\ \bibinfo {pages}
  {1014--1016} (\bibinfo {year} {1980})}\BibitemShut {NoStop}%
\bibitem [{\citenamefont {{Hu}}, \citenamefont {{Denton}},\ and\ \citenamefont
  {{Johnson}}(2010)}]{Hu:2010}%
  \BibitemOpen
  \bibfield  {author} {\bibinfo {author} {\bibfnamefont {Y.}~\bibnamefont
  {{Hu}}}, \bibinfo {author} {\bibfnamefont {R.~E.}\ \bibnamefont {{Denton}}},
  \ and\ \bibinfo {author} {\bibfnamefont {J.~R.}\ \bibnamefont {{Johnson}}},\
  }\bibfield  {title} {\enquote {\bibinfo {title} {{Two-dimensional hybrid code
  simulation of electromagnetic ion cyclotron waves of multi-ion plasmas in a
  dipole magnetic field}},}\ }\href {\doibase 10.1029/2009JA015158} {\bibfield
  {journal} {\bibinfo  {journal} {J.~Geophys.~Res.}\ }\textbf {\bibinfo
  {volume} {115}},\ \bibinfo {eid} {A09218} (\bibinfo {year}
  {2010})}\BibitemShut {NoStop}%
\bibitem [{\citenamefont {{Saikin}}\ \emph {et~al.}(2015)\citenamefont
  {{Saikin}}, \citenamefont {{Zhang}}, \citenamefont {{Allen}}, \citenamefont
  {{Smith}}, \citenamefont {{Kistler}}, \citenamefont {{Spence}}, \citenamefont
  {{Torbert}}, \citenamefont {{Kletzing}},\ and\ \citenamefont
  {{Jordanova}}}]{Saikin:2015}%
  \BibitemOpen
  \bibfield  {author} {\bibinfo {author} {\bibfnamefont {A.~A.}\ \bibnamefont
  {{Saikin}}}, \bibinfo {author} {\bibfnamefont {J.-C.}\ \bibnamefont
  {{Zhang}}}, \bibinfo {author} {\bibfnamefont {R.~C.}\ \bibnamefont
  {{Allen}}}, \bibinfo {author} {\bibfnamefont {C.~W.}\ \bibnamefont
  {{Smith}}}, \bibinfo {author} {\bibfnamefont {L.~M.}\ \bibnamefont
  {{Kistler}}}, \bibinfo {author} {\bibfnamefont {H.~E.}\ \bibnamefont
  {{Spence}}}, \bibinfo {author} {\bibfnamefont {R.~B.}\ \bibnamefont
  {{Torbert}}}, \bibinfo {author} {\bibfnamefont {C.~A.}\ \bibnamefont
  {{Kletzing}}}, \ and\ \bibinfo {author} {\bibfnamefont {V.~K.}\ \bibnamefont
  {{Jordanova}}},\ }\bibfield  {title} {\enquote {\bibinfo {title} {{The
  occurrence and wave properties of H$^{+}$-, He$^{+}$-, and O$^{+}$-band EMIC
  waves observed by the Van Allen Probes}},}\ }\href {\doibase
  10.1002/2015JA021358} {\bibfield  {journal} {\bibinfo  {journal}
  {J.~Geophys.~Res.}\ }\textbf {\bibinfo {volume} {120}},\ \bibinfo {pages}
  {7477--7492} (\bibinfo {year} {2015})}\BibitemShut {NoStop}%
\bibitem [{\citenamefont {{Mithaiwala}}, \citenamefont {{Rudakov}},\ and\
  \citenamefont {{Ganguli}}(2010)}]{Mithaiwala:2010}%
  \BibitemOpen
  \bibfield  {author} {\bibinfo {author} {\bibfnamefont {M.}~\bibnamefont
  {{Mithaiwala}}}, \bibinfo {author} {\bibfnamefont {L.}~\bibnamefont
  {{Rudakov}}}, \ and\ \bibinfo {author} {\bibfnamefont {G.}~\bibnamefont
  {{Ganguli}}},\ }\bibfield  {title} {\enquote {\bibinfo {title} {{Stability of
  an ion-ring distribution in a multi-ion component plasma}},}\ }\href
  {\doibase 10.1063/1.3372842} {\bibfield  {journal} {\bibinfo  {journal}
  {Physics of Plasmas}\ }\textbf {\bibinfo {volume} {17}},\ \bibinfo {eid}
  {042113} (\bibinfo {year} {2010})}\BibitemShut {NoStop}%
\bibitem [{\citenamefont {{Hu}}\ and\ \citenamefont {{Rifai
  Habbal}}(1999)}]{Hu:1999}%
  \BibitemOpen
  \bibfield  {author} {\bibinfo {author} {\bibfnamefont {Y.~Q.}\ \bibnamefont
  {{Hu}}}\ and\ \bibinfo {author} {\bibfnamefont {S.}~\bibnamefont {{Rifai
  Habbal}}},\ }\bibfield  {title} {\enquote {\bibinfo {title} {{Resonant
  acceleration and heating of solar wind ions by dispersive ion cyclotron
  waves}},}\ }\href {\doibase 10.1029/1999JA900193} {\bibfield  {journal}
  {\bibinfo  {journal} {J.~Geophys.~Res.}\ }\textbf {\bibinfo {volume} {104}},\
  \bibinfo {pages} {17045--17056} (\bibinfo {year} {1999})}\BibitemShut
  {NoStop}%
\bibitem [{\citenamefont {{Perrone}}\ \emph {et~al.}(2013)\citenamefont
  {{Perrone}}, \citenamefont {{Valentini}}, \citenamefont {{Servidio}},
  \citenamefont {{Dalena}},\ and\ \citenamefont {{Veltri}}}]{Perrone:2013}%
  \BibitemOpen
  \bibfield  {author} {\bibinfo {author} {\bibfnamefont {D.}~\bibnamefont
  {{Perrone}}}, \bibinfo {author} {\bibfnamefont {F.}~\bibnamefont
  {{Valentini}}}, \bibinfo {author} {\bibfnamefont {S.}~\bibnamefont
  {{Servidio}}}, \bibinfo {author} {\bibfnamefont {S.}~\bibnamefont
  {{Dalena}}}, \ and\ \bibinfo {author} {\bibfnamefont {P.}~\bibnamefont
  {{Veltri}}},\ }\bibfield  {title} {\enquote {\bibinfo {title} {{Vlasov
  Simulations of Multi-ion Plasma Turbulence in the Solar Wind}},}\ }\href
  {\doibase 10.1088/0004-637X/762/2/99} {\bibfield  {journal} {\bibinfo
  {journal} {Astrophys.~J.}\ }\textbf {\bibinfo {volume} {762}},\ \bibinfo
  {eid} {99} (\bibinfo {year} {2013})},\ \Eprint
  {http://arxiv.org/abs/1207.5379} {arXiv:1207.5379 [physics.space-ph]}
  \BibitemShut {NoStop}%
\bibitem [{\citenamefont {{Isenberg}}\ and\ \citenamefont
  {{Hollweg}}(1982)}]{Isenberg:1982}%
  \BibitemOpen
  \bibfield  {author} {\bibinfo {author} {\bibfnamefont {P.~A.}\ \bibnamefont
  {{Isenberg}}}\ and\ \bibinfo {author} {\bibfnamefont {J.~V.}\ \bibnamefont
  {{Hollweg}}},\ }\bibfield  {title} {\enquote {\bibinfo {title} {{Finite
  amplitude Alfven waves in a multi-ion plasma - Propagation, acceleration, and
  heating}},}\ }\href {\doibase 10.1029/JA087iA07p05023} {\bibfield  {journal}
  {\bibinfo  {journal} {J.~Geophys.~Res.}\ }\textbf {\bibinfo {volume} {87}},\
  \bibinfo {pages} {5023--5029} (\bibinfo {year} {1982})}\BibitemShut {NoStop}%
\bibitem [{\citenamefont {{McKenzie}}(1994)}]{McKenzie:1994}%
  \BibitemOpen
  \bibfield  {author} {\bibinfo {author} {\bibfnamefont {J.~F.}\ \bibnamefont
  {{McKenzie}}},\ }\bibfield  {title} {\enquote {\bibinfo {title} {{Interaction
  between Alfven waves and a multicomponent plasma with differential ion
  streaming}},}\ }\href {\doibase 10.1029/93JA02928} {\bibfield  {journal}
  {\bibinfo  {journal} {J.~Geophys.~Res.}\ }\textbf {\bibinfo {volume} {99}},\
  \bibinfo {pages} {4193--4200} (\bibinfo {year} {1994})}\BibitemShut {NoStop}%
\bibitem [{\citenamefont {{Ofman}}\ \emph {et~al.}(2005)\citenamefont
  {{Ofman}}, \citenamefont {{Davila}}, \citenamefont {{Nakariakov}},\ and\
  \citenamefont {{Vi{\~n}As}}}]{Ofman:2005}%
  \BibitemOpen
  \bibfield  {author} {\bibinfo {author} {\bibfnamefont {L.}~\bibnamefont
  {{Ofman}}}, \bibinfo {author} {\bibfnamefont {J.~M.}\ \bibnamefont
  {{Davila}}}, \bibinfo {author} {\bibfnamefont {V.~M.}\ \bibnamefont
  {{Nakariakov}}}, \ and\ \bibinfo {author} {\bibfnamefont {A.-F.}\
  \bibnamefont {{Vi{\~n}As}}},\ }\bibfield  {title} {\enquote {\bibinfo {title}
  {{High-frequency Alfv{\'e}n waves in multi-ion coronal plasma: Observational
  implications}},}\ }\href {\doibase 10.1029/2004JA010969} {\bibfield
  {journal} {\bibinfo  {journal} {J.~Geophys.~Res.}\ }\textbf {\bibinfo
  {volume} {110}},\ \bibinfo {eid} {A09102} (\bibinfo {year}
  {2005})}\BibitemShut {NoStop}%
\bibitem [{\citenamefont {{Steinacker}}\ \emph {et~al.}(1997)\citenamefont
  {{Steinacker}}, \citenamefont {{Meyer}}, \citenamefont {{Steinacker}},\ and\
  \citenamefont {{Reames}}}]{Steinacker:1997}%
  \BibitemOpen
  \bibfield  {author} {\bibinfo {author} {\bibfnamefont {J.}~\bibnamefont
  {{Steinacker}}}, \bibinfo {author} {\bibfnamefont {J.-P.}\ \bibnamefont
  {{Meyer}}}, \bibinfo {author} {\bibfnamefont {A.}~\bibnamefont
  {{Steinacker}}}, \ and\ \bibinfo {author} {\bibfnamefont {D.~V.}\
  \bibnamefont {{Reames}}},\ }\bibfield  {title} {\enquote {\bibinfo {title}
  {{The Helium Valley: Comparison of Impulsive Solar Flare Ion Abundances and
  Gyroresonant Acceleration with Oblique Turbulence in a Hot Multi-Ion
  Plasma}},}\ }\href {\doibase 10.1086/303589} {\bibfield  {journal} {\bibinfo
  {journal} {Astrophys.~J.}\ }\textbf {\bibinfo {volume} {476}},\ \bibinfo
  {pages} {403--427} (\bibinfo {year} {1997})}\BibitemShut {NoStop}%
\bibitem [{\citenamefont {{Eastes}}\ \emph {et~al.}(2017)\citenamefont
  {{Eastes}}, \citenamefont {{McClintock}}, \citenamefont {{Burns}},
  \citenamefont {{Anderson}}, \citenamefont {{Andersson}}, \citenamefont
  {{Codrescu}}, \citenamefont {{Correira}}, \citenamefont {{Daniell}},
  \citenamefont {{England}}, \citenamefont {{Evans}}, \citenamefont {{Harvey}},
  \citenamefont {{Krywonos}}, \citenamefont {{Lumpe}}, \citenamefont
  {{Richmond}}, \citenamefont {{Rusch}}, \citenamefont {{Siegmund}},
  \citenamefont {{Solomon}}, \citenamefont {{Strickland}}, \citenamefont
  {{Woods}}, \citenamefont {{Aksnes}}, \citenamefont {{Budzien}}, \citenamefont
  {{Dymond}}, \citenamefont {{Eparvier}}, \citenamefont {{Martinis}},\ and\
  \citenamefont {{Oberheide}}}]{Eastes:2017}%
  \BibitemOpen
  \bibfield  {author} {\bibinfo {author} {\bibfnamefont {R.~W.}\ \bibnamefont
  {{Eastes}}}, \bibinfo {author} {\bibfnamefont {W.~E.}\ \bibnamefont
  {{McClintock}}}, \bibinfo {author} {\bibfnamefont {A.~G.}\ \bibnamefont
  {{Burns}}}, \bibinfo {author} {\bibfnamefont {D.~N.}\ \bibnamefont
  {{Anderson}}}, \bibinfo {author} {\bibfnamefont {L.}~\bibnamefont
  {{Andersson}}}, \bibinfo {author} {\bibfnamefont {M.}~\bibnamefont
  {{Codrescu}}}, \bibinfo {author} {\bibfnamefont {J.~T.}\ \bibnamefont
  {{Correira}}}, \bibinfo {author} {\bibfnamefont {R.~E.}\ \bibnamefont
  {{Daniell}}}, \bibinfo {author} {\bibfnamefont {S.~L.}\ \bibnamefont
  {{England}}}, \bibinfo {author} {\bibfnamefont {J.~S.}\ \bibnamefont
  {{Evans}}}, \bibinfo {author} {\bibfnamefont {J.}~\bibnamefont {{Harvey}}},
  \bibinfo {author} {\bibfnamefont {A.}~\bibnamefont {{Krywonos}}}, \bibinfo
  {author} {\bibfnamefont {J.~D.}\ \bibnamefont {{Lumpe}}}, \bibinfo {author}
  {\bibfnamefont {A.~D.}\ \bibnamefont {{Richmond}}}, \bibinfo {author}
  {\bibfnamefont {D.~W.}\ \bibnamefont {{Rusch}}}, \bibinfo {author}
  {\bibfnamefont {O.}~\bibnamefont {{Siegmund}}}, \bibinfo {author}
  {\bibfnamefont {S.~C.}\ \bibnamefont {{Solomon}}}, \bibinfo {author}
  {\bibfnamefont {D.~J.}\ \bibnamefont {{Strickland}}}, \bibinfo {author}
  {\bibfnamefont {T.~N.}\ \bibnamefont {{Woods}}}, \bibinfo {author}
  {\bibfnamefont {A.}~\bibnamefont {{Aksnes}}}, \bibinfo {author}
  {\bibfnamefont {S.~A.}\ \bibnamefont {{Budzien}}}, \bibinfo {author}
  {\bibfnamefont {K.~F.}\ \bibnamefont {{Dymond}}}, \bibinfo {author}
  {\bibfnamefont {F.~G.}\ \bibnamefont {{Eparvier}}}, \bibinfo {author}
  {\bibfnamefont {C.~R.}\ \bibnamefont {{Martinis}}}, \ and\ \bibinfo {author}
  {\bibfnamefont {J.}~\bibnamefont {{Oberheide}}},\ }\bibfield  {title}
  {\enquote {\bibinfo {title} {{The Global-Scale Observations of the Limb and
  Disk (GOLD) Mission}},}\ }\href {\doibase 10.1007/s11214-017-0392-2}
  {\bibfield  {journal} {\bibinfo  {journal} {Space Sci.~Rev.}\ }\textbf
  {\bibinfo {volume} {212}},\ \bibinfo {pages} {383--408} (\bibinfo {year}
  {2017})}\BibitemShut {NoStop}%
\bibitem [{\citenamefont {{Immel}}\ \emph {et~al.}(2018)\citenamefont
  {{Immel}}, \citenamefont {{England}}, \citenamefont {{Mende}}, \citenamefont
  {{Heelis}}, \citenamefont {{Englert}}, \citenamefont {{Edelstein}},
  \citenamefont {{Frey}}, \citenamefont {{Korpela}}, \citenamefont {{Taylor}},
  \citenamefont {{Craig}}, \citenamefont {{Harris}}, \citenamefont {{Bester}},
  \citenamefont {{Bust}}, \citenamefont {{Crowley}}, \citenamefont {{Forbes}},
  \citenamefont {{G{\'e}rard}}, \citenamefont {{Harlander}}, \citenamefont
  {{Huba}}, \citenamefont {{Hubert}}, \citenamefont {{Kamalabadi}},
  \citenamefont {{Makela}}, \citenamefont {{Maute}}, \citenamefont {{Meier}},
  \citenamefont {{Raftery}}, \citenamefont {{Rochus}}, \citenamefont
  {{Siegmund}}, \citenamefont {{Stephan}}, \citenamefont {{Swenson}},
  \citenamefont {{Frey}}, \citenamefont {{Hysell}}, \citenamefont {{Saito}},
  \citenamefont {{Rider}},\ and\ \citenamefont {{Sirk}}}]{Immel:2018}%
  \BibitemOpen
  \bibfield  {author} {\bibinfo {author} {\bibfnamefont {T.~J.}\ \bibnamefont
  {{Immel}}}, \bibinfo {author} {\bibfnamefont {S.~L.}\ \bibnamefont
  {{England}}}, \bibinfo {author} {\bibfnamefont {S.~B.}\ \bibnamefont
  {{Mende}}}, \bibinfo {author} {\bibfnamefont {R.~A.}\ \bibnamefont
  {{Heelis}}}, \bibinfo {author} {\bibfnamefont {C.~R.}\ \bibnamefont
  {{Englert}}}, \bibinfo {author} {\bibfnamefont {J.}~\bibnamefont
  {{Edelstein}}}, \bibinfo {author} {\bibfnamefont {H.~U.}\ \bibnamefont
  {{Frey}}}, \bibinfo {author} {\bibfnamefont {E.~J.}\ \bibnamefont
  {{Korpela}}}, \bibinfo {author} {\bibfnamefont {E.~R.}\ \bibnamefont
  {{Taylor}}}, \bibinfo {author} {\bibfnamefont {W.~W.}\ \bibnamefont
  {{Craig}}}, \bibinfo {author} {\bibfnamefont {S.~E.}\ \bibnamefont
  {{Harris}}}, \bibinfo {author} {\bibfnamefont {M.}~\bibnamefont {{Bester}}},
  \bibinfo {author} {\bibfnamefont {G.~S.}\ \bibnamefont {{Bust}}}, \bibinfo
  {author} {\bibfnamefont {G.}~\bibnamefont {{Crowley}}}, \bibinfo {author}
  {\bibfnamefont {J.~M.}\ \bibnamefont {{Forbes}}}, \bibinfo {author}
  {\bibfnamefont {J.-C.}\ \bibnamefont {{G{\'e}rard}}}, \bibinfo {author}
  {\bibfnamefont {J.~M.}\ \bibnamefont {{Harlander}}}, \bibinfo {author}
  {\bibfnamefont {J.~D.}\ \bibnamefont {{Huba}}}, \bibinfo {author}
  {\bibfnamefont {B.}~\bibnamefont {{Hubert}}}, \bibinfo {author}
  {\bibfnamefont {F.}~\bibnamefont {{Kamalabadi}}}, \bibinfo {author}
  {\bibfnamefont {J.~J.}\ \bibnamefont {{Makela}}}, \bibinfo {author}
  {\bibfnamefont {A.~I.}\ \bibnamefont {{Maute}}}, \bibinfo {author}
  {\bibfnamefont {R.~R.}\ \bibnamefont {{Meier}}}, \bibinfo {author}
  {\bibfnamefont {C.}~\bibnamefont {{Raftery}}}, \bibinfo {author}
  {\bibfnamefont {P.}~\bibnamefont {{Rochus}}}, \bibinfo {author}
  {\bibfnamefont {O.~H.~W.}\ \bibnamefont {{Siegmund}}}, \bibinfo {author}
  {\bibfnamefont {A.~W.}\ \bibnamefont {{Stephan}}}, \bibinfo {author}
  {\bibfnamefont {G.~R.}\ \bibnamefont {{Swenson}}}, \bibinfo {author}
  {\bibfnamefont {S.}~\bibnamefont {{Frey}}}, \bibinfo {author} {\bibfnamefont
  {D.~L.}\ \bibnamefont {{Hysell}}}, \bibinfo {author} {\bibfnamefont
  {A.}~\bibnamefont {{Saito}}}, \bibinfo {author} {\bibfnamefont {K.~A.}\
  \bibnamefont {{Rider}}}, \ and\ \bibinfo {author} {\bibfnamefont {M.~M.}\
  \bibnamefont {{Sirk}}},\ }\bibfield  {title} {\enquote {\bibinfo {title}
  {{The Ionospheric Connection Explorer Mission: Mission Goals and Design}},}\
  }\href {\doibase 10.1007/s11214-017-0449-2} {\bibfield  {journal} {\bibinfo
  {journal} {Space Sci.~Rev.}\ }\textbf {\bibinfo {volume} {214}},\ \bibinfo
  {eid} {13} (\bibinfo {year} {2018})}\BibitemShut {NoStop}%
\bibitem [{\citenamefont {{Arnett}}(1996)}]{Arnett:1996}%
  \BibitemOpen
  \bibfield  {author} {\bibinfo {author} {\bibfnamefont {D.}~\bibnamefont
  {{Arnett}}},\ }\href@noop {} {\emph {\bibinfo {title} {Supernovae and
  Nucleosynthesis: An Investigation of the History of Matter, from the Big Bang
  to the Present, by D.~Arnett.~Princeton: Princeton University Press,
  1996.}}}\ (\bibinfo  {publisher} {Princeton University Press},\ \bibinfo
  {year} {1996})\BibitemShut {NoStop}%
\bibitem [{\citenamefont {{Bailey}}\ \emph {et~al.}(2015)\citenamefont
  {{Bailey}}, \citenamefont {{Nagayama}}, \citenamefont {{Loisel}},
  \citenamefont {{Rochau}}, \citenamefont {{Blancard}}, \citenamefont
  {{Colgan}}, \citenamefont {{Cosse}}, \citenamefont {{Faussurier}},
  \citenamefont {{Fontes}}, \citenamefont {{Gilleron}}, \citenamefont
  {{Golovkin}}, \citenamefont {{Hansen}}, \citenamefont {{Iglesias}},
  \citenamefont {{Kilcrease}}, \citenamefont {{Macfarlane}}, \citenamefont
  {{Mancini}}, \citenamefont {{Nahar}}, \citenamefont {{Orban}}, \citenamefont
  {{Pain}}, \citenamefont {{Pradhan}}, \citenamefont {{Sherrill}},\ and\
  \citenamefont {{Wilson}}}]{Bailey:2015}%
  \BibitemOpen
  \bibfield  {author} {\bibinfo {author} {\bibfnamefont {J.~E.}\ \bibnamefont
  {{Bailey}}}, \bibinfo {author} {\bibfnamefont {T.}~\bibnamefont
  {{Nagayama}}}, \bibinfo {author} {\bibfnamefont {G.~P.}\ \bibnamefont
  {{Loisel}}}, \bibinfo {author} {\bibfnamefont {G.~A.}\ \bibnamefont
  {{Rochau}}}, \bibinfo {author} {\bibfnamefont {C.}~\bibnamefont
  {{Blancard}}}, \bibinfo {author} {\bibfnamefont {J.}~\bibnamefont
  {{Colgan}}}, \bibinfo {author} {\bibfnamefont {P.}~\bibnamefont {{Cosse}}},
  \bibinfo {author} {\bibfnamefont {G.}~\bibnamefont {{Faussurier}}}, \bibinfo
  {author} {\bibfnamefont {C.~J.}\ \bibnamefont {{Fontes}}}, \bibinfo {author}
  {\bibfnamefont {F.}~\bibnamefont {{Gilleron}}}, \bibinfo {author}
  {\bibfnamefont {I.}~\bibnamefont {{Golovkin}}}, \bibinfo {author}
  {\bibfnamefont {S.~B.}\ \bibnamefont {{Hansen}}}, \bibinfo {author}
  {\bibfnamefont {C.~A.}\ \bibnamefont {{Iglesias}}}, \bibinfo {author}
  {\bibfnamefont {D.~P.}\ \bibnamefont {{Kilcrease}}}, \bibinfo {author}
  {\bibfnamefont {J.~J.}\ \bibnamefont {{Macfarlane}}}, \bibinfo {author}
  {\bibfnamefont {R.~C.}\ \bibnamefont {{Mancini}}}, \bibinfo {author}
  {\bibfnamefont {S.~N.}\ \bibnamefont {{Nahar}}}, \bibinfo {author}
  {\bibfnamefont {C.}~\bibnamefont {{Orban}}}, \bibinfo {author} {\bibfnamefont
  {J.-C.}\ \bibnamefont {{Pain}}}, \bibinfo {author} {\bibfnamefont {A.~K.}\
  \bibnamefont {{Pradhan}}}, \bibinfo {author} {\bibfnamefont {M.}~\bibnamefont
  {{Sherrill}}}, \ and\ \bibinfo {author} {\bibfnamefont {B.~G.}\ \bibnamefont
  {{Wilson}}},\ }\bibfield  {title} {\enquote {\bibinfo {title} {{A
  higher-than-predicted measurement of iron opacity at solar interior
  temperatures}},}\ }\href {\doibase 10.1038/nature14048} {\bibfield  {journal}
  {\bibinfo  {journal} {Nature}\ }\textbf {\bibinfo {volume} {517}},\ \bibinfo
  {pages} {56--59} (\bibinfo {year} {2015})}\BibitemShut {NoStop}%
\bibitem [{\citenamefont {{Cox}}\ and\ \citenamefont
  {{Stewart}}(1970)}]{Cox:1970}%
  \BibitemOpen
  \bibfield  {author} {\bibinfo {author} {\bibfnamefont {A.~N.}\ \bibnamefont
  {{Cox}}}\ and\ \bibinfo {author} {\bibfnamefont {J.~N.}\ \bibnamefont
  {{Stewart}}},\ }\bibfield  {title} {\enquote {\bibinfo {title} {{Rosseland
  Opacity Tables for Population I Compositions}},}\ }\href {\doibase
  10.1086/190207} {\bibfield  {journal} {\bibinfo  {journal}
  {Astrophys.~J.~Supp.}\ }\textbf {\bibinfo {volume} {19}},\ \bibinfo {pages}
  {243} (\bibinfo {year} {1970})}\BibitemShut {NoStop}%
\bibitem [{\citenamefont {{Shibata}}\ and\ \citenamefont
  {{Uchida}}(1985)}]{Shibata:1985}%
  \BibitemOpen
  \bibfield  {author} {\bibinfo {author} {\bibfnamefont {K.}~\bibnamefont
  {{Shibata}}}\ and\ \bibinfo {author} {\bibfnamefont {Y.}~\bibnamefont
  {{Uchida}}},\ }\bibfield  {title} {\enquote {\bibinfo {title} {{A
  magnetodynamic mechanism for the formation of astrophysical jets. I -
  Dynamical effects of the relaxation of nonlinear magnetic twists}},}\
  }\href@noop {} {\bibfield  {journal} {\bibinfo  {journal}
  {Pub.~Astron.~Soc.~Japan}\ }\textbf {\bibinfo {volume} {37}},\ \bibinfo
  {pages} {31--46} (\bibinfo {year} {1985})}\BibitemShut {NoStop}%
\bibitem [{\citenamefont {{Shibata}}\ and\ \citenamefont
  {{Uchida}}(1986)}]{Shibata:1986}%
  \BibitemOpen
  \bibfield  {author} {\bibinfo {author} {\bibfnamefont {K.}~\bibnamefont
  {{Shibata}}}\ and\ \bibinfo {author} {\bibfnamefont {Y.}~\bibnamefont
  {{Uchida}}},\ }\bibfield  {title} {\enquote {\bibinfo {title} {{A
  magnetodynamic mechanism for the formation of astrophysical jets. II -
  Dynamical processes in the accretion of magnetized mass in rotation}},}\
  }\href@noop {} {\bibfield  {journal} {\bibinfo  {journal}
  {Pub.~Astron.~Soc.~Japan}\ }\textbf {\bibinfo {volume} {38}},\ \bibinfo
  {pages} {631--660} (\bibinfo {year} {1986})}\BibitemShut {NoStop}%
\bibitem [{\citenamefont {{Blandford}}\ and\ \citenamefont
  {{Payne}}(1982)}]{Blandford:1982}%
  \BibitemOpen
  \bibfield  {author} {\bibinfo {author} {\bibfnamefont {R.~D.}\ \bibnamefont
  {{Blandford}}}\ and\ \bibinfo {author} {\bibfnamefont {D.~G.}\ \bibnamefont
  {{Payne}}},\ }\bibfield  {title} {\enquote {\bibinfo {title} {{Hydromagnetic
  flows from accretion discs and the production of radio jets}},}\ }\href
  {\doibase 10.1093/mnras/199.4.883} {\bibfield  {journal} {\bibinfo  {journal}
  {Mon.~Not.~Roy.~Astron.~Soc.}\ }\textbf {\bibinfo {volume} {199}},\ \bibinfo
  {pages} {883--903} (\bibinfo {year} {1982})}\BibitemShut {NoStop}%
\bibitem [{\citenamefont {{Begelman}}, \citenamefont {{Blandford}},\ and\
  \citenamefont {{Rees}}(1984)}]{Begelman:1984}%
  \BibitemOpen
  \bibfield  {author} {\bibinfo {author} {\bibfnamefont {M.~C.}\ \bibnamefont
  {{Begelman}}}, \bibinfo {author} {\bibfnamefont {R.~D.}\ \bibnamefont
  {{Blandford}}}, \ and\ \bibinfo {author} {\bibfnamefont {M.~J.}\ \bibnamefont
  {{Rees}}},\ }\bibfield  {title} {\enquote {\bibinfo {title} {{Theory of
  extragalactic radio sources}},}\ }\href {\doibase 10.1103/RevModPhys.56.255}
  {\bibfield  {journal} {\bibinfo  {journal} {Rev. Mod. Phys.}\ }\textbf
  {\bibinfo {volume} {56}},\ \bibinfo {pages} {255--351} (\bibinfo {year}
  {1984})}\BibitemShut {NoStop}%
\bibitem [{\citenamefont {{McKinney}}(2006)}]{McKinney:2006}%
  \BibitemOpen
  \bibfield  {author} {\bibinfo {author} {\bibfnamefont {J.~C.}\ \bibnamefont
  {{McKinney}}},\ }\bibfield  {title} {\enquote {\bibinfo {title} {{General
  relativistic magnetohydrodynamic simulations of the jet formation and
  large-scale propagation from black hole accretion systems}},}\ }\href
  {\doibase 10.1111/j.1365-2966.2006.10256.x} {\bibfield  {journal} {\bibinfo
  {journal} {Mon.~Not.~Roy.~Astron.~Soc.}\ }\textbf {\bibinfo {volume} {368}},\
  \bibinfo {pages} {1561--1582} (\bibinfo {year} {2006})},\ \Eprint
  {http://arxiv.org/abs/astro-ph/0603045} {astro-ph/0603045} \BibitemShut
  {NoStop}%
\bibitem [{\citenamefont {{Farley}}\ \emph {et~al.}(1999)\citenamefont
  {{Farley}}, \citenamefont {{Estabrook}}, \citenamefont {{Glendinning}},
  \citenamefont {{Glenzer}}, \citenamefont {{Remington}}, \citenamefont
  {{Shigemori}}, \citenamefont {{Stone}}, \citenamefont {{Wallace}},
  \citenamefont {{Zimmerman}},\ and\ \citenamefont {{Harte}}}]{Farley:1999}%
  \BibitemOpen
  \bibfield  {author} {\bibinfo {author} {\bibfnamefont {D.~R.}\ \bibnamefont
  {{Farley}}}, \bibinfo {author} {\bibfnamefont {K.~G.}\ \bibnamefont
  {{Estabrook}}}, \bibinfo {author} {\bibfnamefont {S.~G.}\ \bibnamefont
  {{Glendinning}}}, \bibinfo {author} {\bibfnamefont {S.~H.}\ \bibnamefont
  {{Glenzer}}}, \bibinfo {author} {\bibfnamefont {B.~A.}\ \bibnamefont
  {{Remington}}}, \bibinfo {author} {\bibfnamefont {K.}~\bibnamefont
  {{Shigemori}}}, \bibinfo {author} {\bibfnamefont {J.~M.}\ \bibnamefont
  {{Stone}}}, \bibinfo {author} {\bibfnamefont {R.~J.}\ \bibnamefont
  {{Wallace}}}, \bibinfo {author} {\bibfnamefont {G.~B.}\ \bibnamefont
  {{Zimmerman}}}, \ and\ \bibinfo {author} {\bibfnamefont {J.~A.}\ \bibnamefont
  {{Harte}}},\ }\bibfield  {title} {\enquote {\bibinfo {title} {{Radiative Jet
  Experiments of Astrophysical Interest Using Intense Lasers}},}\ }\href
  {\doibase 10.1103/PhysRevLett.83.1982} {\bibfield  {journal} {\bibinfo
  {journal} {Phys.~Rev.~Lett.}\ }\textbf {\bibinfo {volume} {83}},\ \bibinfo
  {pages} {1982--1985} (\bibinfo {year} {1999})}\BibitemShut {NoStop}%
\bibitem [{\citenamefont {{Pringle}}(1981)}]{Pringle:1981}%
  \BibitemOpen
  \bibfield  {author} {\bibinfo {author} {\bibfnamefont {J.~E.}\ \bibnamefont
  {{Pringle}}},\ }\bibfield  {title} {\enquote {\bibinfo {title} {{Accretion
  discs in astrophysics}},}\ }\href {\doibase
  10.1146/annurev.aa.19.090181.001033} {\bibfield  {journal} {\bibinfo
  {journal} {Ann.~Rev.~Astron.~Astrophys.}\ }\textbf {\bibinfo {volume} {19}},\
  \bibinfo {pages} {137--162} (\bibinfo {year} {1981})}\BibitemShut {NoStop}%
\bibitem [{\citenamefont {{Narayan}}\ and\ \citenamefont
  {{Yi}}(1994)}]{Narayan:1994}%
  \BibitemOpen
  \bibfield  {author} {\bibinfo {author} {\bibfnamefont {R.}~\bibnamefont
  {{Narayan}}}\ and\ \bibinfo {author} {\bibfnamefont {I.}~\bibnamefont
  {{Yi}}},\ }\bibfield  {title} {\enquote {\bibinfo {title}
  {{Advection-dominated accretion: A self-similar solution}},}\ }\href
  {\doibase 10.1086/187381} {\bibfield  {journal} {\bibinfo  {journal}
  {Astrophys.~J.~Lett.}\ }\textbf {\bibinfo {volume} {428}},\ \bibinfo {pages}
  {L13--L16} (\bibinfo {year} {1994})},\ \Eprint
  {http://arxiv.org/abs/astro-ph/9403052} {astro-ph/9403052} \BibitemShut
  {NoStop}%
\bibitem [{\citenamefont {{Blandford}}\ and\ \citenamefont
  {{Begelman}}(1999)}]{Blandford:1999}%
  \BibitemOpen
  \bibfield  {author} {\bibinfo {author} {\bibfnamefont {R.~D.}\ \bibnamefont
  {{Blandford}}}\ and\ \bibinfo {author} {\bibfnamefont {M.~C.}\ \bibnamefont
  {{Begelman}}},\ }\bibfield  {title} {\enquote {\bibinfo {title} {{On the fate
  of gas accreting at a low rate on to a black hole}},}\ }\href {\doibase
  10.1046/j.1365-8711.1999.02358.x} {\bibfield  {journal} {\bibinfo  {journal}
  {Mon.~Not.~Roy.~Astron.~Soc.}\ }\textbf {\bibinfo {volume} {303}},\ \bibinfo
  {pages} {L1--L5} (\bibinfo {year} {1999})},\ \Eprint
  {http://arxiv.org/abs/astro-ph/9809083} {astro-ph/9809083} \BibitemShut
  {NoStop}%
\bibitem [{\citenamefont {{Frank}}, \citenamefont {{King}},\ and\ \citenamefont
  {{Raine}}(2002)}]{Frank:2002}%
  \BibitemOpen
  \bibfield  {author} {\bibinfo {author} {\bibfnamefont {J.}~\bibnamefont
  {{Frank}}}, \bibinfo {author} {\bibfnamefont {A.}~\bibnamefont {{King}}}, \
  and\ \bibinfo {author} {\bibfnamefont {D.~J.}\ \bibnamefont {{Raine}}},\
  }\href@noop {} {\emph {\bibinfo {title} {Accretion Power in Astrophysics, by
  Juhan Frank and Andrew King and Derek Raine, pp.~398.~ISBN
  0521620538.~Cambridge, UK: Cambridge University Press, February 2002.}}}\
  (\bibinfo  {publisher} {Cambridge University Press},\ \bibinfo {year}
  {2002})\ p.\ \bibinfo {pages} {398}\BibitemShut {NoStop}%
\bibitem [{\citenamefont {Balbus}\ and\ \citenamefont
  {Hawley}(1991)}]{Balbus:1991}%
  \BibitemOpen
  \bibfield  {author} {\bibinfo {author} {\bibfnamefont {S.~A.}\ \bibnamefont
  {Balbus}}\ and\ \bibinfo {author} {\bibfnamefont {J.~F.}\ \bibnamefont
  {Hawley}},\ }\bibfield  {title} {\enquote {\bibinfo {title} {A powerful local
  shear instability in weakly magnetized disks. i. linear analysis},}\
  }\href@noop {} {\bibfield  {journal} {\bibinfo  {journal} {Astrophys.~J.}\
  }\textbf {\bibinfo {volume} {376}},\ \bibinfo {pages} {214--222} (\bibinfo
  {year} {1991})}\BibitemShut {NoStop}%
\bibitem [{\citenamefont {{Hawley}}\ and\ \citenamefont
  {{Balbus}}(1991)}]{Hawley:1991}%
  \BibitemOpen
  \bibfield  {author} {\bibinfo {author} {\bibfnamefont {J.~F.}\ \bibnamefont
  {{Hawley}}}\ and\ \bibinfo {author} {\bibfnamefont {S.~A.}\ \bibnamefont
  {{Balbus}}},\ }\bibfield  {title} {\enquote {\bibinfo {title} {{A Powerful
  Local Shear Instability in Weakly Magnetized Disks. II. Nonlinear
  Evolution}},}\ }\href {\doibase 10.1086/170271} {\bibfield  {journal}
  {\bibinfo  {journal} {Astrophys.~J.}\ }\textbf {\bibinfo {volume} {376}},\
  \bibinfo {pages} {223} (\bibinfo {year} {1991})}\BibitemShut {NoStop}%
\bibitem [{\citenamefont {{Balbus}}\ and\ \citenamefont
  {{Hawley}}(1998)}]{Balbus:1998}%
  \BibitemOpen
  \bibfield  {author} {\bibinfo {author} {\bibfnamefont {S.~A.}\ \bibnamefont
  {{Balbus}}}\ and\ \bibinfo {author} {\bibfnamefont {J.~F.}\ \bibnamefont
  {{Hawley}}},\ }\bibfield  {title} {\enquote {\bibinfo {title} {{Instability,
  turbulence, and enhanced transport in accretion disks}},}\ }\href@noop {}
  {\bibfield  {journal} {\bibinfo  {journal} {Reviews of Modern Physics}\
  }\textbf {\bibinfo {volume} {70}},\ \bibinfo {pages} {1--53} (\bibinfo {year}
  {1998})}\BibitemShut {NoStop}%
\bibitem [{\citenamefont {{Quataert}}, \citenamefont {{Dorland}},\ and\
  \citenamefont {{Hammett}}(2002)}]{Quataert:2002}%
  \BibitemOpen
  \bibfield  {author} {\bibinfo {author} {\bibfnamefont {E.}~\bibnamefont
  {{Quataert}}}, \bibinfo {author} {\bibfnamefont {W.}~\bibnamefont
  {{Dorland}}}, \ and\ \bibinfo {author} {\bibfnamefont {G.~W.}\ \bibnamefont
  {{Hammett}}},\ }\bibfield  {title} {\enquote {\bibinfo {title} {{The
  Magnetorotational Instability in a Collisionless Plasma}},}\ }\href {\doibase
  10.1086/342174} {\bibfield  {journal} {\bibinfo  {journal} {Astrophys.~J.}\
  }\textbf {\bibinfo {volume} {577}},\ \bibinfo {pages} {524--533} (\bibinfo
  {year} {2002})},\ \Eprint {http://arxiv.org/abs/astro-ph/0205492}
  {astro-ph/0205492} \BibitemShut {NoStop}%
\bibitem [{\citenamefont {{Sharma}}\ \emph {et~al.}(2006)\citenamefont
  {{Sharma}}, \citenamefont {{Hammett}}, \citenamefont {{Quataert}},\ and\
  \citenamefont {{Stone}}}]{Sharma:2006}%
  \BibitemOpen
  \bibfield  {author} {\bibinfo {author} {\bibfnamefont {P.}~\bibnamefont
  {{Sharma}}}, \bibinfo {author} {\bibfnamefont {G.~W.}\ \bibnamefont
  {{Hammett}}}, \bibinfo {author} {\bibfnamefont {E.}~\bibnamefont
  {{Quataert}}}, \ and\ \bibinfo {author} {\bibfnamefont {J.~M.}\ \bibnamefont
  {{Stone}}},\ }\bibfield  {title} {\enquote {\bibinfo {title} {{Shearing Box
  Simulations of the MRI in a Collisionless Plasma}},}\ }\href {\doibase
  10.1086/498405} {\bibfield  {journal} {\bibinfo  {journal} {Astrophys.~J.}\
  }\textbf {\bibinfo {volume} {637}},\ \bibinfo {pages} {952--967} (\bibinfo
  {year} {2006})},\ \Eprint {http://arxiv.org/abs/astro-ph/0508502}
  {astro-ph/0508502} \BibitemShut {NoStop}%
\bibitem [{\citenamefont {{Sharma}}\ \emph {et~al.}(2007)\citenamefont
  {{Sharma}}, \citenamefont {{Quataert}}, \citenamefont {{Hammett}},\ and\
  \citenamefont {{Stone}}}]{Sharma:2007}%
  \BibitemOpen
  \bibfield  {author} {\bibinfo {author} {\bibfnamefont {P.}~\bibnamefont
  {{Sharma}}}, \bibinfo {author} {\bibfnamefont {E.}~\bibnamefont
  {{Quataert}}}, \bibinfo {author} {\bibfnamefont {G.~W.}\ \bibnamefont
  {{Hammett}}}, \ and\ \bibinfo {author} {\bibfnamefont {J.~M.}\ \bibnamefont
  {{Stone}}},\ }\bibfield  {title} {\enquote {\bibinfo {title} {{Electron
  Heating in Hot Accretion Flows}},}\ }\href {\doibase 10.1086/520800}
  {\bibfield  {journal} {\bibinfo  {journal} {Astrophys.~J.}\ }\textbf
  {\bibinfo {volume} {667}},\ \bibinfo {pages} {714--723} (\bibinfo {year}
  {2007})},\ \Eprint {http://arxiv.org/abs/astro-ph/0703572} {astro-ph/0703572}
  \BibitemShut {NoStop}%
\bibitem [{\citenamefont {{Riquelme}}\ \emph {et~al.}(2012)\citenamefont
  {{Riquelme}}, \citenamefont {{Quataert}}, \citenamefont {{Sharma}},\ and\
  \citenamefont {{Spitkovsky}}}]{Riquelme:2012}%
  \BibitemOpen
  \bibfield  {author} {\bibinfo {author} {\bibfnamefont {M.~A.}\ \bibnamefont
  {{Riquelme}}}, \bibinfo {author} {\bibfnamefont {E.}~\bibnamefont
  {{Quataert}}}, \bibinfo {author} {\bibfnamefont {P.}~\bibnamefont
  {{Sharma}}}, \ and\ \bibinfo {author} {\bibfnamefont {A.}~\bibnamefont
  {{Spitkovsky}}},\ }\bibfield  {title} {\enquote {\bibinfo {title} {{Local
  Two-dimensional Particle-in-cell Simulations of the Collisionless
  Magnetorotational Instability}},}\ }\href {\doibase
  10.1088/0004-637X/755/1/50} {\bibfield  {journal} {\bibinfo  {journal}
  {Astrophys.~J.}\ }\textbf {\bibinfo {volume} {755}},\ \bibinfo {eid} {50}
  (\bibinfo {year} {2012})},\ \Eprint {http://arxiv.org/abs/1201.6407}
  {arXiv:1201.6407 [astro-ph.HE]} \BibitemShut {NoStop}%
\bibitem [{\citenamefont {{Hoshino}}(2013)}]{Hoshino:2013}%
  \BibitemOpen
  \bibfield  {author} {\bibinfo {author} {\bibfnamefont {M.}~\bibnamefont
  {{Hoshino}}},\ }\bibfield  {title} {\enquote {\bibinfo {title} {{Particle
  Acceleration during Magnetorotational Instability in a Collisionless
  Accretion Disk}},}\ }\href {\doibase 10.1088/0004-637X/773/2/118} {\bibfield
  {journal} {\bibinfo  {journal} {Astrophys.~J.}\ }\textbf {\bibinfo {volume}
  {773}},\ \bibinfo {eid} {118} (\bibinfo {year} {2013})},\ \Eprint
  {http://arxiv.org/abs/1306.6720} {arXiv:1306.6720 [astro-ph.HE]} \BibitemShut
  {NoStop}%
\bibitem [{\citenamefont {{Heinemann}}\ and\ \citenamefont
  {{Quataert}}(2014)}]{Heinemann:2014}%
  \BibitemOpen
  \bibfield  {author} {\bibinfo {author} {\bibfnamefont {T.}~\bibnamefont
  {{Heinemann}}}\ and\ \bibinfo {author} {\bibfnamefont {E.}~\bibnamefont
  {{Quataert}}},\ }\bibfield  {title} {\enquote {\bibinfo {title} {{Linear
  Vlasov Theory in the Shearing Sheet Approximation with Application to the
  Magneto-rotational Instability}},}\ }\href {\doibase
  10.1088/0004-637X/792/1/70} {\bibfield  {journal} {\bibinfo  {journal}
  {Astrophys.~J.}\ }\textbf {\bibinfo {volume} {792}},\ \bibinfo {eid} {70}
  (\bibinfo {year} {2014})},\ \Eprint {http://arxiv.org/abs/1405.7698}
  {arXiv:1405.7698 [astro-ph.HE]} \BibitemShut {NoStop}%
\bibitem [{\citenamefont {{Hoshino}}(2015)}]{Hoshino:2015}%
  \BibitemOpen
  \bibfield  {author} {\bibinfo {author} {\bibfnamefont {M.}~\bibnamefont
  {{Hoshino}}},\ }\bibfield  {title} {\enquote {\bibinfo {title} {{Angular
  Momentum Transport and Particle Acceleration During Magnetorotational
  Instability in a Kinetic Accretion Disk}},}\ }\href {\doibase
  10.1103/PhysRevLett.114.061101} {\bibfield  {journal} {\bibinfo  {journal}
  {Phys.~Rev.~Lett.}\ }\textbf {\bibinfo {volume} {114}},\ \bibinfo {eid}
  {061101} (\bibinfo {year} {2015})},\ \Eprint
  {http://arxiv.org/abs/1502.02452} {arXiv:1502.02452 [astro-ph.HE]}
  \BibitemShut {NoStop}%
\bibitem [{\citenamefont {{Sironi}}\ and\ \citenamefont
  {{Narayan}}(2015)}]{Sironi:2015a}%
  \BibitemOpen
  \bibfield  {author} {\bibinfo {author} {\bibfnamefont {L.}~\bibnamefont
  {{Sironi}}}\ and\ \bibinfo {author} {\bibfnamefont {R.}~\bibnamefont
  {{Narayan}}},\ }\bibfield  {title} {\enquote {\bibinfo {title} {{Electron
  Heating by the Ion Cyclotron Instability in Collisionless Accretion Flows. I.
  Compression-driven Instabilities and the Electron Heating Mechanism}},}\
  }\href {\doibase 10.1088/0004-637X/800/2/88} {\bibfield  {journal} {\bibinfo
  {journal} {Astrophys.~J.}\ }\textbf {\bibinfo {volume} {800}},\ \bibinfo
  {eid} {88} (\bibinfo {year} {2015})},\ \Eprint
  {http://arxiv.org/abs/1411.5685} {arXiv:1411.5685 [astro-ph.HE]} \BibitemShut
  {NoStop}%
\bibitem [{\citenamefont {{Sironi}}(2015)}]{Sironi:2015b}%
  \BibitemOpen
  \bibfield  {author} {\bibinfo {author} {\bibfnamefont {L.}~\bibnamefont
  {{Sironi}}},\ }\bibfield  {title} {\enquote {\bibinfo {title} {{Electron
  Heating by the Ion Cyclotron Instability in Collisionless Accretion Flows.
  II. Electron Heating Efficiency as a Function of Flow Conditions}},}\ }\href
  {\doibase 10.1088/0004-637X/800/2/89} {\bibfield  {journal} {\bibinfo
  {journal} {Astrophys.~J.}\ }\textbf {\bibinfo {volume} {800}},\ \bibinfo
  {eid} {89} (\bibinfo {year} {2015})},\ \Eprint
  {http://arxiv.org/abs/1411.6014} {arXiv:1411.6014 [astro-ph.HE]} \BibitemShut
  {NoStop}%
\bibitem [{\citenamefont {{Kunz}}, \citenamefont {{Stone}},\ and\ \citenamefont
  {{Quataert}}(2016)}]{Kunz:2016}%
  \BibitemOpen
  \bibfield  {author} {\bibinfo {author} {\bibfnamefont {M.~W.}\ \bibnamefont
  {{Kunz}}}, \bibinfo {author} {\bibfnamefont {J.~M.}\ \bibnamefont {{Stone}}},
  \ and\ \bibinfo {author} {\bibfnamefont {E.}~\bibnamefont {{Quataert}}},\
  }\bibfield  {title} {\enquote {\bibinfo {title} {{Magnetorotational
  Turbulence and Dynamo in a Collisionless Plasma}},}\ }\href {\doibase
  10.1103/PhysRevLett.117.235101} {\bibfield  {journal} {\bibinfo  {journal}
  {Phys.~Rev.~Lett.}\ }\textbf {\bibinfo {volume} {117}},\ \bibinfo {eid}
  {235101} (\bibinfo {year} {2016})},\ \Eprint
  {http://arxiv.org/abs/1608.07911} {arXiv:1608.07911 [astro-ph.HE]}
  \BibitemShut {NoStop}%
\bibitem [{\citenamefont {{Kunz}}\ and\ \citenamefont
  {{Lesur}}(2013)}]{Kunz:2013}%
  \BibitemOpen
  \bibfield  {author} {\bibinfo {author} {\bibfnamefont {M.~W.}\ \bibnamefont
  {{Kunz}}}\ and\ \bibinfo {author} {\bibfnamefont {G.}~\bibnamefont
  {{Lesur}}},\ }\bibfield  {title} {\enquote {\bibinfo {title} {{Magnetic
  self-organization in Hall-dominated magnetorotational turbulence}},}\ }\href
  {\doibase 10.1093/mnras/stt1171} {\bibfield  {journal} {\bibinfo  {journal}
  {Mon.~Not.~Roy.~Astron.~Soc.}\ }\textbf {\bibinfo {volume} {434}},\ \bibinfo
  {pages} {2295--2312} (\bibinfo {year} {2013})},\ \Eprint
  {http://arxiv.org/abs/1306.5887} {arXiv:1306.5887 [astro-ph.EP]} \BibitemShut
  {NoStop}%
\bibitem [{\citenamefont {{Kunz}}, \citenamefont {{Schekochihin}},\ and\
  \citenamefont {{Stone}}(2014)}]{Kunz:2014}%
  \BibitemOpen
  \bibfield  {author} {\bibinfo {author} {\bibfnamefont {M.~W.}\ \bibnamefont
  {{Kunz}}}, \bibinfo {author} {\bibfnamefont {A.~A.}\ \bibnamefont
  {{Schekochihin}}}, \ and\ \bibinfo {author} {\bibfnamefont {J.~M.}\
  \bibnamefont {{Stone}}},\ }\bibfield  {title} {\enquote {\bibinfo {title}
  {{Firehose and Mirror Instabilities in a Collisionless Shearing Plasma}},}\
  }\href {\doibase 10.1103/PhysRevLett.112.205003} {\bibfield  {journal}
  {\bibinfo  {journal} {Phys.~Rev.~Lett.}\ }\textbf {\bibinfo {volume} {112}},\
  \bibinfo {eid} {205003} (\bibinfo {year} {2014})},\ \Eprint
  {http://arxiv.org/abs/1402.0010} {arXiv:1402.0010 [astro-ph.HE]} \BibitemShut
  {NoStop}%
\bibitem [{\citenamefont {{Quataert}}, \citenamefont {{Heinemann}},\ and\
  \citenamefont {{Spitkovsky}}(2015)}]{Quataert:2015}%
  \BibitemOpen
  \bibfield  {author} {\bibinfo {author} {\bibfnamefont {E.}~\bibnamefont
  {{Quataert}}}, \bibinfo {author} {\bibfnamefont {T.}~\bibnamefont
  {{Heinemann}}}, \ and\ \bibinfo {author} {\bibfnamefont {A.}~\bibnamefont
  {{Spitkovsky}}},\ }\bibfield  {title} {\enquote {\bibinfo {title} {{Linear
  instabilities driven by differential rotation in very weakly magnetized
  plasmas}},}\ }\href {\doibase 10.1093/mnras/stu2483} {\bibfield  {journal}
  {\bibinfo  {journal} {Mon.~Not.~Roy.~Astron.~Soc.}\ }\textbf {\bibinfo
  {volume} {447}},\ \bibinfo {pages} {3328--3341} (\bibinfo {year} {2015})},\
  \Eprint {http://arxiv.org/abs/1412.1097} {arXiv:1412.1097} \BibitemShut
  {NoStop}%
\bibitem [{\citenamefont {{Riquelme}}, \citenamefont {{Quataert}},\ and\
  \citenamefont {{Verscharen}}(2015)}]{Riquelme:2015}%
  \BibitemOpen
  \bibfield  {author} {\bibinfo {author} {\bibfnamefont {M.~A.}\ \bibnamefont
  {{Riquelme}}}, \bibinfo {author} {\bibfnamefont {E.}~\bibnamefont
  {{Quataert}}}, \ and\ \bibinfo {author} {\bibfnamefont {D.}~\bibnamefont
  {{Verscharen}}},\ }\bibfield  {title} {\enquote {\bibinfo {title}
  {{Particle-in-cell Simulations of Continuously Driven Mirror and Ion
  Cyclotron Instabilities in High Beta Astrophysical and Heliospheric
  Plasmas}},}\ }\href {\doibase 10.1088/0004-637X/800/1/27} {\bibfield
  {journal} {\bibinfo  {journal} {Astrophys.~J.}\ }\textbf {\bibinfo {volume}
  {800}},\ \bibinfo {eid} {27} (\bibinfo {year} {2015})},\ \Eprint
  {http://arxiv.org/abs/1402.0014} {arXiv:1402.0014 [astro-ph.HE]} \BibitemShut
  {NoStop}%
\bibitem [{\citenamefont {{Melville}}, \citenamefont {{Schekochihin}},\ and\
  \citenamefont {{Kunz}}(2016)}]{Melville:2016}%
  \BibitemOpen
  \bibfield  {author} {\bibinfo {author} {\bibfnamefont {S.}~\bibnamefont
  {{Melville}}}, \bibinfo {author} {\bibfnamefont {A.~A.}\ \bibnamefont
  {{Schekochihin}}}, \ and\ \bibinfo {author} {\bibfnamefont {M.~W.}\
  \bibnamefont {{Kunz}}},\ }\bibfield  {title} {\enquote {\bibinfo {title}
  {{Pressure-anisotropy-driven microturbulence and magnetic-field evolution in
  shearing, collisionless plasma}},}\ }\href {\doibase 10.1093/mnras/stw793}
  {\bibfield  {journal} {\bibinfo  {journal} {Mon.~Not.~Roy.~Astron.~Soc.}\
  }\textbf {\bibinfo {volume} {459}},\ \bibinfo {pages} {2701--2720} (\bibinfo
  {year} {2016})},\ \Eprint {http://arxiv.org/abs/1512.08131} {arXiv:1512.08131
  [astro-ph.HE]} \BibitemShut {NoStop}%
\bibitem [{\citenamefont {{Colgate}}, \citenamefont {{Li}},\ and\ \citenamefont
  {{Pariev}}(2001)}]{Colgate:2001}%
  \BibitemOpen
  \bibfield  {author} {\bibinfo {author} {\bibfnamefont {S.~A.}\ \bibnamefont
  {{Colgate}}}, \bibinfo {author} {\bibfnamefont {H.}~\bibnamefont {{Li}}}, \
  and\ \bibinfo {author} {\bibfnamefont {V.}~\bibnamefont {{Pariev}}},\
  }\bibfield  {title} {\enquote {\bibinfo {title} {{The origin of the magnetic
  fields of the universe: The plasma astrophysics of the free energy of the
  universe}},}\ }\href {\doibase 10.1063/1.1351827} {\bibfield  {journal}
  {\bibinfo  {journal} {Phys.~Plasmas}\ }\textbf {\bibinfo {volume} {8}},\
  \bibinfo {pages} {2425--2431} (\bibinfo {year} {2001})},\ \Eprint
  {http://arxiv.org/abs/astro-ph/0012484} {astro-ph/0012484} \BibitemShut
  {NoStop}%
\bibitem [{\citenamefont {{Schekochihin}}\ and\ \citenamefont
  {{Cowley}}(2006)}]{Schekochihin:2006}%
  \BibitemOpen
  \bibfield  {author} {\bibinfo {author} {\bibfnamefont {A.~A.}\ \bibnamefont
  {{Schekochihin}}}\ and\ \bibinfo {author} {\bibfnamefont {S.~C.}\
  \bibnamefont {{Cowley}}},\ }\bibfield  {title} {\enquote {\bibinfo {title}
  {{Turbulence, magnetic fields, and plasma physics in clusters of
  galaxies}},}\ }\href {\doibase 10.1063/1.2179053} {\bibfield  {journal}
  {\bibinfo  {journal} {Phys.~Plasmas}\ }\textbf {\bibinfo {volume} {13}},\
  \bibinfo {pages} {056501--056501} (\bibinfo {year} {2006})},\ \Eprint
  {http://arxiv.org/abs/astro-ph/0601246} {astro-ph/0601246} \BibitemShut
  {NoStop}%
\bibitem [{\citenamefont {{Parrish}}\ and\ \citenamefont
  {{Quataert}}(2008)}]{Parrish:2008}%
  \BibitemOpen
  \bibfield  {author} {\bibinfo {author} {\bibfnamefont {I.~J.}\ \bibnamefont
  {{Parrish}}}\ and\ \bibinfo {author} {\bibfnamefont {E.}~\bibnamefont
  {{Quataert}}},\ }\bibfield  {title} {\enquote {\bibinfo {title} {{Nonlinear
  Simulations of the Heat-Flux-driven Buoyancy Instability and Its Implications
  for Galaxy Clusters}},}\ }\href {\doibase 10.1086/587937} {\bibfield
  {journal} {\bibinfo  {journal} {Astrophys.~J.~Lett.}\ }\textbf {\bibinfo
  {volume} {677}},\ \bibinfo {eid} {L9} (\bibinfo {year} {2008})},\ \Eprint
  {http://arxiv.org/abs/0712.3048} {arXiv:0712.3048} \BibitemShut {NoStop}%
\bibitem [{\citenamefont {{McCourt}}\ \emph {et~al.}(2011)\citenamefont
  {{McCourt}}, \citenamefont {{Parrish}}, \citenamefont {{Sharma}},\ and\
  \citenamefont {{Quataert}}}]{McCourt:2011}%
  \BibitemOpen
  \bibfield  {author} {\bibinfo {author} {\bibfnamefont {M.}~\bibnamefont
  {{McCourt}}}, \bibinfo {author} {\bibfnamefont {I.~J.}\ \bibnamefont
  {{Parrish}}}, \bibinfo {author} {\bibfnamefont {P.}~\bibnamefont {{Sharma}}},
  \ and\ \bibinfo {author} {\bibfnamefont {E.}~\bibnamefont {{Quataert}}},\
  }\bibfield  {title} {\enquote {\bibinfo {title} {{Can conduction induce
  convection? On the non-linear saturation of buoyancy instabilities in dilute
  plasmas}},}\ }\href {\doibase 10.1111/j.1365-2966.2011.18216.x} {\bibfield
  {journal} {\bibinfo  {journal} {Mon.~Not.~Roy.~Astron.~Soc.}\ }\textbf
  {\bibinfo {volume} {413}},\ \bibinfo {pages} {1295--1310} (\bibinfo {year}
  {2011})},\ \Eprint {http://arxiv.org/abs/1009.2498} {arXiv:1009.2498
  [astro-ph.CO]} \BibitemShut {NoStop}%
\bibitem [{\citenamefont {{Santos-Lima}}\ \emph {et~al.}(2014)\citenamefont
  {{Santos-Lima}}, \citenamefont {{de Gouveia Dal Pino}}, \citenamefont
  {{Kowal}}, \citenamefont {{Falceta-Gon{\c c}alves}}, \citenamefont
  {{Lazarian}},\ and\ \citenamefont {{Nakwacki}}}]{Santos-Lima:2014}%
  \BibitemOpen
  \bibfield  {author} {\bibinfo {author} {\bibfnamefont {R.}~\bibnamefont
  {{Santos-Lima}}}, \bibinfo {author} {\bibfnamefont {E.~M.}\ \bibnamefont {{de
  Gouveia Dal Pino}}}, \bibinfo {author} {\bibfnamefont {G.}~\bibnamefont
  {{Kowal}}}, \bibinfo {author} {\bibfnamefont {D.}~\bibnamefont
  {{Falceta-Gon{\c c}alves}}}, \bibinfo {author} {\bibfnamefont
  {A.}~\bibnamefont {{Lazarian}}}, \ and\ \bibinfo {author} {\bibfnamefont
  {M.~S.}\ \bibnamefont {{Nakwacki}}},\ }\bibfield  {title} {\enquote {\bibinfo
  {title} {{Magnetic Field Amplification and Evolution in Turbulent
  Collisionless Magnetohydrodynamics: An Application to the Intracluster
  Medium}},}\ }\href {\doibase 10.1088/0004-637X/781/2/84} {\bibfield
  {journal} {\bibinfo  {journal} {Astrophys.~J.}\ }\textbf {\bibinfo {volume}
  {781}},\ \bibinfo {eid} {84} (\bibinfo {year} {2014})},\ \Eprint
  {http://arxiv.org/abs/1305.5654} {arXiv:1305.5654} \BibitemShut {NoStop}%
\bibitem [{\citenamefont {{Rincon}}\ \emph {et~al.}(2016)\citenamefont
  {{Rincon}}, \citenamefont {{Califano}}, \citenamefont {{Schekochihin}},\ and\
  \citenamefont {{Valentini}}}]{Rincon:2016}%
  \BibitemOpen
  \bibfield  {author} {\bibinfo {author} {\bibfnamefont {F.}~\bibnamefont
  {{Rincon}}}, \bibinfo {author} {\bibfnamefont {F.}~\bibnamefont
  {{Califano}}}, \bibinfo {author} {\bibfnamefont {A.~A.}\ \bibnamefont
  {{Schekochihin}}}, \ and\ \bibinfo {author} {\bibfnamefont {F.}~\bibnamefont
  {{Valentini}}},\ }\bibfield  {title} {\enquote {\bibinfo {title} {{Turbulent
  dynamo in a collisionless plasma}},}\ }\href {\doibase
  10.1073/pnas.1525194113} {\bibfield  {journal} {\bibinfo  {journal}
  {Proc.~Nat.~Acad.~Sci.}\ }\textbf {\bibinfo {volume} {113}},\ \bibinfo
  {pages} {3950--3953} (\bibinfo {year} {2016})},\ \Eprint
  {http://arxiv.org/abs/1512.06455} {arXiv:1512.06455} \BibitemShut {NoStop}%
\bibitem [{\citenamefont {{Fabian}}, \citenamefont {{Nulsen}},\ and\
  \citenamefont {{Canizares}}(1984)}]{Fabian:1984}%
  \BibitemOpen
  \bibfield  {author} {\bibinfo {author} {\bibfnamefont {A.~C.}\ \bibnamefont
  {{Fabian}}}, \bibinfo {author} {\bibfnamefont {P.~E.~J.}\ \bibnamefont
  {{Nulsen}}}, \ and\ \bibinfo {author} {\bibfnamefont {C.~R.}\ \bibnamefont
  {{Canizares}}},\ }\bibfield  {title} {\enquote {\bibinfo {title} {{Cooling
  flows in clusters of galaxies}},}\ }\href {\doibase 10.1038/310733a0}
  {\bibfield  {journal} {\bibinfo  {journal} {Nature}\ }\textbf {\bibinfo
  {volume} {310}},\ \bibinfo {pages} {733--740} (\bibinfo {year}
  {1984})}\BibitemShut {NoStop}%
\bibitem [{\citenamefont {{Fabian}}(1994)}]{Fabian:1994}%
  \BibitemOpen
  \bibfield  {author} {\bibinfo {author} {\bibfnamefont {A.~C.}\ \bibnamefont
  {{Fabian}}},\ }\bibfield  {title} {\enquote {\bibinfo {title} {{Cooling Flows
  in Clusters of Galaxies}},}\ }\href {\doibase
  10.1146/annurev.aa.32.090194.001425} {\bibfield  {journal} {\bibinfo
  {journal} {Ann.~Rev.~Astron.~Astrophys.}\ }\textbf {\bibinfo {volume} {32}},\
  \bibinfo {pages} {277--318} (\bibinfo {year} {1994})}\BibitemShut {NoStop}%
\bibitem [{\citenamefont {{Carilli}}\ and\ \citenamefont
  {{Taylor}}(2002)}]{Carilli:2002}%
  \BibitemOpen
  \bibfield  {author} {\bibinfo {author} {\bibfnamefont {C.~L.}\ \bibnamefont
  {{Carilli}}}\ and\ \bibinfo {author} {\bibfnamefont {G.~B.}\ \bibnamefont
  {{Taylor}}},\ }\bibfield  {title} {\enquote {\bibinfo {title} {{Cluster
  Magnetic Fields}},}\ }\href {\doibase 10.1146/annurev.astro.40.060401.093852}
  {\bibfield  {journal} {\bibinfo  {journal} {Ann.~Rev.~Astron.~Astrophys.}\
  }\textbf {\bibinfo {volume} {40}},\ \bibinfo {pages} {319--348} (\bibinfo
  {year} {2002})},\ \Eprint {http://arxiv.org/abs/astro-ph/0110655}
  {astro-ph/0110655} \BibitemShut {NoStop}%
\bibitem [{\citenamefont {{Chandran}}\ and\ \citenamefont
  {{Cowley}}(1998)}]{Chandran:1998}%
  \BibitemOpen
  \bibfield  {author} {\bibinfo {author} {\bibfnamefont {B.~D.~G.}\
  \bibnamefont {{Chandran}}}\ and\ \bibinfo {author} {\bibfnamefont {S.~C.}\
  \bibnamefont {{Cowley}}},\ }\bibfield  {title} {\enquote {\bibinfo {title}
  {{Thermal Conduction in a Tangled Magnetic Field}},}\ }\href {\doibase
  10.1103/PhysRevLett.80.3077} {\bibfield  {journal} {\bibinfo  {journal}
  {Phys.~Rev.~Lett.}\ }\textbf {\bibinfo {volume} {80}},\ \bibinfo {pages}
  {3077--3080} (\bibinfo {year} {1998})}\BibitemShut {NoStop}%
\bibitem [{\citenamefont {{Narayan}}\ and\ \citenamefont
  {{Medvedev}}(2001)}]{Narayan:2001}%
  \BibitemOpen
  \bibfield  {author} {\bibinfo {author} {\bibfnamefont {R.}~\bibnamefont
  {{Narayan}}}\ and\ \bibinfo {author} {\bibfnamefont {M.~V.}\ \bibnamefont
  {{Medvedev}}},\ }\bibfield  {title} {\enquote {\bibinfo {title} {{Thermal
  Conduction in Clusters of Galaxies}},}\ }\href {\doibase 10.1086/338325}
  {\bibfield  {journal} {\bibinfo  {journal} {Astrophys.~J.~Lett.}\ }\textbf
  {\bibinfo {volume} {562}},\ \bibinfo {pages} {L129--L132} (\bibinfo {year}
  {2001})},\ \Eprint {http://arxiv.org/abs/astro-ph/0110567} {astro-ph/0110567}
  \BibitemShut {NoStop}%
\bibitem [{\citenamefont {{Parrish}}, \citenamefont {{Quataert}},\ and\
  \citenamefont {{Sharma}}(2009)}]{Parrish:2009}%
  \BibitemOpen
  \bibfield  {author} {\bibinfo {author} {\bibfnamefont {I.~J.}\ \bibnamefont
  {{Parrish}}}, \bibinfo {author} {\bibfnamefont {E.}~\bibnamefont
  {{Quataert}}}, \ and\ \bibinfo {author} {\bibfnamefont {P.}~\bibnamefont
  {{Sharma}}},\ }\bibfield  {title} {\enquote {\bibinfo {title} {{Anisotropic
  Thermal Conduction and the Cooling Flow Problem in Galaxy Clusters}},}\
  }\href {\doibase 10.1088/0004-637X/703/1/96} {\bibfield  {journal} {\bibinfo
  {journal} {Astrophys.~J.}\ }\textbf {\bibinfo {volume} {703}},\ \bibinfo
  {pages} {96--108} (\bibinfo {year} {2009})},\ \Eprint
  {http://arxiv.org/abs/0905.4500} {arXiv:0905.4500 [astro-ph.CO]} \BibitemShut
  {NoStop}%
\bibitem [{\citenamefont {{Komarov}}\ \emph {et~al.}(2016)\citenamefont
  {{Komarov}}, \citenamefont {{Churazov}}, \citenamefont {{Kunz}},\ and\
  \citenamefont {{Schekochihin}}}]{Komarov:2016}%
  \BibitemOpen
  \bibfield  {author} {\bibinfo {author} {\bibfnamefont {S.~V.}\ \bibnamefont
  {{Komarov}}}, \bibinfo {author} {\bibfnamefont {E.~M.}\ \bibnamefont
  {{Churazov}}}, \bibinfo {author} {\bibfnamefont {M.~W.}\ \bibnamefont
  {{Kunz}}}, \ and\ \bibinfo {author} {\bibfnamefont {A.~A.}\ \bibnamefont
  {{Schekochihin}}},\ }\bibfield  {title} {\enquote {\bibinfo {title} {{Thermal
  conduction in a mirror-unstable plasma}},}\ }\href {\doibase
  10.1093/mnras/stw963} {\bibfield  {journal} {\bibinfo  {journal}
  {Mon.~Not.~Roy.~Astron.~Soc.}\ }\textbf {\bibinfo {volume} {460}},\ \bibinfo
  {pages} {467--477} (\bibinfo {year} {2016})},\ \Eprint
  {http://arxiv.org/abs/1603.00524} {arXiv:1603.00524 [astro-ph.HE]}
  \BibitemShut {NoStop}%
\bibitem [{\citenamefont {{Balbus}}(2000)}]{Balbus:2000}%
  \BibitemOpen
  \bibfield  {author} {\bibinfo {author} {\bibfnamefont {S.~A.}\ \bibnamefont
  {{Balbus}}},\ }\bibfield  {title} {\enquote {\bibinfo {title} {{Stability,
  Instability, and ``Backward'' Transport in Stratified Fluids}},}\ }\href
  {\doibase 10.1086/308732} {\bibfield  {journal} {\bibinfo  {journal}
  {Astrophys.~J.}\ }\textbf {\bibinfo {volume} {534}},\ \bibinfo {pages}
  {420--427} (\bibinfo {year} {2000})},\ \Eprint
  {http://arxiv.org/abs/astro-ph/9906315} {astro-ph/9906315} \BibitemShut
  {NoStop}%
\bibitem [{\citenamefont {{Parrish}}\ and\ \citenamefont
  {{Stone}}(2005)}]{Parrish:2005}%
  \BibitemOpen
  \bibfield  {author} {\bibinfo {author} {\bibfnamefont {I.~J.}\ \bibnamefont
  {{Parrish}}}\ and\ \bibinfo {author} {\bibfnamefont {J.~M.}\ \bibnamefont
  {{Stone}}},\ }\bibfield  {title} {\enquote {\bibinfo {title} {{Nonlinear
  Evolution of the Magnetothermal Instability in Two Dimensions}},}\ }\href
  {\doibase 10.1086/444589} {\bibfield  {journal} {\bibinfo  {journal}
  {Astrophys.~J.}\ }\textbf {\bibinfo {volume} {633}},\ \bibinfo {pages}
  {334--348} (\bibinfo {year} {2005})},\ \Eprint
  {http://arxiv.org/abs/astro-ph/0507212} {astro-ph/0507212} \BibitemShut
  {NoStop}%
\bibitem [{\citenamefont {{Parrish}}\ and\ \citenamefont
  {{Stone}}(2007)}]{Parrish:2007}%
  \BibitemOpen
  \bibfield  {author} {\bibinfo {author} {\bibfnamefont {I.~J.}\ \bibnamefont
  {{Parrish}}}\ and\ \bibinfo {author} {\bibfnamefont {J.~M.}\ \bibnamefont
  {{Stone}}},\ }\bibfield  {title} {\enquote {\bibinfo {title} {{Saturation of
  the Magnetothermal Instability in Three Dimensions}},}\ }\href {\doibase
  10.1086/518881} {\bibfield  {journal} {\bibinfo  {journal} {Astrophys.~J.}\
  }\textbf {\bibinfo {volume} {664}},\ \bibinfo {pages} {135--148} (\bibinfo
  {year} {2007})},\ \Eprint {http://arxiv.org/abs/astro-ph/0612195}
  {astro-ph/0612195} \BibitemShut {NoStop}%
\bibitem [{\citenamefont {{Quataert}}(2008)}]{Quataert:2008}%
  \BibitemOpen
  \bibfield  {author} {\bibinfo {author} {\bibfnamefont {E.}~\bibnamefont
  {{Quataert}}},\ }\bibfield  {title} {\enquote {\bibinfo {title} {{Buoyancy
  Instabilities in Weakly Magnetized Low-Collisionality Plasmas}},}\ }\href
  {\doibase 10.1086/525248} {\bibfield  {journal} {\bibinfo  {journal}
  {Astrophys.~J.}\ }\textbf {\bibinfo {volume} {673}},\ \bibinfo {eid}
  {758-762} (\bibinfo {year} {2008})},\ \Eprint
  {http://arxiv.org/abs/0710.5521} {arXiv:0710.5521} \BibitemShut {NoStop}%
\bibitem [{\citenamefont {{Marsch}}\ \emph {et~al.}(1982)\citenamefont
  {{Marsch}}, \citenamefont {{Schwenn}}, \citenamefont {{Rosenbauer}},
  \citenamefont {{Muehlhaeuser}}, \citenamefont {{Pilipp}},\ and\ \citenamefont
  {{Neubauer}}}]{Marsch:1982}%
  \BibitemOpen
  \bibfield  {author} {\bibinfo {author} {\bibfnamefont {E.}~\bibnamefont
  {{Marsch}}}, \bibinfo {author} {\bibfnamefont {R.}~\bibnamefont {{Schwenn}}},
  \bibinfo {author} {\bibfnamefont {H.}~\bibnamefont {{Rosenbauer}}}, \bibinfo
  {author} {\bibfnamefont {K.-H.}\ \bibnamefont {{Muehlhaeuser}}}, \bibinfo
  {author} {\bibfnamefont {W.}~\bibnamefont {{Pilipp}}}, \ and\ \bibinfo
  {author} {\bibfnamefont {F.~M.}\ \bibnamefont {{Neubauer}}},\ }\bibfield
  {title} {\enquote {\bibinfo {title} {{Solar wind protons - Three-dimensional
  velocity distributions and derived plasma parameters measured between 0.3 and
  1 AU}},}\ }\href@noop {} {\bibfield  {journal} {\bibinfo  {journal}
  {J.~Geophys.~Res.}\ }\textbf {\bibinfo {volume} {87}},\ \bibinfo {pages}
  {52--72} (\bibinfo {year} {1982})}\BibitemShut {NoStop}%
\bibitem [{\citenamefont {{Marsch}}(1991)}]{Marsch:1991}%
  \BibitemOpen
  \bibfield  {author} {\bibinfo {author} {\bibfnamefont {E.}~\bibnamefont
  {{Marsch}}},\ }\bibfield  {title} {\enquote {\bibinfo {title} {{Kinetic
  Physics of the Solar Wind Plasma}},}\ }in\ \href@noop {} {\emph {\bibinfo
  {booktitle} {{Physics of the Inner Heliosphere II. Particles, Waves and
  Turbulence.}}}},\ \bibinfo {editor} {edited by\ \bibinfo {editor}
  {\bibfnamefont {R.}~\bibnamefont {{Schwenn}}}\ and\ \bibinfo {editor}
  {\bibfnamefont {E.}~\bibnamefont {{Marsch}}}}\ (\bibinfo  {publisher}
  {{Springer-Verlag}},\ \bibinfo {address} {{Berlin}},\ \bibinfo {year}
  {1991})\ pp.\ \bibinfo {pages} {45--133}\BibitemShut {NoStop}%
\bibitem [{\citenamefont {{Marsch}}, \citenamefont {{Ao}},\ and\ \citenamefont
  {{Tu}}(2004)}]{Marsch:2004}%
  \BibitemOpen
  \bibfield  {author} {\bibinfo {author} {\bibfnamefont {E.}~\bibnamefont
  {{Marsch}}}, \bibinfo {author} {\bibfnamefont {X.-Z.}\ \bibnamefont {{Ao}}},
  \ and\ \bibinfo {author} {\bibfnamefont {C.-Y.}\ \bibnamefont {{Tu}}},\
  }\bibfield  {title} {\enquote {\bibinfo {title} {{On the temperature
  anisotropy of the core part of the proton velocity distribution function in
  the solar wind}},}\ }\href {\doibase 10.1029/2003JA010330} {\bibfield
  {journal} {\bibinfo  {journal} {J.~Geophys.~Res.}\ }\textbf {\bibinfo
  {volume} {109}},\ \bibinfo {eid} {A04102} (\bibinfo {year}
  {2004})}\BibitemShut {NoStop}%
\bibitem [{\citenamefont {Marsch}(2006)}]{Marsch:2006}%
  \BibitemOpen
  \bibfield  {author} {\bibinfo {author} {\bibfnamefont {E.}~\bibnamefont
  {Marsch}},\ }\bibfield  {title} {\enquote {\bibinfo {title} {Kinetic physics
  of the solar corona and solar wind},}\ }\href
  {http://www.livingreviews.org/lrsp-2006-1} {\bibfield  {journal} {\bibinfo
  {journal} {Living Rev.~Solar Phys.}\ }\textbf {\bibinfo {volume} {3}},\
  \bibinfo {pages} {1} (\bibinfo {year} {2006})}\BibitemShut {NoStop}%
\bibitem [{\citenamefont {{Howes}}(2017)}]{Howes:2017c}%
  \BibitemOpen
  \bibfield  {author} {\bibinfo {author} {\bibfnamefont {G.~G.}\ \bibnamefont
  {{Howes}}},\ }\bibfield  {title} {\enquote {\bibinfo {title} {A prospectus on
  kinetic heliophysics},}\ }\href {\doibase 10.1063/1.4983993} {\bibfield
  {journal} {\bibinfo  {journal} {Phys.~Plasmas}\ }\textbf {\bibinfo {volume}
  {24}},\ \bibinfo {pages} {055907} (\bibinfo {year} {2017})}\BibitemShut
  {NoStop}%
\bibitem [{\citenamefont {{Thomas}}\ \emph
  {et~al.}(2015{\natexlab{b}})\citenamefont {{Thomas}}, \citenamefont
  {{Konopka}}, \citenamefont {{Lynch}}, \citenamefont {{Adams}}, \citenamefont
  {{LeBlanc}}, \citenamefont {{Merlino}},\ and\ \citenamefont
  {{Rosenberg}}}]{Thomas:2015b}%
  \BibitemOpen
  \bibfield  {author} {\bibinfo {author} {\bibfnamefont {E.}~\bibnamefont
  {{Thomas}}}, \bibinfo {author} {\bibfnamefont {U.}~\bibnamefont {{Konopka}}},
  \bibinfo {author} {\bibfnamefont {B.}~\bibnamefont {{Lynch}}}, \bibinfo
  {author} {\bibfnamefont {S.}~\bibnamefont {{Adams}}}, \bibinfo {author}
  {\bibfnamefont {S.}~\bibnamefont {{LeBlanc}}}, \bibinfo {author}
  {\bibfnamefont {R.~L.}\ \bibnamefont {{Merlino}}}, \ and\ \bibinfo {author}
  {\bibfnamefont {M.}~\bibnamefont {{Rosenberg}}},\ }\bibfield  {title}
  {\enquote {\bibinfo {title} {{Quasi-discrete particle motion in an externally
  imposed, ordered structure in a dusty plasma at high magnetic field}},}\
  }\href {\doibase 10.1063/1.4936244} {\bibfield  {journal} {\bibinfo
  {journal} {Physics of Plasmas}\ }\textbf {\bibinfo {volume} {22}},\ \bibinfo
  {eid} {113708} (\bibinfo {year} {2015}{\natexlab{b}})}\BibitemShut {NoStop}%
\bibitem [{\citenamefont {{Thomas}}\ \emph {et~al.}(2016)\citenamefont
  {{Thomas}}, \citenamefont {{Konopka}}, \citenamefont {{Merlino}},\ and\
  \citenamefont {{Rosenberg}}}]{Thomas:2016}%
  \BibitemOpen
  \bibfield  {author} {\bibinfo {author} {\bibfnamefont {E.}~\bibnamefont
  {{Thomas}}}, \bibinfo {author} {\bibfnamefont {U.}~\bibnamefont {{Konopka}}},
  \bibinfo {author} {\bibfnamefont {R.~L.}\ \bibnamefont {{Merlino}}}, \ and\
  \bibinfo {author} {\bibfnamefont {M.}~\bibnamefont {{Rosenberg}}},\
  }\bibfield  {title} {\enquote {\bibinfo {title} {{Initial measurements of
  two- and three-dimensional ordering, waves, and plasma filamentation in the
  Magnetized Dusty Plasma Experiment}},}\ }\href {\doibase 10.1063/1.4943112}
  {\bibfield  {journal} {\bibinfo  {journal} {Phys.~Plasmas}\ }\textbf
  {\bibinfo {volume} {23}},\ \bibinfo {eid} {055701} (\bibinfo {year}
  {2016})}\BibitemShut {NoStop}%
\bibitem [{\citenamefont {{Jaiswal}}\ \emph {et~al.}(2017)\citenamefont
  {{Jaiswal}}, \citenamefont {{Hall}}, \citenamefont {{LeBlanc}}, \citenamefont
  {{Mukherjee}},\ and\ \citenamefont {{Thomas}}}]{Jaiswal:2017}%
  \BibitemOpen
  \bibfield  {author} {\bibinfo {author} {\bibfnamefont {S.}~\bibnamefont
  {{Jaiswal}}}, \bibinfo {author} {\bibfnamefont {T.}~\bibnamefont {{Hall}}},
  \bibinfo {author} {\bibfnamefont {S.}~\bibnamefont {{LeBlanc}}}, \bibinfo
  {author} {\bibfnamefont {R.}~\bibnamefont {{Mukherjee}}}, \ and\ \bibinfo
  {author} {\bibfnamefont {E.}~\bibnamefont {{Thomas}}},\ }\bibfield  {title}
  {\enquote {\bibinfo {title} {{Effect of magnetic field on the phase
  transition in a dusty plasma}},}\ }\href {\doibase 10.1063/1.5003972}
  {\bibfield  {journal} {\bibinfo  {journal} {Phys.~Plasmas}\ }\textbf
  {\bibinfo {volume} {24}},\ \bibinfo {eid} {113703} (\bibinfo {year}
  {2017})},\ \Eprint {http://arxiv.org/abs/1709.03744} {arXiv:1709.03744
  [physics.plasm-ph]} \BibitemShut {NoStop}%
\bibitem [{\citenamefont {Amatucci}\ \emph {et~al.}(1998)\citenamefont
  {Amatucci}, \citenamefont {Walker}, \citenamefont {Ganguli}, \citenamefont
  {Duncan}, \citenamefont {Antoniades}, \citenamefont {Bowles}, \citenamefont
  {Gavrishchaka},\ and\ \citenamefont {Koepke}}]{Amatucci:1998}%
  \BibitemOpen
  \bibfield  {author} {\bibinfo {author} {\bibfnamefont {W.~E.}\ \bibnamefont
  {Amatucci}}, \bibinfo {author} {\bibfnamefont {D.~N.}\ \bibnamefont
  {Walker}}, \bibinfo {author} {\bibfnamefont {G.}~\bibnamefont {Ganguli}},
  \bibinfo {author} {\bibfnamefont {D.}~\bibnamefont {Duncan}}, \bibinfo
  {author} {\bibfnamefont {J.~A.}\ \bibnamefont {Antoniades}}, \bibinfo
  {author} {\bibfnamefont {J.~H.}\ \bibnamefont {Bowles}}, \bibinfo {author}
  {\bibfnamefont {V.}~\bibnamefont {Gavrishchaka}}, \ and\ \bibinfo {author}
  {\bibfnamefont {M.~E.}\ \bibnamefont {Koepke}},\ }\bibfield  {title}
  {\enquote {\bibinfo {title} {Velocity-shear-driven ion-cyclotron waves and
  associated transverse ion heating},}\ }\href@noop {} {\bibfield  {journal}
  {\bibinfo  {journal} {J.~Geophys.~Res.}\ }\textbf {\bibinfo {volume} {103}},\
  \bibinfo {pages} {11711} (\bibinfo {year} {1998})}\BibitemShut {NoStop}%
\bibitem [{\citenamefont {Amatucci}\ \emph {et~al.}(1999)\citenamefont
  {Amatucci}, \citenamefont {Ganguli}, \citenamefont {Walker},\ and\
  \citenamefont {Duncan}}]{Amatucci:1999}%
  \BibitemOpen
  \bibfield  {author} {\bibinfo {author} {\bibfnamefont {W.~E.}\ \bibnamefont
  {Amatucci}}, \bibinfo {author} {\bibfnamefont {G.}~\bibnamefont {Ganguli}},
  \bibinfo {author} {\bibfnamefont {D.~N.}\ \bibnamefont {Walker}}, \ and\
  \bibinfo {author} {\bibfnamefont {D.}~\bibnamefont {Duncan}},\ }\bibfield
  {title} {\enquote {\bibinfo {title} {Wave and joule heating in a rotating
  plasma},}\ }\href {\doibase 10.1063/1.873207} {\bibfield  {journal} {\bibinfo
   {journal} {Phys.~Plasmas}\ }\textbf {\bibinfo {volume} {6}},\ \bibinfo
  {pages} {619--622} (\bibinfo {year} {1999})}\BibitemShut {NoStop}%
\bibitem [{\citenamefont {{Bellan}}, \citenamefont {{You}},\ and\ \citenamefont
  {{Hsu}}(2005)}]{Bellan:2005}%
  \BibitemOpen
  \bibfield  {author} {\bibinfo {author} {\bibfnamefont {P.~M.}\ \bibnamefont
  {{Bellan}}}, \bibinfo {author} {\bibfnamefont {S.}~\bibnamefont {{You}}}, \
  and\ \bibinfo {author} {\bibfnamefont {S.~C.}\ \bibnamefont {{Hsu}}},\
  }\bibfield  {title} {\enquote {\bibinfo {title} {{Simulating Astrophysical
  Jets in Laboratory Experiments}},}\ }\href {\doibase
  10.1007/s10509-005-3933-1} {\bibfield  {journal} {\bibinfo  {journal}
  {Astrophys. Space Sci.}\ }\textbf {\bibinfo {volume} {298}},\ \bibinfo
  {pages} {203--209} (\bibinfo {year} {2005})}\BibitemShut {NoStop}%
\bibitem [{\citenamefont {{Gary}}\ and\ \citenamefont
  {{Lee}}(1994)}]{Gary:1994}%
  \BibitemOpen
  \bibfield  {author} {\bibinfo {author} {\bibfnamefont {S.~P.}\ \bibnamefont
  {{Gary}}}\ and\ \bibinfo {author} {\bibfnamefont {M.~A.}\ \bibnamefont
  {{Lee}}},\ }\bibfield  {title} {\enquote {\bibinfo {title} {{The ion
  cyclotron anisotropy instability and the inverse correlation between proton
  anisotropy and proton beta}},}\ }\href {\doibase 10.1029/94JA00253}
  {\bibfield  {journal} {\bibinfo  {journal} {J.~Geophys.~Res.}\ }\textbf
  {\bibinfo {volume} {99}},\ \bibinfo {pages} {11297--11302} (\bibinfo {year}
  {1994})}\BibitemShut {NoStop}%
\bibitem [{\citenamefont {{Ulibarri}}\ \emph {et~al.}(2017)\citenamefont
  {{Ulibarri}}, \citenamefont {{Han}}, \citenamefont {{Hor{\'a}nyi}},
  \citenamefont {{Munsat}}, \citenamefont {{Wang}}, \citenamefont
  {{Whittall-Scherfee}},\ and\ \citenamefont {{Yeo}}}]{Ulibarri:2017}%
  \BibitemOpen
  \bibfield  {author} {\bibinfo {author} {\bibfnamefont {Z.}~\bibnamefont
  {{Ulibarri}}}, \bibinfo {author} {\bibfnamefont {J.}~\bibnamefont {{Han}}},
  \bibinfo {author} {\bibfnamefont {M.}~\bibnamefont {{Hor{\'a}nyi}}}, \bibinfo
  {author} {\bibfnamefont {T.}~\bibnamefont {{Munsat}}}, \bibinfo {author}
  {\bibfnamefont {X.}~\bibnamefont {{Wang}}}, \bibinfo {author} {\bibfnamefont
  {G.}~\bibnamefont {{Whittall-Scherfee}}}, \ and\ \bibinfo {author}
  {\bibfnamefont {L.~H.}\ \bibnamefont {{Yeo}}},\ }\bibfield  {title} {\enquote
  {\bibinfo {title} {{A large ion beam device for laboratory solar wind
  studies}},}\ }\href {\doibase 10.1063/1.5011785} {\bibfield  {journal}
  {\bibinfo  {journal} {Rev. Sci. Instrum.}\ }\textbf {\bibinfo {volume}
  {88}},\ \bibinfo {eid} {115112} (\bibinfo {year} {2017})}\BibitemShut
  {NoStop}%
\bibitem [{\citenamefont {{Spong}}\ \emph {et~al.}(2017)\citenamefont
  {{Spong}}, \citenamefont {{Heidbrink}}, \citenamefont {C.~{Paz-Soldan}},
  \citenamefont {{Du}}, \citenamefont {{Thome}}, \citenamefont {{Van Zeeland}},
  \citenamefont {{Collins}}, \citenamefont {{Lvovskiy}}, \citenamefont
  {{Moyer}}, \citenamefont {{Brennan}}, \citenamefont {{Liu}}, \citenamefont
  {{Jaeger}},\ and\ \citenamefont {{Lau}}}]{Spong:2018}%
  \BibitemOpen
  \bibfield  {author} {\bibinfo {author} {\bibfnamefont {D.~A.}\ \bibnamefont
  {{Spong}}}, \bibinfo {author} {\bibfnamefont {W.~W.}\ \bibnamefont
  {{Heidbrink}}}, \bibinfo {author} {\bibfnamefont {C.}~\bibnamefont
  {C.~{Paz-Soldan}}}, \bibinfo {author} {\bibfnamefont {X.~D.}\ \bibnamefont
  {{Du}}}, \bibinfo {author} {\bibfnamefont {K.~E.}\ \bibnamefont {{Thome}}},
  \bibinfo {author} {\bibfnamefont {M.~A.}\ \bibnamefont {{Van Zeeland}}},
  \bibinfo {author} {\bibfnamefont {C.}~\bibnamefont {{Collins}}}, \bibinfo
  {author} {\bibfnamefont {A.}~\bibnamefont {{Lvovskiy}}}, \bibinfo {author}
  {\bibfnamefont {R.~A.}\ \bibnamefont {{Moyer}}}, \bibinfo {author}
  {\bibfnamefont {D.~P.}\ \bibnamefont {{Brennan}}}, \bibinfo {author}
  {\bibfnamefont {C.}~\bibnamefont {{Liu}}}, \bibinfo {author} {\bibfnamefont
  {E.~F.}\ \bibnamefont {{Jaeger}}}, \ and\ \bibinfo {author} {\bibfnamefont
  {C.}~\bibnamefont {{Lau}}},\ }\bibfield  {title} {\enquote {\bibinfo {title}
  {{First direct observation of runaway electron-driven whistler waves in
  tokamaks}},}\ }\href@noop {} {\bibfield  {journal} {\bibinfo  {journal}
  {Phys.~Rev.~Lett.}\ } (\bibinfo {year} {2017})},\ \bibinfo {note}
  {submitted}\BibitemShut {NoStop}%
\bibitem [{\citenamefont {{Berisford}}\ \emph {et~al.}(2016)\citenamefont
  {{Berisford}}, \citenamefont {{Foster}}, \citenamefont {{Poston}},\ and\
  \citenamefont {{Hand}}}]{Berisford:2016}%
  \BibitemOpen
  \bibfield  {author} {\bibinfo {author} {\bibfnamefont {D.~F.}\ \bibnamefont
  {{Berisford}}}, \bibinfo {author} {\bibfnamefont {J.}~\bibnamefont
  {{Foster}}}, \bibinfo {author} {\bibfnamefont {M.~J.}\ \bibnamefont
  {{Poston}}}, \ and\ \bibinfo {author} {\bibfnamefont {K.~P.}\ \bibnamefont
  {{Hand}}},\ }\bibfield  {title} {\enquote {\bibinfo {title} {{Cryogenic Ices
  Under Vacuum: Preliminary Tests Related to Sampling Material on Europa's
  Surface}},}\ }in\ \href@noop {} {\emph {\bibinfo {booktitle} {Lunar and
  Planetary Science Conference}}},\ \bibinfo {series} {Lunar and Planetary
  Science Conference}, Vol.~\bibinfo {volume} {47}\ (\bibinfo {year} {2016})\
  p.\ \bibinfo {pages} {2998}\BibitemShut {NoStop}%
\bibitem [{\citenamefont {{Berisford}}\ \emph {et~al.}(2017)\citenamefont
  {{Berisford}}, \citenamefont {{Furst}}, \citenamefont {{Foster}},
  \citenamefont {{Hofmann}},\ and\ \citenamefont {{Hand}}}]{Berisford:2017}%
  \BibitemOpen
  \bibfield  {author} {\bibinfo {author} {\bibfnamefont {D.~F.}\ \bibnamefont
  {{Berisford}}}, \bibinfo {author} {\bibfnamefont {B.}~\bibnamefont
  {{Furst}}}, \bibinfo {author} {\bibfnamefont {J.}~\bibnamefont {{Foster}}},
  \bibinfo {author} {\bibfnamefont {A.}~\bibnamefont {{Hofmann}}}, \ and\
  \bibinfo {author} {\bibfnamefont {K.~P.}\ \bibnamefont {{Hand}}},\ }\bibfield
   {title} {\enquote {\bibinfo {title} {{Penitent Ice on Europa? Laboratory
  Testing of Cryogenic Ices Related to Icy Moon Surfaces}},}\ }in\ \href@noop
  {} {\emph {\bibinfo {booktitle} {Lunar and Planetary Science Conference}}},\
  \bibinfo {series} {Lunar and Planetary Science Conference}, Vol.~\bibinfo
  {volume} {48}\ (\bibinfo {year} {2017})\ p.\ \bibinfo {pages}
  {2581}\BibitemShut {NoStop}%
\bibitem [{\citenamefont {{Thomas}}\ \emph {et~al.}(2017)\citenamefont
  {{Thomas}}, \citenamefont {{Simolka}}, \citenamefont {{DeLuca}},
  \citenamefont {{Hor{\'a}nyi}}, \citenamefont {{Janches}}, \citenamefont
  {{Marshall}}, \citenamefont {{Munsat}}, \citenamefont {{Plane}},\ and\
  \citenamefont {{Sternovsky}}}]{Thomas:2017}%
  \BibitemOpen
  \bibfield  {author} {\bibinfo {author} {\bibfnamefont {E.}~\bibnamefont
  {{Thomas}}}, \bibinfo {author} {\bibfnamefont {J.}~\bibnamefont {{Simolka}}},
  \bibinfo {author} {\bibfnamefont {M.}~\bibnamefont {{DeLuca}}}, \bibinfo
  {author} {\bibfnamefont {M.}~\bibnamefont {{Hor{\'a}nyi}}}, \bibinfo {author}
  {\bibfnamefont {D.}~\bibnamefont {{Janches}}}, \bibinfo {author}
  {\bibfnamefont {R.~A.}\ \bibnamefont {{Marshall}}}, \bibinfo {author}
  {\bibfnamefont {T.}~\bibnamefont {{Munsat}}}, \bibinfo {author}
  {\bibfnamefont {J.~M.~C.}\ \bibnamefont {{Plane}}}, \ and\ \bibinfo {author}
  {\bibfnamefont {Z.}~\bibnamefont {{Sternovsky}}},\ }\bibfield  {title}
  {\enquote {\bibinfo {title} {{Experimental setup for the laboratory
  investigation of micrometeoroid ablation using a dust accelerator}},}\ }\href
  {\doibase 10.1063/1.4977832} {\bibfield  {journal} {\bibinfo  {journal} {Rev.
  Sci. Instrum.}\ }\textbf {\bibinfo {volume} {88}},\ \bibinfo {eid} {034501}
  (\bibinfo {year} {2017})}\BibitemShut {NoStop}%
\bibitem [{\citenamefont {{Nelson}}\ \emph {et~al.}(2016)\citenamefont
  {{Nelson}}, \citenamefont {{Dee}}, \citenamefont {{Gudipati}}, \citenamefont
  {{Hor{\'a}nyi}}, \citenamefont {{James}}, \citenamefont {{Kempf}},
  \citenamefont {{Munsat}}, \citenamefont {{Sternovsky}},\ and\ \citenamefont
  {{Ulibarri}}}]{Nelson:2016}%
  \BibitemOpen
  \bibfield  {author} {\bibinfo {author} {\bibfnamefont {A.~O.}\ \bibnamefont
  {{Nelson}}}, \bibinfo {author} {\bibfnamefont {R.}~\bibnamefont {{Dee}}},
  \bibinfo {author} {\bibfnamefont {M.~S.}\ \bibnamefont {{Gudipati}}},
  \bibinfo {author} {\bibfnamefont {M.}~\bibnamefont {{Hor{\'a}nyi}}}, \bibinfo
  {author} {\bibfnamefont {D.}~\bibnamefont {{James}}}, \bibinfo {author}
  {\bibfnamefont {S.}~\bibnamefont {{Kempf}}}, \bibinfo {author} {\bibfnamefont
  {T.}~\bibnamefont {{Munsat}}}, \bibinfo {author} {\bibfnamefont
  {Z.}~\bibnamefont {{Sternovsky}}}, \ and\ \bibinfo {author} {\bibfnamefont
  {Z.}~\bibnamefont {{Ulibarri}}},\ }\bibfield  {title} {\enquote {\bibinfo
  {title} {{New experimental capability to investigate the hypervelocity
  micrometeoroid bombardment of cryogenic surfaces}},}\ }\href {\doibase
  10.1063/1.4941960} {\bibfield  {journal} {\bibinfo  {journal} {Rev. Sci.
  Instrum.}\ }\textbf {\bibinfo {volume} {87}},\ \bibinfo {eid} {024502}
  (\bibinfo {year} {2016})}\BibitemShut {NoStop}%
\bibitem [{\citenamefont {{Malaska}}\ \emph {et~al.}(2017)\citenamefont
  {{Malaska}}, \citenamefont {{Hodyss}}, \citenamefont {{Lunine}},
  \citenamefont {{Hayes}}, \citenamefont {{Hofgartner}}, \citenamefont
  {{Hollyday}},\ and\ \citenamefont {{Lorenz}}}]{Malaska:2017}%
  \BibitemOpen
  \bibfield  {author} {\bibinfo {author} {\bibfnamefont {M.~J.}\ \bibnamefont
  {{Malaska}}}, \bibinfo {author} {\bibfnamefont {R.}~\bibnamefont {{Hodyss}}},
  \bibinfo {author} {\bibfnamefont {J.~I.}\ \bibnamefont {{Lunine}}}, \bibinfo
  {author} {\bibfnamefont {A.~G.}\ \bibnamefont {{Hayes}}}, \bibinfo {author}
  {\bibfnamefont {J.~D.}\ \bibnamefont {{Hofgartner}}}, \bibinfo {author}
  {\bibfnamefont {G.}~\bibnamefont {{Hollyday}}}, \ and\ \bibinfo {author}
  {\bibfnamefont {R.~D.}\ \bibnamefont {{Lorenz}}},\ }\bibfield  {title}
  {\enquote {\bibinfo {title} {{Laboratory measurements of nitrogen dissolution
  in Titan lake fluids}},}\ }\href {\doibase 10.1016/j.icarus.2017.01.033}
  {\bibfield  {journal} {\bibinfo  {journal} {Icarus}\ }\textbf {\bibinfo
  {volume} {289}},\ \bibinfo {pages} {94--105} (\bibinfo {year}
  {2017})}\BibitemShut {NoStop}%
\bibitem [{\citenamefont {{Vu}}\ \emph {et~al.}(2017)\citenamefont {{Vu}},
  \citenamefont {{Hodyss}}, \citenamefont {{Johnson}},\ and\ \citenamefont
  {{Choukroun}}}]{Vu:2017}%
  \BibitemOpen
  \bibfield  {author} {\bibinfo {author} {\bibfnamefont {T.~H.}\ \bibnamefont
  {{Vu}}}, \bibinfo {author} {\bibfnamefont {R.}~\bibnamefont {{Hodyss}}},
  \bibinfo {author} {\bibfnamefont {P.~V.}\ \bibnamefont {{Johnson}}}, \ and\
  \bibinfo {author} {\bibfnamefont {M.}~\bibnamefont {{Choukroun}}},\
  }\bibfield  {title} {\enquote {\bibinfo {title} {{Preferential formation of
  sodium salts from frozen sodium-ammonium-chloride-carbonate brines -
  Implications for Ceres' bright spots}},}\ }\href {\doibase
  10.1016/j.pss.2017.04.014} {\bibfield  {journal} {\bibinfo  {journal}
  {Planet. Space Sci.}\ }\textbf {\bibinfo {volume} {141}},\ \bibinfo {pages}
  {73--77} (\bibinfo {year} {2017})}\BibitemShut {NoStop}%
\bibitem [{\citenamefont {{Mahjoub}}\ \emph {et~al.}(2017)\citenamefont
  {{Mahjoub}}, \citenamefont {{Poston}}, \citenamefont {{Blacksberg}},
  \citenamefont {{Eiler}}, \citenamefont {{Brown}}, \citenamefont {{Ehlmann}},
  \citenamefont {{Hodyss}}, \citenamefont {{Hand}}, \citenamefont {{Carlson}},\
  and\ \citenamefont {{Choukroun}}}]{Mahjoub:2017}%
  \BibitemOpen
  \bibfield  {author} {\bibinfo {author} {\bibfnamefont {A.}~\bibnamefont
  {{Mahjoub}}}, \bibinfo {author} {\bibfnamefont {M.~J.}\ \bibnamefont
  {{Poston}}}, \bibinfo {author} {\bibfnamefont {J.}~\bibnamefont
  {{Blacksberg}}}, \bibinfo {author} {\bibfnamefont {J.~M.}\ \bibnamefont
  {{Eiler}}}, \bibinfo {author} {\bibfnamefont {M.~E.}\ \bibnamefont
  {{Brown}}}, \bibinfo {author} {\bibfnamefont {B.~L.}\ \bibnamefont
  {{Ehlmann}}}, \bibinfo {author} {\bibfnamefont {R.}~\bibnamefont {{Hodyss}}},
  \bibinfo {author} {\bibfnamefont {K.~P.}\ \bibnamefont {{Hand}}}, \bibinfo
  {author} {\bibfnamefont {R.}~\bibnamefont {{Carlson}}}, \ and\ \bibinfo
  {author} {\bibfnamefont {M.}~\bibnamefont {{Choukroun}}},\ }\bibfield
  {title} {\enquote {\bibinfo {title} {{Production of Sulfur Allotropes in
  Electron Irradiated Jupiter Trojans Ice Analogs}},}\ }\href {\doibase
  10.3847/1538-4357/aa85e0} {\bibfield  {journal} {\bibinfo  {journal}
  {Astrophys.~J.}\ }\textbf {\bibinfo {volume} {846}},\ \bibinfo {eid} {148}
  (\bibinfo {year} {2017})}\BibitemShut {NoStop}%
\bibitem [{\citenamefont {{Henderson}}\ and\ \citenamefont
  {{Dahlstrom}}(1939)}]{Henderson:1939}%
  \BibitemOpen
  \bibfield  {author} {\bibinfo {author} {\bibfnamefont {J.~E.}\ \bibnamefont
  {{Henderson}}}\ and\ \bibinfo {author} {\bibfnamefont {R.~K.}\ \bibnamefont
  {{Dahlstrom}}},\ }\bibfield  {title} {\enquote {\bibinfo {title} {{The Energy
  Distribution in Field Emission}},}\ }\href {\doibase 10.1103/PhysRev.55.473}
  {\bibfield  {journal} {\bibinfo  {journal} {Phys.~Rev.}\ }\textbf {\bibinfo
  {volume} {55}},\ \bibinfo {pages} {473--481} (\bibinfo {year}
  {1939})}\BibitemShut {NoStop}%
\bibitem [{\citenamefont {{Young}}\ and\ \citenamefont
  {{M{\"u}ller}}(1959)}]{Young:1959}%
  \BibitemOpen
  \bibfield  {author} {\bibinfo {author} {\bibfnamefont {R.~D.}\ \bibnamefont
  {{Young}}}\ and\ \bibinfo {author} {\bibfnamefont {E.~W.}\ \bibnamefont
  {{M{\"u}ller}}},\ }\bibfield  {title} {\enquote {\bibinfo {title}
  {{Experimental Measurement of the Total-Energy Distribution of Field-Emitted
  Electrons}},}\ }\href {\doibase 10.1103/PhysRev.113.115} {\bibfield
  {journal} {\bibinfo  {journal} {Phys.~Rev.}\ }\textbf {\bibinfo {volume}
  {113}},\ \bibinfo {pages} {115--120} (\bibinfo {year} {1959})}\BibitemShut
  {NoStop}%
\bibitem [{\citenamefont {{Conway}}, \citenamefont {{Perry}},\ and\
  \citenamefont {{Boswell}}(1998)}]{Conway:1998}%
  \BibitemOpen
  \bibfield  {author} {\bibinfo {author} {\bibfnamefont {G.~D.}\ \bibnamefont
  {{Conway}}}, \bibinfo {author} {\bibfnamefont {A.~J.}\ \bibnamefont
  {{Perry}}}, \ and\ \bibinfo {author} {\bibfnamefont {R.~W.}\ \bibnamefont
  {{Boswell}}},\ }\bibfield  {title} {\enquote {\bibinfo {title} {{Evolution of
  ion and electron energy distributions in pulsed helicon plasma
  discharges}},}\ }\href {\doibase 10.1088/0963-0252/7/3/012} {\bibfield
  {journal} {\bibinfo  {journal} {Plasma Sources Sci. Tech.}\ }\textbf
  {\bibinfo {volume} {7}},\ \bibinfo {pages} {337--347} (\bibinfo {year}
  {1998})}\BibitemShut {NoStop}%
\bibitem [{\citenamefont {{Rudakov}}\ \emph {et~al.}(1999)\citenamefont
  {{Rudakov}}, \citenamefont {{Shats}}, \citenamefont {{Boswell}},
  \citenamefont {{Charles}},\ and\ \citenamefont {{Howard}}}]{Rudakov:1999}%
  \BibitemOpen
  \bibfield  {author} {\bibinfo {author} {\bibfnamefont {D.~L.}\ \bibnamefont
  {{Rudakov}}}, \bibinfo {author} {\bibfnamefont {M.~G.}\ \bibnamefont
  {{Shats}}}, \bibinfo {author} {\bibfnamefont {R.~W.}\ \bibnamefont
  {{Boswell}}}, \bibinfo {author} {\bibfnamefont {C.}~\bibnamefont
  {{Charles}}}, \ and\ \bibinfo {author} {\bibfnamefont {J.}~\bibnamefont
  {{Howard}}},\ }\bibfield  {title} {\enquote {\bibinfo {title} {{Overview of
  probe diagnostics on the H-1 heliac}},}\ }\href {\doibase 10.1063/1.1149483}
  {\bibfield  {journal} {\bibinfo  {journal} {Rev. Sci. Instrum.}\ }\textbf
  {\bibinfo {volume} {70}},\ \bibinfo {pages} {476--479} (\bibinfo {year}
  {1999})}\BibitemShut {NoStop}%
\bibitem [{\citenamefont {{Charles}}, \citenamefont {{Boswell}},\ and\
  \citenamefont {{Porteous}}(1992)}]{Charles:1992}%
  \BibitemOpen
  \bibfield  {author} {\bibinfo {author} {\bibfnamefont {C.}~\bibnamefont
  {{Charles}}}, \bibinfo {author} {\bibfnamefont {R.~W.}\ \bibnamefont
  {{Boswell}}}, \ and\ \bibinfo {author} {\bibfnamefont {R.~K.}\ \bibnamefont
  {{Porteous}}},\ }\bibfield  {title} {\enquote {\bibinfo {title} {{Measurement
  and modeling of ion energy distribution functions in a low pressure argon
  plasma diffusing from a 13.56 MHz helicon source}},}\ }\href@noop {}
  {\bibfield  {journal} {\bibinfo  {journal} {J. Vac. Sci. Tech.}\ }\textbf
  {\bibinfo {volume} {10}},\ \bibinfo {pages} {398--403} (\bibinfo {year}
  {1992})}\BibitemShut {NoStop}%
\bibitem [{\citenamefont {{Charles}}(1993)}]{Charles:1993}%
  \BibitemOpen
  \bibfield  {author} {\bibinfo {author} {\bibfnamefont {C.}~\bibnamefont
  {{Charles}}},\ }\bibfield  {title} {\enquote {\bibinfo {title} {{Ion energy
  distribution functions in a multipole confined argon plasma diffusing from a
  13.56-MHz helicon source}},}\ }\href@noop {} {\bibfield  {journal} {\bibinfo
  {journal} {J. Vac. Sci. Tech.}\ }\textbf {\bibinfo {volume} {11}},\ \bibinfo
  {pages} {157--163} (\bibinfo {year} {1993})}\BibitemShut {NoStop}%
\bibitem [{\citenamefont {{Charles}}\ \emph {et~al.}(2000)\citenamefont
  {{Charles}}, \citenamefont {{Degeling}}, \citenamefont {{Sheridan}},
  \citenamefont {{Harris}}, \citenamefont {{Lieberman}},\ and\ \citenamefont
  {{Boswell}}}]{Charles:2000}%
  \BibitemOpen
  \bibfield  {author} {\bibinfo {author} {\bibfnamefont {C.}~\bibnamefont
  {{Charles}}}, \bibinfo {author} {\bibfnamefont {A.~W.}\ \bibnamefont
  {{Degeling}}}, \bibinfo {author} {\bibfnamefont {T.~E.}\ \bibnamefont
  {{Sheridan}}}, \bibinfo {author} {\bibfnamefont {J.~H.}\ \bibnamefont
  {{Harris}}}, \bibinfo {author} {\bibfnamefont {M.~A.}\ \bibnamefont
  {{Lieberman}}}, \ and\ \bibinfo {author} {\bibfnamefont {R.~W.}\ \bibnamefont
  {{Boswell}}},\ }\bibfield  {title} {\enquote {\bibinfo {title} {{Absolute
  measurements and modeling of radio frequency electric fields using a
  retarding field energy analyzer}},}\ }\href {\doibase 10.1063/1.1322557}
  {\bibfield  {journal} {\bibinfo  {journal} {Phys.~Plasmas}\ }\textbf
  {\bibinfo {volume} {7}},\ \bibinfo {pages} {5232--5241} (\bibinfo {year}
  {2000})}\BibitemShut {NoStop}%
\bibitem [{\citenamefont {{Enloe}}\ \emph {et~al.}(2015)\citenamefont
  {{Enloe}}, \citenamefont {{Amatucci}}, \citenamefont {{McHarg}},\ and\
  \citenamefont {{Balthazor}}}]{Enloe:2015}%
  \BibitemOpen
  \bibfield  {author} {\bibinfo {author} {\bibfnamefont {C.~L.}\ \bibnamefont
  {{Enloe}}}, \bibinfo {author} {\bibfnamefont {W.~E.}\ \bibnamefont
  {{Amatucci}}}, \bibinfo {author} {\bibfnamefont {M.~G.}\ \bibnamefont
  {{McHarg}}}, \ and\ \bibinfo {author} {\bibfnamefont {R.~L.}\ \bibnamefont
  {{Balthazor}}},\ }\bibfield  {title} {\enquote {\bibinfo {title} {{Screens
  versus microarrays for ruggedized retarding potential analyzers}},}\ }\href
  {\doibase 10.1063/1.4929532} {\bibfield  {journal} {\bibinfo  {journal} {Rev.
  Sci. Instrum.}\ }\textbf {\bibinfo {volume} {86}},\ \bibinfo {eid} {093302}
  (\bibinfo {year} {2015})}\BibitemShut {NoStop}%
\bibitem [{\citenamefont {{Scime}}\ \emph {et~al.}(2016)\citenamefont
  {{Scime}}, \citenamefont {{Keesee}}, \citenamefont {{Dugas}}, \citenamefont
  {{Ellison}}, \citenamefont {{Tersteeg}}, \citenamefont {{Wagner}},
  \citenamefont {{Barrie}}, \citenamefont {{Rager}},\ and\ \citenamefont
  {{Elliott}}}]{Scime:2016}%
  \BibitemOpen
  \bibfield  {author} {\bibinfo {author} {\bibfnamefont {E.~E.}\ \bibnamefont
  {{Scime}}}, \bibinfo {author} {\bibfnamefont {A.~M.}\ \bibnamefont
  {{Keesee}}}, \bibinfo {author} {\bibfnamefont {M.}~\bibnamefont {{Dugas}}},
  \bibinfo {author} {\bibfnamefont {S.}~\bibnamefont {{Ellison}}}, \bibinfo
  {author} {\bibfnamefont {J.}~\bibnamefont {{Tersteeg}}}, \bibinfo {author}
  {\bibfnamefont {G.}~\bibnamefont {{Wagner}}}, \bibinfo {author}
  {\bibfnamefont {A.}~\bibnamefont {{Barrie}}}, \bibinfo {author}
  {\bibfnamefont {A.}~\bibnamefont {{Rager}}}, \ and\ \bibinfo {author}
  {\bibfnamefont {D.}~\bibnamefont {{Elliott}}},\ }\bibfield  {title} {\enquote
  {\bibinfo {title} {{A micro-scale plasma spectrometer for space and plasma
  edge applications (invited)}},}\ }\href {\doibase 10.1063/1.4960145}
  {\bibfield  {journal} {\bibinfo  {journal} {Rev. Sci. Instrum.}\ }\textbf
  {\bibinfo {volume} {87}},\ \bibinfo {eid} {11D302} (\bibinfo {year}
  {2016})}\BibitemShut {NoStop}%
\bibitem [{\citenamefont {{Skiff}}, \citenamefont {{Anderegg}},\ and\
  \citenamefont {{Tran}}(1987)}]{Skiff:1987}%
  \BibitemOpen
  \bibfield  {author} {\bibinfo {author} {\bibfnamefont {F.}~\bibnamefont
  {{Skiff}}}, \bibinfo {author} {\bibfnamefont {F.}~\bibnamefont {{Anderegg}}},
  \ and\ \bibinfo {author} {\bibfnamefont {M.~Q.}\ \bibnamefont {{Tran}}},\
  }\bibfield  {title} {\enquote {\bibinfo {title} {{Stochastic particle
  acceleration in an electrostatic wave}},}\ }\href {\doibase
  10.1103/PhysRevLett.58.1430} {\bibfield  {journal} {\bibinfo  {journal}
  {Phys.~Rev.~Lett.}\ }\textbf {\bibinfo {volume} {58}},\ \bibinfo {pages}
  {1430--1433} (\bibinfo {year} {1987})}\BibitemShut {NoStop}%
\bibitem [{\citenamefont {{Sarfaty}}, \citenamefont {{De Souza-Machado}},\ and\
  \citenamefont {{Skiff}}(1996)}]{Sarfaty:1996}%
  \BibitemOpen
  \bibfield  {author} {\bibinfo {author} {\bibfnamefont {M.}~\bibnamefont
  {{Sarfaty}}}, \bibinfo {author} {\bibfnamefont {S.}~\bibnamefont {{De
  Souza-Machado}}}, \ and\ \bibinfo {author} {\bibfnamefont {F.}~\bibnamefont
  {{Skiff}}},\ }\bibfield  {title} {\enquote {\bibinfo {title} {{Direct
  determination of ion wave fields in a hot magnetized and weakly collisional
  plasma}},}\ }\href {\doibase 10.1063/1.871581} {\bibfield  {journal}
  {\bibinfo  {journal} {Phys.~Plasmas}\ }\textbf {\bibinfo {volume} {3}},\
  \bibinfo {pages} {4316--4324} (\bibinfo {year} {1996})}\BibitemShut {NoStop}%
\bibitem [{\citenamefont {{Scime}}\ \emph {et~al.}(1998)\citenamefont
  {{Scime}}, \citenamefont {{Keiter}}, \citenamefont {{Zintl}}, \citenamefont
  {{Balkey}}, \citenamefont {{Kline}},\ and\ \citenamefont
  {{Koepke}}}]{Scime:1998}%
  \BibitemOpen
  \bibfield  {author} {\bibinfo {author} {\bibfnamefont {E.~E.}\ \bibnamefont
  {{Scime}}}, \bibinfo {author} {\bibfnamefont {P.~A.}\ \bibnamefont
  {{Keiter}}}, \bibinfo {author} {\bibfnamefont {M.~W.}\ \bibnamefont
  {{Zintl}}}, \bibinfo {author} {\bibfnamefont {M.~M.}\ \bibnamefont
  {{Balkey}}}, \bibinfo {author} {\bibfnamefont {J.~L.}\ \bibnamefont
  {{Kline}}}, \ and\ \bibinfo {author} {\bibfnamefont {M.~E.}\ \bibnamefont
  {{Koepke}}},\ }\bibfield  {title} {\enquote {\bibinfo {title} {{Control of
  ion temperature anisotropy in a helicon plasma}},}\ }\href {\doibase
  10.1088/0963-0252/7/2/013} {\bibfield  {journal} {\bibinfo  {journal} {Plasma
  Sources Sci. Tech.}\ }\textbf {\bibinfo {volume} {7}},\ \bibinfo {pages}
  {186--191} (\bibinfo {year} {1998})}\BibitemShut {NoStop}%
\bibitem [{\citenamefont {{Boivin}}\ and\ \citenamefont
  {{Scime}}(2003)}]{Boivin:2003}%
  \BibitemOpen
  \bibfield  {author} {\bibinfo {author} {\bibfnamefont {R.~F.}\ \bibnamefont
  {{Boivin}}}\ and\ \bibinfo {author} {\bibfnamefont {E.~E.}\ \bibnamefont
  {{Scime}}},\ }\bibfield  {title} {\enquote {\bibinfo {title} {{Laser induced
  fluorescence in Ar and He plasmas with a tunable diode laser}},}\ }\href
  {\doibase 10.1063/1.1606095} {\bibfield  {journal} {\bibinfo  {journal} {Rev.
  Sci. Instrum.}\ }\textbf {\bibinfo {volume} {74}},\ \bibinfo {pages}
  {4352--4360} (\bibinfo {year} {2003})}\BibitemShut {NoStop}%
\bibitem [{\citenamefont {{Keesee}}, \citenamefont {{Scime}},\ and\
  \citenamefont {{Boivin}}(2004)}]{Keesee:2004}%
  \BibitemOpen
  \bibfield  {author} {\bibinfo {author} {\bibfnamefont {A.~M.}\ \bibnamefont
  {{Keesee}}}, \bibinfo {author} {\bibfnamefont {E.~E.}\ \bibnamefont
  {{Scime}}}, \ and\ \bibinfo {author} {\bibfnamefont {R.~F.}\ \bibnamefont
  {{Boivin}}},\ }\bibfield  {title} {\enquote {\bibinfo {title} {{Laser-induced
  fluorescence measurements of three plasma species with a tunable diode
  laser}},}\ }\href {\doibase 10.1063/1.1787166} {\bibfield  {journal}
  {\bibinfo  {journal} {Rev. Sci. Instrum.}\ }\textbf {\bibinfo {volume}
  {75}},\ \bibinfo {pages} {4091--4093} (\bibinfo {year} {2004})}\BibitemShut
  {NoStop}%
\bibitem [{\citenamefont {{Biloiu}}\ \emph {et~al.}(2005)\citenamefont
  {{Biloiu}}, \citenamefont {{Sun}}, \citenamefont {{Choueiri}}, \citenamefont
  {{Doss}}, \citenamefont {{Scime}}, \citenamefont {{Heard}}, \citenamefont
  {{Spektor}},\ and\ \citenamefont {{Ventura}}}]{Biloiu:2005}%
  \BibitemOpen
  \bibfield  {author} {\bibinfo {author} {\bibfnamefont {C.}~\bibnamefont
  {{Biloiu}}}, \bibinfo {author} {\bibfnamefont {X.}~\bibnamefont {{Sun}}},
  \bibinfo {author} {\bibfnamefont {E.}~\bibnamefont {{Choueiri}}}, \bibinfo
  {author} {\bibfnamefont {F.}~\bibnamefont {{Doss}}}, \bibinfo {author}
  {\bibfnamefont {E.}~\bibnamefont {{Scime}}}, \bibinfo {author} {\bibfnamefont
  {J.}~\bibnamefont {{Heard}}}, \bibinfo {author} {\bibfnamefont
  {R.}~\bibnamefont {{Spektor}}}, \ and\ \bibinfo {author} {\bibfnamefont
  {D.}~\bibnamefont {{Ventura}}},\ }\bibfield  {title} {\enquote {\bibinfo
  {title} {{Evolution of the parallel and perpendicular ion velocity
  distribution functions in pulsed helicon plasma sources obtained by time
  resolved laser induced fluorescence}},}\ }\href {\doibase
  10.1088/0963-0252/14/4/016} {\bibfield  {journal} {\bibinfo  {journal}
  {Plasma Sources Sci. Tech.}\ }\textbf {\bibinfo {volume} {14}},\ \bibinfo
  {pages} {766--776} (\bibinfo {year} {2005})}\BibitemShut {NoStop}%
\bibitem [{\citenamefont {{Scime}}\ \emph {et~al.}(2005)\citenamefont
  {{Scime}}, \citenamefont {{Biloiu}}, \citenamefont {{Compton}}, \citenamefont
  {{Doss}}, \citenamefont {{Venture}}, \citenamefont {{Heard}}, \citenamefont
  {{Choueiri}},\ and\ \citenamefont {{Spektor}}}]{Scime:2005}%
  \BibitemOpen
  \bibfield  {author} {\bibinfo {author} {\bibfnamefont {E.}~\bibnamefont
  {{Scime}}}, \bibinfo {author} {\bibfnamefont {C.}~\bibnamefont {{Biloiu}}},
  \bibinfo {author} {\bibfnamefont {C.}~\bibnamefont {{Compton}}}, \bibinfo
  {author} {\bibfnamefont {F.}~\bibnamefont {{Doss}}}, \bibinfo {author}
  {\bibfnamefont {D.}~\bibnamefont {{Venture}}}, \bibinfo {author}
  {\bibfnamefont {J.}~\bibnamefont {{Heard}}}, \bibinfo {author} {\bibfnamefont
  {E.}~\bibnamefont {{Choueiri}}}, \ and\ \bibinfo {author} {\bibfnamefont
  {R.}~\bibnamefont {{Spektor}}},\ }\bibfield  {title} {\enquote {\bibinfo
  {title} {{Laser induced fluorescence in a pulsed argon plasma}},}\ }\href
  {\doibase 10.1063/1.1848491} {\bibfield  {journal} {\bibinfo  {journal} {Rev.
  Sci. Instrum.}\ }\textbf {\bibinfo {volume} {76}},\ \bibinfo {eid}
  {026107-026107-3} (\bibinfo {year} {2005})}\BibitemShut {NoStop}%
\bibitem [{\citenamefont {{Biloiu}}, \citenamefont {{Sun}},\ and\ \citenamefont
  {{Scime}}(2006)}]{Biloiu:2006}%
  \BibitemOpen
  \bibfield  {author} {\bibinfo {author} {\bibfnamefont {I.~A.}\ \bibnamefont
  {{Biloiu}}}, \bibinfo {author} {\bibfnamefont {X.}~\bibnamefont {{Sun}}}, \
  and\ \bibinfo {author} {\bibfnamefont {E.~E.}\ \bibnamefont {{Scime}}},\
  }\bibfield  {title} {\enquote {\bibinfo {title} {{High time resolution laser
  induced fluorescence in pulsed argon plasma}},}\ }\href {\doibase
  10.1063/1.2217919} {\bibfield  {journal} {\bibinfo  {journal} {Rev. Sci.
  Instrum.}\ }\textbf {\bibinfo {volume} {77}},\ \bibinfo {eid}
  {10F301-10F301-3} (\bibinfo {year} {2006})}\BibitemShut {NoStop}%
\bibitem [{\citenamefont {{Uzun-Kaymak}}\ and\ \citenamefont
  {{Skiff}}(2006)}]{UzunKaymak:2006}%
  \BibitemOpen
  \bibfield  {author} {\bibinfo {author} {\bibfnamefont {I.~{\"U}.}\
  \bibnamefont {{Uzun-Kaymak}}}\ and\ \bibinfo {author} {\bibfnamefont
  {F.}~\bibnamefont {{Skiff}}},\ }\bibfield  {title} {\enquote {\bibinfo
  {title} {{Observation of coherent nonlinear interactions in the ion velocity
  distribution function}},}\ }\href {\doibase 10.1063/1.2387143} {\bibfield
  {journal} {\bibinfo  {journal} {Phys.~Plasmas}\ }\textbf {\bibinfo {volume}
  {13}},\ \bibinfo {eid} {112108} (\bibinfo {year} {2006})}\BibitemShut
  {NoStop}%
\bibitem [{\citenamefont {{Mattingly}}\ \emph {et~al.}(2013)\citenamefont
  {{Mattingly}}, \citenamefont {{Berumen}}, \citenamefont {{Chu}},
  \citenamefont {{Hood}},\ and\ \citenamefont {{Skiff}}}]{Mattingly:2013}%
  \BibitemOpen
  \bibfield  {author} {\bibinfo {author} {\bibfnamefont {S.~W.}\ \bibnamefont
  {{Mattingly}}}, \bibinfo {author} {\bibfnamefont {J.}~\bibnamefont
  {{Berumen}}}, \bibinfo {author} {\bibfnamefont {F.}~\bibnamefont {{Chu}}},
  \bibinfo {author} {\bibfnamefont {R.}~\bibnamefont {{Hood}}}, \ and\ \bibinfo
  {author} {\bibfnamefont {F.}~\bibnamefont {{Skiff}}},\ }\bibfield  {title}
  {\enquote {\bibinfo {title} {{Measurement and interpretation of the velocity
  space correlation of a laboratory plasma fluctuation with laser induced
  fluorescence}},}\ }\href {\doibase 10.1088/1748-0221/8/11/C11015} {\bibfield
  {journal} {\bibinfo  {journal} {J. Instrumentation}\ }\textbf {\bibinfo
  {volume} {8}},\ \bibinfo {eid} {C11015} (\bibinfo {year} {2013})}\BibitemShut
  {NoStop}%
\bibitem [{\citenamefont {{Thompson}}\ \emph {et~al.}(2017)\citenamefont
  {{Thompson}}, \citenamefont {{Henriquez}}, \citenamefont {{Scime}},\ and\
  \citenamefont {{Good}}}]{Thompson:2017}%
  \BibitemOpen
  \bibfield  {author} {\bibinfo {author} {\bibfnamefont {D.~S.}\ \bibnamefont
  {{Thompson}}}, \bibinfo {author} {\bibfnamefont {M.~F.}\ \bibnamefont
  {{Henriquez}}}, \bibinfo {author} {\bibfnamefont {E.~E.}\ \bibnamefont
  {{Scime}}}, \ and\ \bibinfo {author} {\bibfnamefont {T.~N.}\ \bibnamefont
  {{Good}}},\ }\bibfield  {title} {\enquote {\bibinfo {title} {{Confocal laser
  induced fluorescence with comparable spatial localization to the conventional
  method}},}\ }\href {\doibase 10.1063/1.4991637} {\bibfield  {journal}
  {\bibinfo  {journal} {Rev. Sci. Instrum.}\ }\textbf {\bibinfo {volume}
  {88}},\ \bibinfo {eid} {103506} (\bibinfo {year} {2017})}\BibitemShut
  {NoStop}%
\bibitem [{\citenamefont {{Harvey}}\ \emph {et~al.}(2008)\citenamefont
  {{Harvey}}, \citenamefont {{Chakraborty Thakur}}, \citenamefont {{Hansen}},
  \citenamefont {{Hardin}}, \citenamefont {{Przybysz}},\ and\ \citenamefont
  {{Scime}}}]{Harvey:2008}%
  \BibitemOpen
  \bibfield  {author} {\bibinfo {author} {\bibfnamefont {Z.}~\bibnamefont
  {{Harvey}}}, \bibinfo {author} {\bibfnamefont {S.}~\bibnamefont {{Chakraborty
  Thakur}}}, \bibinfo {author} {\bibfnamefont {A.}~\bibnamefont {{Hansen}}},
  \bibinfo {author} {\bibfnamefont {R.}~\bibnamefont {{Hardin}}}, \bibinfo
  {author} {\bibfnamefont {W.~S.}\ \bibnamefont {{Przybysz}}}, \ and\ \bibinfo
  {author} {\bibfnamefont {E.~E.}\ \bibnamefont {{Scime}}},\ }\bibfield
  {title} {\enquote {\bibinfo {title} {{Comparison of gridded energy analyzer
  and laser induced fluorescence measurements of a two-component ion
  distributiona)}},}\ }\href {\doibase 10.1063/1.2953411} {\bibfield  {journal}
  {\bibinfo  {journal} {Rev. Sci. Instrum.}\ }\textbf {\bibinfo {volume}
  {79}},\ \bibinfo {eid} {10F314-10F314-3} (\bibinfo {year}
  {2008})}\BibitemShut {NoStop}%
\bibitem [{\citenamefont {{Gulbrandsen}}\ \emph {et~al.}(2015)\citenamefont
  {{Gulbrandsen}}, \citenamefont {{Fredriksen}}, \citenamefont {{Carr}},\ and\
  \citenamefont {{Scime}}}]{Gulbrandsen:2015}%
  \BibitemOpen
  \bibfield  {author} {\bibinfo {author} {\bibfnamefont {N.}~\bibnamefont
  {{Gulbrandsen}}}, \bibinfo {author} {\bibfnamefont {{\AA}.}~\bibnamefont
  {{Fredriksen}}}, \bibinfo {author} {\bibfnamefont {J.}~\bibnamefont
  {{Carr}}}, \ and\ \bibinfo {author} {\bibfnamefont {E.}~\bibnamefont
  {{Scime}}},\ }\bibfield  {title} {\enquote {\bibinfo {title} {{A comparison
  of ion beam measurements by retarding field energy analyzer and laser induced
  fluorescence in helicon plasma devices}},}\ }\href {\doibase
  10.1063/1.4913990} {\bibfield  {journal} {\bibinfo  {journal}
  {Phys.~Plasmas}\ }\textbf {\bibinfo {volume} {22}},\ \bibinfo {eid} {033505}
  (\bibinfo {year} {2015})}\BibitemShut {NoStop}%
\bibitem [{\citenamefont {{Skiff}}, \citenamefont {{Boyd}},\ and\ \citenamefont
  {{Colborn}}(1993)}]{Skiff:1993}%
  \BibitemOpen
  \bibfield  {author} {\bibinfo {author} {\bibfnamefont {F.}~\bibnamefont
  {{Skiff}}}, \bibinfo {author} {\bibfnamefont {D.~A.}\ \bibnamefont {{Boyd}}},
  \ and\ \bibinfo {author} {\bibfnamefont {J.~A.}\ \bibnamefont {{Colborn}}},\
  }\bibfield  {title} {\enquote {\bibinfo {title} {{Measurements of electron
  parallel-momentum distributions using cyclotron wave transmission}},}\ }\href
  {\doibase 10.1063/1.860729} {\bibfield  {journal} {\bibinfo  {journal}
  {Phys.~Fluids B}\ }\textbf {\bibinfo {volume} {5}},\ \bibinfo {pages}
  {2445--2450} (\bibinfo {year} {1993})}\BibitemShut {NoStop}%
\bibitem [{\citenamefont {{Skiff}}(2006)}]{Skiff:2006}%
  \BibitemOpen
  \bibfield  {author} {\bibinfo {author} {\bibfnamefont {F.}~\bibnamefont
  {{Skiff}}},\ }\bibfield  {title} {\enquote {\bibinfo {title} {{Diagnostics of
  Collisionless Processes in Plasma}},}\ }\href {\doibase
  10.1109/TPS.2006.878359} {\bibfield  {journal} {\bibinfo  {journal} {IEEE
  Trans. Plasma Sci.}\ }\textbf {\bibinfo {volume} {34}},\ \bibinfo {pages}
  {1548--1552} (\bibinfo {year} {2006})}\BibitemShut {NoStop}%
\bibitem [{\citenamefont {{Fonck}}, \citenamefont {{Darrow}},\ and\
  \citenamefont {{Jaehnig}}(1984)}]{Fonck:1984}%
  \BibitemOpen
  \bibfield  {author} {\bibinfo {author} {\bibfnamefont {R.~J.}\ \bibnamefont
  {{Fonck}}}, \bibinfo {author} {\bibfnamefont {D.~S.}\ \bibnamefont
  {{Darrow}}}, \ and\ \bibinfo {author} {\bibfnamefont {K.~P.}\ \bibnamefont
  {{Jaehnig}}},\ }\bibfield  {title} {\enquote {\bibinfo {title}
  {{Determination of plasma-ion velocity distribution via charge-exchange
  recombination spectroscopy}},}\ }\href {\doibase 10.1103/PhysRevA.29.3288}
  {\bibfield  {journal} {\bibinfo  {journal} {Phys.~Rev.~A}\ }\textbf {\bibinfo
  {volume} {29}},\ \bibinfo {pages} {3288--3309} (\bibinfo {year}
  {1984})}\BibitemShut {NoStop}%
\bibitem [{\citenamefont {{Den Hartog}}\ \emph {et~al.}(2006)\citenamefont
  {{Den Hartog}}, \citenamefont {{Craig}}, \citenamefont {{Ennis}},
  \citenamefont {{Fiksel}}, \citenamefont {{Gangadhara}}, \citenamefont
  {{Holly}}, \citenamefont {{Reardon}}, \citenamefont {{Davydenko}},
  \citenamefont {{Ivanov}}, \citenamefont {{Lizunov}}, \citenamefont
  {{O'Mullane}},\ and\ \citenamefont {{Summers}}}]{DenHartog:2006}%
  \BibitemOpen
  \bibfield  {author} {\bibinfo {author} {\bibfnamefont {D.~J.}\ \bibnamefont
  {{Den Hartog}}}, \bibinfo {author} {\bibfnamefont {D.}~\bibnamefont
  {{Craig}}}, \bibinfo {author} {\bibfnamefont {D.~A.}\ \bibnamefont
  {{Ennis}}}, \bibinfo {author} {\bibfnamefont {G.}~\bibnamefont {{Fiksel}}},
  \bibinfo {author} {\bibfnamefont {S.}~\bibnamefont {{Gangadhara}}}, \bibinfo
  {author} {\bibfnamefont {D.~J.}\ \bibnamefont {{Holly}}}, \bibinfo {author}
  {\bibfnamefont {J.~C.}\ \bibnamefont {{Reardon}}}, \bibinfo {author}
  {\bibfnamefont {V.~I.}\ \bibnamefont {{Davydenko}}}, \bibinfo {author}
  {\bibfnamefont {A.~A.}\ \bibnamefont {{Ivanov}}}, \bibinfo {author}
  {\bibfnamefont {A.~A.}\ \bibnamefont {{Lizunov}}}, \bibinfo {author}
  {\bibfnamefont {M.~G.}\ \bibnamefont {{O'Mullane}}}, \ and\ \bibinfo {author}
  {\bibfnamefont {H.~P.}\ \bibnamefont {{Summers}}},\ }\bibfield  {title}
  {\enquote {\bibinfo {title} {{Advances in neutral-beam-based diagnostics on
  the Madison Symmetric Torus reversed-field pinch (invited)}},}\ }\href
  {\doibase 10.1063/1.2217920} {\bibfield  {journal} {\bibinfo  {journal} {Rev.
  Sci. Instrum.}\ }\textbf {\bibinfo {volume} {77}},\ \bibinfo {eid}
  {10F122-10F122-8} (\bibinfo {year} {2006})}\BibitemShut {NoStop}%
\bibitem [{\citenamefont {{Podest{\`a}}}\ \emph {et~al.}(2008)\citenamefont
  {{Podest{\`a}}}, \citenamefont {{Heidbrink}}, \citenamefont {{Bell}},\ and\
  \citenamefont {{Feder}}}]{Podesta:2008}%
  \BibitemOpen
  \bibfield  {author} {\bibinfo {author} {\bibfnamefont {M.}~\bibnamefont
  {{Podest{\`a}}}}, \bibinfo {author} {\bibfnamefont {W.~W.}\ \bibnamefont
  {{Heidbrink}}}, \bibinfo {author} {\bibfnamefont {R.~E.}\ \bibnamefont
  {{Bell}}}, \ and\ \bibinfo {author} {\bibfnamefont {R.}~\bibnamefont
  {{Feder}}},\ }\bibfield  {title} {\enquote {\bibinfo {title} {{The NSTX
  fast-ion D-alpha diagnostica)}},}\ }\href {\doibase 10.1063/1.2956744}
  {\bibfield  {journal} {\bibinfo  {journal} {Rev. Sci. Instrum.}\ }\textbf
  {\bibinfo {volume} {79}},\ \bibinfo {eid} {10E521-10E521-5} (\bibinfo {year}
  {2008})}\BibitemShut {NoStop}%
\bibitem [{\citenamefont {{Heidbrink}}\ \emph {et~al.}(2008)\citenamefont
  {{Heidbrink}}, \citenamefont {{Luo}}, \citenamefont {{Muscatello}},
  \citenamefont {{Zhu}},\ and\ \citenamefont {{Burrell}}}]{Heidbrink:2008}%
  \BibitemOpen
  \bibfield  {author} {\bibinfo {author} {\bibfnamefont {W.~W.}\ \bibnamefont
  {{Heidbrink}}}, \bibinfo {author} {\bibfnamefont {Y.}~\bibnamefont {{Luo}}},
  \bibinfo {author} {\bibfnamefont {C.~M.}\ \bibnamefont {{Muscatello}}},
  \bibinfo {author} {\bibfnamefont {Y.}~\bibnamefont {{Zhu}}}, \ and\ \bibinfo
  {author} {\bibfnamefont {K.~H.}\ \bibnamefont {{Burrell}}},\ }\bibfield
  {title} {\enquote {\bibinfo {title} {{A new fast-ion D$_{α}$ diagnostic for
  DIII-Da)}},}\ }\href {\doibase 10.1063/1.2956828} {\bibfield  {journal}
  {\bibinfo  {journal} {Review of Scientific Instruments}\ }\textbf {\bibinfo
  {volume} {79}},\ \bibinfo {eid} {10E520-10E520-4} (\bibinfo {year}
  {2008})}\BibitemShut {NoStop}%
\bibitem [{\citenamefont {{Heidbrink}}(2010)}]{Heidbrink:2010}%
  \BibitemOpen
  \bibfield  {author} {\bibinfo {author} {\bibfnamefont {W.~W.}\ \bibnamefont
  {{Heidbrink}}},\ }\bibfield  {title} {\enquote {\bibinfo {title} {{Fast-ion
  D{$\alpha$} measurements of the fast-ion distribution (invited)a)}},}\ }\href
  {\doibase 10.1063/1.3478739} {\bibfield  {journal} {\bibinfo  {journal} {Rev.
  Sci. Instrum.}\ }\textbf {\bibinfo {volume} {81}},\ \bibinfo {eid}
  {10D727-10D727-8} (\bibinfo {year} {2010})}\BibitemShut {NoStop}%
\bibitem [{\citenamefont {{Chakraborty Thakur}}\ \emph
  {et~al.}(2012)\citenamefont {{Chakraborty Thakur}}, \citenamefont
  {{McCarren}}, \citenamefont {{Carr}},\ and\ \citenamefont
  {{Scime}}}]{ChakrabortyThakur:2012}%
  \BibitemOpen
  \bibfield  {author} {\bibinfo {author} {\bibfnamefont {S.}~\bibnamefont
  {{Chakraborty Thakur}}}, \bibinfo {author} {\bibfnamefont {D.}~\bibnamefont
  {{McCarren}}}, \bibinfo {author} {\bibfnamefont {J.}~\bibnamefont {{Carr}}},
  \ and\ \bibinfo {author} {\bibfnamefont {E.~E.}\ \bibnamefont {{Scime}}},\
  }\bibfield  {title} {\enquote {\bibinfo {title} {{Continuous wave cavity ring
  down spectroscopy measurements of velocity distribution functions of argon
  ions in a helicon plasma}},}\ }\href {\doibase 10.1063/1.3687429} {\bibfield
  {journal} {\bibinfo  {journal} {Rev. Sci. Instrum.}\ }\textbf {\bibinfo
  {volume} {83}},\ \bibinfo {eid} {023508-023508-8} (\bibinfo {year}
  {2012})}\BibitemShut {NoStop}%
\bibitem [{\citenamefont {{McCarren}}\ and\ \citenamefont
  {{Scime}}(2015)}]{McCarren:2015}%
  \BibitemOpen
  \bibfield  {author} {\bibinfo {author} {\bibfnamefont {D.}~\bibnamefont
  {{McCarren}}}\ and\ \bibinfo {author} {\bibfnamefont {E.}~\bibnamefont
  {{Scime}}},\ }\bibfield  {title} {\enquote {\bibinfo {title} {{Continuous
  wave cavity ring-down spectroscopy for velocity distribution measurements in
  plasma}},}\ }\href {\doibase 10.1063/1.4932313} {\bibfield  {journal}
  {\bibinfo  {journal} {Rev. Sci. Instrum.}\ }\textbf {\bibinfo {volume}
  {86}},\ \bibinfo {eid} {103505} (\bibinfo {year} {2015})}\BibitemShut
  {NoStop}%
\bibitem [{\citenamefont {{Skiff}}\ and\ \citenamefont
  {{Curry}}(1995)}]{Skiff:1995}%
  \BibitemOpen
  \bibfield  {author} {\bibinfo {author} {\bibfnamefont {F.}~\bibnamefont
  {{Skiff}}}\ and\ \bibinfo {author} {\bibfnamefont {J.~J.}\ \bibnamefont
  {{Curry}}},\ }\bibfield  {title} {\enquote {\bibinfo {title} {{Nonlinear
  optical tagging and laser induced fluorescence}},}\ }\href {\doibase
  10.1063/1.1146308} {\bibfield  {journal} {\bibinfo  {journal} {Rev. Sci.
  Instrum.}\ }\textbf {\bibinfo {volume} {66}},\ \bibinfo {pages} {629--631}
  (\bibinfo {year} {1995})}\BibitemShut {NoStop}%
\bibitem [{\citenamefont {{Claire}}\ \emph {et~al.}(2001)\citenamefont
  {{Claire}}, \citenamefont {{Dindelegan}}, \citenamefont {{Bachet}},\ and\
  \citenamefont {{Skiff}}}]{Claire:2001}%
  \BibitemOpen
  \bibfield  {author} {\bibinfo {author} {\bibfnamefont {N.}~\bibnamefont
  {{Claire}}}, \bibinfo {author} {\bibfnamefont {M.}~\bibnamefont
  {{Dindelegan}}}, \bibinfo {author} {\bibfnamefont {G.}~\bibnamefont
  {{Bachet}}}, \ and\ \bibinfo {author} {\bibfnamefont {F.}~\bibnamefont
  {{Skiff}}},\ }\bibfield  {title} {\enquote {\bibinfo {title} {{Nonlinear
  optical tagging diagnostic for the measurement of Fokker-Planck diffusion and
  electric fields}},}\ }\href {\doibase 10.1063/1.1419221} {\bibfield
  {journal} {\bibinfo  {journal} {Rev. Sci. Instrum.}\ }\textbf {\bibinfo
  {volume} {72}},\ \bibinfo {pages} {4372--4376} (\bibinfo {year}
  {2001})}\BibitemShut {NoStop}%
\bibitem [{\citenamefont {{Klein}}\ and\ \citenamefont
  {{Howes}}(2016)}]{Klein:2016a}%
  \BibitemOpen
  \bibfield  {author} {\bibinfo {author} {\bibfnamefont {K.~G.}\ \bibnamefont
  {{Klein}}}\ and\ \bibinfo {author} {\bibfnamefont {G.~G.}\ \bibnamefont
  {{Howes}}},\ }\bibfield  {title} {\enquote {\bibinfo {title} {{Measuring
  Collisionless Damping in Heliospheric Plasmas using Field-Particle
  Correlations}},}\ }\href {\doibase 10.3847/2041-8205/826/2/L30} {\bibfield
  {journal} {\bibinfo  {journal} {Astrophys.~J.~Lett.}\ }\textbf {\bibinfo
  {volume} {826}},\ \bibinfo {eid} {L30} (\bibinfo {year} {2016})},\ \Eprint
  {http://arxiv.org/abs/1607.01738} {arXiv:1607.01738 [physics.space-ph]}
  \BibitemShut {NoStop}%
\bibitem [{\citenamefont {{Howes}}, \citenamefont {{Klein}},\ and\
  \citenamefont {{Li}}(2017)}]{Howes:2017a}%
  \BibitemOpen
  \bibfield  {author} {\bibinfo {author} {\bibfnamefont {G.~G.}\ \bibnamefont
  {{Howes}}}, \bibinfo {author} {\bibfnamefont {K.~G.}\ \bibnamefont
  {{Klein}}}, \ and\ \bibinfo {author} {\bibfnamefont {T.~C.}\ \bibnamefont
  {{Li}}},\ }\bibfield  {title} {\enquote {\bibinfo {title} {{Diagnosing
  collisionless energy transfer using field-particle correlations:
  Vlasov-Poisson plasmas}},}\ }\href {\doibase 10.1017/S0022377816001197}
  {\bibfield  {journal} {\bibinfo  {journal} {J.~Plasma Phys.}\ }\textbf
  {\bibinfo {volume} {83}},\ \bibinfo {eid} {705830102} (\bibinfo {year}
  {2017})}\BibitemShut {NoStop}%
\bibitem [{\citenamefont {{Klein}}, \citenamefont {{Howes}},\ and\
  \citenamefont {{TenBarge}}(2017)}]{Klein:2017b}%
  \BibitemOpen
  \bibfield  {author} {\bibinfo {author} {\bibfnamefont {K.~G.}\ \bibnamefont
  {{Klein}}}, \bibinfo {author} {\bibfnamefont {G.~G.}\ \bibnamefont
  {{Howes}}}, \ and\ \bibinfo {author} {\bibfnamefont {J.~M.}\ \bibnamefont
  {{TenBarge}}},\ }\bibfield  {title} {\enquote {\bibinfo {title} {{Diagnosing
  collisionless energy transfer using field-particle correlations: gyrokinetic
  turbulence}},}\ }\href {\doibase 10.1017/S0022377817000563} {\bibfield
  {journal} {\bibinfo  {journal} {J.~Plasma Phys.}\ }\textbf {\bibinfo {volume}
  {83}},\ \bibinfo {eid} {535830401} (\bibinfo {year} {2017})},\ \Eprint
  {http://arxiv.org/abs/1705.06385} {arXiv:1705.06385 [physics.plasm-ph]}
  \BibitemShut {NoStop}%
\bibitem [{\citenamefont {{Klein}}(2017)}]{Klein:2017a}%
  \BibitemOpen
  \bibfield  {author} {\bibinfo {author} {\bibfnamefont {K.~G.}\ \bibnamefont
  {{Klein}}},\ }\bibfield  {title} {\enquote {\bibinfo {title} {{Characterizing
  fluid and kinetic instabilities using field-particle correlations on
  single-point time series}},}\ }\href {\doibase 10.1063/1.4977465} {\bibfield
  {journal} {\bibinfo  {journal} {Phys.~Plasmas}\ }\textbf {\bibinfo {volume}
  {24}},\ \bibinfo {eid} {055901} (\bibinfo {year} {2017})},\ \Eprint
  {http://arxiv.org/abs/1701.03687} {arXiv:1701.03687 [physics.plasm-ph]}
  \BibitemShut {NoStop}%
\bibitem [{\citenamefont {{Howes}}, \citenamefont {{McCubbin}},\ and\
  \citenamefont {{Klein}}(2018{\natexlab{b}})}]{Howes:2018}%
  \BibitemOpen
  \bibfield  {author} {\bibinfo {author} {\bibfnamefont {G.~G.}\ \bibnamefont
  {{Howes}}}, \bibinfo {author} {\bibfnamefont {A.~J.}\ \bibnamefont
  {{McCubbin}}}, \ and\ \bibinfo {author} {\bibfnamefont {K.~G.}\ \bibnamefont
  {{Klein}}},\ }\bibfield  {title} {\enquote {\bibinfo {title} {{Spatial
  Localization of Particle Enegization in Current Sheets Produced by Alfv\'en
  Wave Collisions}},}\ }\href@noop {} {\bibfield  {journal} {\bibinfo
  {journal} {J.~Plasma Phys.}\ }\textbf {\bibinfo {volume} {84}},\ \bibinfo
  {pages} {905840105} (\bibinfo {year} {2018}{\natexlab{b}})}\BibitemShut
  {NoStop}%
\bibitem [{\citenamefont {{Watanabe}}\ and\ \citenamefont
  {{Sugama}}(2004)}]{Watanabe:2004}%
  \BibitemOpen
  \bibfield  {author} {\bibinfo {author} {\bibfnamefont {T.-H.}\ \bibnamefont
  {{Watanabe}}}\ and\ \bibinfo {author} {\bibfnamefont {H.}~\bibnamefont
  {{Sugama}}},\ }\bibfield  {title} {\enquote {\bibinfo {title} {{Kinetic
  simulation of steady states of ion temperature gradient driven turbulence
  with weak collisionality}},}\ }\href {\doibase 10.1063/1.1669393} {\bibfield
  {journal} {\bibinfo  {journal} {Phys.~Plasmas}\ }\textbf {\bibinfo {volume}
  {11}},\ \bibinfo {pages} {1476--1483} (\bibinfo {year} {2004})}\BibitemShut
  {NoStop}%
\bibitem [{\citenamefont {{Zocco}}\ and\ \citenamefont
  {{Schekochihin}}(2011)}]{Zocco:2011}%
  \BibitemOpen
  \bibfield  {author} {\bibinfo {author} {\bibfnamefont {A.}~\bibnamefont
  {{Zocco}}}\ and\ \bibinfo {author} {\bibfnamefont {A.~A.}\ \bibnamefont
  {{Schekochihin}}},\ }\bibfield  {title} {\enquote {\bibinfo {title} {{Reduced
  fluid-kinetic equations for low-frequency dynamics, magnetic reconnection,
  and electron heating in low-beta plasmas}},}\ }\href {\doibase
  10.1063/1.3628639} {\bibfield  {journal} {\bibinfo  {journal}
  {Phys.~Plasmas}\ }\textbf {\bibinfo {volume} {18}},\ \bibinfo {pages}
  {102309} (\bibinfo {year} {2011})},\ \Eprint {http://arxiv.org/abs/1104.4622}
  {arXiv:1104.4622 [physics.plasm-ph]} \BibitemShut {NoStop}%
\bibitem [{\citenamefont {{Hatch}}\ \emph {et~al.}(2013)\citenamefont
  {{Hatch}}, \citenamefont {{Jenko}}, \citenamefont {{Ba{\~n}{\'o}n Navarro}},\
  and\ \citenamefont {{Bratanov}}}]{Hatch:2013}%
  \BibitemOpen
  \bibfield  {author} {\bibinfo {author} {\bibfnamefont {D.~R.}\ \bibnamefont
  {{Hatch}}}, \bibinfo {author} {\bibfnamefont {F.}~\bibnamefont {{Jenko}}},
  \bibinfo {author} {\bibfnamefont {A.}~\bibnamefont {{Ba{\~n}{\'o}n
  Navarro}}}, \ and\ \bibinfo {author} {\bibfnamefont {V.}~\bibnamefont
  {{Bratanov}}},\ }\bibfield  {title} {\enquote {\bibinfo {title} {{Transition
  between Saturation Regimes of Gyrokinetic Turbulence}},}\ }\href {\doibase
  10.1103/PhysRevLett.111.175001} {\bibfield  {journal} {\bibinfo  {journal}
  {Phys.~Rev.~Lett.}\ }\textbf {\bibinfo {volume} {111}},\ \bibinfo {eid}
  {175001} (\bibinfo {year} {2013})}\BibitemShut {NoStop}%
\bibitem [{\citenamefont {{Loureiro}}, \citenamefont {{Schekochihin}},\ and\
  \citenamefont {{Zocco}}(2013)}]{Loureiro:2013}%
  \BibitemOpen
  \bibfield  {author} {\bibinfo {author} {\bibfnamefont {N.~F.}\ \bibnamefont
  {{Loureiro}}}, \bibinfo {author} {\bibfnamefont {A.~A.}\ \bibnamefont
  {{Schekochihin}}}, \ and\ \bibinfo {author} {\bibfnamefont {A.}~\bibnamefont
  {{Zocco}}},\ }\bibfield  {title} {\enquote {\bibinfo {title} {{Fast
  Collisionless Reconnection and Electron Heating in Strongly Magnetized
  Plasmas}},}\ }\href {\doibase 10.1103/PhysRevLett.111.025002} {\bibfield
  {journal} {\bibinfo  {journal} {Phys.~Rev.~Lett.}\ }\textbf {\bibinfo
  {volume} {111}},\ \bibinfo {eid} {025002} (\bibinfo {year} {2013})},\ \Eprint
  {http://arxiv.org/abs/1301.0338} {arXiv:1301.0338 [physics.plasm-ph]}
  \BibitemShut {NoStop}%
\bibitem [{\citenamefont {{Hatch}}\ \emph {et~al.}(2014)\citenamefont
  {{Hatch}}, \citenamefont {{Jenko}}, \citenamefont {{Bratanov}}, \citenamefont
  {{Navarro}},\ and\ \citenamefont {{Navarro}}}]{Hatch:2014}%
  \BibitemOpen
  \bibfield  {author} {\bibinfo {author} {\bibfnamefont {D.~R.}\ \bibnamefont
  {{Hatch}}}, \bibinfo {author} {\bibfnamefont {F.}~\bibnamefont {{Jenko}}},
  \bibinfo {author} {\bibfnamefont {V.}~\bibnamefont {{Bratanov}}}, \bibinfo
  {author} {\bibfnamefont {A.~B.}\ \bibnamefont {{Navarro}}}, \ and\ \bibinfo
  {author} {\bibnamefont {{Navarro}}},\ }\bibfield  {title} {\enquote {\bibinfo
  {title} {{Phase space scales of free energy dissipation in gradient-driven
  gyrokinetic turbulence}},}\ }\href {\doibase 10.1017/S0022377814000154}
  {\bibfield  {journal} {\bibinfo  {journal} {Journal of Plasma Physics}\
  }\textbf {\bibinfo {volume} {80}},\ \bibinfo {pages} {531--551} (\bibinfo
  {year} {2014})}\BibitemShut {NoStop}%
\bibitem [{\citenamefont {{Kanekar}}\ \emph {et~al.}(2015)\citenamefont
  {{Kanekar}}, \citenamefont {{Schekochihin}}, \citenamefont {{Dorland}},\ and\
  \citenamefont {{Loureiro}}}]{Kanekar:2015b}%
  \BibitemOpen
  \bibfield  {author} {\bibinfo {author} {\bibfnamefont {A.}~\bibnamefont
  {{Kanekar}}}, \bibinfo {author} {\bibfnamefont {A.~A.}\ \bibnamefont
  {{Schekochihin}}}, \bibinfo {author} {\bibfnamefont {W.}~\bibnamefont
  {{Dorland}}}, \ and\ \bibinfo {author} {\bibfnamefont {N.~F.}\ \bibnamefont
  {{Loureiro}}},\ }\bibfield  {title} {\enquote {\bibinfo {title}
  {{Fluctuation-dissipation relations for a plasma-kinetic Langevin
  equation}},}\ }\href {\doibase 10.1017/S0022377814000622} {\bibfield
  {journal} {\bibinfo  {journal} {J.~Plasma Phys.}\ }\textbf {\bibinfo {volume}
  {81}},\ \bibinfo {eid} {305810104} (\bibinfo {year} {2015})},\ \Eprint
  {http://arxiv.org/abs/1403.6257} {arXiv:1403.6257 [physics.plasm-ph]}
  \BibitemShut {NoStop}%
\bibitem [{\citenamefont {{Parker}}\ and\ \citenamefont
  {{Dellar}}(2015)}]{Parker:2015}%
  \BibitemOpen
  \bibfield  {author} {\bibinfo {author} {\bibfnamefont {J.~T.}\ \bibnamefont
  {{Parker}}}\ and\ \bibinfo {author} {\bibfnamefont {P.~J.}\ \bibnamefont
  {{Dellar}}},\ }\bibfield  {title} {\enquote {\bibinfo {title}
  {{Fourier-Hermite spectral representation for the Vlasov-Poisson system in
  the weakly collisional limit}},}\ }\href {\doibase 10.1017/S0022377814001287}
  {\bibfield  {journal} {\bibinfo  {journal} {J.~Plasma Phys.}\ }\textbf
  {\bibinfo {volume} {81}},\ \bibinfo {eid} {305810203} (\bibinfo {year}
  {2015})},\ \Eprint {http://arxiv.org/abs/1407.1932} {arXiv:1407.1932
  [physics.plasm-ph]} \BibitemShut {NoStop}%
\bibitem [{\citenamefont {{Pezzi}}, \citenamefont {{Camporeale}},\ and\
  \citenamefont {{Valentini}}(2016)}]{Pezzi:2016}%
  \BibitemOpen
  \bibfield  {author} {\bibinfo {author} {\bibfnamefont {O.}~\bibnamefont
  {{Pezzi}}}, \bibinfo {author} {\bibfnamefont {E.}~\bibnamefont
  {{Camporeale}}}, \ and\ \bibinfo {author} {\bibfnamefont {F.}~\bibnamefont
  {{Valentini}}},\ }\bibfield  {title} {\enquote {\bibinfo {title}
  {{Collisional effects on the numerical recurrence in Vlasov-Poisson
  simulations}},}\ }\href {\doibase 10.1063/1.4940963} {\bibfield  {journal}
  {\bibinfo  {journal} {Phys.~Plasmas}\ }\textbf {\bibinfo {volume} {23}},\
  \bibinfo {eid} {022103} (\bibinfo {year} {2016})},\ \Eprint
  {http://arxiv.org/abs/1601.05240} {arXiv:1601.05240 [physics.plasm-ph]}
  \BibitemShut {NoStop}%
\bibitem [{\citenamefont {{Schekochihin}}\ \emph {et~al.}(2016)\citenamefont
  {{Schekochihin}}, \citenamefont {{Parker}}, \citenamefont {{Highcock}},
  \citenamefont {{Dellar}}, \citenamefont {{Dorland}},\ and\ \citenamefont
  {{Hammett}}}]{Schekochihin:2016}%
  \BibitemOpen
  \bibfield  {author} {\bibinfo {author} {\bibfnamefont {A.~A.}\ \bibnamefont
  {{Schekochihin}}}, \bibinfo {author} {\bibfnamefont {J.~T.}\ \bibnamefont
  {{Parker}}}, \bibinfo {author} {\bibfnamefont {E.~G.}\ \bibnamefont
  {{Highcock}}}, \bibinfo {author} {\bibfnamefont {P.~J.}\ \bibnamefont
  {{Dellar}}}, \bibinfo {author} {\bibfnamefont {W.}~\bibnamefont {{Dorland}}},
  \ and\ \bibinfo {author} {\bibfnamefont {G.~W.}\ \bibnamefont {{Hammett}}},\
  }\bibfield  {title} {\enquote {\bibinfo {title} {{Phase mixing versus
  nonlinear advection in drift-kinetic plasma turbulence}},}\ }\href {\doibase
  10.1017/S0022377816000374} {\bibfield  {journal} {\bibinfo  {journal}
  {J.~Plasma Phys.}\ }\textbf {\bibinfo {volume} {82}},\ \bibinfo {eid}
  {905820212} (\bibinfo {year} {2016})},\ \Eprint
  {http://arxiv.org/abs/1508.05988} {arXiv:1508.05988 [physics.plasm-ph]}
  \BibitemShut {NoStop}%
\bibitem [{\citenamefont {{Parker}}\ \emph {et~al.}(2016)\citenamefont
  {{Parker}}, \citenamefont {{Highcock}}, \citenamefont {{Schekochihin}},\ and\
  \citenamefont {{Dellar}}}]{Parker:2016}%
  \BibitemOpen
  \bibfield  {author} {\bibinfo {author} {\bibfnamefont {J.~T.}\ \bibnamefont
  {{Parker}}}, \bibinfo {author} {\bibfnamefont {E.~G.}\ \bibnamefont
  {{Highcock}}}, \bibinfo {author} {\bibfnamefont {A.~A.}\ \bibnamefont
  {{Schekochihin}}}, \ and\ \bibinfo {author} {\bibfnamefont {P.~J.}\
  \bibnamefont {{Dellar}}},\ }\bibfield  {title} {\enquote {\bibinfo {title}
  {{Suppression of phase mixing in drift-kinetic plasma turbulence}},}\ }\href
  {\doibase 10.1063/1.4958954} {\bibfield  {journal} {\bibinfo  {journal}
  {Phys.~Plasmas}\ }\textbf {\bibinfo {volume} {23}},\ \bibinfo {eid} {070703}
  (\bibinfo {year} {2016})},\ \Eprint {http://arxiv.org/abs/1603.06968}
  {arXiv:1603.06968 [physics.plasm-ph]} \BibitemShut {NoStop}%
\bibitem [{\citenamefont {{Gro{\v s}elj}}\ \emph {et~al.}(2017)\citenamefont
  {{Gro{\v s}elj}}, \citenamefont {{Cerri}}, \citenamefont {{Ba{\~n}{\'o}n
  Navarro}}, \citenamefont {{Willmott}}, \citenamefont {{Told}}, \citenamefont
  {{Loureiro}}, \citenamefont {{Califano}},\ and\ \citenamefont
  {{Jenko}}}]{Groselj:2017}%
  \BibitemOpen
  \bibfield  {author} {\bibinfo {author} {\bibfnamefont {D.}~\bibnamefont
  {{Gro{\v s}elj}}}, \bibinfo {author} {\bibfnamefont {S.~S.}\ \bibnamefont
  {{Cerri}}}, \bibinfo {author} {\bibfnamefont {A.}~\bibnamefont
  {{Ba{\~n}{\'o}n Navarro}}}, \bibinfo {author} {\bibfnamefont
  {C.}~\bibnamefont {{Willmott}}}, \bibinfo {author} {\bibfnamefont
  {D.}~\bibnamefont {{Told}}}, \bibinfo {author} {\bibfnamefont {N.~F.}\
  \bibnamefont {{Loureiro}}}, \bibinfo {author} {\bibfnamefont
  {F.}~\bibnamefont {{Califano}}}, \ and\ \bibinfo {author} {\bibfnamefont
  {F.}~\bibnamefont {{Jenko}}},\ }\bibfield  {title} {\enquote {\bibinfo
  {title} {{Fully Kinetic versus Reduced-kinetic Modeling of Collisionless
  Plasma Turbulence}},}\ }\href {\doibase 10.3847/1538-4357/aa894d} {\bibfield
  {journal} {\bibinfo  {journal} {Astrophys.~J.}\ }\textbf {\bibinfo {volume}
  {847}},\ \bibinfo {eid} {28} (\bibinfo {year} {2017})},\ \Eprint
  {http://arxiv.org/abs/1706.02652} {arXiv:1706.02652 [physics.plasm-ph]}
  \BibitemShut {NoStop}%
\bibitem [{\citenamefont {{White}}\ and\ \citenamefont
  {{Hazeltine}}(2017)}]{White:2017}%
  \BibitemOpen
  \bibfield  {author} {\bibinfo {author} {\bibfnamefont {R.~L.}\ \bibnamefont
  {{White}}}\ and\ \bibinfo {author} {\bibfnamefont {R.~D.}\ \bibnamefont
  {{Hazeltine}}},\ }\bibfield  {title} {\enquote {\bibinfo {title} {{Analysis
  of the Hermite spectrum in plasma turbulence}},}\ }\href {\doibase
  10.1063/1.5000518} {\bibfield  {journal} {\bibinfo  {journal}
  {Phys.~Plasmas}\ }\textbf {\bibinfo {volume} {24}},\ \bibinfo {eid} {102315}
  (\bibinfo {year} {2017})}\BibitemShut {NoStop}%
\bibitem [{\citenamefont {{Servidio}}\ \emph {et~al.}(2017)\citenamefont
  {{Servidio}}, \citenamefont {{Chasapis}}, \citenamefont {{Matthaeus}},
  \citenamefont {{Perrone}}, \citenamefont {{Valentini}}, \citenamefont
  {{Parashar}}, \citenamefont {{Veltri}}, \citenamefont {{Gershman}},
  \citenamefont {{Russell}}, \citenamefont {{Giles}}, \citenamefont
  {{Fuselier}}, \citenamefont {{Phan}},\ and\ \citenamefont
  {{Burch}}}]{Servidio:2017}%
  \BibitemOpen
  \bibfield  {author} {\bibinfo {author} {\bibfnamefont {S.}~\bibnamefont
  {{Servidio}}}, \bibinfo {author} {\bibfnamefont {A.}~\bibnamefont
  {{Chasapis}}}, \bibinfo {author} {\bibfnamefont {W.~H.}\ \bibnamefont
  {{Matthaeus}}}, \bibinfo {author} {\bibfnamefont {D.}~\bibnamefont
  {{Perrone}}}, \bibinfo {author} {\bibfnamefont {F.}~\bibnamefont
  {{Valentini}}}, \bibinfo {author} {\bibfnamefont {T.~N.}\ \bibnamefont
  {{Parashar}}}, \bibinfo {author} {\bibfnamefont {P.}~\bibnamefont
  {{Veltri}}}, \bibinfo {author} {\bibfnamefont {D.}~\bibnamefont
  {{Gershman}}}, \bibinfo {author} {\bibfnamefont {C.~T.}\ \bibnamefont
  {{Russell}}}, \bibinfo {author} {\bibfnamefont {B.}~\bibnamefont {{Giles}}},
  \bibinfo {author} {\bibfnamefont {S.~A.}\ \bibnamefont {{Fuselier}}},
  \bibinfo {author} {\bibfnamefont {T.~D.}\ \bibnamefont {{Phan}}}, \ and\
  \bibinfo {author} {\bibfnamefont {J.}~\bibnamefont {{Burch}}},\ }\bibfield
  {title} {\enquote {\bibinfo {title} {{Magnetospheric Multiscale Observation
  of Plasma Velocity-Space Cascade: Hermite Representation and Theory}},}\
  }\href {\doibase 10.1103/PhysRevLett.119.205101} {\bibfield  {journal}
  {\bibinfo  {journal} {Phys.~Rev.~Lett.}\ }\textbf {\bibinfo {volume} {119}},\
  \bibinfo {eid} {205101} (\bibinfo {year} {2017})},\ \Eprint
  {http://arxiv.org/abs/1707.08180} {arXiv:1707.08180 [physics.plasm-ph]}
  \BibitemShut {NoStop}%
\bibitem [{\citenamefont {{Bellan}}(2012)}]{Bellan:2012}%
  \BibitemOpen
  \bibfield  {author} {\bibinfo {author} {\bibfnamefont {P.~M.}\ \bibnamefont
  {{Bellan}}},\ }\bibfield  {title} {\enquote {\bibinfo {title} {{Improved
  basis set for low frequency plasma waves}},}\ }\href {\doibase
  10.1029/2012JA017856} {\bibfield  {journal} {\bibinfo  {journal}
  {J.~Geophys.~Res.}\ }\textbf {\bibinfo {volume} {117}},\ \bibinfo {eid}
  {A12219} (\bibinfo {year} {2012})}\BibitemShut {NoStop}%
\bibitem [{\citenamefont {{Bellan}}(2016)}]{Bellan:2016}%
  \BibitemOpen
  \bibfield  {author} {\bibinfo {author} {\bibfnamefont {P.~M.}\ \bibnamefont
  {{Bellan}}},\ }\bibfield  {title} {\enquote {\bibinfo {title} {Revised
  single-spacecraft method for determining wavevector k and resolving
  space-time ambiguity},}\ }\href@noop {} {\bibfield  {journal} {\bibinfo
  {journal} {J.~Geophys.~Res.}\ } (\bibinfo {year} {2016})},\ \bibinfo {note}
  {submitted}\BibitemShut {NoStop}%
\bibitem [{\citenamefont {{Gershman}}\ \emph {et~al.}(2017)\citenamefont
  {{Gershman}}, \citenamefont {{F-Vi{\~n}as}}, \citenamefont {{Dorelli}},
  \citenamefont {{Boardsen}}, \citenamefont {{Avanov}}, \citenamefont
  {{Bellan}}, \citenamefont {{Schwartz}}, \citenamefont {{Lavraud}},
  \citenamefont {{Coffey}}, \citenamefont {{Chandler}}, \citenamefont
  {{Saito}}, \citenamefont {{Paterson}}, \citenamefont {{Fuselier}},
  \citenamefont {{Ergun}}, \citenamefont {{Strangeway}}, \citenamefont
  {{Russell}}, \citenamefont {{Giles}}, \citenamefont {{Pollock}},
  \citenamefont {{Torbert}},\ and\ \citenamefont {{Burch}}}]{Gershman:2017}%
  \BibitemOpen
  \bibfield  {author} {\bibinfo {author} {\bibfnamefont {D.~J.}\ \bibnamefont
  {{Gershman}}}, \bibinfo {author} {\bibfnamefont {A.}~\bibnamefont
  {{F-Vi{\~n}as}}}, \bibinfo {author} {\bibfnamefont {J.~C.}\ \bibnamefont
  {{Dorelli}}}, \bibinfo {author} {\bibfnamefont {S.~A.}\ \bibnamefont
  {{Boardsen}}}, \bibinfo {author} {\bibfnamefont {L.~A.}\ \bibnamefont
  {{Avanov}}}, \bibinfo {author} {\bibfnamefont {P.~M.}\ \bibnamefont
  {{Bellan}}}, \bibinfo {author} {\bibfnamefont {S.~J.}\ \bibnamefont
  {{Schwartz}}}, \bibinfo {author} {\bibfnamefont {B.}~\bibnamefont
  {{Lavraud}}}, \bibinfo {author} {\bibfnamefont {V.~N.}\ \bibnamefont
  {{Coffey}}}, \bibinfo {author} {\bibfnamefont {M.~O.}\ \bibnamefont
  {{Chandler}}}, \bibinfo {author} {\bibfnamefont {Y.}~\bibnamefont {{Saito}}},
  \bibinfo {author} {\bibfnamefont {W.~R.}\ \bibnamefont {{Paterson}}},
  \bibinfo {author} {\bibfnamefont {S.~A.}\ \bibnamefont {{Fuselier}}},
  \bibinfo {author} {\bibfnamefont {R.~E.}\ \bibnamefont {{Ergun}}}, \bibinfo
  {author} {\bibfnamefont {R.~J.}\ \bibnamefont {{Strangeway}}}, \bibinfo
  {author} {\bibfnamefont {C.~T.}\ \bibnamefont {{Russell}}}, \bibinfo {author}
  {\bibfnamefont {B.~L.}\ \bibnamefont {{Giles}}}, \bibinfo {author}
  {\bibfnamefont {C.~J.}\ \bibnamefont {{Pollock}}}, \bibinfo {author}
  {\bibfnamefont {R.~B.}\ \bibnamefont {{Torbert}}}, \ and\ \bibinfo {author}
  {\bibfnamefont {J.~L.}\ \bibnamefont {{Burch}}},\ }\bibfield  {title}
  {\enquote {\bibinfo {title} {{Wave-particle energy exchange directly observed
  in a kinetic Alfv{\'e}n-branch wave}},}\ }\href {\doibase
  10.1038/ncomms14719} {\bibfield  {journal} {\bibinfo  {journal} {Nature
  Comm.}\ }\textbf {\bibinfo {volume} {8}},\ \bibinfo {eid} {14719} (\bibinfo
  {year} {2017})}\BibitemShut {NoStop}%
\bibitem [{\citenamefont {{Klein}}\ \emph {et~al.}(2017)\citenamefont
  {{Klein}}, \citenamefont {{Kasper}}, \citenamefont {{Korreck}},\ and\
  \citenamefont {{Stevens}}}]{Klein:2017c}%
  \BibitemOpen
  \bibfield  {author} {\bibinfo {author} {\bibfnamefont {K.~G.}\ \bibnamefont
  {{Klein}}}, \bibinfo {author} {\bibfnamefont {J.~C.}\ \bibnamefont
  {{Kasper}}}, \bibinfo {author} {\bibfnamefont {K.~E.}\ \bibnamefont
  {{Korreck}}}, \ and\ \bibinfo {author} {\bibfnamefont {M.~L.}\ \bibnamefont
  {{Stevens}}},\ }\bibfield  {title} {\enquote {\bibinfo {title} {Applying
  nyquist's method for stability determination to solar wind observations},}\
  }\href {\doibase 10.1002/2017JA024486} {\bibfield  {journal} {\bibinfo
  {journal} {J.~Geophys.~Res.}\ } (\bibinfo {year} {2017}),\
  10.1002/2017JA024486},\ \bibinfo {note} {in press}\BibitemShut {NoStop}%
\bibitem [{\citenamefont {{Bale}}\ \emph {et~al.}(2016)\citenamefont {{Bale}},
  \citenamefont {{Goetz}}, \citenamefont {{Harvey}}, \citenamefont {{Turin}},
  \citenamefont {{Bonnell}}, \citenamefont {{Dudok de Wit}}, \citenamefont
  {{Ergun}}, \citenamefont {{MacDowall}}, \citenamefont {{Pulupa}},
  \citenamefont {{Andre}}, \citenamefont {{Bolton}}, \citenamefont
  {{Bougeret}}, \citenamefont {{Bowen}}, \citenamefont {{Burgess}},
  \citenamefont {{Cattell}}, \citenamefont {{Chandran}}, \citenamefont
  {{Chaston}}, \citenamefont {{Chen}}, \citenamefont {{Choi}}, \citenamefont
  {{Connerney}}, \citenamefont {{Cranmer}}, \citenamefont {{Diaz-Aguado}},
  \citenamefont {{Donakowski}}, \citenamefont {{Drake}}, \citenamefont
  {{Farrell}}, \citenamefont {{Fergeau}}, \citenamefont {{Fermin}},
  \citenamefont {{Fischer}}, \citenamefont {{Fox}}, \citenamefont {{Glaser}},
  \citenamefont {{Goldstein}}, \citenamefont {{Gordon}}, \citenamefont
  {{Hanson}}, \citenamefont {{Harris}}, \citenamefont {{Hayes}}, \citenamefont
  {{Hinze}}, \citenamefont {{Hollweg}}, \citenamefont {{Horbury}},
  \citenamefont {{Howard}}, \citenamefont {{Hoxie}}, \citenamefont {{Jannet}},
  \citenamefont {{Karlsson}}, \citenamefont {{Kasper}}, \citenamefont
  {{Kellogg}}, \citenamefont {{Kien}}, \citenamefont {{Klimchuk}},
  \citenamefont {{Krasnoselskikh}}, \citenamefont {{Krucker}}, \citenamefont
  {{Lynch}}, \citenamefont {{Maksimovic}}, \citenamefont {{Malaspina}},
  \citenamefont {{Marker}}, \citenamefont {{Martin}}, \citenamefont
  {{Martinez-Oliveros}}, \citenamefont {{McCauley}}, \citenamefont {{McComas}},
  \citenamefont {{McDonald}}, \citenamefont {{Meyer-Vernet}}, \citenamefont
  {{Moncuquet}}, \citenamefont {{Monson}}, \citenamefont {{Mozer}},
  \citenamefont {{Murphy}}, \citenamefont {{Odom}}, \citenamefont
  {{Oliverson}}, \citenamefont {{Olson}}, \citenamefont {{Parker}},
  \citenamefont {{Pankow}}, \citenamefont {{Phan}}, \citenamefont {{Quataert}},
  \citenamefont {{Quinn}}, \citenamefont {{Ruplin}}, \citenamefont {{Salem}},
  \citenamefont {{Seitz}}, \citenamefont {{Sheppard}}, \citenamefont {{Siy}},
  \citenamefont {{Stevens}}, \citenamefont {{Summers}}, \citenamefont
  {{Szabo}}, \citenamefont {{Timofeeva}}, \citenamefont {{Vaivads}},
  \citenamefont {{Velli}}, \citenamefont {{Yehle}}, \citenamefont
  {{Werthimer}},\ and\ \citenamefont {{Wygant}}}]{Bale:2016}%
  \BibitemOpen
  \bibfield  {author} {\bibinfo {author} {\bibfnamefont {S.~D.}\ \bibnamefont
  {{Bale}}}, \bibinfo {author} {\bibfnamefont {K.}~\bibnamefont {{Goetz}}},
  \bibinfo {author} {\bibfnamefont {P.~R.}\ \bibnamefont {{Harvey}}}, \bibinfo
  {author} {\bibfnamefont {P.}~\bibnamefont {{Turin}}}, \bibinfo {author}
  {\bibfnamefont {J.~W.}\ \bibnamefont {{Bonnell}}}, \bibinfo {author}
  {\bibfnamefont {T.}~\bibnamefont {{Dudok de Wit}}}, \bibinfo {author}
  {\bibfnamefont {R.~E.}\ \bibnamefont {{Ergun}}}, \bibinfo {author}
  {\bibfnamefont {R.~J.}\ \bibnamefont {{MacDowall}}}, \bibinfo {author}
  {\bibfnamefont {M.}~\bibnamefont {{Pulupa}}}, \bibinfo {author}
  {\bibfnamefont {M.}~\bibnamefont {{Andre}}}, \bibinfo {author} {\bibfnamefont
  {M.}~\bibnamefont {{Bolton}}}, \bibinfo {author} {\bibfnamefont {J.-L.}\
  \bibnamefont {{Bougeret}}}, \bibinfo {author} {\bibfnamefont {T.~A.}\
  \bibnamefont {{Bowen}}}, \bibinfo {author} {\bibfnamefont {D.}~\bibnamefont
  {{Burgess}}}, \bibinfo {author} {\bibfnamefont {C.~A.}\ \bibnamefont
  {{Cattell}}}, \bibinfo {author} {\bibfnamefont {B.~D.~G.}\ \bibnamefont
  {{Chandran}}}, \bibinfo {author} {\bibfnamefont {C.~C.}\ \bibnamefont
  {{Chaston}}}, \bibinfo {author} {\bibfnamefont {C.~H.~K.}\ \bibnamefont
  {{Chen}}}, \bibinfo {author} {\bibfnamefont {M.~K.}\ \bibnamefont {{Choi}}},
  \bibinfo {author} {\bibfnamefont {J.~E.}\ \bibnamefont {{Connerney}}},
  \bibinfo {author} {\bibfnamefont {S.}~\bibnamefont {{Cranmer}}}, \bibinfo
  {author} {\bibfnamefont {M.}~\bibnamefont {{Diaz-Aguado}}}, \bibinfo {author}
  {\bibfnamefont {W.}~\bibnamefont {{Donakowski}}}, \bibinfo {author}
  {\bibfnamefont {J.~F.}\ \bibnamefont {{Drake}}}, \bibinfo {author}
  {\bibfnamefont {W.~M.}\ \bibnamefont {{Farrell}}}, \bibinfo {author}
  {\bibfnamefont {P.}~\bibnamefont {{Fergeau}}}, \bibinfo {author}
  {\bibfnamefont {J.}~\bibnamefont {{Fermin}}}, \bibinfo {author}
  {\bibfnamefont {J.}~\bibnamefont {{Fischer}}}, \bibinfo {author}
  {\bibfnamefont {N.}~\bibnamefont {{Fox}}}, \bibinfo {author} {\bibfnamefont
  {D.}~\bibnamefont {{Glaser}}}, \bibinfo {author} {\bibfnamefont
  {M.}~\bibnamefont {{Goldstein}}}, \bibinfo {author} {\bibfnamefont
  {D.}~\bibnamefont {{Gordon}}}, \bibinfo {author} {\bibfnamefont
  {E.}~\bibnamefont {{Hanson}}}, \bibinfo {author} {\bibfnamefont {S.~E.}\
  \bibnamefont {{Harris}}}, \bibinfo {author} {\bibfnamefont {L.~M.}\
  \bibnamefont {{Hayes}}}, \bibinfo {author} {\bibfnamefont {J.~J.}\
  \bibnamefont {{Hinze}}}, \bibinfo {author} {\bibfnamefont {J.~V.}\
  \bibnamefont {{Hollweg}}}, \bibinfo {author} {\bibfnamefont {T.~S.}\
  \bibnamefont {{Horbury}}}, \bibinfo {author} {\bibfnamefont {R.~A.}\
  \bibnamefont {{Howard}}}, \bibinfo {author} {\bibfnamefont {V.}~\bibnamefont
  {{Hoxie}}}, \bibinfo {author} {\bibfnamefont {G.}~\bibnamefont {{Jannet}}},
  \bibinfo {author} {\bibfnamefont {M.}~\bibnamefont {{Karlsson}}}, \bibinfo
  {author} {\bibfnamefont {J.~C.}\ \bibnamefont {{Kasper}}}, \bibinfo {author}
  {\bibfnamefont {P.~J.}\ \bibnamefont {{Kellogg}}}, \bibinfo {author}
  {\bibfnamefont {M.}~\bibnamefont {{Kien}}}, \bibinfo {author} {\bibfnamefont
  {J.~A.}\ \bibnamefont {{Klimchuk}}}, \bibinfo {author} {\bibfnamefont
  {V.~V.}\ \bibnamefont {{Krasnoselskikh}}}, \bibinfo {author} {\bibfnamefont
  {S.}~\bibnamefont {{Krucker}}}, \bibinfo {author} {\bibfnamefont {J.~J.}\
  \bibnamefont {{Lynch}}}, \bibinfo {author} {\bibfnamefont {M.}~\bibnamefont
  {{Maksimovic}}}, \bibinfo {author} {\bibfnamefont {D.~M.}\ \bibnamefont
  {{Malaspina}}}, \bibinfo {author} {\bibfnamefont {S.}~\bibnamefont
  {{Marker}}}, \bibinfo {author} {\bibfnamefont {P.}~\bibnamefont {{Martin}}},
  \bibinfo {author} {\bibfnamefont {J.}~\bibnamefont {{Martinez-Oliveros}}},
  \bibinfo {author} {\bibfnamefont {J.}~\bibnamefont {{McCauley}}}, \bibinfo
  {author} {\bibfnamefont {D.~J.}\ \bibnamefont {{McComas}}}, \bibinfo {author}
  {\bibfnamefont {T.}~\bibnamefont {{McDonald}}}, \bibinfo {author}
  {\bibfnamefont {N.}~\bibnamefont {{Meyer-Vernet}}}, \bibinfo {author}
  {\bibfnamefont {M.}~\bibnamefont {{Moncuquet}}}, \bibinfo {author}
  {\bibfnamefont {S.~J.}\ \bibnamefont {{Monson}}}, \bibinfo {author}
  {\bibfnamefont {F.~S.}\ \bibnamefont {{Mozer}}}, \bibinfo {author}
  {\bibfnamefont {S.~D.}\ \bibnamefont {{Murphy}}}, \bibinfo {author}
  {\bibfnamefont {J.}~\bibnamefont {{Odom}}}, \bibinfo {author} {\bibfnamefont
  {R.}~\bibnamefont {{Oliverson}}}, \bibinfo {author} {\bibfnamefont
  {J.}~\bibnamefont {{Olson}}}, \bibinfo {author} {\bibfnamefont {E.~N.}\
  \bibnamefont {{Parker}}}, \bibinfo {author} {\bibfnamefont {D.}~\bibnamefont
  {{Pankow}}}, \bibinfo {author} {\bibfnamefont {T.}~\bibnamefont {{Phan}}},
  \bibinfo {author} {\bibfnamefont {E.}~\bibnamefont {{Quataert}}}, \bibinfo
  {author} {\bibfnamefont {T.}~\bibnamefont {{Quinn}}}, \bibinfo {author}
  {\bibfnamefont {S.~W.}\ \bibnamefont {{Ruplin}}}, \bibinfo {author}
  {\bibfnamefont {C.}~\bibnamefont {{Salem}}}, \bibinfo {author} {\bibfnamefont
  {D.}~\bibnamefont {{Seitz}}}, \bibinfo {author} {\bibfnamefont {D.~A.}\
  \bibnamefont {{Sheppard}}}, \bibinfo {author} {\bibfnamefont
  {A.}~\bibnamefont {{Siy}}}, \bibinfo {author} {\bibfnamefont
  {K.}~\bibnamefont {{Stevens}}}, \bibinfo {author} {\bibfnamefont
  {D.}~\bibnamefont {{Summers}}}, \bibinfo {author} {\bibfnamefont
  {A.}~\bibnamefont {{Szabo}}}, \bibinfo {author} {\bibfnamefont
  {M.}~\bibnamefont {{Timofeeva}}}, \bibinfo {author} {\bibfnamefont
  {A.}~\bibnamefont {{Vaivads}}}, \bibinfo {author} {\bibfnamefont
  {M.}~\bibnamefont {{Velli}}}, \bibinfo {author} {\bibfnamefont
  {A.}~\bibnamefont {{Yehle}}}, \bibinfo {author} {\bibfnamefont
  {D.}~\bibnamefont {{Werthimer}}}, \ and\ \bibinfo {author} {\bibfnamefont
  {J.~R.}\ \bibnamefont {{Wygant}}},\ }\bibfield  {title} {\enquote {\bibinfo
  {title} {{The FIELDS Instrument Suite for Solar Probe Plus. Measuring the
  Coronal Plasma and Magnetic Field, Plasma Waves and Turbulence, and Radio
  Signatures of Solar Transients}},}\ }\href {\doibase
  10.1007/s11214-016-0244-5} {\bibfield  {journal} {\bibinfo  {journal} {Space
  Sci.~Rev.}\ }\textbf {\bibinfo {volume} {204}},\ \bibinfo {pages} {49--82}
  (\bibinfo {year} {2016})}\BibitemShut {NoStop}%
\bibitem [{\citenamefont {{Kasper}}\ \emph {et~al.}(2016)\citenamefont
  {{Kasper}}, \citenamefont {{Abiad}}, \citenamefont {{Austin}}, \citenamefont
  {{Balat-Pichelin}}, \citenamefont {{Bale}}, \citenamefont {{Belcher}},
  \citenamefont {{Berg}}, \citenamefont {{Bergner}}, \citenamefont
  {{Berthomier}}, \citenamefont {{Bookbinder}}, \citenamefont {{Brodu}},
  \citenamefont {{Caldwell}}, \citenamefont {{Case}}, \citenamefont
  {{Chandran}}, \citenamefont {{Cheimets}}, \citenamefont {{Cirtain}},
  \citenamefont {{Cranmer}}, \citenamefont {{Curtis}}, \citenamefont
  {{Daigneau}}, \citenamefont {{Dalton}}, \citenamefont {{Dasgupta}},
  \citenamefont {{DeTomaso}}, \citenamefont {{Diaz-Aguado}}, \citenamefont
  {{Djordjevic}}, \citenamefont {{Donaskowski}}, \citenamefont {{Effinger}},
  \citenamefont {{Florinski}}, \citenamefont {{Fox}}, \citenamefont
  {{Freeman}}, \citenamefont {{Gallagher}}, \citenamefont {{Gary}},
  \citenamefont {{Gauron}}, \citenamefont {{Gates}}, \citenamefont
  {{Goldstein}}, \citenamefont {{Golub}}, \citenamefont {{Gordon}},
  \citenamefont {{Gurnee}}, \citenamefont {{Guth}}, \citenamefont {{Halekas}},
  \citenamefont {{Hatch}}, \citenamefont {{Heerikuisen}}, \citenamefont {{Ho}},
  \citenamefont {{Hu}}, \citenamefont {{Johnson}}, \citenamefont {{Jordan}},
  \citenamefont {{Korreck}}, \citenamefont {{Larson}}, \citenamefont
  {{Lazarus}}, \citenamefont {{Li}}, \citenamefont {{Livi}}, \citenamefont
  {{Ludlam}}, \citenamefont {{Maksimovic}}, \citenamefont {{McFadden}},
  \citenamefont {{Marchant}}, \citenamefont {{Maruca}}, \citenamefont
  {{McComas}}, \citenamefont {{Messina}}, \citenamefont {{Mercer}},
  \citenamefont {{Park}}, \citenamefont {{Peddie}}, \citenamefont
  {{Pogorelov}}, \citenamefont {{Reinhart}}, \citenamefont {{Richardson}},
  \citenamefont {{Robinson}}, \citenamefont {{Rosen}}, \citenamefont {{Skoug}},
  \citenamefont {{Slagle}}, \citenamefont {{Steinberg}}, \citenamefont
  {{Stevens}}, \citenamefont {{Szabo}}, \citenamefont {{Taylor}}, \citenamefont
  {{Tiu}}, \citenamefont {{Turin}}, \citenamefont {{Velli}}, \citenamefont
  {{Webb}}, \citenamefont {{Whittlesey}}, \citenamefont {{Wright}},
  \citenamefont {{Wu}},\ and\ \citenamefont {{Zank}}}]{Kasper:2016}%
  \BibitemOpen
  \bibfield  {author} {\bibinfo {author} {\bibfnamefont {J.~C.}\ \bibnamefont
  {{Kasper}}}, \bibinfo {author} {\bibfnamefont {R.}~\bibnamefont {{Abiad}}},
  \bibinfo {author} {\bibfnamefont {G.}~\bibnamefont {{Austin}}}, \bibinfo
  {author} {\bibfnamefont {M.}~\bibnamefont {{Balat-Pichelin}}}, \bibinfo
  {author} {\bibfnamefont {S.~D.}\ \bibnamefont {{Bale}}}, \bibinfo {author}
  {\bibfnamefont {J.~W.}\ \bibnamefont {{Belcher}}}, \bibinfo {author}
  {\bibfnamefont {P.}~\bibnamefont {{Berg}}}, \bibinfo {author} {\bibfnamefont
  {H.}~\bibnamefont {{Bergner}}}, \bibinfo {author} {\bibfnamefont
  {M.}~\bibnamefont {{Berthomier}}}, \bibinfo {author} {\bibfnamefont
  {J.}~\bibnamefont {{Bookbinder}}}, \bibinfo {author} {\bibfnamefont
  {E.}~\bibnamefont {{Brodu}}}, \bibinfo {author} {\bibfnamefont
  {D.}~\bibnamefont {{Caldwell}}}, \bibinfo {author} {\bibfnamefont {A.~W.}\
  \bibnamefont {{Case}}}, \bibinfo {author} {\bibfnamefont {B.~D.~G.}\
  \bibnamefont {{Chandran}}}, \bibinfo {author} {\bibfnamefont
  {P.}~\bibnamefont {{Cheimets}}}, \bibinfo {author} {\bibfnamefont {J.~W.}\
  \bibnamefont {{Cirtain}}}, \bibinfo {author} {\bibfnamefont {S.~R.}\
  \bibnamefont {{Cranmer}}}, \bibinfo {author} {\bibfnamefont {D.~W.}\
  \bibnamefont {{Curtis}}}, \bibinfo {author} {\bibfnamefont {P.}~\bibnamefont
  {{Daigneau}}}, \bibinfo {author} {\bibfnamefont {G.}~\bibnamefont
  {{Dalton}}}, \bibinfo {author} {\bibfnamefont {B.}~\bibnamefont
  {{Dasgupta}}}, \bibinfo {author} {\bibfnamefont {D.}~\bibnamefont
  {{DeTomaso}}}, \bibinfo {author} {\bibfnamefont {M.}~\bibnamefont
  {{Diaz-Aguado}}}, \bibinfo {author} {\bibfnamefont {B.}~\bibnamefont
  {{Djordjevic}}}, \bibinfo {author} {\bibfnamefont {B.}~\bibnamefont
  {{Donaskowski}}}, \bibinfo {author} {\bibfnamefont {M.}~\bibnamefont
  {{Effinger}}}, \bibinfo {author} {\bibfnamefont {V.}~\bibnamefont
  {{Florinski}}}, \bibinfo {author} {\bibfnamefont {N.}~\bibnamefont {{Fox}}},
  \bibinfo {author} {\bibfnamefont {M.}~\bibnamefont {{Freeman}}}, \bibinfo
  {author} {\bibfnamefont {D.}~\bibnamefont {{Gallagher}}}, \bibinfo {author}
  {\bibfnamefont {S.~P.}\ \bibnamefont {{Gary}}}, \bibinfo {author}
  {\bibfnamefont {T.}~\bibnamefont {{Gauron}}}, \bibinfo {author}
  {\bibfnamefont {R.}~\bibnamefont {{Gates}}}, \bibinfo {author} {\bibfnamefont
  {M.}~\bibnamefont {{Goldstein}}}, \bibinfo {author} {\bibfnamefont
  {L.}~\bibnamefont {{Golub}}}, \bibinfo {author} {\bibfnamefont {D.~A.}\
  \bibnamefont {{Gordon}}}, \bibinfo {author} {\bibfnamefont {R.}~\bibnamefont
  {{Gurnee}}}, \bibinfo {author} {\bibfnamefont {G.}~\bibnamefont {{Guth}}},
  \bibinfo {author} {\bibfnamefont {J.}~\bibnamefont {{Halekas}}}, \bibinfo
  {author} {\bibfnamefont {K.}~\bibnamefont {{Hatch}}}, \bibinfo {author}
  {\bibfnamefont {J.}~\bibnamefont {{Heerikuisen}}}, \bibinfo {author}
  {\bibfnamefont {G.}~\bibnamefont {{Ho}}}, \bibinfo {author} {\bibfnamefont
  {Q.}~\bibnamefont {{Hu}}}, \bibinfo {author} {\bibfnamefont {G.}~\bibnamefont
  {{Johnson}}}, \bibinfo {author} {\bibfnamefont {S.~P.}\ \bibnamefont
  {{Jordan}}}, \bibinfo {author} {\bibfnamefont {K.~E.}\ \bibnamefont
  {{Korreck}}}, \bibinfo {author} {\bibfnamefont {D.}~\bibnamefont {{Larson}}},
  \bibinfo {author} {\bibfnamefont {A.~J.}\ \bibnamefont {{Lazarus}}}, \bibinfo
  {author} {\bibfnamefont {G.}~\bibnamefont {{Li}}}, \bibinfo {author}
  {\bibfnamefont {R.}~\bibnamefont {{Livi}}}, \bibinfo {author} {\bibfnamefont
  {M.}~\bibnamefont {{Ludlam}}}, \bibinfo {author} {\bibfnamefont
  {M.}~\bibnamefont {{Maksimovic}}}, \bibinfo {author} {\bibfnamefont {J.~P.}\
  \bibnamefont {{McFadden}}}, \bibinfo {author} {\bibfnamefont
  {W.}~\bibnamefont {{Marchant}}}, \bibinfo {author} {\bibfnamefont {B.~A.}\
  \bibnamefont {{Maruca}}}, \bibinfo {author} {\bibfnamefont {D.~J.}\
  \bibnamefont {{McComas}}}, \bibinfo {author} {\bibfnamefont {L.}~\bibnamefont
  {{Messina}}}, \bibinfo {author} {\bibfnamefont {T.}~\bibnamefont {{Mercer}}},
  \bibinfo {author} {\bibfnamefont {S.}~\bibnamefont {{Park}}}, \bibinfo
  {author} {\bibfnamefont {A.~M.}\ \bibnamefont {{Peddie}}}, \bibinfo {author}
  {\bibfnamefont {N.}~\bibnamefont {{Pogorelov}}}, \bibinfo {author}
  {\bibfnamefont {M.~J.}\ \bibnamefont {{Reinhart}}}, \bibinfo {author}
  {\bibfnamefont {J.~D.}\ \bibnamefont {{Richardson}}}, \bibinfo {author}
  {\bibfnamefont {M.}~\bibnamefont {{Robinson}}}, \bibinfo {author}
  {\bibfnamefont {I.}~\bibnamefont {{Rosen}}}, \bibinfo {author} {\bibfnamefont
  {R.~M.}\ \bibnamefont {{Skoug}}}, \bibinfo {author} {\bibfnamefont
  {A.}~\bibnamefont {{Slagle}}}, \bibinfo {author} {\bibfnamefont {J.~T.}\
  \bibnamefont {{Steinberg}}}, \bibinfo {author} {\bibfnamefont {M.~L.}\
  \bibnamefont {{Stevens}}}, \bibinfo {author} {\bibfnamefont {A.}~\bibnamefont
  {{Szabo}}}, \bibinfo {author} {\bibfnamefont {E.~R.}\ \bibnamefont
  {{Taylor}}}, \bibinfo {author} {\bibfnamefont {C.}~\bibnamefont {{Tiu}}},
  \bibinfo {author} {\bibfnamefont {P.}~\bibnamefont {{Turin}}}, \bibinfo
  {author} {\bibfnamefont {M.}~\bibnamefont {{Velli}}}, \bibinfo {author}
  {\bibfnamefont {G.}~\bibnamefont {{Webb}}}, \bibinfo {author} {\bibfnamefont
  {P.}~\bibnamefont {{Whittlesey}}}, \bibinfo {author} {\bibfnamefont
  {K.}~\bibnamefont {{Wright}}}, \bibinfo {author} {\bibfnamefont {S.~T.}\
  \bibnamefont {{Wu}}}, \ and\ \bibinfo {author} {\bibfnamefont
  {G.}~\bibnamefont {{Zank}}},\ }\bibfield  {title} {\enquote {\bibinfo {title}
  {{Solar Wind Electrons Alphas and Protons (SWEAP) Investigation: Design of
  the Solar Wind and Coronal Plasma Instrument Suite for Solar Probe Plus}},}\
  }\href {\doibase 10.1007/s11214-015-0206-3} {\bibfield  {journal} {\bibinfo
  {journal} {Space Sci.~Rev.}\ }\textbf {\bibinfo {volume} {204}},\ \bibinfo
  {pages} {131--186} (\bibinfo {year} {2016})}\BibitemShut {NoStop}%
\bibitem [{\citenamefont {{M{\"u}ller}}\ \emph {et~al.}(2013)\citenamefont
  {{M{\"u}ller}}, \citenamefont {{Marsden}}, \citenamefont {{St.~Cyr}},\ and\
  \citenamefont {{Gilbert}}}]{Muller:2013}%
  \BibitemOpen
  \bibfield  {author} {\bibinfo {author} {\bibfnamefont {D.}~\bibnamefont
  {{M{\"u}ller}}}, \bibinfo {author} {\bibfnamefont {R.~G.}\ \bibnamefont
  {{Marsden}}}, \bibinfo {author} {\bibfnamefont {O.~C.}\ \bibnamefont
  {{St.~Cyr}}}, \ and\ \bibinfo {author} {\bibfnamefont {H.~R.}\ \bibnamefont
  {{Gilbert}}},\ }\bibfield  {title} {\enquote {\bibinfo {title} {{Solar
  Orbiter . Exploring the Sun-Heliosphere Connection}},}\ }\href {\doibase
  10.1007/s11207-012-0085-7} {\bibfield  {journal} {\bibinfo  {journal}
  {Sol.~Phys.}\ }\textbf {\bibinfo {volume} {285}},\ \bibinfo {pages} {25--70}
  (\bibinfo {year} {2013})},\ \Eprint {http://arxiv.org/abs/1207.4579}
  {arXiv:1207.4579 [astro-ph.SR]} \BibitemShut {NoStop}%
\bibitem [{\citenamefont {{Vaivads}}\ \emph {et~al.}(2016)\citenamefont
  {{Vaivads}}, \citenamefont {{Retin{\`o}}}, \citenamefont {{Soucek}},
  \citenamefont {{Khotyaintsev}}, \citenamefont {{Valentini}}, \citenamefont
  {{Escoubet}}, \citenamefont {{Alexandrova}}, \citenamefont {{Andr{\'e}}},
  \citenamefont {{Bale}}, \citenamefont {{Balikhin}}, \citenamefont
  {{Burgess}}, \citenamefont {{Camporeale}}, \citenamefont {{Caprioli}},
  \citenamefont {{Chen}}, \citenamefont {{Clacey}}, \citenamefont {{Cully}},
  \citenamefont {{de Keyser}}, \citenamefont {{Eastwood}}, \citenamefont
  {{Fazakerley}}, \citenamefont {{Eriksson}}, \citenamefont {{Goldstein}},
  \citenamefont {{Graham}}, \citenamefont {{Haaland}}, \citenamefont
  {{Hoshino}}, \citenamefont {{Ji}}, \citenamefont {{Karimabadi}},
  \citenamefont {{Kucharek}}, \citenamefont {{Lavraud}}, \citenamefont
  {{Marcucci}}, \citenamefont {{Matthaeus}}, \citenamefont {{Moore}},
  \citenamefont {{Nakamura}}, \citenamefont {{Narita}}, \citenamefont
  {{Nemecek}}, \citenamefont {{Norgren}}, \citenamefont {{Opgenoorth}},
  \citenamefont {{Palmroth}}, \citenamefont {{Perrone}}, \citenamefont {{Pin{\c
  c}on}}, \citenamefont {{Rathsman}}, \citenamefont {{Rothkaehl}},
  \citenamefont {{Sahraoui}}, \citenamefont {{Servidio}}, \citenamefont
  {{Sorriso-Valvo}}, \citenamefont {{Vainio}}, \citenamefont
  {{V{\"o}r{\"o}s}},\ and\ \citenamefont
  {{Wimmer-Schweingruber}}}]{Vaivads:2016}%
  \BibitemOpen
  \bibfield  {author} {\bibinfo {author} {\bibfnamefont {A.}~\bibnamefont
  {{Vaivads}}}, \bibinfo {author} {\bibfnamefont {A.}~\bibnamefont
  {{Retin{\`o}}}}, \bibinfo {author} {\bibfnamefont {J.}~\bibnamefont
  {{Soucek}}}, \bibinfo {author} {\bibfnamefont {Y.~V.}\ \bibnamefont
  {{Khotyaintsev}}}, \bibinfo {author} {\bibfnamefont {F.}~\bibnamefont
  {{Valentini}}}, \bibinfo {author} {\bibfnamefont {C.~P.}\ \bibnamefont
  {{Escoubet}}}, \bibinfo {author} {\bibfnamefont {O.}~\bibnamefont
  {{Alexandrova}}}, \bibinfo {author} {\bibfnamefont {M.}~\bibnamefont
  {{Andr{\'e}}}}, \bibinfo {author} {\bibfnamefont {S.~D.}\ \bibnamefont
  {{Bale}}}, \bibinfo {author} {\bibfnamefont {M.}~\bibnamefont {{Balikhin}}},
  \bibinfo {author} {\bibfnamefont {D.}~\bibnamefont {{Burgess}}}, \bibinfo
  {author} {\bibfnamefont {E.}~\bibnamefont {{Camporeale}}}, \bibinfo {author}
  {\bibfnamefont {D.}~\bibnamefont {{Caprioli}}}, \bibinfo {author}
  {\bibfnamefont {C.~H.~K.}\ \bibnamefont {{Chen}}}, \bibinfo {author}
  {\bibfnamefont {E.}~\bibnamefont {{Clacey}}}, \bibinfo {author}
  {\bibfnamefont {C.~M.}\ \bibnamefont {{Cully}}}, \bibinfo {author}
  {\bibfnamefont {J.}~\bibnamefont {{de Keyser}}}, \bibinfo {author}
  {\bibfnamefont {J.~P.}\ \bibnamefont {{Eastwood}}}, \bibinfo {author}
  {\bibfnamefont {A.~N.}\ \bibnamefont {{Fazakerley}}}, \bibinfo {author}
  {\bibfnamefont {S.}~\bibnamefont {{Eriksson}}}, \bibinfo {author}
  {\bibfnamefont {M.~L.}\ \bibnamefont {{Goldstein}}}, \bibinfo {author}
  {\bibfnamefont {D.~B.}\ \bibnamefont {{Graham}}}, \bibinfo {author}
  {\bibfnamefont {S.}~\bibnamefont {{Haaland}}}, \bibinfo {author}
  {\bibfnamefont {M.}~\bibnamefont {{Hoshino}}}, \bibinfo {author}
  {\bibfnamefont {H.}~\bibnamefont {{Ji}}}, \bibinfo {author} {\bibfnamefont
  {H.}~\bibnamefont {{Karimabadi}}}, \bibinfo {author} {\bibfnamefont
  {H.}~\bibnamefont {{Kucharek}}}, \bibinfo {author} {\bibfnamefont
  {B.}~\bibnamefont {{Lavraud}}}, \bibinfo {author} {\bibfnamefont
  {F.}~\bibnamefont {{Marcucci}}}, \bibinfo {author} {\bibfnamefont {W.~H.}\
  \bibnamefont {{Matthaeus}}}, \bibinfo {author} {\bibfnamefont {T.~E.}\
  \bibnamefont {{Moore}}}, \bibinfo {author} {\bibfnamefont {R.}~\bibnamefont
  {{Nakamura}}}, \bibinfo {author} {\bibfnamefont {Y.}~\bibnamefont
  {{Narita}}}, \bibinfo {author} {\bibfnamefont {Z.}~\bibnamefont {{Nemecek}}},
  \bibinfo {author} {\bibfnamefont {C.}~\bibnamefont {{Norgren}}}, \bibinfo
  {author} {\bibfnamefont {H.}~\bibnamefont {{Opgenoorth}}}, \bibinfo {author}
  {\bibfnamefont {M.}~\bibnamefont {{Palmroth}}}, \bibinfo {author}
  {\bibfnamefont {D.}~\bibnamefont {{Perrone}}}, \bibinfo {author}
  {\bibfnamefont {J.-L.}\ \bibnamefont {{Pin{\c c}on}}}, \bibinfo {author}
  {\bibfnamefont {P.}~\bibnamefont {{Rathsman}}}, \bibinfo {author}
  {\bibfnamefont {H.}~\bibnamefont {{Rothkaehl}}}, \bibinfo {author}
  {\bibfnamefont {F.}~\bibnamefont {{Sahraoui}}}, \bibinfo {author}
  {\bibfnamefont {S.}~\bibnamefont {{Servidio}}}, \bibinfo {author}
  {\bibfnamefont {L.}~\bibnamefont {{Sorriso-Valvo}}}, \bibinfo {author}
  {\bibfnamefont {R.}~\bibnamefont {{Vainio}}}, \bibinfo {author}
  {\bibfnamefont {Z.}~\bibnamefont {{V{\"o}r{\"o}s}}}, \ and\ \bibinfo {author}
  {\bibfnamefont {R.~F.}\ \bibnamefont {{Wimmer-Schweingruber}}},\ }\bibfield
  {title} {\enquote {\bibinfo {title} {{Turbulence Heating ObserveR - satellite
  mission proposal}},}\ }\href {\doibase 10.1017/S0022377816000775} {\bibfield
  {journal} {\bibinfo  {journal} {J.~Plasma Phys.}\ }\textbf {\bibinfo {volume}
  {82}},\ \bibinfo {eid} {905820501} (\bibinfo {year} {2016})}\BibitemShut
  {NoStop}%
\end{thebibliography}

%

\end{document}